\documentclass[twocolumn,traditabstract,longauth]{aa}
\usepackage{graphicx}
\usepackage[breaklinks, colorlinks, citecolor=blue]{hyperref}
\usepackage{natbib}
\usepackage{subfigure}
\usepackage{amsmath}
\usepackage{epsfig}
\usepackage{multirow}
\usepackage{array}
\usepackage{color}
\usepackage{txfonts}
\newcommand{\ie}{{i.e.}}

\usepackage{url}
\usepackage{psfrag}
\usepackage[T1]{fontenc}
\usepackage{rotating}

\newcolumntype{L}[1]{>{\raggedright\let\newline\\\arraybackslash\hspace{0pt}}m{#1}}
\newcolumntype{C}[1]{>{\centering\let\newline\\\arraybackslash\hspace{0pt}}m{#1}}
\newcolumntype{R}[1]{>{\raggedleft\let\newline\\\arraybackslash\hspace{0pt}}m{#1}}

\def\setsymbol#1#2{\expandafter\def\csname #1\endcsname{#2}}
\def\getsymbol#1{\csname #1\endcsname}

\def\Planck{\textit{Planck}}

\def\all2013resultspapers{\nocite{planck2013-p01, planck2013-p02, planck2013-p02a, planck2013-p02d, planck2013-p02b, planck2013-p03, planck2013-p03c, planck2013-p03f, planck2013-p03d, planck2013-p03e, planck2013-p01a, planck2013-p06, planck2013-p03a, planck2013-pip88, planck2013-p08, planck2013-p11, planck2013-p12, planck2013-p13, planck2013-p14, planck2013-p15, planck2013-p05b, planck2013-p17, planck2013-p09, planck2013-p09a, planck2013-p20, planck2013-p19, planck2013-pipaberration, planck2013-p05, planck2013-p05a, planck2013-pip56, planck2013-p06b}}

\newbox\tablebox    \newdimen\tablewidth
\def\leaderfil{\leaders\hbox to 5pt{\hss.\hss}\hfil}
\def\endPlancktable{\tablewidth=\columnwidth 
    $$\hss\copy\tablebox\hss$$
    \vskip-\lastskip\vskip -2pt}

\def\tablenote#1 #2\par{\begingroup \parindent=0.8em
    \abovedisplayshortskip=0pt\belowdisplayshortskip=0pt
    \noindent
    $$\hss\vbox{\hsize\tablewidth \hangindent=\parindent \hangafter=1 \noindent
    \hbox to \parindent{$^#1$\hss}\strut#2\strut\par}\hss$$
    \endgroup}
\def\doubleline{\vskip 3pt\hrule \vskip 1.5pt \hrule \vskip 5pt}

\def\L2{\ifmmode L_2\else $L_2$\fi}

\def\DeltaT{\ifmmode \Delta T\else $\Delta T$\fi}
\def\deltat{\ifmmode \Delta t\else $\Delta t$\fi}
\def\fknee{\ifmmode f_{\rm knee}\else $f_{\rm knee}$\fi}
\def\Fmax{\ifmmode F_{\rm max}\else $F_{\rm max}$\fi}
\def\solar{\ifmmode{\rm M}_{\mathord\odot}\else${\rm M}_{\mathord\odot}$\fi}
\def\Msolar{\ifmmode{\rm M}_{\mathord\odot}\else${\rm M}_{\mathord\odot}$\fi}
\def\Lsolar{\ifmmode{\rm L}_{\mathord\odot}\else${\rm L}_{\mathord\odot}$\fi}

\def\inv{\ifmmode^{-1}\else$^{-1}$\fi}
\def\mo{\ifmmode^{-1}\else$^{-1}$\fi}
\def\sup#1{\ifmmode ^{\rm #1}\else $^{\rm #1}$\fi}
\def\expo#1{\ifmmode \times 10^{#1}\else $\times 10^{#1}$\fi}
\def\,{\thinspace}
\def\lsim{\mathrel{\raise .4ex\hbox{\rlap{$<$}\lower 1.2ex\hbox{$\sim$}}}}
\def\gsim{\mathrel{\raise .4ex\hbox{\rlap{$>$}\lower 1.2ex\hbox{$\sim$}}}}

\def\simprop{\mathrel{\raise .4ex\hbox{\rlap{$\propto$}\lower 1.2ex\hbox{$\sim$}}}}
\def\deg{\ifmmode^\circ\else$^\circ$\fi}
\def\pdeg{\ifmmode $\setbox0=\hbox{$^{\circ}$}\rlap{\hskip.11\wd0 .}$^{\circ}
          \else \setbox0=\hbox{$^{\circ}$}\rlap{\hskip.11\wd0 .}$^{\circ}$\fi}
\def\arcs{\ifmmode {^{\scriptstyle\prime\prime}}
          \else $^{\scriptstyle\prime\prime}$\fi}
\def\arcm{\ifmmode {^{\scriptstyle\prime}}
          \else $^{\scriptstyle\prime}$\fi}
\newdimen\sa  \newdimen\sb
\def\parcs{\sa=.07em \sb=.03em
     \ifmmode \hbox{\rlap{.}}^{\scriptstyle\prime\kern -\sb\prime}\hbox{\kern -\sa}
     \else \rlap{.}$^{\scriptstyle\prime\kern -\sb\prime}$\kern -\sa\fi}
\def\parcm{\sa=.08em \sb=.03em
     \ifmmode \hbox{\rlap{.}\kern\sa}^{\scriptstyle\prime}\hbox{\kern-\sb}
     \else \rlap{.}\kern\sa$^{\scriptstyle\prime}$\kern-\sb\fi}
\def\ra[#1 #2 #3.#4]{#1\sup{h}#2\sup{m}#3\sup{s}\llap.#4}
\def\dec[#1 #2 #3.#4]{#1\deg#2\arcm#3\arcs\llap.#4}
\def\deco[#1 #2 #3]{#1\deg#2\arcm#3\arcs}
\def\rra[#1 #2]{#1\sup{h}#2\sup{m}}

\def\dots{\relax\ifmmode \ldots\else $\ldots$\fi}
\def\WHzsr{\ifmmode $W\,Hz\mo\,sr\mo$\else W\,Hz\mo\,sr\mo\fi}
\def\mHz{\ifmmode $\,mHz$\else \,mHz\fi}
\def\GHz{\ifmmode $\,GHz$\else \,GHz\fi}
\def\mKs{\ifmmode $\,mK\,s$^{1/2}\else \,mK\,s$^{1/2}$\fi}
\def\muKs{\ifmmode \,\mu$K\,s$^{1/2}\else \,$\mu$K\,s$^{1/2}$\fi}
\def\muKRJs{\ifmmode \,\mu$K$_{\rm RJ}$\,s$^{1/2}\else \,$\mu$K$_{\rm RJ}$\,s$^{1/2}$\fi}
\def\muKHz{\ifmmode \,\mu$K\,Hz$^{-1/2}\else \,$\mu$K\,Hz$^{-1/2}$\fi}
\def\MJysr{\ifmmode \,$MJy\,sr\mo$\else \,MJy\,sr\mo\fi}
\def\MJysrmK{\ifmmode \,$MJy\,sr\mo$\,mK$_{\rm CMB}\mo\else \,MJy\,sr\mo\,mK$_{\rm CMB}\mo$\fi}
\def\microns{\ifmmode \,\mu$m$\else \,$\mu$m\fi}

\def\muK{\ifmmode \,\mu$K$\else \,$\mu$\hbox{K}\fi}
\def\microK{\ifmmode \,\mu$K$\else \,$\mu$\hbox{K}\fi}
\def\muW{\ifmmode \,\mu$W$\else \,$\mu$\hbox{W}\fi}
\def\kms{\ifmmode $\,km\,s$^{-1}\else \,km\,s$^{-1}$\fi}
\def\kmsMpc{\ifmmode $\,\kms\,Mpc\mo$\else \,\kms\,Mpc\mo\fi}
\setsymbol{LFI:center:frequency:70GHz:units}{70.3\,GHz}
\setsymbol{LFI:center:frequency:44GHz:units}{44.1\,GHz}
\setsymbol{LFI:center:frequency:30GHz:units}{28.5\,GHz}

\setsymbol{LFI:center:frequency:70GHz}{70.3}
\setsymbol{LFI:center:frequency:44GHz}{44.1}
\setsymbol{LFI:center:frequency:30GHz}{28.5}

\setsymbol{LFI:center:frequency:LFI18:Rad:M:units}{71.7\GHz}
\setsymbol{LFI:center:frequency:LFI19:Rad:M:units}{67.5\GHz}
\setsymbol{LFI:center:frequency:LFI20:Rad:M:units}{69.2\GHz}
\setsymbol{LFI:center:frequency:LFI21:Rad:M:units}{70.4\GHz}
\setsymbol{LFI:center:frequency:LFI22:Rad:M:units}{71.5\GHz}
\setsymbol{LFI:center:frequency:LFI23:Rad:M:units}{70.8\GHz}
\setsymbol{LFI:center:frequency:LFI24:Rad:M:units}{44.4\GHz}
\setsymbol{LFI:center:frequency:LFI25:Rad:M:units}{44.0\GHz}
\setsymbol{LFI:center:frequency:LFI26:Rad:M:units}{43.9\GHz}
\setsymbol{LFI:center:frequency:LFI27:Rad:M:units}{28.3\GHz}
\setsymbol{LFI:center:frequency:LFI28:Rad:M:units}{28.8\GHz}
\setsymbol{LFI:center:frequency:LFI18:Rad:S:units}{70.1\GHz}
\setsymbol{LFI:center:frequency:LFI19:Rad:S:units}{69.6\GHz}
\setsymbol{LFI:center:frequency:LFI20:Rad:S:units}{69.5\GHz}
\setsymbol{LFI:center:frequency:LFI21:Rad:S:units}{69.5\GHz}
\setsymbol{LFI:center:frequency:LFI22:Rad:S:units}{72.8\GHz}
\setsymbol{LFI:center:frequency:LFI23:Rad:S:units}{71.3\GHz}
\setsymbol{LFI:center:frequency:LFI24:Rad:S:units}{44.1\GHz}
\setsymbol{LFI:center:frequency:LFI25:Rad:S:units}{44.1\GHz}
\setsymbol{LFI:center:frequency:LFI26:Rad:S:units}{44.1\GHz}
\setsymbol{LFI:center:frequency:LFI27:Rad:S:units}{28.5\GHz}
\setsymbol{LFI:center:frequency:LFI28:Rad:S:units}{28.2\GHz}

\setsymbol{LFI:center:frequency:LFI18:Rad:M}{71.7}
\setsymbol{LFI:center:frequency:LFI19:Rad:M}{67.5}
\setsymbol{LFI:center:frequency:LFI20:Rad:M}{69.2}
\setsymbol{LFI:center:frequency:LFI21:Rad:M}{70.4}
\setsymbol{LFI:center:frequency:LFI22:Rad:M}{71.5}
\setsymbol{LFI:center:frequency:LFI23:Rad:M}{70.8}
\setsymbol{LFI:center:frequency:LFI24:Rad:M}{44.4}
\setsymbol{LFI:center:frequency:LFI25:Rad:M}{44.0}
\setsymbol{LFI:center:frequency:LFI26:Rad:M}{43.9}
\setsymbol{LFI:center:frequency:LFI27:Rad:M}{28.3}
\setsymbol{LFI:center:frequency:LFI28:Rad:M}{28.8}
\setsymbol{LFI:center:frequency:LFI18:Rad:S}{70.1}
\setsymbol{LFI:center:frequency:LFI19:Rad:S}{69.6}
\setsymbol{LFI:center:frequency:LFI20:Rad:S}{69.5}
\setsymbol{LFI:center:frequency:LFI21:Rad:S}{69.5}
\setsymbol{LFI:center:frequency:LFI22:Rad:S}{72.8}
\setsymbol{LFI:center:frequency:LFI23:Rad:S}{71.3}
\setsymbol{LFI:center:frequency:LFI24:Rad:S}{44.1}
\setsymbol{LFI:center:frequency:LFI25:Rad:S}{44.1}
\setsymbol{LFI:center:frequency:LFI26:Rad:S}{44.1}
\setsymbol{LFI:center:frequency:LFI27:Rad:S}{28.5}
\setsymbol{LFI:center:frequency:LFI28:Rad:S}{28.2}

\setsymbol{LFI:white:noise:sensitivity:70GHz:units}{134.7\muKs}
\setsymbol{LFI:white:noise:sensitivity:44GHz:units}{164.7\muKs}
\setsymbol{LFI:white:noise:sensitivity:30GHz:units}{143.4\muKs}

\setsymbol{LFI:white:noise:sensitivity:70GHz}{134.7}
\setsymbol{LFI:white:noise:sensitivity:44GHz}{164.7}
\setsymbol{LFI:white:noise:sensitivity:30GHz}{143.4}

\setsymbol{LFI:white:noise:sensitivity:LFI18:Rad:M:units}{512.0\muKs}
\setsymbol{LFI:white:noise:sensitivity:LFI19:Rad:M:units}{581.4\muKs}
\setsymbol{LFI:white:noise:sensitivity:LFI20:Rad:M:units}{590.8\muKs}
\setsymbol{LFI:white:noise:sensitivity:LFI21:Rad:M:units}{455.2\muKs}
\setsymbol{LFI:white:noise:sensitivity:LFI22:Rad:M:units}{492.0\muKs}
\setsymbol{LFI:white:noise:sensitivity:LFI23:Rad:M:units}{507.7\muKs}
\setsymbol{LFI:white:noise:sensitivity:LFI24:Rad:M:units}{462.2\muKs}
\setsymbol{LFI:white:noise:sensitivity:LFI25:Rad:M:units}{413.6\muKs}
\setsymbol{LFI:white:noise:sensitivity:LFI26:Rad:M:units}{478.6\muKs}
\setsymbol{LFI:white:noise:sensitivity:LFI27:Rad:M:units}{277.7\muKs}
\setsymbol{LFI:white:noise:sensitivity:LFI28:Rad:M:units}{312.3\muKs}
\setsymbol{LFI:white:noise:sensitivity:LFI18:Rad:S:units}{465.7\muKs}
\setsymbol{LFI:white:noise:sensitivity:LFI19:Rad:S:units}{555.6\muKs}
\setsymbol{LFI:white:noise:sensitivity:LFI20:Rad:S:units}{623.2\muKs}
\setsymbol{LFI:white:noise:sensitivity:LFI21:Rad:S:units}{564.1\muKs}
\setsymbol{LFI:white:noise:sensitivity:LFI22:Rad:S:units}{534.4\muKs}
\setsymbol{LFI:white:noise:sensitivity:LFI23:Rad:S:units}{542.4\muKs}
\setsymbol{LFI:white:noise:sensitivity:LFI24:Rad:S:units}{399.2\muKs}
\setsymbol{LFI:white:noise:sensitivity:LFI25:Rad:S:units}{392.6\muKs}
\setsymbol{LFI:white:noise:sensitivity:LFI26:Rad:S:units}{418.6\muKs}
\setsymbol{LFI:white:noise:sensitivity:LFI27:Rad:S:units}{302.9\muKs}
\setsymbol{LFI:white:noise:sensitivity:LFI28:Rad:S:units}{285.3\muKs}

\setsymbol{LFI:white:noise:sensitivity:LFI18:Rad:M}{512.0}
\setsymbol{LFI:white:noise:sensitivity:LFI19:Rad:M}{581.4}
\setsymbol{LFI:white:noise:sensitivity:LFI20:Rad:M}{590.8}
\setsymbol{LFI:white:noise:sensitivity:LFI21:Rad:M}{455.2}
\setsymbol{LFI:white:noise:sensitivity:LFI22:Rad:M}{492.0}
\setsymbol{LFI:white:noise:sensitivity:LFI23:Rad:M}{507.7}
\setsymbol{LFI:white:noise:sensitivity:LFI24:Rad:M}{462.2}
\setsymbol{LFI:white:noise:sensitivity:LFI25:Rad:M}{413.6}
\setsymbol{LFI:white:noise:sensitivity:LFI26:Rad:M}{478.6}
\setsymbol{LFI:white:noise:sensitivity:LFI27:Rad:M}{277.7}
\setsymbol{LFI:white:noise:sensitivity:LFI28:Rad:M}{312.3}
\setsymbol{LFI:white:noise:sensitivity:LFI18:Rad:S}{465.7}
\setsymbol{LFI:white:noise:sensitivity:LFI19:Rad:S}{555.6}
\setsymbol{LFI:white:noise:sensitivity:LFI20:Rad:S}{623.2}
\setsymbol{LFI:white:noise:sensitivity:LFI21:Rad:S}{564.1}
\setsymbol{LFI:white:noise:sensitivity:LFI22:Rad:S}{534.4}
\setsymbol{LFI:white:noise:sensitivity:LFI23:Rad:S}{542.4}
\setsymbol{LFI:white:noise:sensitivity:LFI24:Rad:S}{399.2}
\setsymbol{LFI:white:noise:sensitivity:LFI25:Rad:S}{392.6}
\setsymbol{LFI:white:noise:sensitivity:LFI26:Rad:S}{418.6}
\setsymbol{LFI:white:noise:sensitivity:LFI27:Rad:S}{302.9}
\setsymbol{LFI:white:noise:sensitivity:LFI28:Rad:S}{285.3}

\setsymbol{LFI:knee:frequency:70GHz:units}{29.5\mHz}
\setsymbol{LFI:knee:frequency:44GHz:units}{56.2\mHz}
\setsymbol{LFI:knee:frequency:30GHz:units}{113.7\mHz}

\setsymbol{LFI:knee:frequency:70GHz}{29.5}
\setsymbol{LFI:knee:frequency:44GHz}{56.2}
\setsymbol{LFI:knee:frequency:30GHz}{113.7}

\setsymbol{LFI:knee:frequency:LFI18:Rad:M:units}{16.3\mHz}
\setsymbol{LFI:knee:frequency:LFI19:Rad:M:units}{15.1\mHz}
\setsymbol{LFI:knee:frequency:LFI20:Rad:M:units}{18.7\mHz}
\setsymbol{LFI:knee:frequency:LFI21:Rad:M:units}{37.2\mHz}
\setsymbol{LFI:knee:frequency:LFI22:Rad:M:units}{12.7\mHz}
\setsymbol{LFI:knee:frequency:LFI23:Rad:M:units}{34.6\mHz}
\setsymbol{LFI:knee:frequency:LFI24:Rad:M:units}{46.2\mHz}
\setsymbol{LFI:knee:frequency:LFI25:Rad:M:units}{24.9\mHz}
\setsymbol{LFI:knee:frequency:LFI26:Rad:M:units}{67.6\mHz}
\setsymbol{LFI:knee:frequency:LFI27:Rad:M:units}{187.4\mHz}
\setsymbol{LFI:knee:frequency:LFI28:Rad:M:units}{122.2\mHz}
\setsymbol{LFI:knee:frequency:LFI18:Rad:S:units}{17.7\mHz}
\setsymbol{LFI:knee:frequency:LFI19:Rad:S:units}{22.0\mHz}
\setsymbol{LFI:knee:frequency:LFI20:Rad:S:units}{8.7\mHz}
\setsymbol{LFI:knee:frequency:LFI21:Rad:S:units}{25.9\mHz}
\setsymbol{LFI:knee:frequency:LFI22:Rad:S:units}{15.8\mHz}
\setsymbol{LFI:knee:frequency:LFI23:Rad:S:units}{129.8\mHz}
\setsymbol{LFI:knee:frequency:LFI24:Rad:S:units}{100.9\mHz}
\setsymbol{LFI:knee:frequency:LFI25:Rad:S:units}{38.9\mHz}
\setsymbol{LFI:knee:frequency:LFI26:Rad:S:units}{58.9\mHz}
\setsymbol{LFI:knee:frequency:LFI27:Rad:S:units}{104.4\mHz}
\setsymbol{LFI:knee:frequency:LFI28:Rad:S:units}{40.7\mHz}

\setsymbol{LFI:knee:frequency:LFI18:Rad:M}{16.3}
\setsymbol{LFI:knee:frequency:LFI19:Rad:M}{15.1}
\setsymbol{LFI:knee:frequency:LFI20:Rad:M}{18.7}
\setsymbol{LFI:knee:frequency:LFI21:Rad:M}{37.2}
\setsymbol{LFI:knee:frequency:LFI22:Rad:M}{12.7}
\setsymbol{LFI:knee:frequency:LFI23:Rad:M}{34.6}
\setsymbol{LFI:knee:frequency:LFI24:Rad:M}{46.2}
\setsymbol{LFI:knee:frequency:LFI25:Rad:M}{24.9}
\setsymbol{LFI:knee:frequency:LFI26:Rad:M}{67.6}
\setsymbol{LFI:knee:frequency:LFI27:Rad:M}{187.4}
\setsymbol{LFI:knee:frequency:LFI28:Rad:M}{122.2}
\setsymbol{LFI:knee:frequency:LFI18:Rad:S}{17.7}
\setsymbol{LFI:knee:frequency:LFI19:Rad:S}{22.0}
\setsymbol{LFI:knee:frequency:LFI20:Rad:S}{8.7}
\setsymbol{LFI:knee:frequency:LFI21:Rad:S}{25.9}
\setsymbol{LFI:knee:frequency:LFI22:Rad:S}{15.8}
\setsymbol{LFI:knee:frequency:LFI23:Rad:S}{129.8}
\setsymbol{LFI:knee:frequency:LFI24:Rad:S}{100.9}
\setsymbol{LFI:knee:frequency:LFI25:Rad:S}{38.9}
\setsymbol{LFI:knee:frequency:LFI26:Rad:S}{58.9}
\setsymbol{LFI:knee:frequency:LFI27:Rad:S}{104.4}
\setsymbol{LFI:knee:frequency:LFI28:Rad:S}{40.7}

\setsymbol{LFI:slope:70GHz:units}{$-1.03$\mHz}
\setsymbol{LFI:slope:44GHz:units}{$-0.89$\mHz}
\setsymbol{LFI:slope:30GHz:units}{$-0.87$\mHz}

\setsymbol{LFI:slope:70GHz}{$-1.03$}
\setsymbol{LFI:slope:44GHz}{$-0.89$}
\setsymbol{LFI:slope:30GHz}{$-0.87$}

\setsymbol{LFI:slope:LFI18:Rad:M:units}{$-1.04$\mHz}
\setsymbol{LFI:slope:LFI19:Rad:M:units}{$-1.09$\mHz}
\setsymbol{LFI:slope:LFI20:Rad:M:units}{$-0.69$\mHz}
\setsymbol{LFI:slope:LFI21:Rad:M:units}{$-1.56$\mHz}
\setsymbol{LFI:slope:LFI22:Rad:M:units}{$-1.01$\mHz}
\setsymbol{LFI:slope:LFI23:Rad:M:units}{$-0.96$\mHz}
\setsymbol{LFI:slope:LFI24:Rad:M:units}{$-0.83$\mHz}
\setsymbol{LFI:slope:LFI25:Rad:M:units}{$-0.91$\mHz}
\setsymbol{LFI:slope:LFI26:Rad:M:units}{$-0.95$\mHz}
\setsymbol{LFI:slope:LFI27:Rad:M:units}{$-0.87$\mHz}
\setsymbol{LFI:slope:LFI28:Rad:M:units}{$-0.88$\mHz}
\setsymbol{LFI:slope:LFI18:Rad:S:units}{$-1.15$\mHz}
\setsymbol{LFI:slope:LFI19:Rad:S:units}{$-1.00$\mHz}
\setsymbol{LFI:slope:LFI20:Rad:S:units}{$-0.95$\mHz}
\setsymbol{LFI:slope:LFI21:Rad:S:units}{$-0.92$\mHz}
\setsymbol{LFI:slope:LFI22:Rad:S:units}{$-1.01$\mHz}
\setsymbol{LFI:slope:LFI23:Rad:S:units}{$-0.95$\mHz}
\setsymbol{LFI:slope:LFI24:Rad:S:units}{$-0.73$\mHz}
\setsymbol{LFI:slope:LFI25:Rad:S:units}{$-1.16$\mHz}
\setsymbol{LFI:slope:LFI26:Rad:S:units}{$-0.79$\mHz}
\setsymbol{LFI:slope:LFI27:Rad:S:units}{$-0.82$\mHz}
\setsymbol{LFI:slope:LFI28:Rad:S:units}{$-0.91$\mHz}

\setsymbol{LFI:slope:LFI18:Rad:M}{$-1.04$}
\setsymbol{LFI:slope:LFI19:Rad:M}{$-1.09$}
\setsymbol{LFI:slope:LFI20:Rad:M}{$-0.69$}
\setsymbol{LFI:slope:LFI21:Rad:M}{$-1.56$}
\setsymbol{LFI:slope:LFI22:Rad:M}{$-1.01$}
\setsymbol{LFI:slope:LFI23:Rad:M}{$-0.96$}
\setsymbol{LFI:slope:LFI24:Rad:M}{$-0.83$}
\setsymbol{LFI:slope:LFI25:Rad:M}{$-0.91$}
\setsymbol{LFI:slope:LFI26:Rad:M}{$-0.95$}
\setsymbol{LFI:slope:LFI27:Rad:M}{$-0.87$}
\setsymbol{LFI:slope:LFI28:Rad:M}{$-0.88$}
\setsymbol{LFI:slope:LFI18:Rad:S}{$-1.15$}
\setsymbol{LFI:slope:LFI19:Rad:S}{$-1.00$}
\setsymbol{LFI:slope:LFI20:Rad:S}{$-0.95$}
\setsymbol{LFI:slope:LFI21:Rad:S}{$-0.92$}
\setsymbol{LFI:slope:LFI22:Rad:S}{$-1.01$}
\setsymbol{LFI:slope:LFI23:Rad:S}{$-0.95$}
\setsymbol{LFI:slope:LFI24:Rad:S}{$-0.73$}
\setsymbol{LFI:slope:LFI25:Rad:S}{$-1.16$}
\setsymbol{LFI:slope:LFI26:Rad:S}{$-0.79$}
\setsymbol{LFI:slope:LFI27:Rad:S}{$-0.82$}
\setsymbol{LFI:slope:LFI28:Rad:S}{$-0.91$}

\setsymbol{LFI:FWHM:70GHz:units}{13\parcm01}
\setsymbol{LFI:FWHM:44GHz:units}{27\parcm92}
\setsymbol{LFI:FWHM:30GHz:units}{32\parcm65}

\setsymbol{LFI:FWHM:70GHz}{13.01}
\setsymbol{LFI:FWHM:44GHz}{27.92}
\setsymbol{LFI:FWHM:30GHz}{32.65}

\setsymbol{LFI:FWHM:LFI18:units}{13\parcm39}
\setsymbol{LFI:FWHM:LFI19:units}{13\parcm01}
\setsymbol{LFI:FWHM:LFI20:units}{12\parcm75}
\setsymbol{LFI:FWHM:LFI21:units}{12\parcm74}
\setsymbol{LFI:FWHM:LFI22:units}{12\parcm87}
\setsymbol{LFI:FWHM:LFI23:units}{13\parcm27}
\setsymbol{LFI:FWHM:LFI24:units}{22\parcm98}
\setsymbol{LFI:FWHM:LFI25:units}{30\parcm46}
\setsymbol{LFI:FWHM:LFI26:units}{30\parcm31}
\setsymbol{LFI:FWHM:LFI27:units}{32\parcm65}
\setsymbol{LFI:FWHM:LFI28:units}{32\parcm66}

\setsymbol{LFI:FWHM:LFI18}{13.39}
\setsymbol{LFI:FWHM:LFI19}{13.01}
\setsymbol{LFI:FWHM:LFI20}{12.75}
\setsymbol{LFI:FWHM:LFI21}{12.74}
\setsymbol{LFI:FWHM:LFI22}{12.87}
\setsymbol{LFI:FWHM:LFI23}{13.27}
\setsymbol{LFI:FWHM:LFI24}{22.98}
\setsymbol{LFI:FWHM:LFI25}{30.46}
\setsymbol{LFI:FWHM:LFI26}{30.31}
\setsymbol{LFI:FWHM:LFI27}{32.65}
\setsymbol{LFI:FWHM:LFI28}{32.66}

\setsymbol{LFI:FWHM:uncertainty:LFI18:units}{0.170\arcm}
\setsymbol{LFI:FWHM:uncertainty:LFI19:units}{0.174\arcm}
\setsymbol{LFI:FWHM:uncertainty:LFI20:units}{0.170\arcm}
\setsymbol{LFI:FWHM:uncertainty:LFI21:units}{0.156\arcm}
\setsymbol{LFI:FWHM:uncertainty:LFI22:units}{0.164\arcm}
\setsymbol{LFI:FWHM:uncertainty:LFI23:units}{0.171\arcm}
\setsymbol{LFI:FWHM:uncertainty:LFI24:units}{0.652\arcm}
\setsymbol{LFI:FWHM:uncertainty:LFI25:units}{1.075\arcm}
\setsymbol{LFI:FWHM:uncertainty:LFI26:units}{1.131\arcm}
\setsymbol{LFI:FWHM:uncertainty:LFI27:units}{1.266\arcm}
\setsymbol{LFI:FWHM:uncertainty:LFI28:units}{1.287\arcm}

\setsymbol{LFI:FWHM:uncertainty:LFI18}{0.170}
\setsymbol{LFI:FWHM:uncertainty:LFI19}{0.174}
\setsymbol{LFI:FWHM:uncertainty:LFI20}{0.170}
\setsymbol{LFI:FWHM:uncertainty:LFI21}{0.156}
\setsymbol{LFI:FWHM:uncertainty:LFI22}{0.164}
\setsymbol{LFI:FWHM:uncertainty:LFI23}{0.171}
\setsymbol{LFI:FWHM:uncertainty:LFI24}{0.652}
\setsymbol{LFI:FWHM:uncertainty:LFI25}{1.075}
\setsymbol{LFI:FWHM:uncertainty:LFI26}{1.131}
\setsymbol{LFI:FWHM:uncertainty:LFI27}{1.266}
\setsymbol{LFI:FWHM:uncertainty:LFI28}{1.287}

\setsymbol{HFI:center:frequency:100GHz:units}{100\,GHz}
\setsymbol{HFI:center:frequency:143GHz:units}{143\,GHz}
\setsymbol{HFI:center:frequency:217GHz:units}{217\,GHz}
\setsymbol{HFI:center:frequency:353GHz:units}{353\,GHz}
\setsymbol{HFI:center:frequency:545GHz:units}{545\,GHz}
\setsymbol{HFI:center:frequency:857GHz:units}{857\,GHz}

\setsymbol{HFI:center:frequency:100GHz}{100}
\setsymbol{HFI:center:frequency:143GHz}{143}
\setsymbol{HFI:center:frequency:217GHz}{217}
\setsymbol{HFI:center:frequency:353GHz}{353}
\setsymbol{HFI:center:frequency:545GHz}{545}
\setsymbol{HFI:center:frequency:857GHz}{857}

\setsymbol{HFI:Ndetectors:100GHz}{8}
\setsymbol{HFI:Ndetectors:143GHz}{11}
\setsymbol{HFI:Ndetectors:217GHz}{12}
\setsymbol{HFI:Ndetectors:353GHz}{12}
\setsymbol{HFI:Ndetectors:545GHz}{3}
\setsymbol{HFI:Ndetectors:857GHz}{4}

\setsymbol{HFI:FWHM:Maps:100GHz:units}{9\parcm88}
\setsymbol{HFI:FWHM:Maps:143GHz:units}{7\parcm18}
\setsymbol{HFI:FWHM:Maps:217GHz:units}{4\parcm87}
\setsymbol{HFI:FWHM:Maps:353GHz:units}{4\parcm65}
\setsymbol{HFI:FWHM:Maps:545GHz:units}{4\parcm72}
\setsymbol{HFI:FWHM:Maps:857GHz:units}{4\parcm39}
\setsymbol{HFI:FWHM:Maps:100GHz}{9.88}
\setsymbol{HFI:FWHM:Maps:143GHz}{7.18}
\setsymbol{HFI:FWHM:Maps:217GHz}{4.87}
\setsymbol{HFI:FWHM:Maps:353GHz}{4.65}
\setsymbol{HFI:FWHM:Maps:545GHz}{4.72}
\setsymbol{HFI:FWHM:Maps:857GHz}{4.39}

\setsymbol{HFI:beam:ellipticity:Maps:100GHz}{1.15}
\setsymbol{HFI:beam:ellipticity:Maps:143GHz}{1.01}
\setsymbol{HFI:beam:ellipticity:Maps:217GHz}{1.06}
\setsymbol{HFI:beam:ellipticity:Maps:353GHz}{1.05}
\setsymbol{HFI:beam:ellipticity:Maps:545GHz}{1.14}
\setsymbol{HFI:beam:ellipticity:Maps:857GHz}{1.19}

\setsymbol{HFI:FWHM:Mars:100GHz:units}{9\parcm37}
\setsymbol{HFI:FWHM:Mars:143GHz:units}{7\parcm04}
\setsymbol{HFI:FWHM:Mars:217GHz:units}{4\parcm68}
\setsymbol{HFI:FWHM:Mars:353GHz:units}{4\parcm43}
\setsymbol{HFI:FWHM:Mars:545GHz:units}{3\parcm80}
\setsymbol{HFI:FWHM:Mars:857GHz:units}{3\parcm67}

\setsymbol{HFI:FWHM:Mars:100GHz}{9.37}
\setsymbol{HFI:FWHM:Mars:143GHz}{7.04}
\setsymbol{HFI:FWHM:Mars:217GHz}{4.68}
\setsymbol{HFI:FWHM:Mars:353GHz}{4.43}
\setsymbol{HFI:FWHM:Mars:545GHz}{3.80}
\setsymbol{HFI:FWHM:Mars:857GHz}{3.67}

\setsymbol{HFI:beam:ellipticity:Mars:100GHz}{1.18}
\setsymbol{HFI:beam:ellipticity:Mars:143GHz}{1.03}
\setsymbol{HFI:beam:ellipticity:Mars:217GHz}{1.14}
\setsymbol{HFI:beam:ellipticity:Mars:353GHz}{1.09}
\setsymbol{HFI:beam:ellipticity:Mars:545GHz}{1.25}
\setsymbol{HFI:beam:ellipticity:Mars:857GHz}{1.03}

\setsymbol{HFI:CMB:relative:calibration:100GHz}{$\lsim 1\%$}
\setsymbol{HFI:CMB:relative:calibration:143GHz}{$\lsim 1\%$}
\setsymbol{HFI:CMB:relative:calibration:217GHz}{$\lsim 1\%$}
\setsymbol{HFI:CMB:relative:calibration:353GHz}{$\lsim 1\%$}
\setsymbol{HFI:CMB:relative:calibration:545GHz}{}
\setsymbol{HFI:CMB:relative:calibration:857GHz}{}

\setsymbol{HFI:CMB:absolute:calibration:100GHz}{$\lsim 2\%$}
\setsymbol{HFI:CMB:absolute:calibration:143GHz}{$\lsim 2\%$}
\setsymbol{HFI:CMB:absolute:calibration:217GHz}{$\lsim 2\%$}
\setsymbol{HFI:CMB:absolute:calibration:353GHz}{$\lsim 2\%$}
\setsymbol{HFI:CMB:absolute:calibration:545GHz}{}
\setsymbol{HFI:CMB:absolute:calibration:857GHz}{}

\setsymbol{HFI:FIRAS:gain:calibration:accuracy:statistical:100GHz}{}
\setsymbol{HFI:FIRAS:gain:calibration:accuracy:statistical:143GHz}{}
\setsymbol{HFI:FIRAS:gain:calibration:accuracy:statistical:217GHz}{}
\setsymbol{HFI:FIRAS:gain:calibration:accuracy:statistical:353GHz}{2.5\%}
\setsymbol{HFI:FIRAS:gain:calibration:accuracy:statistical:545GHz}{1\%}
\setsymbol{HFI:FIRAS:gain:calibration:accuracy:statistical:857GHz}{0.5\%}

\setsymbol{HFI:FIRAS:gain:calibration:accuracy:systematic:100GHz}{}
\setsymbol{HFI:FIRAS:gain:calibration:accuracy:systematic:143GHz}{}
\setsymbol{HFI:FIRAS:gain:calibration:accuracy:systematic:217GHz}{}
\setsymbol{HFI:FIRAS:gain:calibration:accuracy:systematic:353GHz}{}
\setsymbol{HFI:FIRAS:gain:calibration:accuracy:systematic:545GHz}{7\%}
\setsymbol{HFI:FIRAS:gain:calibration:accuracy:systematic:857GHz}{7\%}

\setsymbol{HFI:FIRAS:zero:point:accuracy:100GHz:units}{0.8\MJysr}
\setsymbol{HFI:FIRAS:zero:point:accuracy:143GHz:units}{}
\setsymbol{HFI:FIRAS:zero:point:accuracy:217GHz:units}{}
\setsymbol{HFI:FIRAS:zero:point:accuracy:353GHz:units}{1.4\MJysr}
\setsymbol{HFI:FIRAS:zero:point:accuracy:545GHz:units}{2.2\MJysr}
\setsymbol{HFI:FIRAS:zero:point:accuracy:857GHz:units}{1.7\MJysr}

\setsymbol{HFI:FIRAS:zero:point:accuracy:100GHz}{0.8}
\setsymbol{HFI:FIRAS:zero:point:accuracy:143GHz}{}
\setsymbol{HFI:FIRAS:zero:point:accuracy:217GHz}{}
\setsymbol{HFI:FIRAS:zero:point:accuracy:353GHz}{1.4}
\setsymbol{HFI:FIRAS:zero:point:accuracy:545GHz}{2.2}
\setsymbol{HFI:FIRAS:zero:point:accuracy:857GHz}{1.7}

\setsymbol{HFI:unit:conversion:100GHz:units}{0.2415\MJysrmK}
\setsymbol{HFI:unit:conversion:143GHz:units}{0.3694\MJysrmK}
\setsymbol{HFI:unit:conversion:217GHz:units}{0.4811\MJysrmK}
\setsymbol{HFI:unit:conversion:353GHz:units}{0.2883\MJysrmK}
\setsymbol{HFI:unit:conversion:545GHz:units}{0.05826\MJysrmK}
\setsymbol{HFI:unit:conversion:857GHz:units}{0.002238\MJysrmK}

\setsymbol{HFI:unit:conversion:100GHz}{0.2415}
\setsymbol{HFI:unit:conversion:143GHz}{0.3694}
\setsymbol{HFI:unit:conversion:217GHz}{0.4811}
\setsymbol{HFI:unit:conversion:353GHz}{0.2883}
\setsymbol{HFI:unit:conversion:545GHz}{0.05826}
\setsymbol{HFI:unit:conversion:857GHz}{0.002238}

\setsymbol{HFI:colour:correction:alpha=-2:V1.01:100GHz}{0.9893}
\setsymbol{HFI:colour:correction:alpha=-2:V1.01:143GHz}{0.9759}
\setsymbol{HFI:colour:correction:alpha=-2:V1.01:217GHz}{1.0007}
\setsymbol{HFI:colour:correction:alpha=-2:V1.01:353GHz}{1.0028}
\setsymbol{HFI:colour:correction:alpha=-2:V1.01:545GHz}{1.0019}
\setsymbol{HFI:colour:correction:alpha=-2:V1.01:857GHz}{0.9889}

\setsymbol{HFI:colour:correction:alpha=0:V1.01:100GHz}{1.0008}
\setsymbol{HFI:colour:correction:alpha=0:V1.01:143GHz}{1.0148}
\setsymbol{HFI:colour:correction:alpha=0:V1.01:217GHz}{0.9909}
\setsymbol{HFI:colour:correction:alpha=0:V1.01:353GHz}{0.9888}
\setsymbol{HFI:colour:correction:alpha=0:V1.01:545GHz}{0.9878}
\setsymbol{HFI:colour:correction:alpha=0:V1.01:857GHz}{1.0014}

\providecommand{\sorthelp}[1]{}

\begin{document}

\topmargin=-1cm
\oddsidemargin=-1cm
\evensidemargin=-1cm
\textwidth=17cm
\textheight=25cm
\raggedbottom
\sloppy

\definecolor{Blue}{rgb}{0.,0.,1.}
\definecolor{LightSkyBlue}{rgb}{0.691,0.827,1.}
\definecolor{Red}{rgb}{1.,0.,0.}
\definecolor{Green}{rgb}{0.,1.,0.}
\definecolor{Purple}{rgb}{0.5, 0., 0.5}
\definecolor{Try}{rgb}{0.15,0.,1}
\definecolor{Black}{rgb}{0., 0., 0.}

\title{\textit{Planck} intermediate results. XXXIX. \\
 The Planck list of high-redshift source candidates}

\author{\small
Planck Collaboration: P.~A.~R.~Ade\inst{80}
\and
N.~Aghanim\inst{55}
\and
M.~Arnaud\inst{69}
\and
J.~Aumont\inst{55}
\and
C.~Baccigalupi\inst{79}
\and
A.~J.~Banday\inst{87, 8}
\and
R.~B.~Barreiro\inst{60}
\and
N.~Bartolo\inst{26, 61}
\and
E.~Battaner\inst{88, 89}
\and
K.~Benabed\inst{56, 86}
% ...
}
\institute{\small
APC, AstroParticule et Cosmologie, Universit\'{e} Paris Diderot, CNRS/IN2P3, CEA/lrfu, Observatoire de Paris, Sorbonne Paris Cit\'{e}, 10, rue Alice Domon et L\'{e}onie Duquet, 75205 Paris Cedex 13, France\goodbreak
\and
African Institute for Mathematical Sciences, 6-8 Melrose Road, Muizenberg, Cape Town, South Africa\goodbreak
\and
Agenzia Spaziale Italiana Science Data Center, Via del Politecnico snc, 00133, Roma, Italy\goodbreak
\and
Aix Marseille Universit\'{e}, CNRS, LAM (Laboratoire d'Astrophysique de Marseille) UMR 7326, 13388, Marseille, France\goodbreak
\and
Astrophysics Group, Cavendish Laboratory, University of Cambridge, J J Thomson Avenue, Cambridge CB3 0HE, U.K.\goodbreak
\and
Astrophysics \& Cosmology Research Unit, School of Mathematics, Statistics \& Computer Science, University of KwaZulu-Natal, Westville Campus, Private Bag X54001, Durban 4000, South Africa\goodbreak
\and
CITA, University of Toronto, 60 St. George St., Toronto, ON M5S 3H8, Canada\goodbreak
\and
CNRS, IRAP, 9 Av. colonel Roche, BP 44346, F-31028 Toulouse cedex 4, France\goodbreak
\and
California Institute of Technology, Pasadena, California, U.S.A.\goodbreak
\and
Centro de Estudios de F\'{i}sica del Cosmos de Arag\'{o}n (CEFCA), Plaza San Juan, 1, planta 2, E-44001, Teruel, Spain\goodbreak
\and
Computational Cosmology Center, Lawrence Berkeley National Laboratory, Berkeley, California, U.S.A.\goodbreak
\and
DSM/Irfu/SPP, CEA-Saclay, F-91191 Gif-sur-Yvette Cedex, France\goodbreak
\and
DTU Space, National Space Institute, Technical University of Denmark, Elektrovej 327, DK-2800 Kgs. Lyngby, Denmark\goodbreak
\and
D\'{e}partement de Physique Th\'{e}orique, Universit\'{e} de Gen\`{e}ve, 24, Quai E. Ansermet,1211 Gen\`{e}ve 4, Switzerland\goodbreak
\and
Departamento de Astrof\'{i}sica, Universidad de La Laguna (ULL), E-38206 La Laguna, Tenerife, Spain\goodbreak
\and
Departamento de F\'{\i}sica, Universidad de Oviedo, Avda. Calvo Sotelo s/n, Oviedo, Spain\goodbreak
\and
Department of Astrophysics/IMAPP, Radboud University Nijmegen, P.O. Box 9010, 6500 GL Nijmegen, The Netherlands\goodbreak
\and
Department of Physics \& Astronomy, University of British Columbia, 6224 Agricultural Road, Vancouver, British Columbia, Canada\goodbreak
\and
Department of Physics and Astronomy, Dana and David Dornsife College of Letter, Arts and Sciences, University of Southern California, Los Angeles, CA 90089, U.S.A.\goodbreak
\and
Department of Physics and Astronomy, University College London, London WC1E 6BT, U.K.\goodbreak
\and
Department of Physics, Florida State University, Keen Physics Building, 77 Chieftan Way, Tallahassee, Florida, U.S.A.\goodbreak
\and
Department of Physics, Gustaf H\"{a}llstr\"{o}min katu 2a, University of Helsinki, Helsinki, Finland\goodbreak
\and
Department of Physics, Princeton University, Princeton, New Jersey, U.S.A.\goodbreak
\and
Department of Physics, University of California, Santa Barbara, California, U.S.A.\goodbreak
\and
Department of Physics, University of Illinois at Urbana-Champaign, 1110 West Green Street, Urbana, Illinois, U.S.A.\goodbreak
\and
Dipartimento di Fisica e Astronomia G. Galilei, Universit\`{a} degli Studi di Padova, via Marzolo 8, 35131 Padova, Italy\goodbreak
\and
Dipartimento di Fisica e Scienze della Terra, Universit\`{a} di Ferrara, Via Saragat 1, 44122 Ferrara, Italy\goodbreak
\and
Dipartimento di Fisica, Universit\`{a} La Sapienza, P. le A. Moro 2, Roma, Italy\goodbreak
\and
Dipartimento di Fisica, Universit\`{a} degli Studi di Milano, Via Celoria, 16, Milano, Italy\goodbreak
\and
Dipartimento di Fisica, Universit\`{a} degli Studi di Trieste, via A. Valerio 2, Trieste, Italy\goodbreak
\and
Dipartimento di Matematica, Universit\`{a} di Roma Tor Vergata, Via della Ricerca Scientifica, 1, Roma, Italy\goodbreak
\and
Discovery Center, Niels Bohr Institute, Blegdamsvej 17, Copenhagen, Denmark\goodbreak
\and
European Space Agency, ESAC, Planck Science Office, Camino bajo del Castillo, s/n, Urbanizaci\'{o}n Villafranca del Castillo, Villanueva de la Ca\~{n}ada, Madrid, Spain\goodbreak
\and
European Space Agency, ESTEC, Keplerlaan 1, 2201 AZ Noordwijk, The Netherlands\goodbreak
\and
Facolt\`{a} di Ingegneria, Universit\`{a} degli Studi e-Campus, Via Isimbardi 10, Novedrate (CO), 22060, Italy\goodbreak
\and
Gran Sasso Science Institute, INFN, viale F. Crispi 7, 67100 L'Aquila, Italy\goodbreak
\and
HGSFP and University of Heidelberg, Theoretical Physics Department, Philosophenweg 16, 69120, Heidelberg, Germany\goodbreak
\and
Haverford College Astronomy Department, 370 Lancaster Avenue, Haverford, Pennsylvania, U.S.A.\goodbreak
\and
Helsinki Institute of Physics, Gustaf H\"{a}llstr\"{o}min katu 2, University of Helsinki, Helsinki, Finland\goodbreak
\and
INAF - Osservatorio Astrofisico di Catania, Via S. Sofia 78, Catania, Italy\goodbreak
\and
INAF - Osservatorio Astronomico di Padova, Vicolo dell'Osservatorio 5, Padova, Italy\goodbreak
\and
INAF - Osservatorio Astronomico di Roma, via di Frascati 33, Monte Porzio Catone, Italy\goodbreak
\and
INAF - Osservatorio Astronomico di Trieste, Via G.B. Tiepolo 11, Trieste, Italy\goodbreak
\and
INAF/IASF Bologna, Via Gobetti 101, Bologna, Italy\goodbreak
\and
INAF/IASF Milano, Via E. Bassini 15, Milano, Italy\goodbreak
\and
INFN, Sezione di Bologna, Via Irnerio 46, I-40126, Bologna, Italy\goodbreak
\and
INFN, Sezione di Roma 1, Universit\`{a} di Roma Sapienza, Piazzale Aldo Moro 2, 00185, Roma, Italy\goodbreak
\and
INFN, Sezione di Roma 2, Universit\`{a} di Roma Tor Vergata, Via della Ricerca Scientifica, 1, Roma, Italy\goodbreak
\and
INFN/National Institute for Nuclear Physics, Via Valerio 2, I-34127 Trieste, Italy\goodbreak
\and
ISDC, Department of Astronomy, University of Geneva, ch. d'Ecogia 16, 1290 Versoix, Switzerland\goodbreak
\and
IUCAA, Post Bag 4, Ganeshkhind, Pune University Campus, Pune 411 007, India\goodbreak
\and
Imperial College London, Astrophysics group, Blackett Laboratory, Prince Consort Road, London, SW7 2AZ, U.K.\goodbreak
\and
Infrared Processing and Analysis Center, California Institute of Technology, Pasadena, CA 91125, U.S.A.\goodbreak
\and
Institut Universitaire de France, 103, bd Saint-Michel, 75005, Paris, France\goodbreak
\and
Institut d'Astrophysique Spatiale, CNRS (UMR8617) Universit\'{e} Paris-Sud 11, B\^{a}timent 121, Orsay, France\goodbreak
\and
Institut d'Astrophysique de Paris, CNRS (UMR7095), 98 bis Boulevard Arago, F-75014, Paris, France\goodbreak
\and
Institute of Astronomy, University of Cambridge, Madingley Road, Cambridge CB3 0HA, U.K.\goodbreak
\and
Institute of Theoretical Astrophysics, University of Oslo, Blindern, Oslo, Norway\goodbreak
\and
Instituto de Astrof\'{\i}sica de Canarias, C/V\'{\i}a L\'{a}ctea s/n, La Laguna, Tenerife, Spain\goodbreak
\and
Instituto de F\'{\i}sica de Cantabria (CSIC-Universidad de Cantabria), Avda. de los Castros s/n, Santander, Spain\goodbreak
\and
Istituto Nazionale di Fisica Nucleare, Sezione di Padova, via Marzolo 8, I-35131 Padova, Italy\goodbreak
\and
Jet Propulsion Laboratory, California Institute of Technology, 4800 Oak Grove Drive, Pasadena, California, U.S.A.\goodbreak
\and
Jodrell Bank Centre for Astrophysics, Alan Turing Building, School of Physics and Astronomy, The University of Manchester, Oxford Road, Manchester, M13 9PL, U.K.\goodbreak
\and
Kavli Institute for Cosmological Physics, University of Chicago, Chicago, IL 60637, USA\goodbreak
\and
Kavli Institute for Cosmology Cambridge, Madingley Road, Cambridge, CB3 0HA, U.K.\goodbreak
\and
Kazan Federal University, 18 Kremlyovskaya St., Kazan, 420008, Russia\goodbreak
\and
LAL, Universit\'{e} Paris-Sud, CNRS/IN2P3, Orsay, France\goodbreak
\and
LERMA, CNRS, Observatoire de Paris, 61 Avenue de l'Observatoire, Paris, France\goodbreak
\and
Laboratoire AIM, IRFU/Service d'Astrophysique - CEA/DSM - CNRS - Universit\'{e} Paris Diderot, B\^{a}t. 709, CEA-Saclay, F-91191 Gif-sur-Yvette Cedex, France\goodbreak
\and
Laboratoire de Physique Subatomique et Cosmologie, Universit\'{e} Grenoble-Alpes, CNRS/IN2P3, 53, rue des Martyrs, 38026 Grenoble Cedex, France\goodbreak
\and
Laboratoire de Physique Th\'{e}orique, Universit\'{e} Paris-Sud 11 \& CNRS, B\^{a}timent 210, 91405 Orsay, France\goodbreak
\and
Lawrence Berkeley National Laboratory, Berkeley, California, U.S.A.\goodbreak
\and
Lebedev Physical Institute of the Russian Academy of Sciences, Astro Space Centre, 84/32 Profsoyuznaya st., Moscow, GSP-7, 117997, Russia\goodbreak
\and
Max-Planck-Institut f\"{u}r Astrophysik, Karl-Schwarzschild-Str. 1, 85741 Garching, Germany\goodbreak
\and
National University of Ireland, Department of Experimental Physics, Maynooth, Co. Kildare, Ireland\goodbreak
\and
Nicolaus Copernicus Astronomical Center, Bartycka 18, 00-716 Warsaw, Poland\goodbreak
\and
Niels Bohr Institute, Blegdamsvej 17, Copenhagen, Denmark\goodbreak
\and
Optical Science Laboratory, University College London, Gower Street, London, U.K.\goodbreak
\and
SISSA, Astrophysics Sector, via Bonomea 265, 34136, Trieste, Italy\goodbreak
\and
School of Physics and Astronomy, Cardiff University, Queens Buildings, The Parade, Cardiff, CF24 3AA, U.K.\goodbreak
\and
Sorbonne Universit\'{e}-UPMC, UMR7095, Institut d'Astrophysique de Paris, 98 bis Boulevard Arago, F-75014, Paris, France\goodbreak
\and
Space Research Institute (IKI), Russian Academy of Sciences, Profsoyuznaya Str, 84/32, Moscow, 117997, Russia\goodbreak
\and
Space Sciences Laboratory, University of California, Berkeley, California, U.S.A.\goodbreak
\and
Special Astrophysical Observatory, Russian Academy of Sciences, Nizhnij Arkhyz, Zelenchukskiy region, Karachai-Cherkessian Republic, 369167, Russia\goodbreak
\and
Sub-Department of Astrophysics, University of Oxford, Keble Road, Oxford OX1 3RH, U.K.\goodbreak
\and
UPMC Univ Paris 06, UMR7095, 98 bis Boulevard Arago, F-75014, Paris, France\goodbreak
\and
Universit\'{e} de Toulouse, UPS-OMP, IRAP, F-31028 Toulouse cedex 4, France\goodbreak
\and
University of Granada, Departamento de F\'{\i}sica Te\'{o}rica y del Cosmos, Facultad de Ciencias, Granada, Spain\goodbreak
\and
University of Granada, Instituto Carlos I de F\'{\i}sica Te\'{o}rica y Computacional, Granada, Spain\goodbreak
% ...
}

\abstract{
  The \Planck\ mission, thanks to its large frequency range and
  all-sky coverage, has a unique potential for systematically detecting
  the brightest, and rarest, submillimetre sources on the sky,
  including distant objects in the high-redshift Universe traced by their dust
  emission. A novel method, based on a component-separation procedure
  using a combination of \Planck\ and IRAS
  data, has been applied to select the most luminous cold submillimetre
  sources with spectral energy distributions peaking between 353 and 857\,GHz
  at 5{\arcmin} resolution.
  A total of 2151 \Planck\ high-$z$ source candidates (the ``PHZ'') have been
  detected in the cleanest 26\,\% of the sky,
  with flux density at 545\,GHz above 500\,mJy. Embedded in the cosmic
  infrared background close to the confusion limit, 
  these high-$z$ candidates exhibit colder colours than their surroundings,
  consistent with redshifts $z>2$, 
  assuming a dust temperature of $T_{\rm{xgal}}=35$\,K and a spectral
  index of $\beta_{\rm{xgal}}=1.5$. 
  First follow-up observations obtained from optical to submillimetre
  wavelengths have confirmed that this list consists of
  two distinct populations. A small fraction (around 3\,\%) of the sources
  have been identified as strongly gravitationally lensed star-forming
  galaxies, which are amongst the brightest submm lensed objects (with flux
  density at 545\,GHz ranging from 350\,mJy  up to 1\,Jy) 
  at redshift 2 to 4. However, the vast majority of the PHZ sources 
  appear as overdensities of dusty star-forming galaxies, having colours
  consistent with $z>2$, 
  and may be considered as proto-cluster candidates.
  The PHZ provides an original sample, which is complementary to the 
  Planck Sunyaev-Zeldovich Catalogue (PSZ2); by extending the population
  of the virialized massive galaxy clusters detected with their SZ signal
  below $z<1.5$ to a population of sources at $z>1.5$, 
  the PHZ may contain the progenitors of today's clusters.
  Hence the Planck List of High-redshift Source Candidates
  opens a new window on the study of the early ages of structure formation, and 
  the understanding of the intensively star-forming phase at high-$z$.
  }


\authorrunning{Planck Collaboration}
\titlerunning{The PHZ}
\maketitle

\section{Introduction}
\label{sec:intro}

Developing an understanding of the birth and growth of the large-scale
structures in the Universe
enables us to build a bridge between cosmology and astrophysics. 
The formation of structures in the nonlinear regime is still poorly constrained, because of the
complex interplay between dark matter halos and baryonic cooling at early times, during this transition from
the epoch of first galaxy formation to the virialization of massive halos.
Hence the analysis of a large sample of high-redshift ($z>2$) objects is crucial for placing new constraints on both cosmological and
astrophysical models. 

Galaxy clusters, as the largest virialized structures in the Universe, are ideal laboratories for studying the intense star-formation occurring in dark
matter halos, and providing observational constraints on galaxy assembly,
quenching, and evolution, driven by the halos' environment.
The first discoveries of strongly gravitationally lensed galaxies at very high redshift \citep[e.g.,][]{Walsh1979,Soucail1987}
opened another window onto the early stages of these intensively star-forming galaxies, and provided
new information on the early star-formation phase \citep{Danielson2011, Swinbank2011,Combes2012}, 
allowing us to probe spatial details at scales well below 1\,kpc \citep[e.g.,][]{Swinbank2010, Swinbank2011}.
From the cosmological point of view, galaxy clusters, considered as the direct descendants of primordial fluctuations on Mpc scales, 
provide a powerful tool for probing structure formation within the $\Lambda$CDM model
\citep{Brodwin2010,Hutsi2010,Williamson2011,Harrison2012, Holz2012,Waizmann2012,Trindade2013}.
More specifically, \citet{planck2013-p11}, \citet{planck2013-p15}, and \citet{planck2014-a30} recently highlighted 
some tension between the cosmological and astrophysical results concerning
 the determination of the $\Omega_{\mathrm{M}}$ and $\sigma_8$ parameters, which still needs to be resolved and properly understood.
 
Galaxy clusters in the local Universe can be efficiently traced by their
dominant red sequence galaxies \citep[e.g.,][]{Gladders2005, Olsen2007}, 
by their diffuse X-ray emission from the hot gas of the intra-cluster medium
\citep[e.g.,][]{Ebeling2001,Fassbender2011} 
or by the Sunyaev-Zeldovich effect \citep[e.g.,][]{Foley2011, Menanteau2012,
planck2012-I, planck2013-p05a, Brodwin2015} up to $z\,{\simeq}\,1.5$.
The standard methods to search for clusters have yielded only a handful of
objects at $z > 1.5$ \citep[e.g.,][]{Henry2010, Tanaka2010, Santos2011}, 
consistent with the prediction of the concordance model that cluster-size
objects virialize late.  Searching for high-$z$ large-scale structures means
we are looking at the progenitors of local galaxy clusters, the so-called
proto-clusters, at the early stages of their evolution, where not enough
processed baryonic material was available to be detected by standard methods.
These proto-clusters, likely lying at $z>2$, are assumed to be in an active
star-forming phase, but not yet fully virialized.  To investigate these
earlier evolutionary stages we need different approaches, such as the one
presented in this paper.
 
During the past decade, more and more proto-cluster candidates have been
detected through different techniques, using X-ray signatures, stellar mass
overdensities, Ly$\alpha$ emission, and association with radio galaxies
\citep[e.g.,][]{Brodwin2005,Miley2006,Nesvadba2006,Doherty2010,
Papovich2010,Hatch2011,Gobat2011,Stanford2012,Santos2011,Santos2013, Santos2014,Brodwin2010, Brodwin2011,Brodwin2013}.
However, only a few detections have been done in ``random'' fields
 \citep[e.g.,][]{Steidel1998,Steidel2005,Toshikawa2012,rettura2014}, and 
most of these detections are biased towards radio galaxies or quasars 
\citep[][]{pentericci2000,Kurk2000,Kurk2004, venemans2002,
venemans2004,venemans2007,galametz2010,galametz2013,rigby2013,wylezalek2013a,
trainor2012,cooke2014}, or obtained over very limited fractions of the sky,
e.g., in the COSMOS field, which is 1.65\,$\rm{deg}^{2}$
\citep[][]{capak2011,Cucciati2014, Chiang2014} and in the
{\it Hubble Space Telescope} Ultra-Deep Field \citep{Beckwith2006,mei2014}
with its $200{\arcsec}\times200{\arcsec}$ area. 
Since the expected surface density of such strongly lensed high-$z$ galaxies
or massive proto-clusters is fairly small, a few times
$10^{-2}$\,$\rm{deg}^{-2}$ \citep{Negrello2007, Negrello2010,
paciga2009,lima2010,Bethermin2011, Hezaveh2012}, performing an unbiased
analysis of this population of sources requires us to explore much larger
regions of the sky.  This has been initiated, for example with the
{\it Spitzer} SPT Deep Field survey covering 94\,$\rm{deg}^2$ 
and yielding the detection of 300 galaxy cluster candidates with redshifts
$1.3<z<2$ \citep{rettura2014}. 

The submm and mm sky has proved to be an efficient window
onto star-forming galaxies with redshifts between 1and 6, since it allows us
to detect the redshifted modified blackbody emission coming from the warm
dust in galaxies.  Taking advantage of the so-called ``negative k-correction''
in the submm \citep{Franceschini1991}, 
which compensates for the cosmological dimming at high redshift in the submm, 
many samples of high-$z$ galaxies and also proto-cluster candidates have been
identified or discovered in this frequency range in the last two decades
\citep[e.g.,][]{Lagache2005, Beelen2008, Smail2014}.  This process has been
accelerated with the observations of larger fields in the submm and mm range.
The South Pole Telescope experiment \citep{carlstrom2011} which covers
1300\,$\rm{deg}^2$ at 1.4 and 2\,mm has built a unique sample of high-$z$
dusty star-forming objects\citep{Vieira2010}, 
which have been shown to be strongly lensed galaxies at a median redshift
$z\simeq3.5$ \citep{Vieira2013, Weiss2013, Hezaveh2013}. 
A population of 38 dusty galaxies at $z>4$ has been discovered
by \citet{Dowell2014} in the HerMES survey (26\,$\rm{deg}^{2}$) with the
{\it Herschel}-SPIRE instrument \citep{Griffin2010}.
Furthermore, as predicted by \citet{Negrello2005}, \citet{Clements2014}
showed that the proto-cluster population can be efficiently detected
in the submm as overdensities of dusty star-forming galaxies. 
 
The \Planck\ satellite\footnote{\Planck\ (\url{http://www.esa.int/Planck}) is
a project of the European Space Agency (ESA) with instruments provided by two
scientific consortia funded by ESA member states and led by Principal
Investigators from France and Italy, telescope reflectors provided through a
collaboration between ESA and a scientific consortium led and funded by
Denmark, and additional contributions from NASA (USA).} combines two of the
main requirements for efficiently detecting high-$z$ sources, namely the
spatial and spectral coverage.  \Planck's combination of the High Frequency
Instrument (HFI) and Low Frequency Instrument (LFI) provides full-sky maps
from 857 down to 30\,GHz,\footnote{Although we use frequency units
for \Planck\ channels here, since most of the relevant literature for
submillimetre spectra uses wavelengths, we will typically discuss the
bands in order of decreasing frequency.}
which allows coverage of the redshifted spectral energy distribution (SED)
of potential dusty star-forming galaxies over a large fraction of the sky. 
The moderate resolution (5{\arcmin} to 10{\arcmin} in the HFI bands) 
of \Planck, compared to other submm experiments, such as {\it Herschel}-SPIRE 
(18{\arcsec} to 36{\arcsec}) or SCUBA-2 (15{\arcsec} at 353\,GHz), 
appears as a benefit when searching for clustered structures at high redshift:
a 5{\arcmin} beam corresponds to a physical size of 2.5\,Mpc at $z=2$,
which matches the expected typical size of proto-clusters in their early
stages.  

We present in this work the Planck List of High-redshift Source Candidates
(the ``PHZ''),
which includes 2151 sources distributed over 26\,\% of the sky, with redshifts
likely to be greater than 2. 
This list is complementary to the Planck Catalogue of Compact Sources
\citep[PCCS2;][]{planck2014-a35}, which has been built in each of the
\Planck-HFI and LFI bands.  The PHZ takes advantage of the spectral
coverage in the HFI bands, between 857 and 353\,GHz, 
to track the redshifted emission from dusty galaxies using an appropriate
colour-cleaning method \citep{Montier2010} and colour-colour selection.
It also covers a different population of sources than the galaxy clusters
of the Planck Sunyaev-Zeldovich Catalogue \citep[PSZ2;][]{planck2013-p05a}, 
with redshifts likely below 1.5, by tracking the dust emission from the
galaxies instead of searching for a signature of the hot intracluster gas.
Because of the limited sensitivity and resolution of \Planck, the PHZ entries 
will point to the rarest and brightest submm excess spots in the extragalactic
sky, which could be either statistical fluctuations of the cosmic infrared
background, single strongly-lensed galaxies, or overdensities of bright
star-forming galaxies in the early Universe.
This list of source candidates may provide important information on the
evolution of the star-formation rate in dense environments: 
the submm luminosity of proto-clusters will obviously be larger if the
star-formation in member galaxies is
synchronous and the abundance of protoclusters detected at submm wavelengths
depends on the duration of the active star-formation phase.

The data that we use and an overview of the processing are presented in
Sect.~\ref{sec:datacat}.  The component separation and point source detection
steps are then detailed in Sects.~\ref{sec:compsep} and \ref{sec:ps_detection},
respectively.  The statistical quality of the selection algorithm is
characterized in Sect.~\ref{sec:mcqa}. The 
final PHZ is described in Sect.~\ref{sec:content}, followed by a discussion
on the nature of the PHZ sources in Sect.~\ref{sec:conclusion}.

\begin{figure*}
\center
\vspace{-0.8cm}
\includegraphics[width=0.6\textwidth, angle=90,viewport=0 0 440 730]{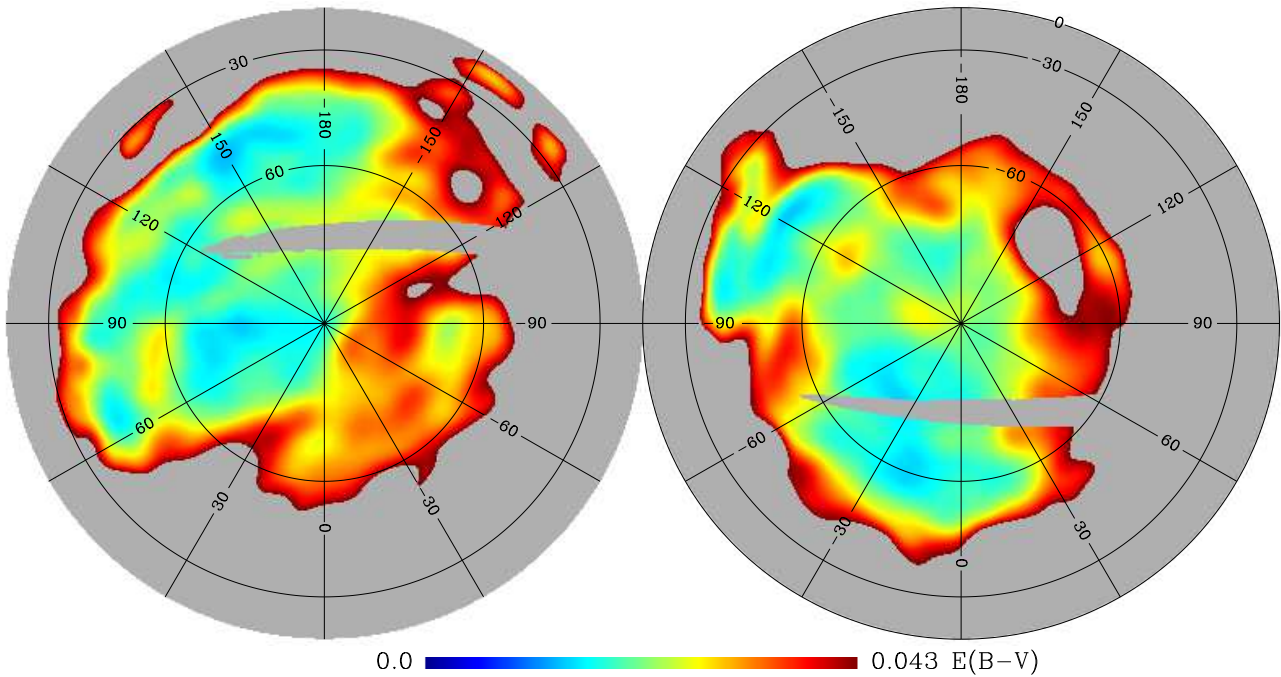} 
\caption{All-sky Galactic map in orthographic projections of the regions at high latitude used for analysis in this paper, with the
masked area built on the \Planck\ extinction map \citep{planck2013-p06b}, 
using the criterion $E(B-V)_{\rm{xgal}}<0.0432$, which is equivalent to 
$N_{\rm{H}}<3\times10^{20}\mathrm{cm^{-2}}$. Poorly defined stripes in the
IRAS data are also rejected.  }
\label{fig:mask_ebv}
\end{figure*}

\section{Data and processing overview}
\label{sec:datacat}

\subsection{Data}
\label{sec:datacat_data}

This paper is based on the \Planck\ 2015 release products corresponding to the
full mission of HFI, i.e., five full-sky surveys.  We refer to
\citet{planck2014-a08} and \citet{planck2014-a09} for the generic scheme 
of time-ordered information (TOI) processing and mapmaking, as well as for
the technical characterization of the \Planck\ frequency maps. The \Planck\
channel maps are provided in {\tt HEALPix} \citep{Gorski2005} format,
at $N_{\rm{side}}=2048$ resolution. 
Here we approximate the \Planck\ beams by effective circular Gaussians
\citep{planck2014-a08}, reported in Table~\ref{tab:fwhm}. 
The noise in the channel maps is assumed to be Gaussian, with a standard 
deviation of 8.8, 9.1, 8.5, and 4.2 kJy sr$^{-1}$ at 857, 545, 353,
and 217\,GHz, respectively \citep{planck2014-a09}.  The absolute gain
calibration of HFI maps is known to better than
5.4 and 5.1\,\% at 857 and 545\,GHz, and 0.78 and 0.16\,\% at 353
and 217\,GHz \citep[see table 6 in][]{planck2014-a09}. 
The mean level of the CIB emission has already been included in the \Planck\
frequency maps of the 2015 release, based on theoretical modelling by
\citet{Bethermin2012}, so that the zero-levels of these maps are compatible
with extragalactic studies.  For further details on the data reduction and
calibration scheme, see \citet{planck2014-a08} and \citet{planck2014-a09}.
In this work we make use of the ``half-ring maps,'' which
correspond to two sets of maps built with only half of the data
as described in \citet{planck2014-a09}.  These can be used to obtain an
estimate of the data noise by computing the half-ring difference maps.

We combine the \Planck-HFI data at 857, 545, 353, and 217\,GHz
with the 3\,THz IRIS data \citep{Miville2005}, the new processing of the
IRAS 3\,THz data \citep{Neugebauer1984}. 
All maps are smoothed at a common FWHM of 5{\arcmin}.

\begin{table}
\caption{FWHM of the effective beam of the IRIS \citep{Miville2005} and \Planck\ \citep{planck2014-a08} maps .}
\label{tab:fwhm}
\nointerlineskip
\setbox\tablebox=\vbox{
\newdimen\digitwidth 
\setbox0=\hbox{\rm 0} 
\digitwidth=\wd0 
\catcode`*=\active 
\def*{\kern\digitwidth} 
\newdimen\signwidth 
\setbox0=\hbox{+} 
\signwidth=\wd0 
\catcode`!=\active 
\def!{\kern\signwidth} 
\newdimen\pointwidth 
\setbox0=\hbox{.} 
\pointwidth=\wd0 
\catcode`?=\active 
\def?{\kern\pointwidth} 
\halign{\tabskip=0pt\hfil#\hfil\tabskip=1em&
\hfil#\hfil\tabskip=1em&
\hfil#\hfil\tabskip=1em&
\hfil#\hfil\tabskip=1em&
\hfil#\hfil\tabskip=0pt\cr
\noalign{\doubleline}
\multispan2 \hfil Band\hfil& FWHM& $\Omega$& $\sigma_{\Omega}$\cr
 [GHz]& [$\mu$m]& [arcmin]& [arcmin$^2$]& [arcmin$^2$]\cr
\noalign{\vskip 3pt\hrule\vskip 4pt}
 3000& *100& 4.3*& 21.04& 1.96\cr
 *857& *350& 4.64& 24.37& 0.02\cr
 *545& *550& 4.83& 26.44& 0.02\cr
 *353& *850& 4.94& 27.69& 0.02\cr
 *217& 1380& 5.02& 28.57& 0.04\cr
\noalign{\vskip 3pt\hrule\vskip 4pt}
}}
\endPlancktable
\end{table}

\subsection{Mask}
\label{sec:datacat_mask}

We define a mask at high Galactic latitude
to minimize the contamination by Galactic dusty structures and to focus on the fraction of the sky dominated by CIB emission.
As recommended in \citet{planck2013-p06b}, we used the $E(B-V)_{\rm{xgal}}$ map, 
released in 2013 in the Planck Legacy Archive,\footnote{\url{http://www.cosmos.esa.int/web/planck/pla}}
as an optimal tracer of the neutral hydrogen column density in diffuse regions. After convolving with a FWHM of $5^\circ$,
we selected regions of the sky with a column density $N_{\rm{H}}<3\times10^{20}\mathrm{cm^{-2}}$, which translates into
$E(B-V)_{\rm{xgal}}<0.0432$. 

We also reject the stripes over the sky that were not covered by the {\it IRAS} satellite, 
and which are filled-in in the IRIS version of the data using an extrapolation of the DIRBE data at lower resolution \citep[see][]{Miville2005}.
These undefined regions of the IRAS map have been masked to avoid spurious detections when combining with the \Planck\ maps.

The resulting mask leaves out the cleanest 25.8\,\% of the sky, approximately equally divided between the northern and southern Galactic hemispheres. 
As shown in Fig.~\ref{fig:mask_ebv}, this fraction of the sky remains heterogeneous, due to elongated Galactic structures with low column density.

\subsection{Data processing overview}
\label{sec:datacat_method}

The purpose of this work is to find extragalactic sources traced by their dust emission in the submillimetre range (submm).
The further away these sources are located, the more redshifted their dust spectral energy distribution (SED) will be, 
or equivalently the colder they appear. The challenge is to separate this 
redshifted dust emission from various foreground or background signals and to extract these sources from the fluctuations of the
cosmic infrared background (CIB) itself.

The data processing is divided into two main steps. The first one is a
component separation on the \Planck\ and IRAS maps (see
Sect.~\ref{sec:compsep}), and the second deals with the compact source
detection and selection (see Sect.~\ref{sec:ps_detection}). 
The full processing can be summarized in the following
steps:\begin{description} \item (i) {\it CMB cleaning} -- we clean maps to
remove the CMB signal in all submm bands using a CMB template (see
Sect.~\ref{sec:compsep_cmb}); \item (ii) {\it Galactic cirrus cleaning} --
we clean maps at 857 to 217\,GHz for Galactic cirrus emission using a Galactic
template combined with the local colour of the maps (see
Sect.~\ref{sec:compsep_cirrus}); \item (iii) {\it excess map at 545\,GHz} --
looking for sources with redshifted SEDs and peaking in the submm range, we
construct an excess map at 545\,GHz, revealing the cold emission of high-$z$
sources, using an optimized combination of all cleaned maps (see
Sect.~\ref{sec:compsep_excess}); \item (iv) {\it point source detection in
the 545\,GHz excess map} -- the point source detection is applied on the
excess map at 545\,GHz (see Sect.~\ref{sec:ps_detection_method});
\item (v) {\it multi frequency detection in the cleaned 857, 545, and 353\,GHz
maps} -- simultaneous detections in the cleaned maps at 857, 545, and 353\,GHz
are also required to consolidate the detection and enable photometry estimates
in these bands (see Sect.~\ref{sec:ps_detection_method});
\item (vi) {\it colour-colour selection} -- complementary to the map-processing
aimed at emphasizing the cold emission from high redshifted sources, we apply a
colour-colour selection based on the photometry (see
Sect.~\ref{sec:ps_detection_colcol});
\item (vii) {\it flux density cut} -- a last selection criterion is applied on
the flux density to deal with the flux boosting affecting our photometry
estimates (see Sect.~\ref{sec:ps_detection_colcol}).
\end{description}

Notice that the first two steps, i.e., CMB and Galactic cleaning, are also
applied independently on the first and last half-ring maps
\citep{planck2014-a09} in all bands, to provide robust estimates of the noise 
in the cleaned maps, which are then used during the photometry processing.

After carrying out this full processing on the \Planck\ and IRAS maps, 
we end-up with a list of 2151 \Planck\ high-$z$ source candidates, distributed 
over the cleanest 25.8\,\% of the sky. We detail in the following sections the
various steps of the processing, the construction of the final list 
and a statistical validation of its quality.

\section{Component separation}
\label{sec:compsep}

\subsection{Astrophysical emissions}
\label{sec:compsep_components}

\begin{figure}[t]
\vspace{-0.4cm}
\hspace{-0.7cm}
\psfrag{----xtitle----}{$\nu\, \rm{[GHz]}$}
\psfrag{-----ytitle-----}{$I_{\nu}\, \rm{[MJy}\,\rm{sr}^{-1}\rm{]}$}
\includegraphics[width=0.55\textwidth]{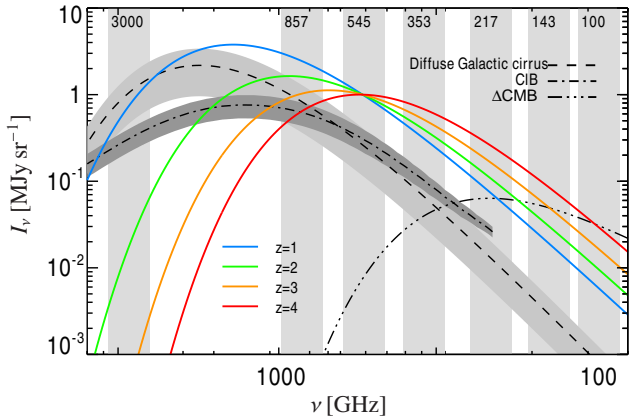} 
\caption{Surface brightness $I_{\nu}$ of the main astrophysical
components of the submm and mm sky at high Galactic latitude, i.e., 
Galactic cirrus, CIB fluctuations, and CMB anisotropies. Typical SEDs of
sources at intermediate and high redshift, i.e., $z=1$--$4$, are 
modelled by a modified blackbody emission law (with $T_{\rm{xgal}}=30$\,K
and $\beta_{\rm{xgal}}=1.5$) and plotted in colours, from blue ($z=1$) to
red ($z=4$).  The $\pm2\,\sigma$ levels of Galactic cirrus and CIB
fluctuations are shown as light and dark grey shaded areas, respectively. 
The bandwidths of the 3\,THz IRIS and the six \Planck-HFI bands are shown as
light grey vertical bands.}
\label{fig:sed_all}
\end{figure}

Owing to the negative k-correction, high-{\it z}
sources (typically $z=1$--$4$) have very ``red'' submillimetre colours. 
Superimposed onto the emission from these sources are other
astrophysical signals, such as the CIB
fluctuations, the CMB anisotropies, and the Galactic foreground dust emission,
each with a different spectral energy distribution (SED). 
A broad frequency coverage from the submm
to mm range is thus mandatory in order to separate these astrophysical
components, so that we can extract faint emission from high-$z$ 
candidates. Combined with IRAS \citep{Neugebauer1984} data at 3\,THz,
the \Planck-HFI data, which spans a wide spectral range from 100 to 857\,GHz, 
represents a unique set of data that is particularly efficient for separating Galactic from
extragalactic and CMB components, as illustrated in Fig.~\ref{fig:sed_all}. 

The Galactic cirrus emission
at high latitude is modelled with a modified blackbody, with a dust temperature
of $T_{\rm{d}}=17.9\,\rm{K}$ and a spectral index $\beta=1.8$ \citep{planck2011-7.12}. 
The SED of Galactic cirrus is normalized at 3\,THz using an averaged 
emissivity \citep[estimated by][at high Galactic latitude]{planck2011-7.12} of
$\epsilon_{100}=0.5\,\rm{MJy}\,\rm{sr}^{-1}/10^{20}\rm{cm}^{-2}$ and a mean column density of
$N_{\rm{H}} =2\times 10^{20}\,\rm{cm}^{-2}$. The grey shaded region in Fig.~\ref{fig:sed_all} 
shows the $\pm2\,\sigma$ domain of the Galactic cirrus fluctuations 
estimated at 3\,THz by computing the integral of the power spectrum $P(k)$ 
over the IRAS maps between multipoles $\ell =200$ and $\ell =2000$, as done in \citet{planck2011-6.6} for the CIB.
This procedure gives $\sigma_{\rm{gal}}^2=2\pi \int{P(k)\times kdk}$,
where $P(k)$ is the 2-D power spectrum obtained in small patches of 100\,deg$^2$, leading to
a value of $\sigma_{\rm{gal}}=0.28\,\rm{MJy}\,\rm{sr}^{-1}$ at 3\,THz.

The CIB emission is given by the model of \citet{Bethermin2011}, with
2$\,\sigma$ values taken from \citet{planck2011-6.6} and defined for spatial
scales of $200<\ell<2000$.  The anisotropies of the CMB, $\Delta$CMB, have been 
normalized at 143\,GHz to correspond to a 2$\,\sigma$ level fluctuations,
with $\sigma_{\rm{CMB}}=65 \ \mu\rm{K_{CMB}}$, 
equivalent to $0.05\,\rm{MJy}\,\rm{sr}^{-1}$.
 
Typical SEDs of extragalactic sources are also indicated on
Fig.~\ref{fig:sed_all} using a modified blackbody emission
law with a temperature of 30\,K and a dust spectral index of 1.5, and 
redshifted to $z=1$,~2,~3, and 4. All SEDs have been normalized to a common
 brightness at 545\,GHz, equivalent to a flux density of 1\,Jy for objects as large as 5{\arcmin} FWHM.
 
As shown in Fig.~\ref{fig:sed_all},
the Galactic cirrus emission, which appears warmer than the other components and peaks at around 2\,THz, 
is well traced by the 3\,THz band of IRAS, as well as the CIB emission which peaks around 1\,THz. 
The CMB anisotropies are well mapped by the low
frequency bands of \Planck-HFI, at 100 and 143\,GHz. 
Finally the emission from high-$z$ sources is dominant in the four bands, from 857 to 217\,GHz, covered by HFI.
This illustrates that the IRAS plus \Planck-HFI bands are well matched to
separate the far-IR emission of high-$z$ ULIRGs from that of the
CMB, Galactic cirrus, and CIB fluctuations.

Because of the special nature of the compact high-$z$ sources, presenting SEDs 
peaking between the Galactic dust component, the CIB component, and the CMB signal,
we had to develop a dedicated approach to component separation, which is detailed below.
This algorithm enables us to clean first for the CMB component, then for the Galactic
 and low-$z$ CIB component, and finally to optimize the excess at 545\,GHz.

\subsection{CMB cleaning}
\label{sec:compsep_cmb}

The CMB component is removed from the 3000, 857, 545, 353, and 217\,GHz IRIS and \Planck\ maps
using a CMB template, which is extrapolated to the other bands according to a CMB spectrum. 
To do this we take into account the spectral bandpass
of each channel, as described in \citet{planck2013-p03d}. 
The cleaning is performed in the {\tt HEALPix} pixel basis, so that the intensity of each pixel after CMB cleaning is given by
\begin{equation}
I_{\nu}^{\rm C} = I^{}_{\nu} - I^{}_{\rm{CMB}} \times \Delta B_{\nu}^{\rm{CMB}} \, , 
\end{equation}
where $I_{\nu}$ is the intensity of a pixel of the input map at frequency $\nu$, $I_{\nu}^{\rm C}$ is the intensity after CMB cleaning, $I_{\rm{CMB}}$ 
is the intensity of the CMB template, and $\Delta B_{\nu}^{\rm{CMB}}$ is the intensity of the CMB fluctuations integrated over the spectral bandpass of the band at frequency $\nu$.

The choice of the CMB template has been driven by the aim of working as close
as possible to the native 5{\arcmin} resolution of the \Planck\ high frequency
maps, in order to match as well as possible the expected physical size of
proto-clusters at high redshift, i.e., around 1 to 2\,Mpc at $z=2$.
Among the four methods applied to the \Planck\ data to produce all-sky
foreground-cleaned CMB maps \citep{planck2014-a11}, only two of them
provide temperature CMB maps at 5{\arcmin} resolution, i.e., {\tt NILC}
\citep{Basak2012, Basak2013} and {\tt SMICA} \citep{Cardoso2008}. 
Since the latter has been shown in \citet{planck2014-a11} to be the least
contaminated by foregrounds for high $\ell$ ($\ell> 2000$), 
it has been chosen as the CMB template in this work. 
The overall agreement between all four methods on the temperature CMB maps is
very good, with an amplitude of pairwise difference maps below
5\,$\muK_{\rm{CMB}}$ over most of the sky on large scales, and below
1$\,\sigma$ at high $\ell$.

However, it is clearly stated that the these maps are not fully cleaned of
high-$\ell$ foregrounds, such as extragalactic point sources or 
Sunyaev-Zeldovich (SZ) emission. Hence this CMB template may be used to
clean efficiently the \Planck\ and IRIS maps for CMB signal at large and
intermediate scales, but not at small scales.
Actually, these residual emission components -- including synchrotron emission
from strong radio sources, thermal emission from Galactic cold dust,
or SZ signal from galaxy clusters -- in the CMB template are extrapolated to
the IRIS and \Planck\ bands with a CMB spectrum during the CMB-cleaning
procedure, which may impact the rest of the analysis,
as we investigate in Sects.~\ref{sec:impact_cleaning_highz} and
\ref{sec:impact_cleaning_foregrounds}.

In order to avoid such issues, it would have been possible to use the 143\,GHz
\Planck\ map as a CMB template.  In this case, the presence of non-CMB signal
could have been more easily quantified; however, the common resolution of all
IRIS and \Planck\ maps would then have to have been degraded
to the 7.3{\arcmin} resolution of the 143\,GHz map, which is
not convenient when looking for compact objects.  As a test case, we have
performed a comparison between the two CMB cleaning options at 8{\arcmin}
resolution to study the impact on the flux density estimates towards PHZ
sources, see Appendix~\ref{sec:cmbclean}.
 
\begin{figure*}
\hspace{-0.5cm}
\begin{tabular}{cccccc}
\quad\quad\quad3\,THz& \quad\quad\quad857\,GHz& \quad\quad\quad 545\,GHz&
 \quad\quad\quad 353\,GHz& \quad\quad\quad 217\,GHz& \quad\quad\quad 5{\arcmin} CMB \\ \\
\includegraphics[width=2.4cm,viewport=100 10 400 400]{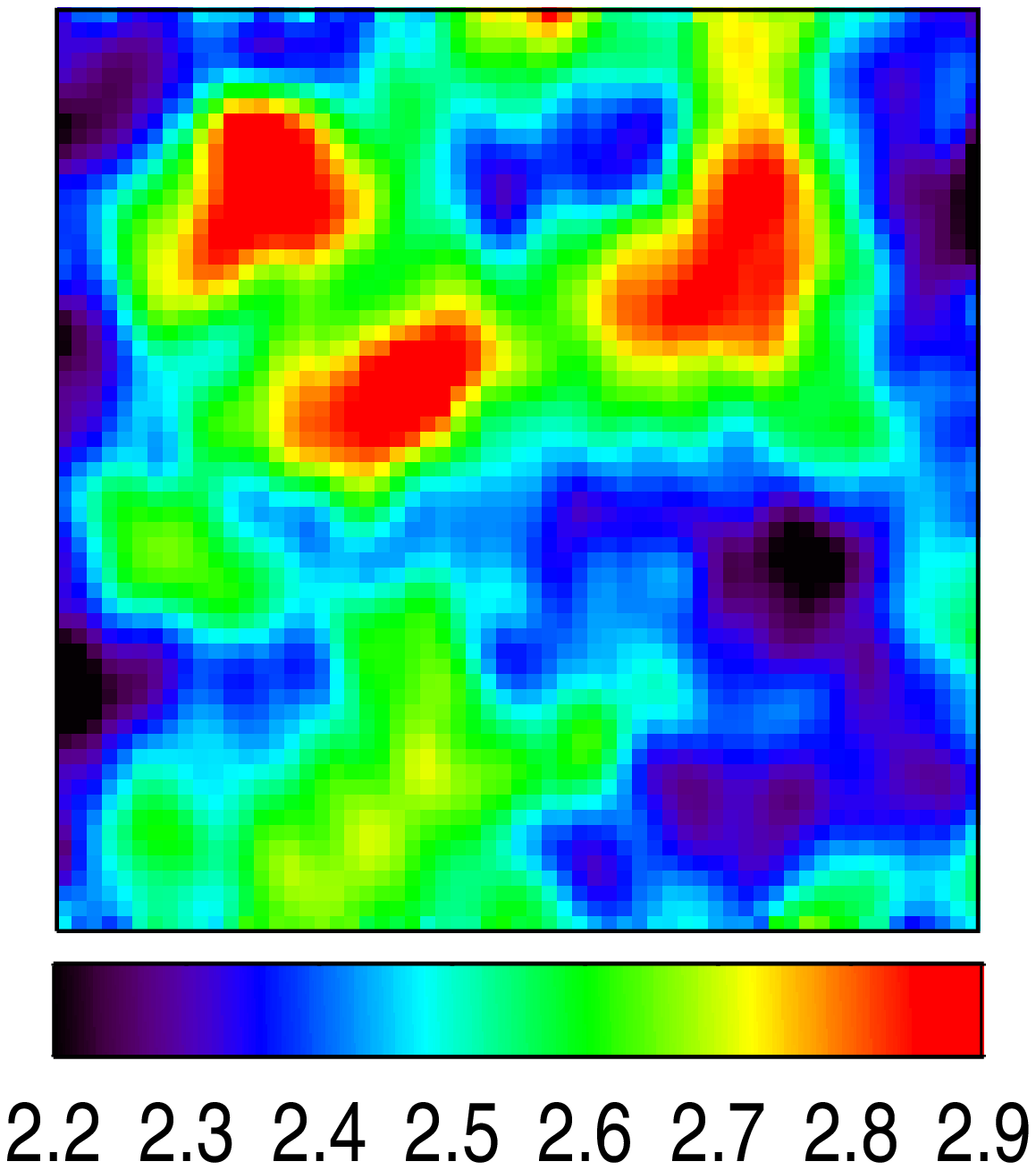}& 
\includegraphics[width=2.4cm,viewport=100 10 400 400]{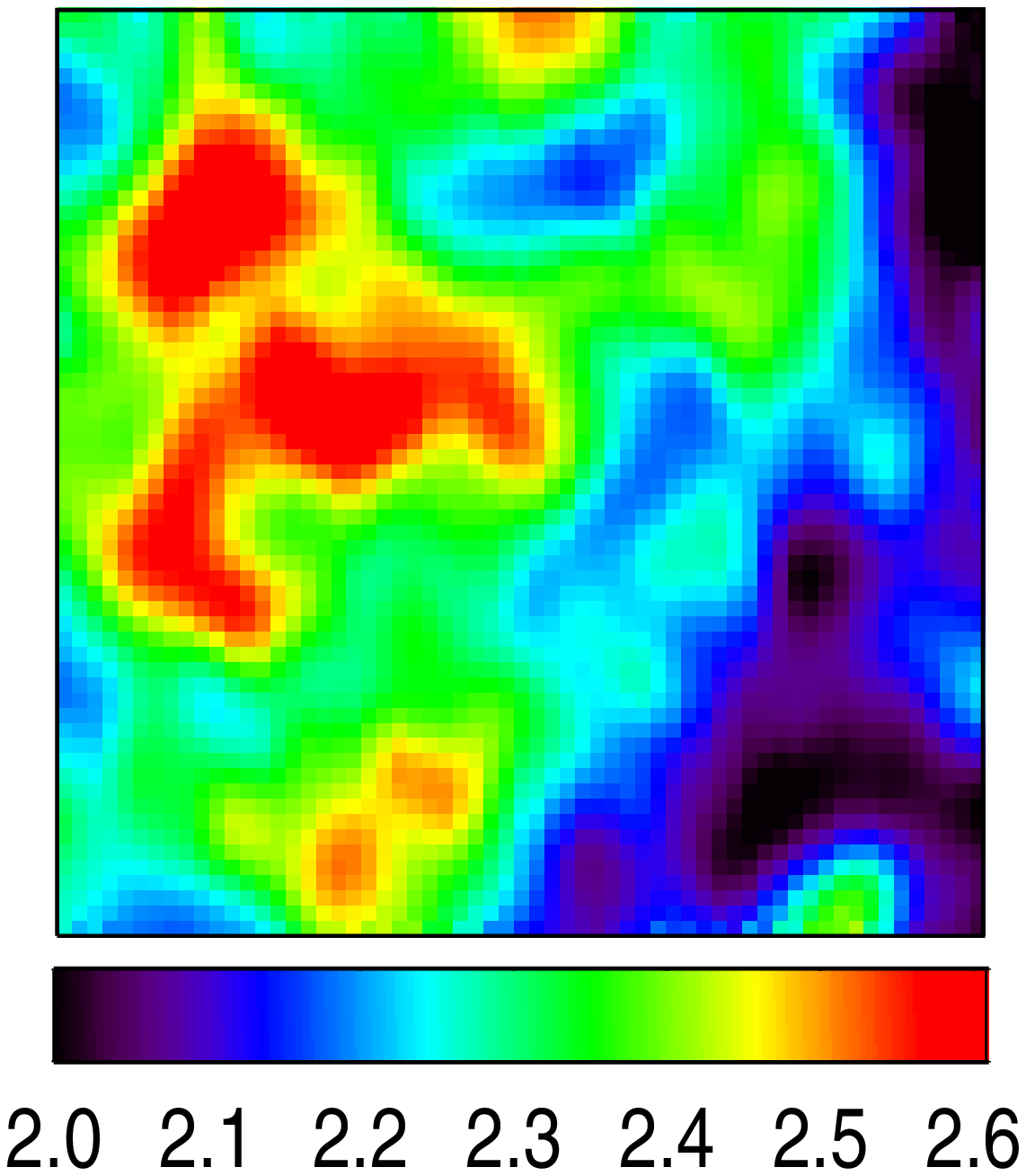}& 
\includegraphics[width=2.4cm,viewport=100 10 400 400]{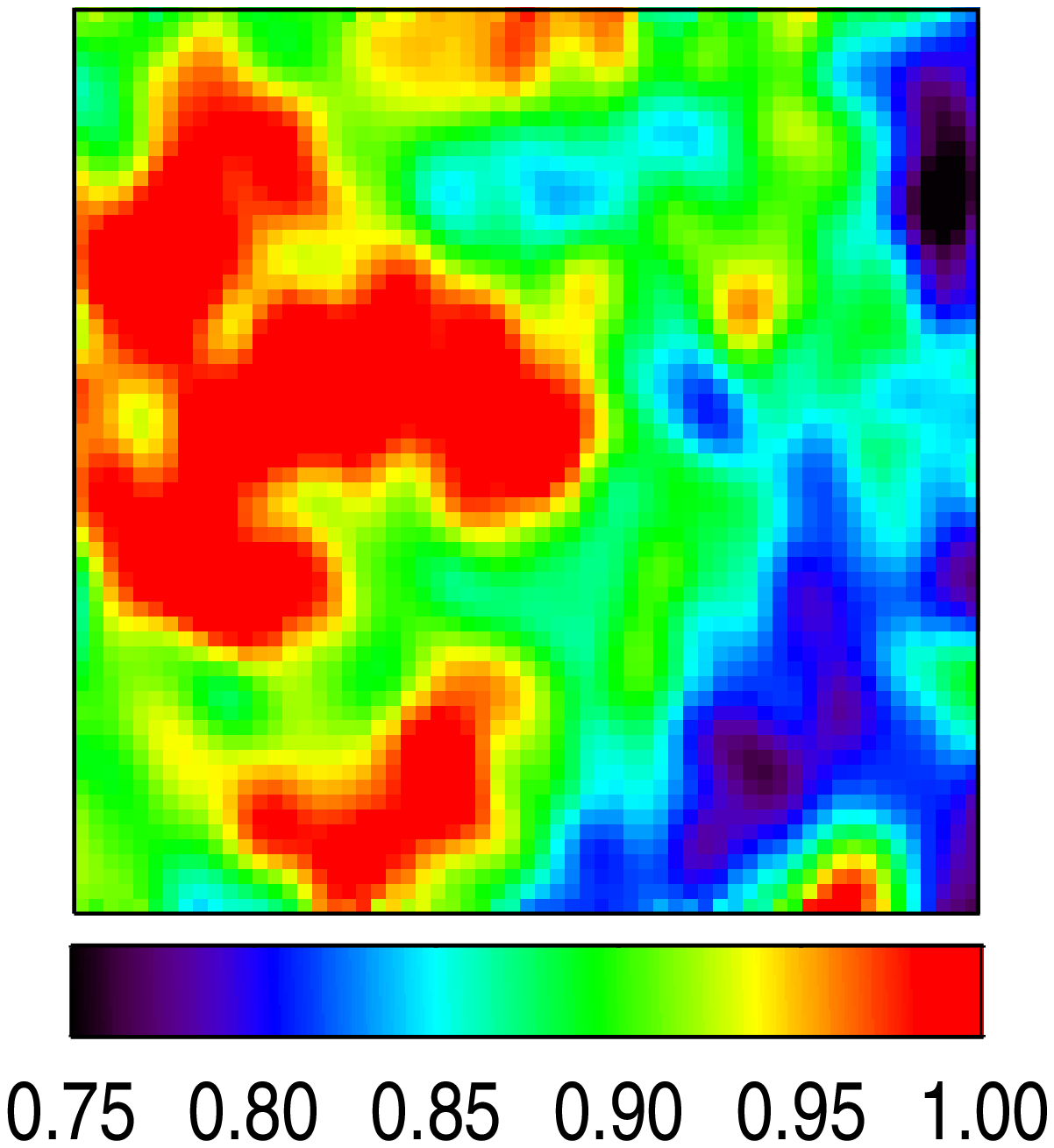}& 
\includegraphics[width=2.4cm,viewport=100 10 400 400]{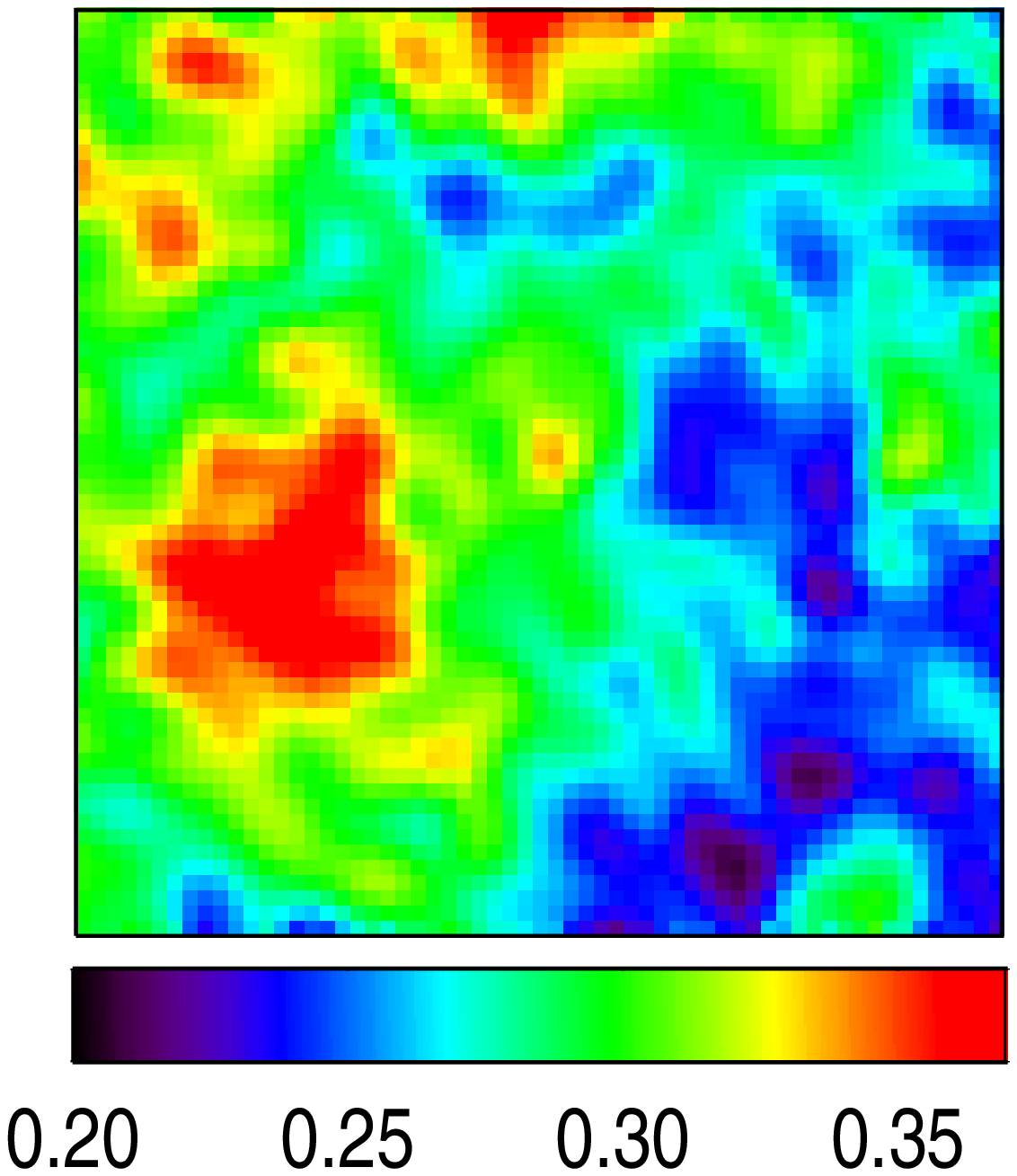}& 
\includegraphics[width=2.4cm,viewport=100 10 400 400]{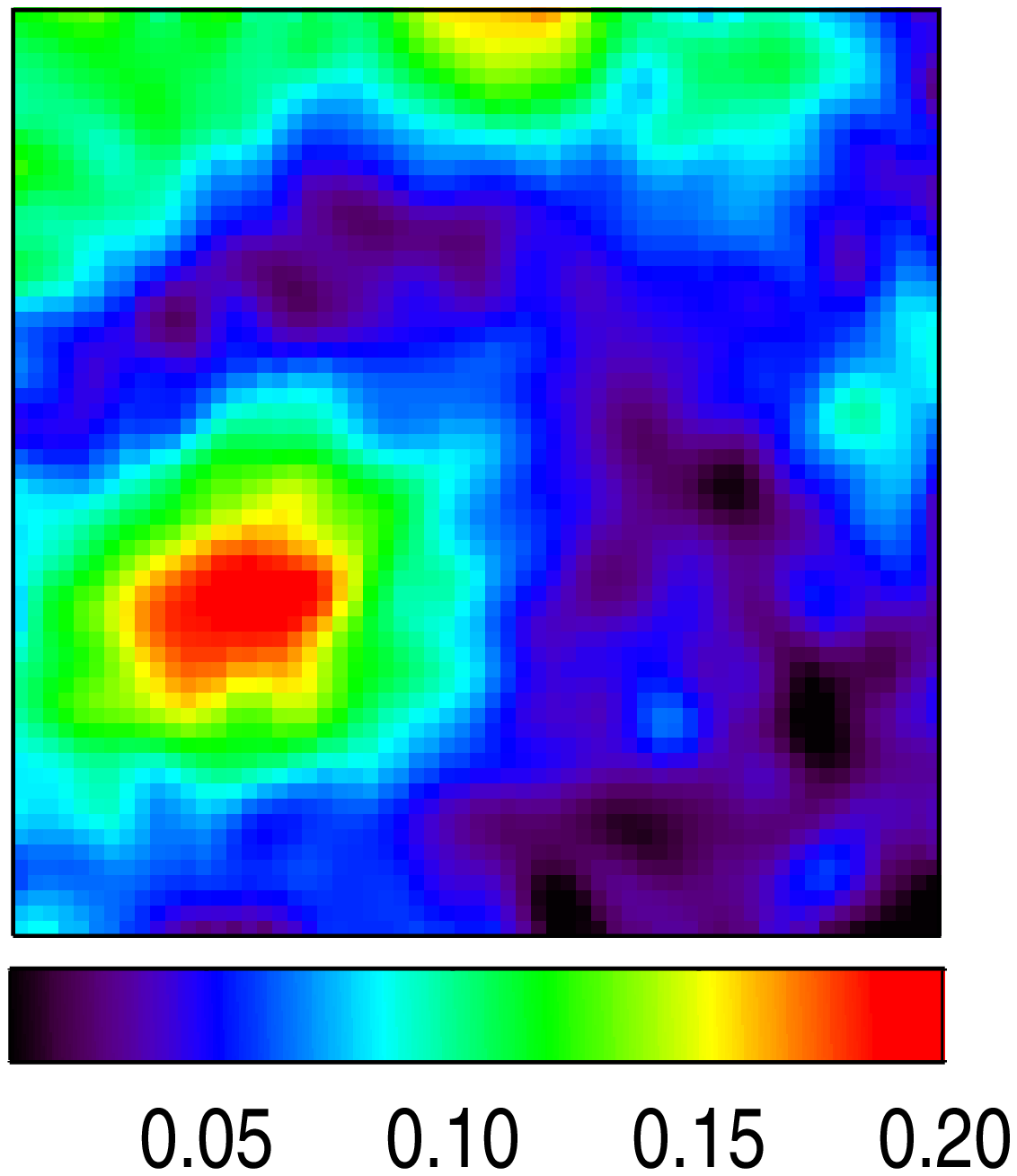}& 
\includegraphics[width=2.4cm,viewport=100 10 400 400]{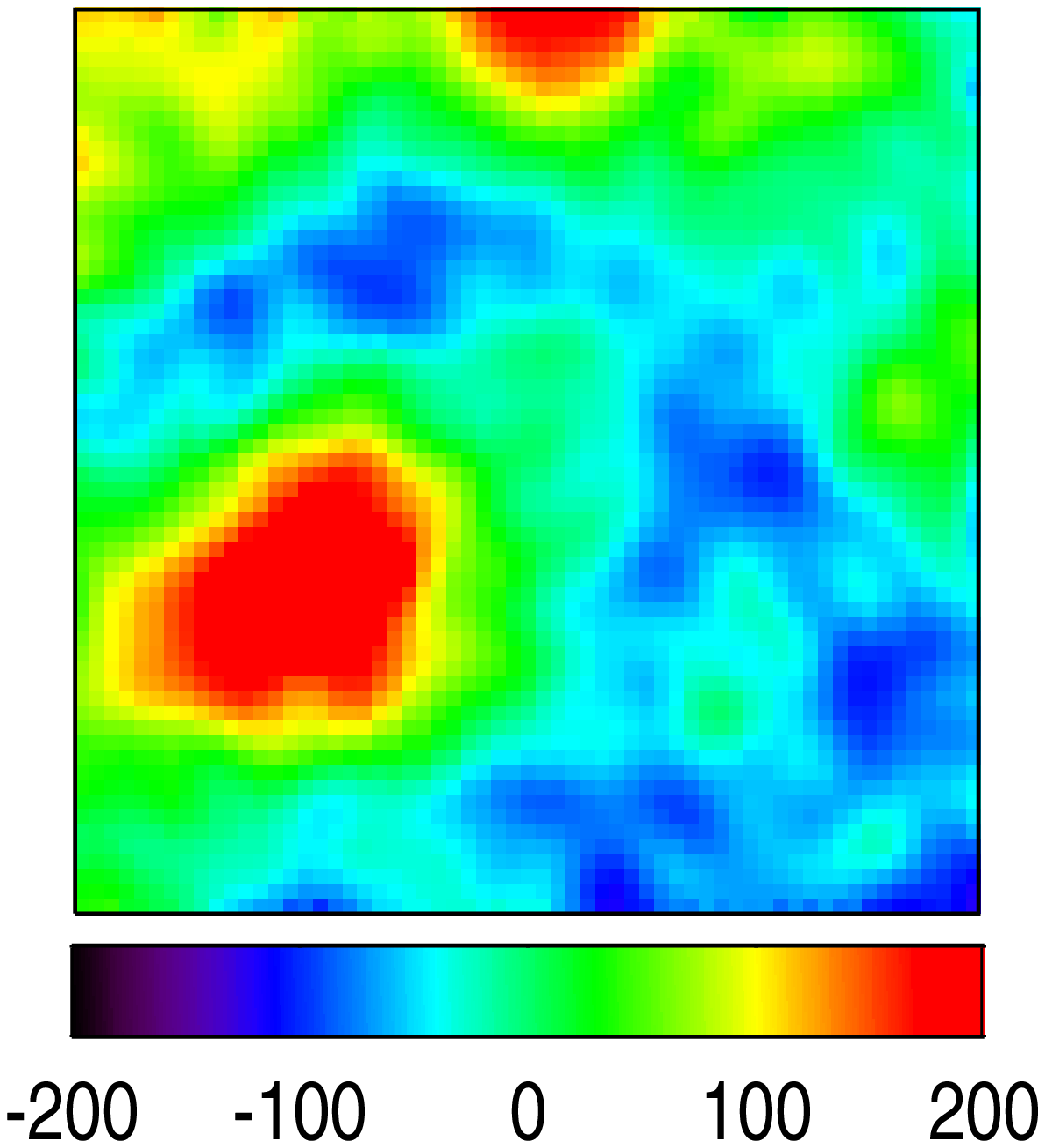}\\
\includegraphics[width=2.4cm,viewport=100 10 400 400]{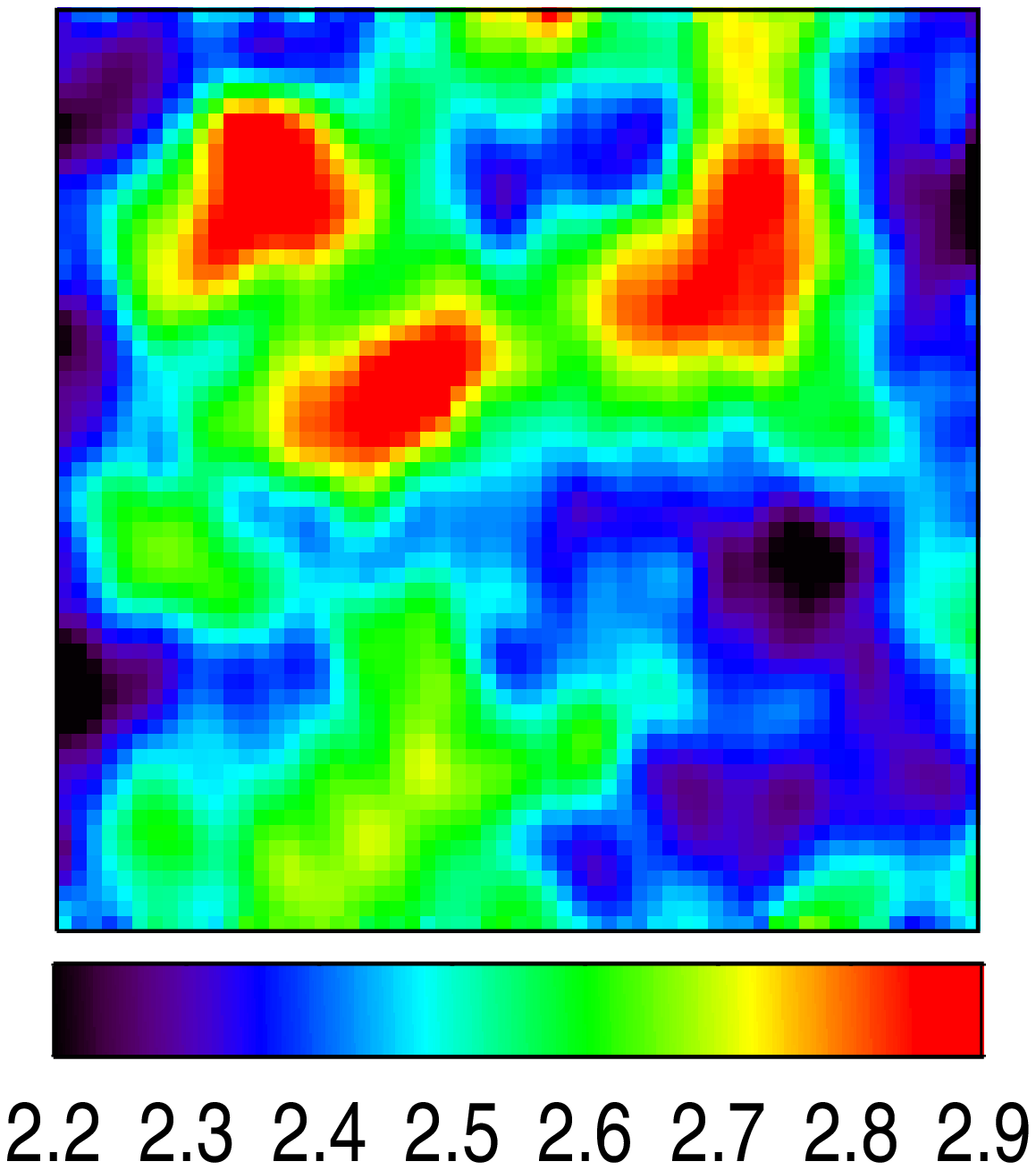}& 
\includegraphics[width=2.4cm,viewport=100 10 400 400]{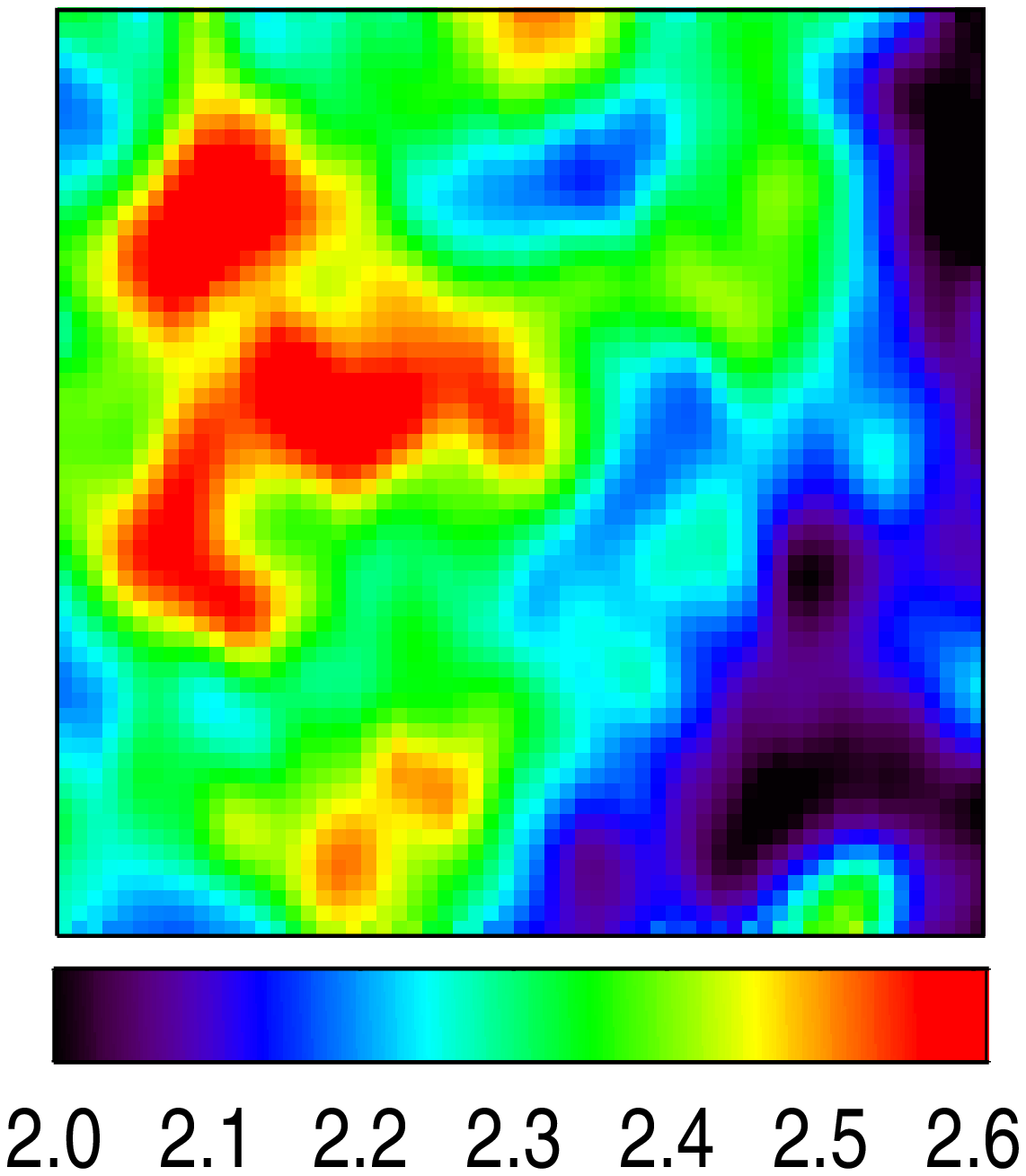}& 
\includegraphics[width=2.4cm,viewport=100 10 400 400]{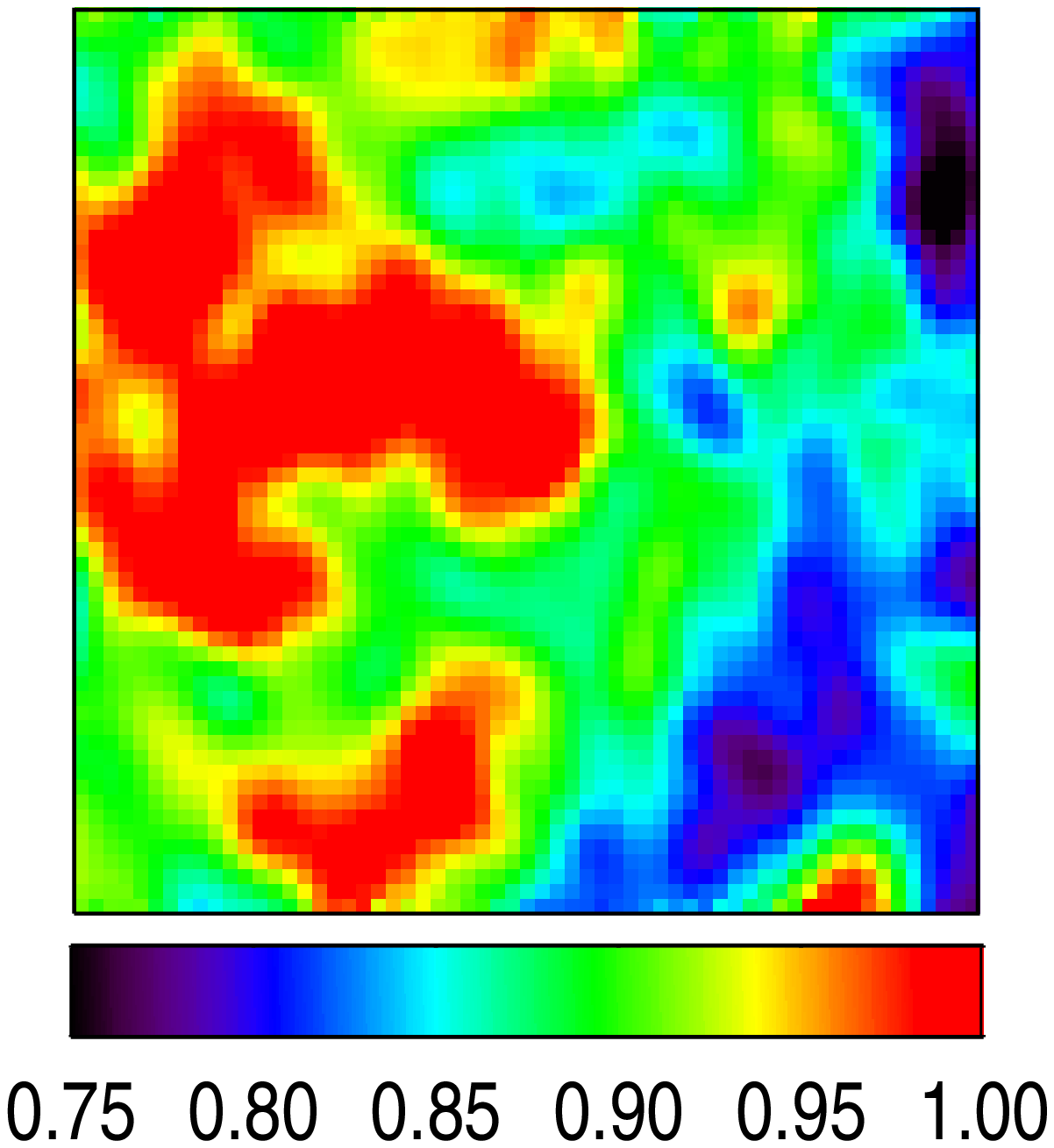}& 
\includegraphics[width=2.4cm,viewport=100 10 400 400]{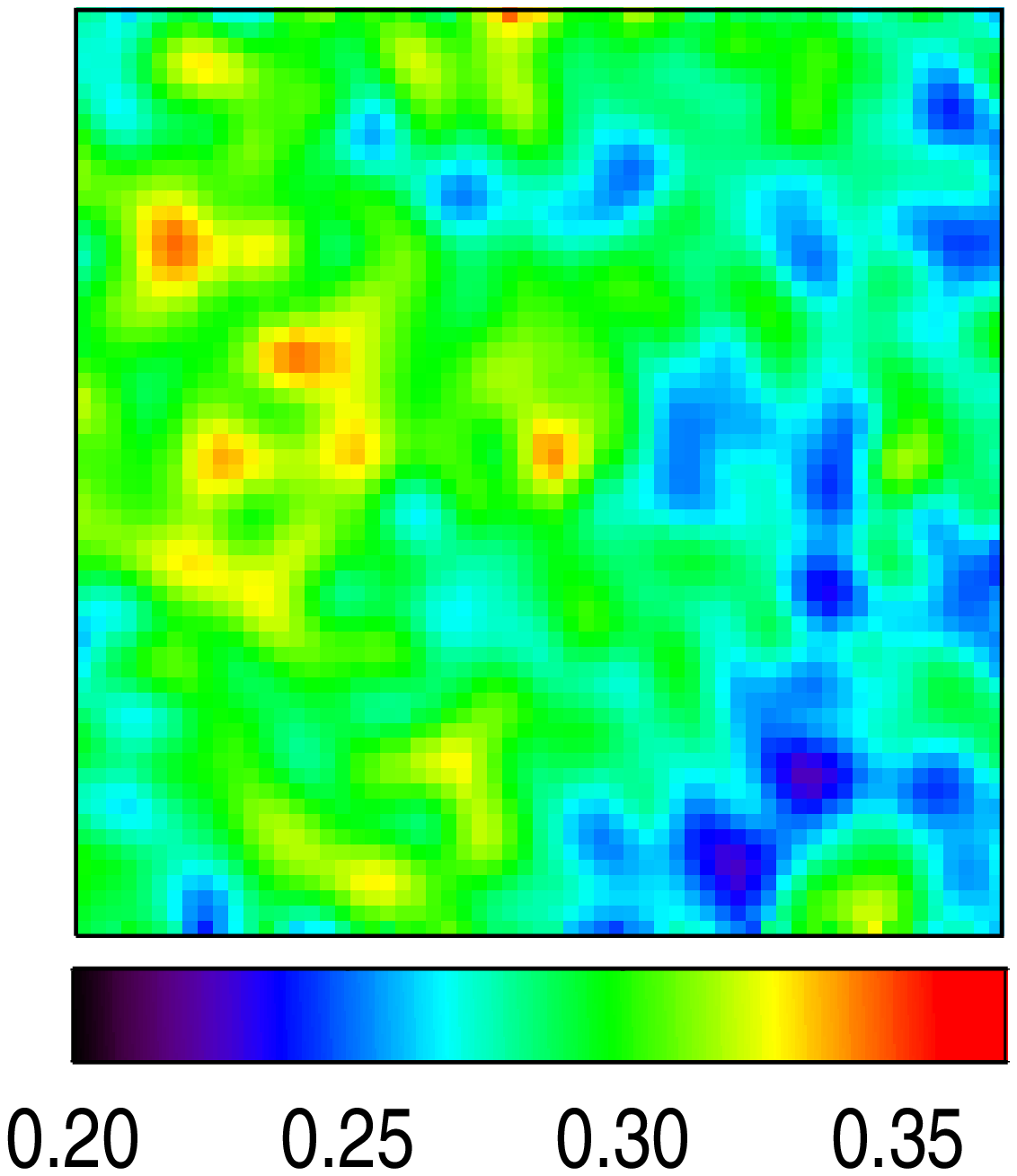}& 
\includegraphics[width=2.4cm,viewport=100 10 400 400]{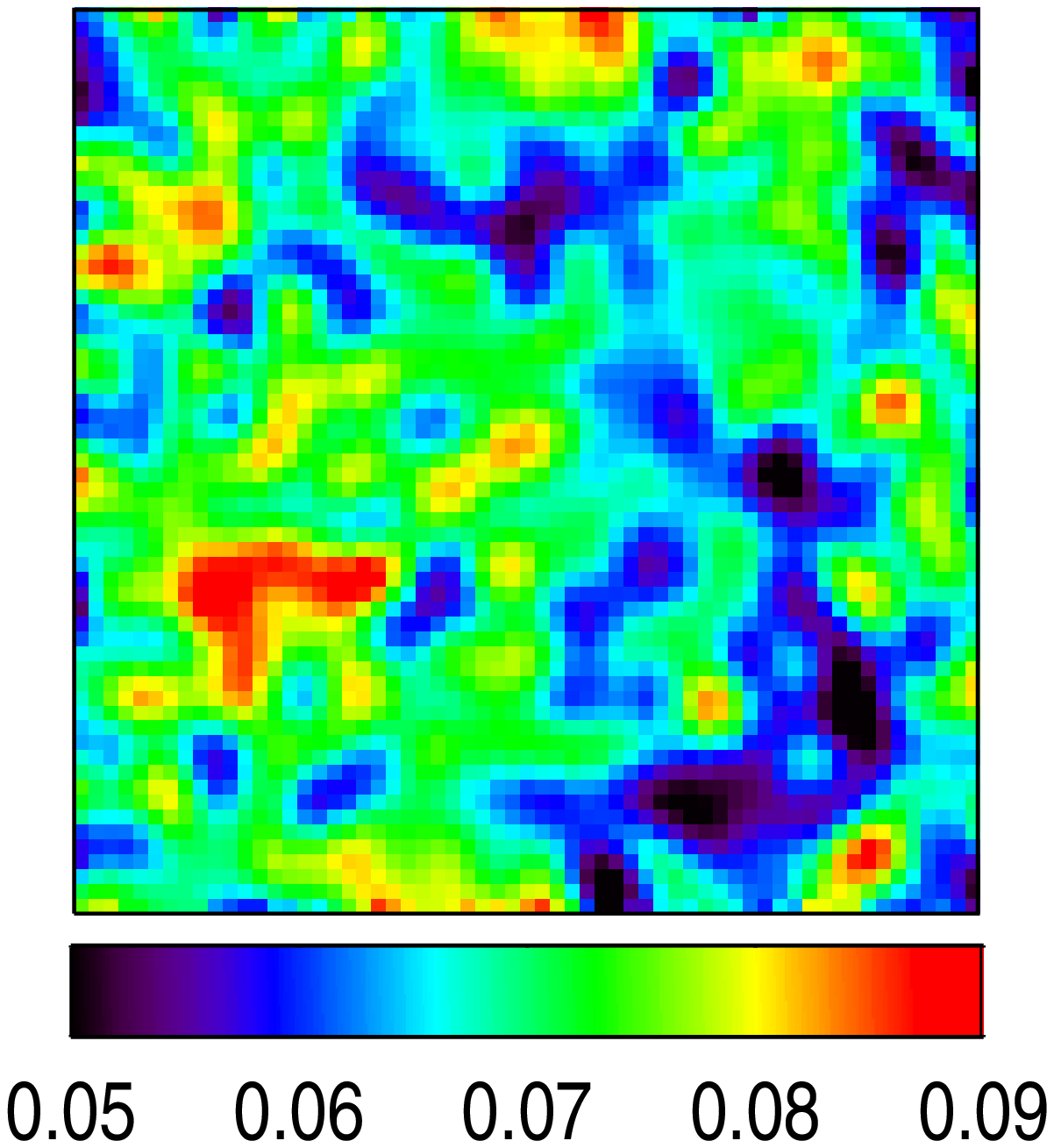}& 
\begin{minipage}[c]{0.05\linewidth} \vspace{-2.5cm} CMB\\ cleaning \end{minipage} \\ & 
\includegraphics[width=2.4cm,viewport=100 10 400 400]{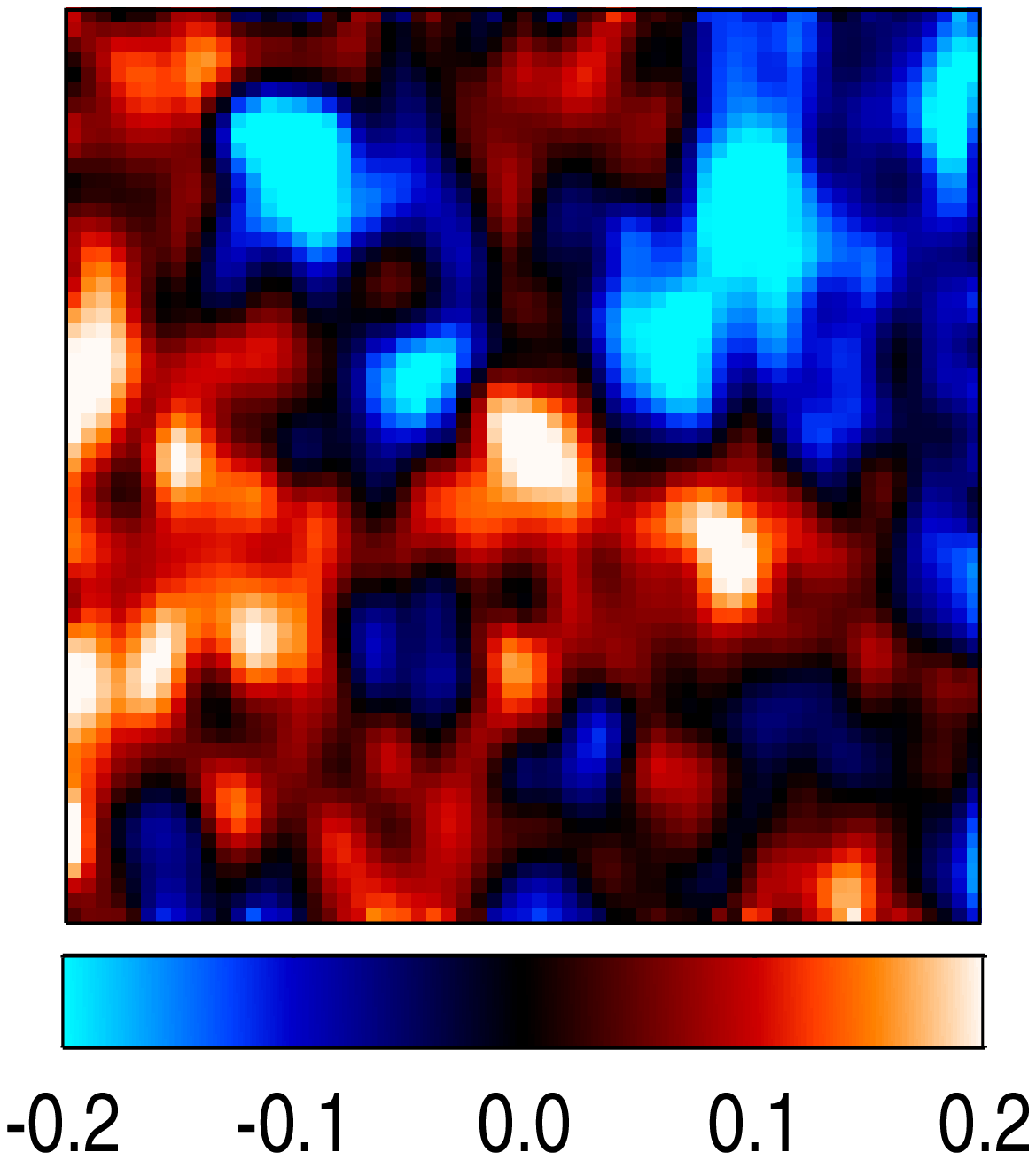}& 
\includegraphics[width=2.4cm,viewport=100 10 400 400]{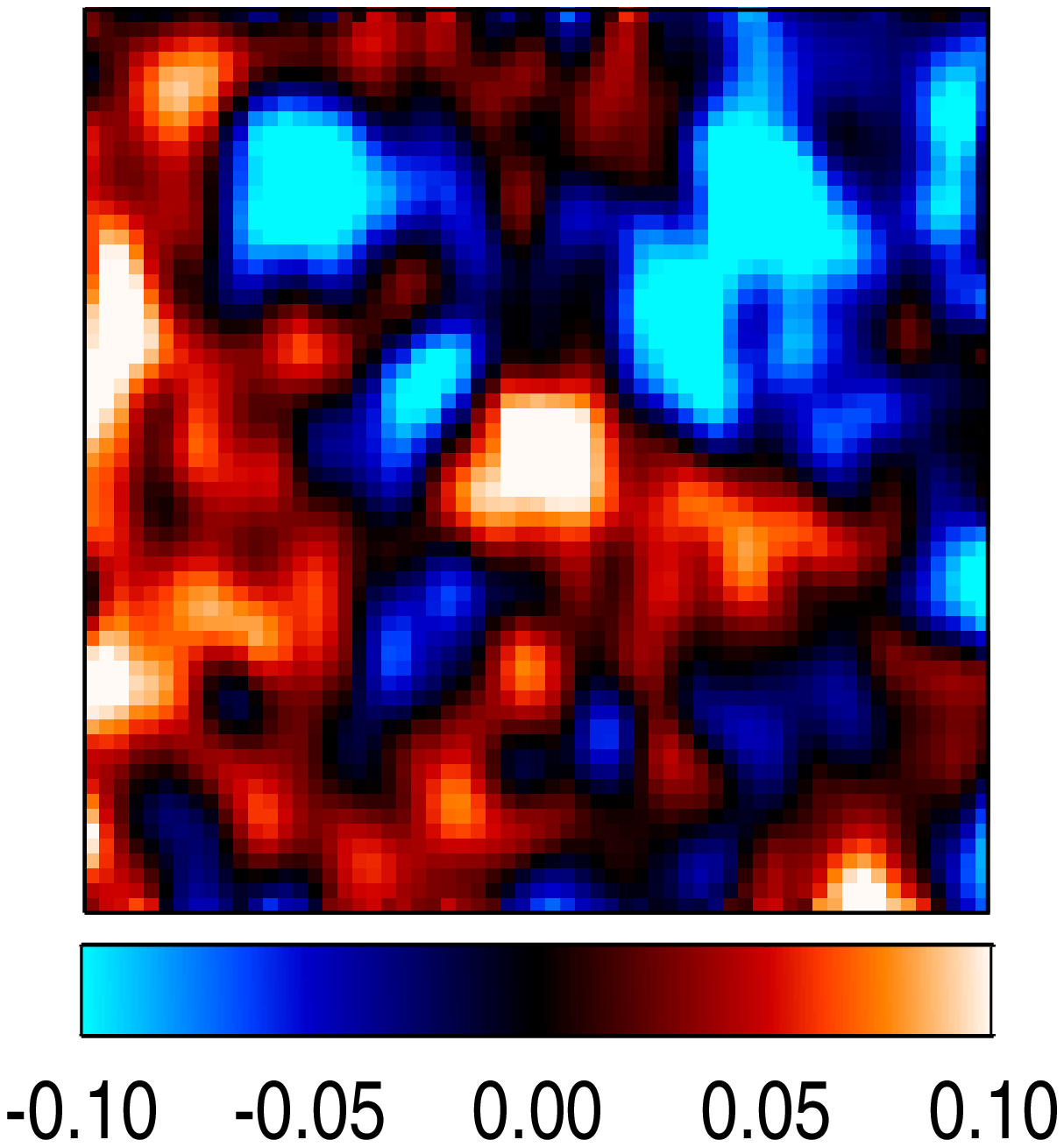}& 
\includegraphics[width=2.4cm,viewport=100 10 400 400]{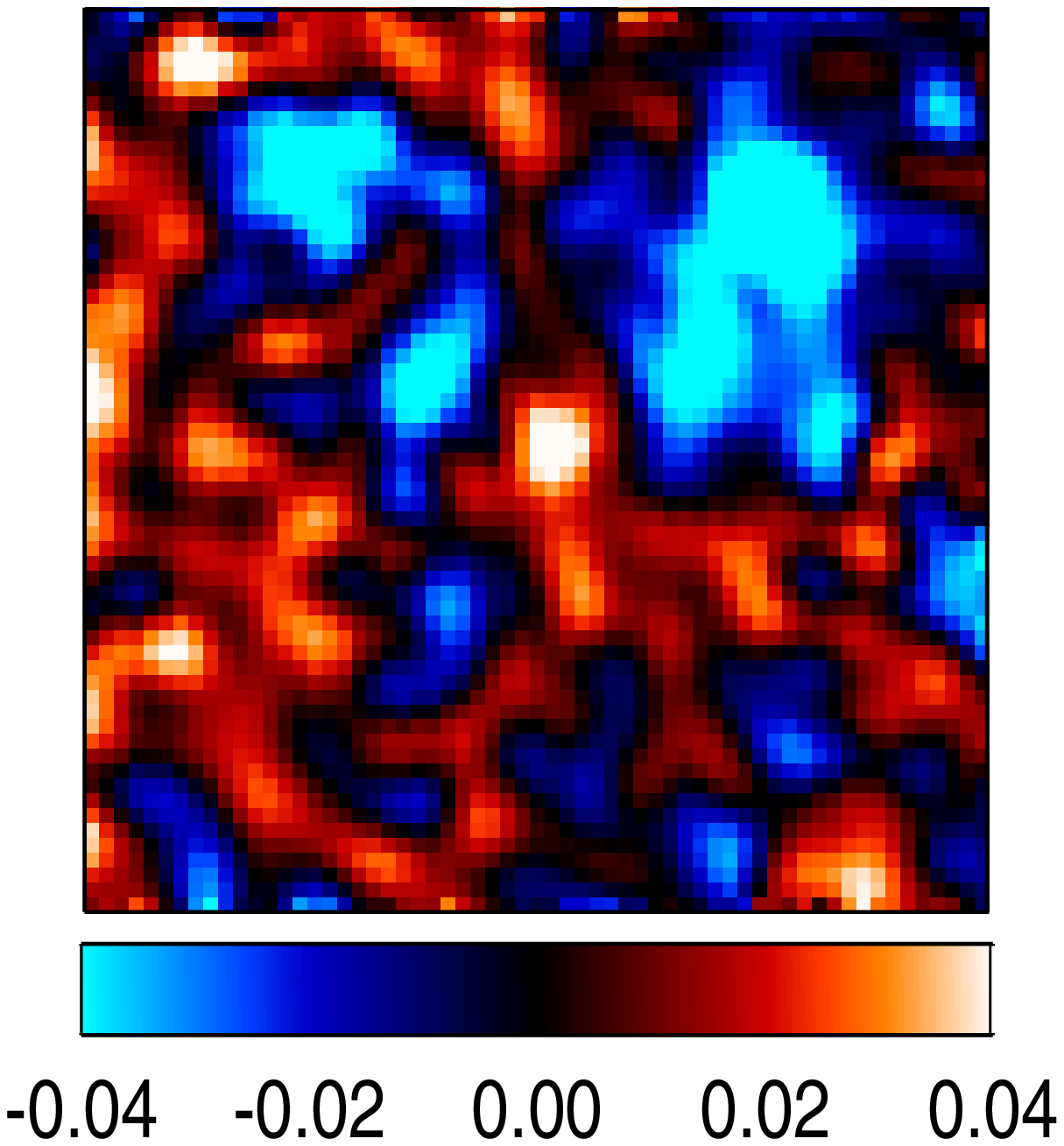}& 
\includegraphics[width=2.4cm,viewport=100 10 400 400]{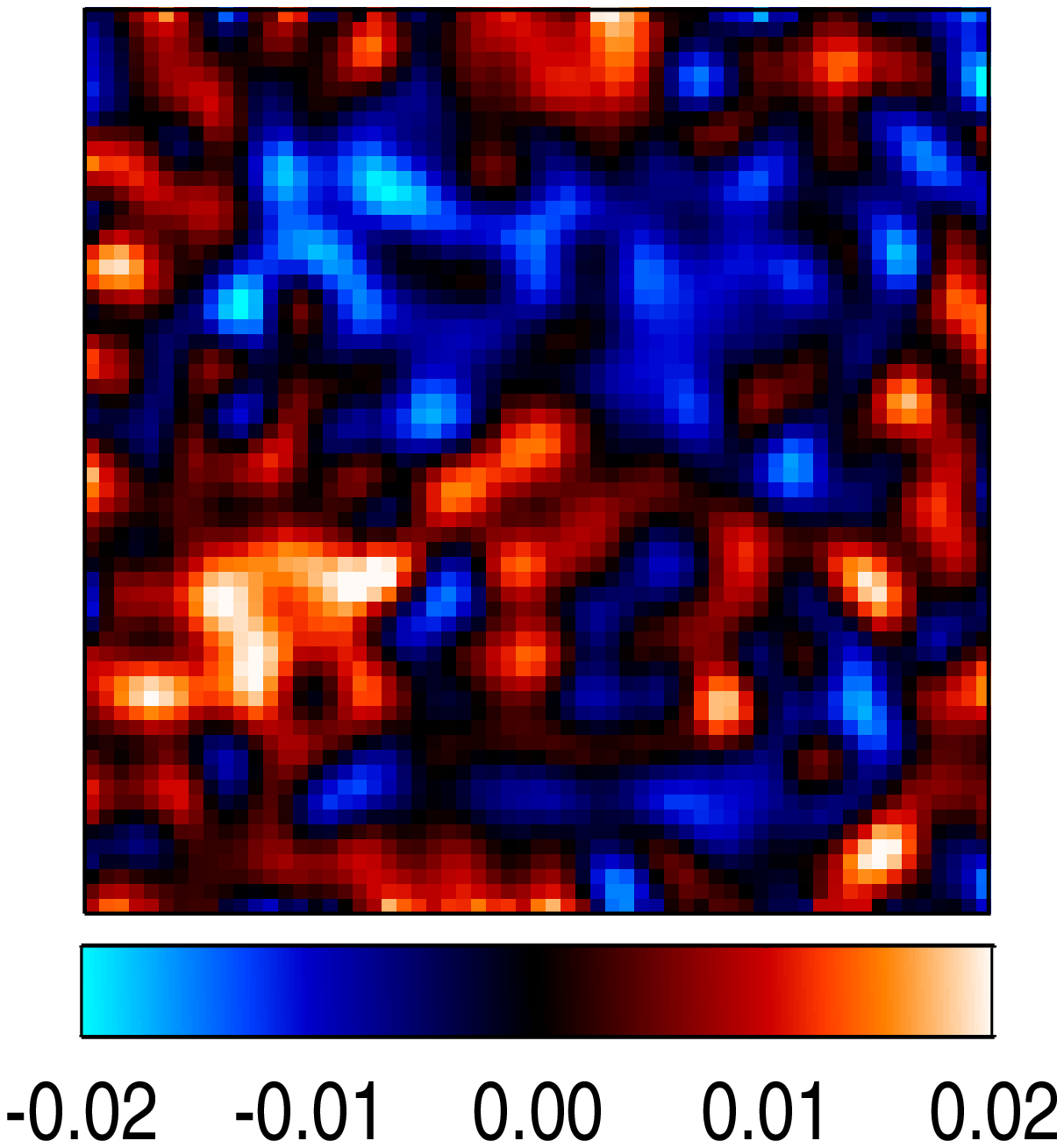}& 
\begin{minipage}[c]{0.05\linewidth} \vspace{-2.5cm} Galactic \\ cirrus \\ cleaning \end{minipage} \\ & & 
\includegraphics[width=2.4cm,viewport=100 10 400 400]{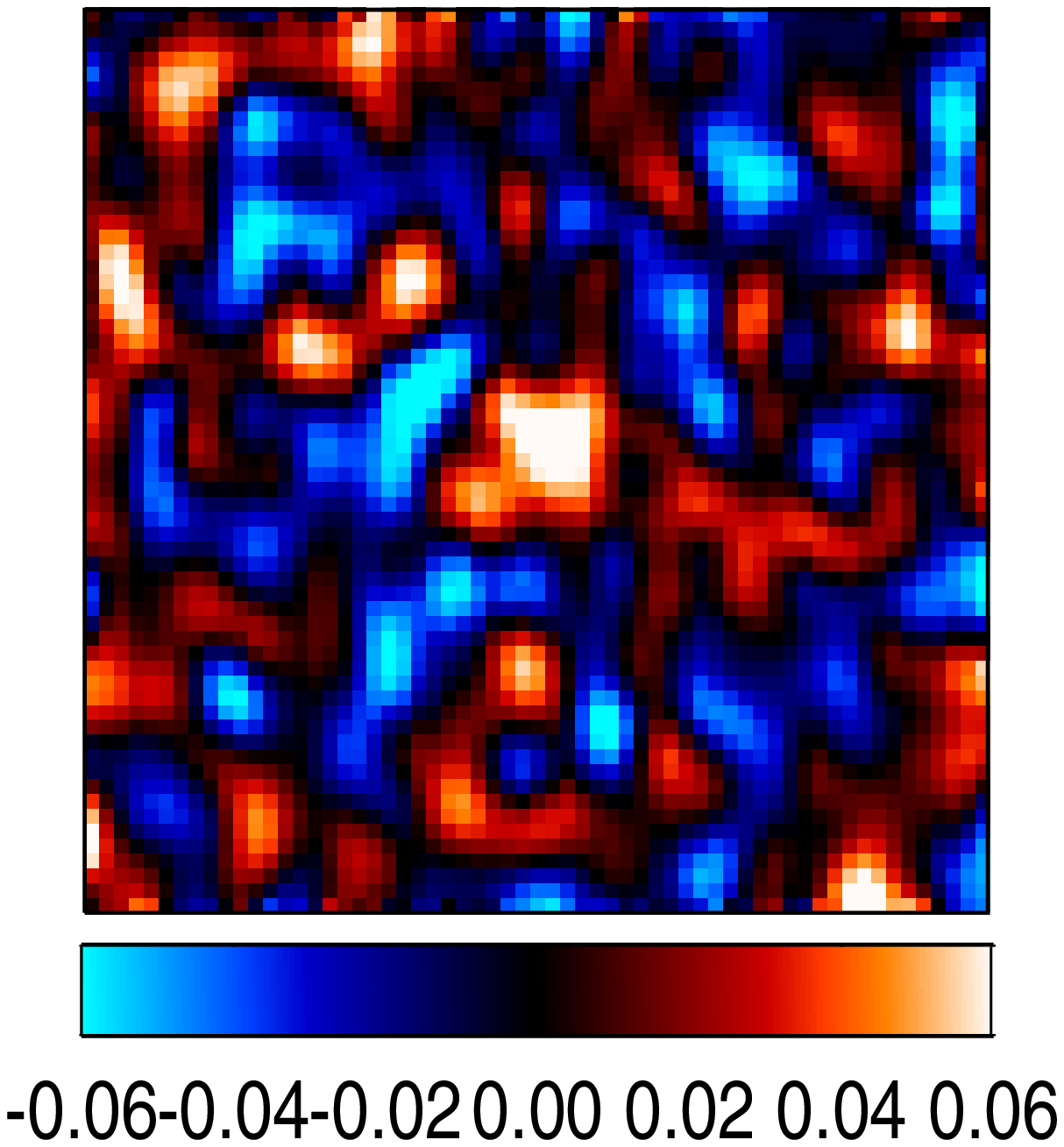}& 
\begin{minipage}[c]{0.05\linewidth} \vspace{-2.5cm} Excess map \end{minipage}& &
\end{tabular}
\caption{Cutouts ($1^{\circ}\times1^{\circ}$) in Galactic coordinates of the IRIS and \Planck\ maps
centred on the source PHZ~G095.50$-$61.59, after the various steps of the
cleaning processing.  {\it First row}: original maps at 5\arcm\ plus the
\Planck\ 5\arcm\ CMB template.  {\it Second row}: maps after CMB cleaning. 
{\it Third row}: maps at 857, 545, 353, and 217\,GHz after Galactic cirrus
cleaning. Fourth row: excess map at 545\,GHz. For the last two rows, 
the colour scale has been chosen so that positive residuals appear in red and
negative residuals in blue.  Units are expressed in ${\rm MJy}\,{\rm sr}^{-1}$,
except for the CMB 5\arcm\ template map, which is expressed in
$\mu$$\rm{K}_{\rm{CMB}}$.}
\label{fig:minimaps}
\end{figure*}

\subsection{Galactic cirrus cleaning}
\label{sec:compsep_cirrus}

In order to clean the Galactic cirrus emission at high latitude,
we apply the colour-cleaning method introduced by \citet{Montier2010}.
In this method, the 3\,THz IRIS map, considered as a template of the Galactic
dust emission, is extrapolated to the lower frequencies using the local
colour around each pixel and is removed from the current map.  Hence the
intensity of a pixel in the output map at frequency ${\nu}$ is given by
\begin{equation}
I_{\nu}^{\rm D} = I_{\nu}^{\rm C} - I_{3000}^{\rm C} \times
 \left< \frac{I_{\nu}^{\rm C}}{I_{3000}^{\rm C}} \right>_{R_{\rm{cirrus}}} , 
\end{equation}
where $I_{\nu}^{\rm C}$ and $I_{\nu}^{\rm D}$ are the intensities of the pixel
after the CMB and Galactic cirrus cleaning, respectively, 
and the $\big< \big>_{R_{\rm{cirrus}}}$ operator is the median estimate over
a ring between radius $R_{\rm{cirrus}}^{\rm{in}}=20${\arcmin} and
$R_{\rm{cirrus}}^{\rm{out}}=30${\arcmin} around the central pixel. 
The extension of the ring has been chosen, following the prescriptions of
\citet{Montier2010}, to maximize (at a beam scale) the signal of pixels with
abnormal colours compared to the background, i.e., 
by cleaning structures larger than 20{\arcmin} using 
the local colour of the background estimated up to 30{\arcmin}.
The ratio $I_{\nu}^{\rm C}/I_{3000}^{\rm C}$ is defined as the colour index.

More generally, this method of cleaning the Galactic dust emission at high
latitude allows us to subtract all the ``warm'' dust components present in the
3\,THz map, compared to the ``cold'' dust components, 
which will preferentially peak at lower frequencies. 
A structure with the same colour index as the average
background within a 30{\arcmin} radius will vanish from the cleaned map.
A structure appearing colder than the background will present a colour index
larger than the average background, and will produce a positive residual.
On the other hand, a structure warmer than the average will be characterized
by a negative residual after this colour cleaning. 
We stress that the definition of ``warm'' or ``cold'' at any frequency
is determined relative to the local background colour, which is a mixture
of Galactic cirrus emission and CIB emission at this location.
Where the emission is dominated by Galactic cirrus, this method will mainly
clean the ``warm'' Galactic dust emission; where the sky is dominated by
CIB emission, it will clean the low-$z$ component of the CIB and 
it will enhance the high-$z$ part as positive emission. 

Notice also that real strong ``warm'' sources present in the 3\,THz map will produce 
extremely negative residuals in the cleaned maps, so that, more generally, the statistics of the negative pixels in the cleaned maps 
 should not be correlated with these of positive pixels, both tracing different phases of the observed sky.

\subsection{Excess maps}
\label{sec:compsep_excess}

The SEDs of sources located at high redshift will exhibit an excess of power
at lower frequencies, located at their dust emission peak.
In order to enhance this effect, we build the excess map at 545\,GHz by
subtracting from the cleaned map at 545\,GHz a linear interpolation between
the two surrounding bands, i.e., the 857 and 353\,GHz maps, as written below:

\begin{eqnarray}
I_{545}^{\rm X} = I_{545}^{\rm D}
 - \Bigg\{ \big< I_{857}^{\rm D} \big>_{R_{\rm{x}}} + 
 \qquad\qquad\qquad\qquad\qquad\qquad\nonumber\\
\qquad\qquad\qquad  \left(\big< I_{353}^{\rm D} \big>_{R_{\rm{x}}}
 - \big< I_{857}^{\rm D} \big>_{R_{\rm{x}}} \right) \cdot
 \frac{ (545 -857)}{(353 - 857)} \Bigg\} \, ,
\end{eqnarray}
where $I_{545}^{\rm X}$ is the intensity in the excess map at 545\,GHz, 
$I_{\nu}^{\rm D}$ is the intensity after CMB and Galactic cirrus cleaning at
frequency $\nu$, and the $\big< \big>_{R_{\rm{x}}}$ operator here is the
median estimate over a disk of radius ${R_{\rm{x}}}=6${\arcmin}.
The value of the radius $R_{\rm{x}}$ has been determined on simulations to
optimize the signal-to-noise ratio of the output signal in the excess map.
The full process of cleaning is illustrated in Fig.~\ref{fig:minimaps} for the
\Planck\ high-$z$ candidate PHZ~G095.50$-$61.59, which has been confirmed
by spectroscopic follow-up as a proto-cluster candidate
\citep{Florescacho2015}.  In Fig.~\ref{fig:minimaps} each row corresponds
to a step in the cleaning, from original maps smoothed at 5{\arcmin}
(first row), to CMB-cleaned maps (second), Galactic cirrus-cleaned maps
(third), and finally yielding the excess map at 545\,GHz (fourth).

\subsection{Impact of cleaning on high-$z$ candidates}
\label{sec:impact_cleaning_highz}

The cleaning process allows us to perform an efficient component separation to 
isolate the extragalactic point sources, but also impacts the original SEDs of
these high-$z$ sources.  The fraction of emission coming from extragalactic
sources present in the CMB and 3\,THz templates are extrapolated and subtracted
from the other bands. Concerning the CMB template, since the amount of residual
emission coming from the extragalactic sources remains unknown, we
bracket the impact by making two extreme assumptions. On the one hand, the
CMB template is assumed to be perfect, i.e., without any foreground residual
emission.  In that case the CMB cleaning has no impact on the cleaned SEDs. 
On the other hand, since the CMB template is mainly dominated by the signal
of the 143-GHz band (where the signal-to-noise ratio of the CMB is the
strongest compared to the other astrophysical components), we assume in the
worse case that it includes a residual emission equivalent to the expected
intensity at 143\,GHz of the extragalactic high-$z$ source.

\begin{figure}[t]
\vspace{-0.4cm}
\hspace{-0.7cm}
\psfrag{----xtitle----}{$\nu\, \rm{[GHz]}$}
\psfrag{-----ytitle-----}{$I_{\nu}\, \rm{[MJy}\,\rm{sr}^{-1}\rm{]}$}
\includegraphics[width=0.55\textwidth]{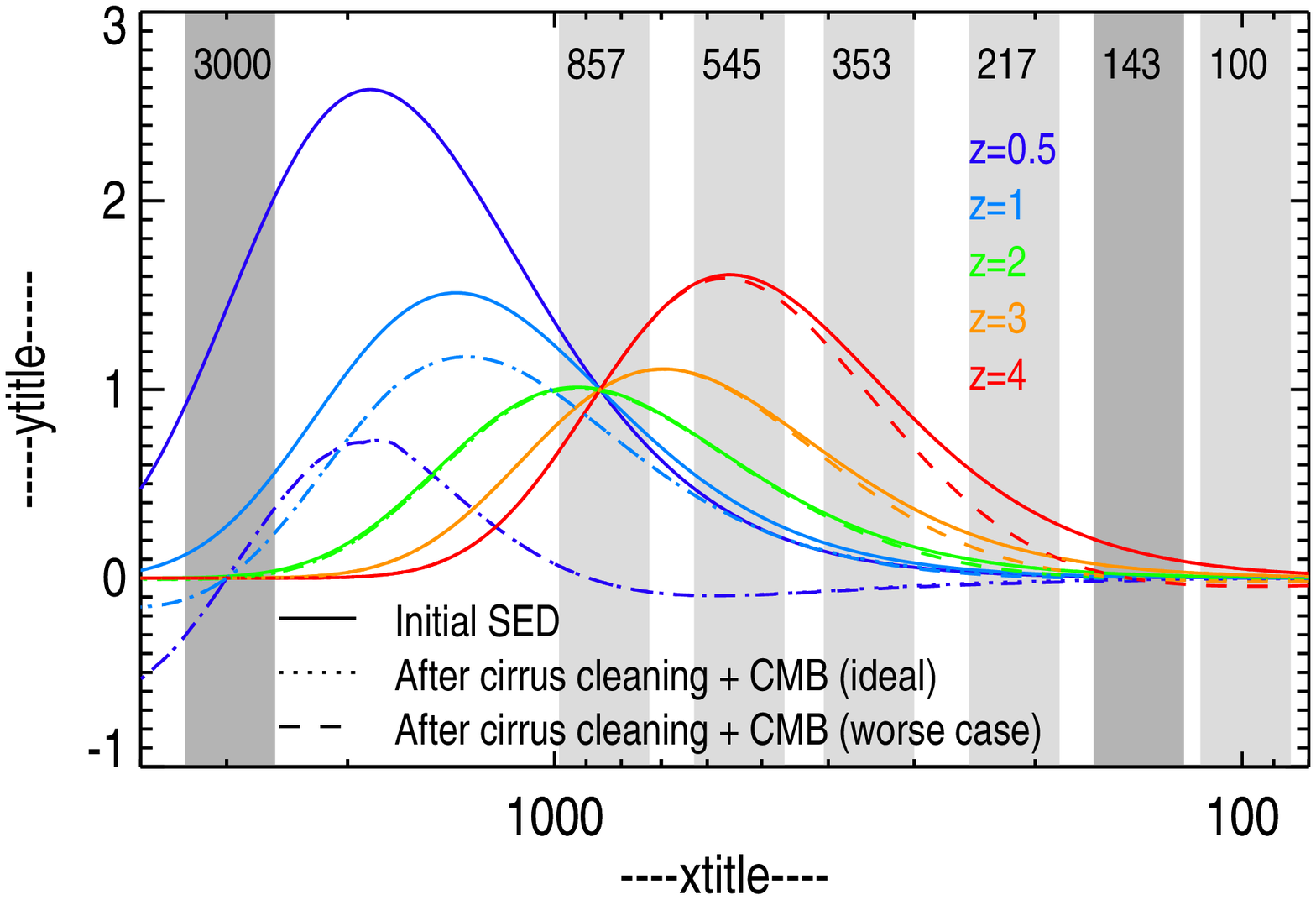} 
\caption{Impact of the cleaning process on the SED of high-$z$ dusty sources.
The SED of the extragalactic sources is modelled by a modified blackbody with
$T_{\rm{xgal}}=30$\,K and a spectral index $\beta_{\rm{xgal}}=1.5$, for
five different redshifts from 0.5 to 4. The original SEDs (solid line) are
normalized to be 1\,MJy\,$\rm{sr}^{-1}$ at 857\,GHz.  The SEDs at various
redshifts after the cleaning process are shown with dotted and dashed lines
when the CMB is assumed to be ideal or highly contaminated by foreground
emission (e.g., SZ clusters), respectively.  Note that the dotted and dashed
lines may be overplotted in some cases.  When not visible at all, those
lines are mixed to the solid line case. }
\label{fig:impact_cleaning_hz}
\end{figure}

The impact of cleaning is illustrated in Fig.~\ref{fig:impact_cleaning_hz},
where the SEDs of extragalactic sources at five redshifts (from 0.5 to 4)
are modelled by modified blackbody emission with a temperature of
$T_{\rm{xgal}}=30$\,K, and a spectral index $\beta_{\rm{xgal}}=1.5$, 
normalized at 1\,MJy\,$\rm{sr}^{-1}$ for 857\,GHz.
The Galactic cirrus cleaning has been performed assuming a balanced mixture
of CIB and Galactic dust emission.  Cleaned SEDs are shown shown for the two
cases of CMB template quality, i.e., ideal or highly foreground-contaminated.
The SEDs of low-$z$ ($<1$) sources are strongly affected by the cleaning
from Galactic cirrus, as expected, while the SEDs at higher redshifts
($z=4$) are potentially more affected by the CMB cleaning.
This has to be kept in mind when computing the photometry for any such
sources detected in the cleaned \Planck\ maps.

\begin{figure}[t]
\vspace{-0.4cm}
\begin{tabular}{c}
\vspace{-0.4cm}
\hspace{-1cm}
\psfrag{xx}{$z$}
\psfrag{----ytitle----}{$A_{\nu}^{\rm{clean}}\,\rm{[\%]}$}
\includegraphics[width=0.55\textwidth]{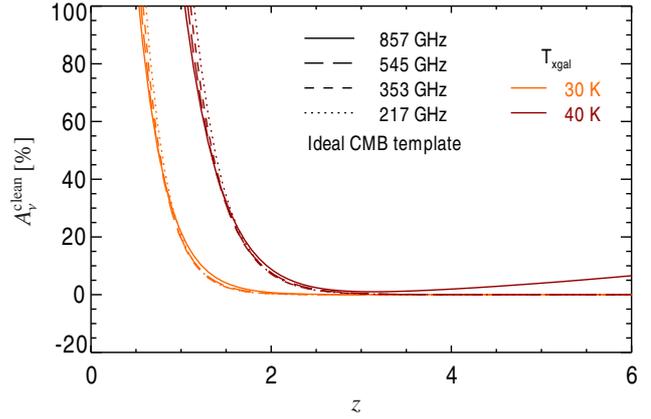} \\
\vspace{-0.4cm}
\hspace{-1cm}
\psfrag{xx}{$z$}
\psfrag{----ytitle----}{$A_{\nu}^{\rm{clean}}\,\rm{[\%]}$}
\includegraphics[width=0.55\textwidth]{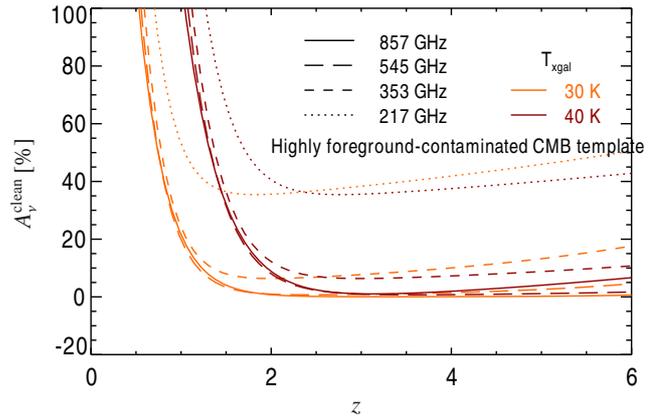} 
\end{tabular}
\caption{Relative attenuation coefficient $A_{\nu}^{\rm{clean}}$ due to the
cleaning process, as a function of redshift for each \Planck-HFI band.
This is computed for two input SEDs modelled by a modified blackbody
($T_{\rm{xgal}}=30$\,K and 40\,K, with $\beta_{\rm{xgal}}=1.5$) 
embedded in foreground emission equally balanced between CIB and Galactic
cirrus. It is shown for two extreme cases of the CMB template quality, i.e.,
ideal (upper panel) or highly contaminated by extragalactic foregrounds at
high $\ell$ (lower panel).}
\label{fig:impact_cleaning_attenuation}
\end{figure}

Note that cleaning will tend to remove some of the flux of real sources.
We define the relative attenuation coefficient in each \Planck-HFI band
due to the cleaning process, $A_{\nu}^{\rm{clean}}$, as
\begin{equation}
A_{\nu}^{\rm{clean}} = \frac{I_{\nu}-I_{\nu}^{\rm{D}}}{I_{\nu}} \, .
\end{equation}
Again this attenuation coefficient ranges between two extreme cases,
depending on the level of contamination by extragalactic foregrounds in the
CMB template.  An estimate of this relative attenuation coefficient is shown
in Fig.~\ref{fig:impact_cleaning_attenuation} 
as a function of redshift for the 857, 545, 353, and 217\,GHz \Planck\ bands. 
We observe that, in the worse case (lower panel), flux densities at 857 and
545\,GHz are barely impacted by the cleaning for redshifts $z>2$, 
while for the 353-GHz band the attenuation reaches 5\,\% to 20\,\%.
The attenuation for the 217-GHz band is much larger, ranging between 30\,\%
and 40\,\%.  When the CMB template is assumed to be ideal (upper panel), the
attenuation remains small for $z>2$ in all bands.
At low redshifts ($<1$), the attenuation coefficient reaches 100\,\% in both
cases, which means that the cleaning 
process fully removes these sources from the maps. In the intermediate range of 
redshifts ($1<z<2$), the situation is less clear and requires
more realistic simulations to provide a reliable assessment of the detection
of such sources, as performed in Sect.~\ref{sec:mcqa}.

We emphasize that this attenuation coefficient strongly 
depends on the SED type and the redshift of each source. 
Simply changing the temperature of the source $T_{\rm{xgal}}$ from 30\,K to
40\,K shifts the transition zone from redshift 1--2 to 2--3
(see Fig.~\ref{fig:impact_cleaning_attenuation}), 
making it hard to predict the actual attenuation coefficients.

\subsection{Contamination by foreground astrophysical sources}
\label{sec:impact_cleaning_foregrounds}

\subsubsection{Thermal emission from cold Galactic dust}
\label{sec:compsep_dust}

Because of the degeneracy between the temperature of a source and its redshift,
cold clouds at high latitude represent an important contaminant for the
detection of high-$z$ sources, Indeed, the SED of a Galactic cold source 
modelled by a modified blackbody with a temperature $T_{\rm{dust}}=10$\,K (blue curve of Fig.~\ref{fig:impact_cleaning_foreground})
will mimic the same spectral trend in the submm range as
the SED of a warm source ($T_{\rm{xgal}}=30$\,K) redshifted to $z=2$ (green curve of Fig.~\ref{fig:impact_cleaning_hz}).
This can only be disentangled by taking into account other properties of such Galactic sources, such as the \ion{H}{i} column density 
or the structure of its surroundings, which may be associated with Galactic components. 
For this reason, with each detection there will be associated an estimate of the local extinction at the
 source location and in the background, as a tracer of the local \ion{H}{i} column density.
 This is further discussed in Sect.~\ref{sec:ancillary_internal_xcat}, in the analysis of the cross-correlation
 between the list of high-$z$ source candidates and the catalogue of Planck Galactic Cold Clumps \citep[PGCC;][]{planck2014-a37}.

\subsubsection{Synchrotron emission from radio sources}
\label{sec:compsep_radio}

The typical SED of the synchrotron emission from radio sources is observed in the submm 
range as a power law with a spectral index around --0.5 \citep{planck2011-6.1}, as shown as a red solid line in Fig.~\ref{fig:impact_cleaning_foreground}. 
While the slope of the cleaned SED is accentuated by the cleaning when the CMB template is assumed to be ideal (red dotted line), 
if the CMB template is highly contaminated by extragalactic foregrounds at high $\ell$, the cleaned SED (red dashed line) exhibits a positive bump in the 
857 and 545-GHz bands, and a deficit in the 353- and 217-GHz bands, 
which mimics the excess at 545\,GHz expected for the high-$z$ sources.
This artefact can be identified by looking
 at the intensity in the 100-GHz band, which remains strongly positive in the case of synchrotron emission,
 compared to the expected emission of high-$z$ dusty galaxies at 100\,GHz.
 For this reason we provide a systematic estimate of the 100\,GHz flux density and compare it to the flux densities at higher frequencies
 in order to reject spurious detections of radio sources.

\subsubsection{SZ emission from galaxy clusters}
\label{sec:compsep_sz}

\begin{figure}[t]
\vspace{-0.4cm}
\hspace{-0.7cm}
\psfrag{----xtitle----}{$\nu\, \rm{[GHz]}$}
\psfrag{-----ytitle-----}{$I_{\nu}\, \rm{[MJy}\,\rm{sr}^{-1}\rm{]}$}
\includegraphics[width=0.55\textwidth]{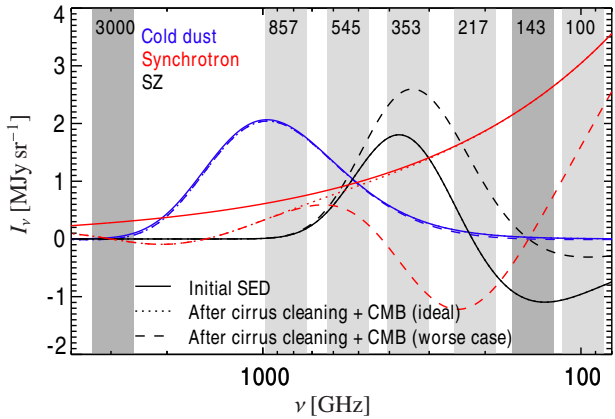} 
\caption{Impact of the cleaning process on the SED of foreground astrophysical
sources: cold Galactic sources (blue); SZ signal from galaxy clusters (black);
and radio sources (red). The SEDs are shown before (solid line) and after
cleaning, assuming two levels of CMB template quality, i.e., 
ideal (dotted line) or highly contaminated by extragalactic foregrounds at
high ${\ell}$ (dashed line).}
\label{fig:impact_cleaning_foreground}
\end{figure}

The SZ effect \citep{Sunyaev70} is a distortion of
the CMB due to the inverse Compton scattering induced by hot electrons
of the intra-cluster medium.  It generates a loss of power at frequencies below
217\,GHz, and a gain above this frequency. An SZ spectrum after removal of the CMB monopole spectrum is shown 
as a black solid line in Fig.~\ref{fig:impact_cleaning_foreground}, 
using a typical integrated Compton parameter $Y_{500}=10^{-3}$ \citep[see][]{planck2013-p05a,planck2014-a36}.
Along the direction towards galaxy clusters, if the CMB template is not fully cleaned for SZ emission, the CMB cleaning method will
 artificially enhance the signal of the resulting SED (black dashed line) by subtracting
the (negative) SZ signal at 143\,GHz. This produces a clear bump of the cleaned SED in the 353-GHz band,
as expected for the SED of a dusty source ($T_{\rm{xgal}}=30$\,K) at $z=7.5$, 
which is not likely to be detected at 5{\arcmin} resolution.
 
Hence the SZ SED does not properly reproduce the expected colours of the dusty galaxies at high $z$, 
and should not be detected by our algorithm. However, it may represent an important contaminant if a 
galaxy cluster and a high-$z$ dusty source lie along the same line of sight. This is addressed 
in Sect.~\ref{sec:ancillary_internal_xcat}, in the analysis of the cross-correlation 
between this list and the Planck Catalogue of SZ sources \citep[PSZ;][]{planck2014-a36}.

\section{Point source detection}
\label{sec:ps_detection}

We describe in this section how the point source detection is performed and the photometry estimates are obtained.
We also detail the final selection process, based on both a colour-colour analysis and a flux density threshold.

\subsection{Detection method}
\label{sec:ps_detection_method}

The point source detection algorithm requires positive detections simultaneously within a 5{\arcmin} radius in the
545\,GHz excess map, and the 857, 545, and 353\,GHz cleaned maps. It also requires a non-detection in the 
100\,GHz cleaned maps, which traces emission from synchrotron sources. 

As already mentioned in Sect.~\ref{sec:compsep_cirrus}, negative pixels in the cleaned and excess maps represent
the locally warmer phase of the high-latitude sky, which may statistically strongly differ from the one of
 the positive pixels tracing the colder phase. For this reason negative pixels are masked afterwards, so that 
 we characterize the significance of a detection by comparing the value of each pixel to the statistics of positive pixels only.
 Hence the local noise is estimated as the median absolute deviation estimate 
over the positive pixels of each map within a radius $R_{\rm{det}}=60${\arcmin} around each pixel.
A disk of $1^{\circ}$ radius covers about 150 times the beam of 5{\arcmin}, providing enough 
statistics to obtain a reliable estimate of the standard deviation. It also covers twice the typical scale of any 
Galactic cirrus structures filtered by the cleaning process (using a radius $R_{\rm{cirrus}}=30${\arcmin}, see Sect.~\ref{sec:compsep_cirrus}).

A detection is then defined as a local maximum of the signal-to-noise ratio
(S/N) above a given threshold in each map, with a spatial separation of at
least 5{\arcmin} being required between two local maxima. 
A threshold of $\rm{S/N}>5$ is adopted for detections in the 545\,GHz excess
map, while this is slightly relaxed to $\rm{S/N}>3$ for detections in the
cleaned maps, because the constraint imposed by the spatial consistency
between detections in all three bands is expected to reinforce the robustness
of a simultaneous detection. Concerning the 100-GHz band, we 
adopt a similar threshold by requiring the absence of 
any local maximum with $\rm{S/N}>3$ within a radius of 5{\arcmin}.
Notice also that this criterion is applied on the 100\,GHz map, which is
only cleaned from CMB after convolving the CMB template and 
100\,GHz maps at a common 10{\arcmin} resolution.
A detection is finally defined by the following simultaneous criteria:
\begin{equation}
\left\{
\begin{array}{l c l}
I_{545}^{\rm{X}}/\sigma^{\rm{X}}_{545}& >& 5\, ; \\[0.5em]
I_{\nu}^{\rm{D}}\,\,/\,\,\sigma^{\rm{D}}_{\nu} \,\,& >& 3\, , \quad {\rm for}
 \, \nu=857, 545{\rm{, \, and}}\, 353\,{\rm{GHz}} \, ; \\ [0.5em]
I_{100}^{\rm{C}}/\sigma^{\rm{C}}_{100}& <& 3\, .
\end{array} \right.
\end{equation}

\subsection{Photometry}
\label{sec:ps_detection_photometry}

The photometry is computed at the location of the detections in the cleaned
857, 545, 353, and 217\,GHz maps.  It is performed in two steps:
(i) determination of the extension of the source in the 545\,GHz cleaned map;
and (ii) aperture photometry in all bands in the cleaned maps.  We perform an
elliptical Gaussian fit in the 545\,GHz cleaned map at the location of the
detection in order to find the exact centroid coordinates, the major and
minor axis FWHM values, and the position angle, with associated uncertainties.
Flux densities, $S_{\nu}^{\rm{D}}$, are obtained consistently in all four
bands via an aperture photometry procedure using the elliptical Gaussian
parameters derived above in the cleaned maps.  The accuracy of the flux
densities, $\sigma_{\nu}^{S}$, can be decomposed into three components: 
$\sigma_{\nu}^{\rm{geom}}$ comes from the uncertainty of the elliptical
Gaussian fit; $\sigma_{\nu}^{\rm{sky}}$ represents the level of the local
CIB fluctuations that dominate the signal at high latitude; and
$\sigma_{\nu}^{\rm{data}}$ is due to the noise measurement of the \Planck\
data and estimated using half-ring maps. 

An estimate of the elliptical Gaussian fit accuracy,
$\sigma_{\nu}^{\rm{geom}}$, is obtained by repeating the aperture photometry
in 1000 Monte Carlo simulations, where the elliptical Gaussian 
parameters are allowed to vary within a normal distribution centred on the
best-fit parameters and a $\sigma$-dispersion provided by the fit.
The uncertainty $\sigma_{\nu}^{\rm{geom}}$ is defined as the mean absolute
deviation over the 1000 flux density estimates.
 
We use the first and last half-ring maps, which have been 
cleaned following the same process as the full maps, to obtain an estimate of
the accuracy of the photometry related to the noise in the data. This is
computed as the absolute half difference of the photometry estimates,
$S_{\nu}^{\rm{first}}$ and $S_{\nu}^{\rm{last}}$, obtained from the 
first and last half-ring cleaned maps, respectively.
Since this quantity follows a half-normal distribution, the estimate of the
noise measurement in the full survey is finally given by
\begin{equation}
\sigma_{\nu}^{\rm{data}} = \sqrt{\frac{\pi}{2}}\, \left|
 \frac{S_{\nu}^{\rm{first}} - S_{\nu}^{\rm{last}}}{2} \right| \, .
\end{equation}

The local level of the CIB fluctuations, $\sigma_{\nu}^{\rm{sky}}$, is
obtained by computing the standard deviation over 400 flux density estimates
obtained by an aperture photometry with the nominal elliptical Gaussian shape
parameters in the cleaned maps at 400 random locations within a radius of
$1^{\circ}$ around the centroid coordinates.  Those random locations are
chosen among the positive pixels of the excess maps, for the same reason
as given in Sect.~\ref{sec:ps_detection_method}, i.e., to explore the same
statistics as the detection pixels.  Notice that this estimate of
$\sigma_{\nu}^{\rm{sky}}$ also includes the noise of the data, 
even if the latter is shown to be low compared to the CIB fluctuation level.
 
We stress that the flux densities are computed using the cleaned maps,
since their S/N values are higher than in the original maps, where the
high-$z$ source candidates are embedded in Galactic cirrus, CIB structures,
and CMB fluctuations.  Nevertheless they still suffer from several potential
systematic effects: (1) attenuation due to the
cleaning; (2) contamination by the Sunyaev-Zeldovich effect (SZ) discussed in 
Sect.~\ref{sec:ancillary_internal_xcat}; and (3) the flux boosting effect
presented in Sect.~\ref{sec:mcqa_photometry}.

\subsection{Colour-colour selection and flux cut}
\label{sec:ps_detection_colcol}

A colour-colour selection is applied to the cleaned flux densities in order
to keep only reliable high-$z$ candidates. This aims to reject 
Galactic cold clumps and radio sources, if still present in the detected
sample.  We use the three highest frequency \Planck\ bands in which
detections at $\rm{S/N}>3$ are simultaneously required. 
The colour-colour space is thus defined by the 
$S_{545}/S_{857}$ and $S_{353}/S_{545}$ colours.

Firstly, we require $S_{545}/S_{857}>0.5$, to
reject potential Galactic cold sources, which exhibit colour ratios
ranging from 0.2 to 0.5 for dust temperatures ranging between 20\,K
and 10\,K (with a spectral index equal to 2).  It is found that 98.5\,\%
of the cold clumps in the PGCC
catalogue \citep{planck2014-a37} have a colour $S_{545}/S_{857}<0.5$. 
We emphasize that this criterion can be safely applied to the colour ratio
$S_{545}^{\rm{D}}/S_{857}^{\rm{D}} $ obtained on cleaned maps, as quantified
with Monte Carlo simulations (see Sect.~\ref{sec:mcqa_photometry}). 

Secondly, it is common to constrain $S_{353}/S_{545}$ to be less than 1 in
order to avoid contamination from radio sources, which have negative spectral
indices \citep[e.g., see][]{planck2014-a37}.  However, this criterion has to
be adapted when using the photometry based on the cleaned maps.
As already mentioned in Sect.~\ref{sec:compsep_radio}, typical SEDs of
radio sources are transformed after cleaning,
so that they no longer have $S_{353}/S_{545}>1$. While SEDs of extremely
redshifted dusty galaxies may present colour ratios larger than 1, 
their cleaned SEDs will be strongly affected by the cleaning process,
so that their colour ratio goes below 0.9 whatever the redshift
(as discussed in Sect.~\ref{sec:mcqa_photometry}). This remains the case
for galaxy clusters with an SZ signature, which produces an excess of the
flux density at 353\,GHz after the cleaning process, so that 
this colour ratio would be larger than 1.
Hence the criterion is finally set to $S_{353}^{\rm{D}}/S_{545}^{\rm{D}}<0.9$, 
so that dusty galaxies are not rejected, but SZ contamination is.

In order to properly propagate the uncertainties of the flux density estimates in all three bands during the colour-colour 
selection process, we construct for each source the probability for the two colour ratios to lie within the high-$z$ 
domain, given the 1$\,\sigma$ error bars associated with the flux densities:
\begin{equation}
\label{eq:prob}
\mathcal{P} \left(\,\, \frac{S_{545}^{\rm{D}}}{S_{857}^{\rm{D}}} \, > \, 0.5
 \quad {\rm and}\ \quad \frac{S_{353}^{\rm{D}}}{S_{545}^{\rm{D}}}\, < \,
 0.9\,\, \right) \, .
\end{equation}
This probability is built numerically by simulating for each source
100\,000 flux densities including noise in the 857-, 545-, and 353-GHz bands,
($S^{\rm I}_{857}$, $S^{\rm I}_{545}$, and $S^{\rm I}_{353}$),
using the cleaned flux density estimates and their 1$\,\sigma$ uncertainties.
The flux density uncertainties used to build these noise realizations are
defined as the quadratic sum of the data noise, $\sigma_{\nu}^{\rm{data}}$,
and the elliptical Gaussian fit accuracy, $\sigma_{\nu}^{\rm{geom}}$,
so that only proper noise components of the uncertainty are included, 
but not the confusion level from CIB fluctuations. The probability estimate
$\mathcal{P}$ for each source is then defined as the ratio 
between the number of occurrences satisfying the two colour criteria of
Eq.~(\ref{eq:prob}) and the total number of realizations.
The colour-colour selection criterion has been finally set up as the condition
$\mathcal{P}>0.9$, based on the Monte Carlo analysis described in
Sect.~\ref{sec:mcqa}.  This approach is far more robust than a simple cut
based on the two colour criteria. It also enables us to reject sources that
might satisfy the criteria owing to poor photometry alone.

\section{Monte Carlo quality assessment}
\label{sec:mcqa}

\subsection{Monte Carlo simulations}
\label{sec:mcqa_method}

In order to assess the impact of the cleaning method on the recovered flux
densities of the \Planck\ high-$z$ candidates and to explore the selection
function of the algorithm, we have performed Monte Carlo simulations.
A total of 90 sets of mock IRIS plus \Planck\ maps have been built by
injecting 10\,000 simulated high-$z$ point sources into the original \Planck\
and IRIS maps, yielding a total of 900\,000 fake injected sources.
The SEDs of these sources are modelled via modified blackbody emission with
a spectral index $\beta_{\rm{xgal}}=1.5$, and four equally probable values
of the temperature, $T_{\rm{xgal}}=20$, 30, 40, and 50\,K. The redshift of
these sources is uniformly sampled between $z=0$ and $z=5$. 
The flux density distribution follows a
power law with an index equal to the Euclidean value ($-2.5$) between
200\,mJy and 5\,Jy at 545\,GHz. Each source is modelled as an elliptical
Gaussian with a FWHM varying uniformly between 5{\arcmin} and 8{\arcmin},
and a ratio between the major and minor axes ranging uniformly between 
1 and 2. The point sources are then injected into the real IRIS and
\Planck\ maps (already convolved at 5{\arcmin} resolution),
excluding the regions within 5{\arcmin} 
of true detections of high-$z$ source candidates.

The full cleaning, extraction, photometry, and colour-colour selection
processing described in Sects.~\ref{sec:compsep} and \ref{sec:ps_detection}
is performed on this set of mock maps, yielding a sub-sample of about 70\,000
detected sources from the 900\,000 injected.
Notice that the cut on the 545\,GHz flux density has been omitted in this analysis in order to explore the completeness of the detection algorithm
beyond this flux density limit. Furthermore, we have tested two options of the CMB template during the cleaning processing: 
an ideal template, which consists in the {\tt SMICA} 5{\arcmin} CMB map; and a highly contaminated template, which has been built by injecting the expected
flux densities at 143\,GHz into the {\tt SMICA} 5{\arcmin} CMB template before cleaning, assuming here that the signal from 
the extragalactic source is still fully included in the CMB template.
This allowed us to quantify the maximum impact of the 
uncleaned foregrounds present in the CMB template we use for the official cleaning.
Finally, we stress that the fraction of total detections over the total number 
of injected sources cannot be considered as an estimate of the overall recovery rate of the algorithm, 
because of the unrealistic statistics of the injected population in terms of temperature, redshift or flux density.
However, these mock simulations allow us to build the a posteriori uncertainties on the properties of the recovered sources, 
and the selection function due to the detection algorithm.

\subsection{Geometry accuracy}
\label{sec:mcqa_position_accuracy}

\begin{figure}[t]
\vspace{-0.4cm}
\psfrag{yy}{$N$}
\includegraphics[width=0.5\textwidth]{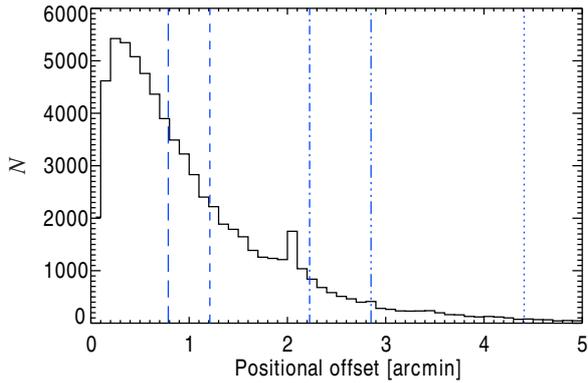}
\caption{Positional recovery. Histogram of the positional offsets between the
centroid coordinates of the recovered sources and the initial coordinates.
Vertical lines give levels of the maximum positional offset for the following
lower percentiles: 50\,\% (long dashed); 68\,\% (dashed); 90\,\% (dash-dotted); 
95\,\% (dash-dot-dot-dotted); and 99\,\% (dotted).}
\label{fig:mcqa_accuracy_position}
\end{figure}

\begin{figure}[t]
\vspace{0.38cm}
\hspace{-0.9cm}
\begin{tabular}{cc}
\psfrag{------xtitle------}{S/N $\quad S_{545}^{\rm{D}} $}
\psfrag{------ytitle------}{Rec./In.}
$\quad\,$\includegraphics[width=0.17\textwidth, viewport=100 0 450 380]{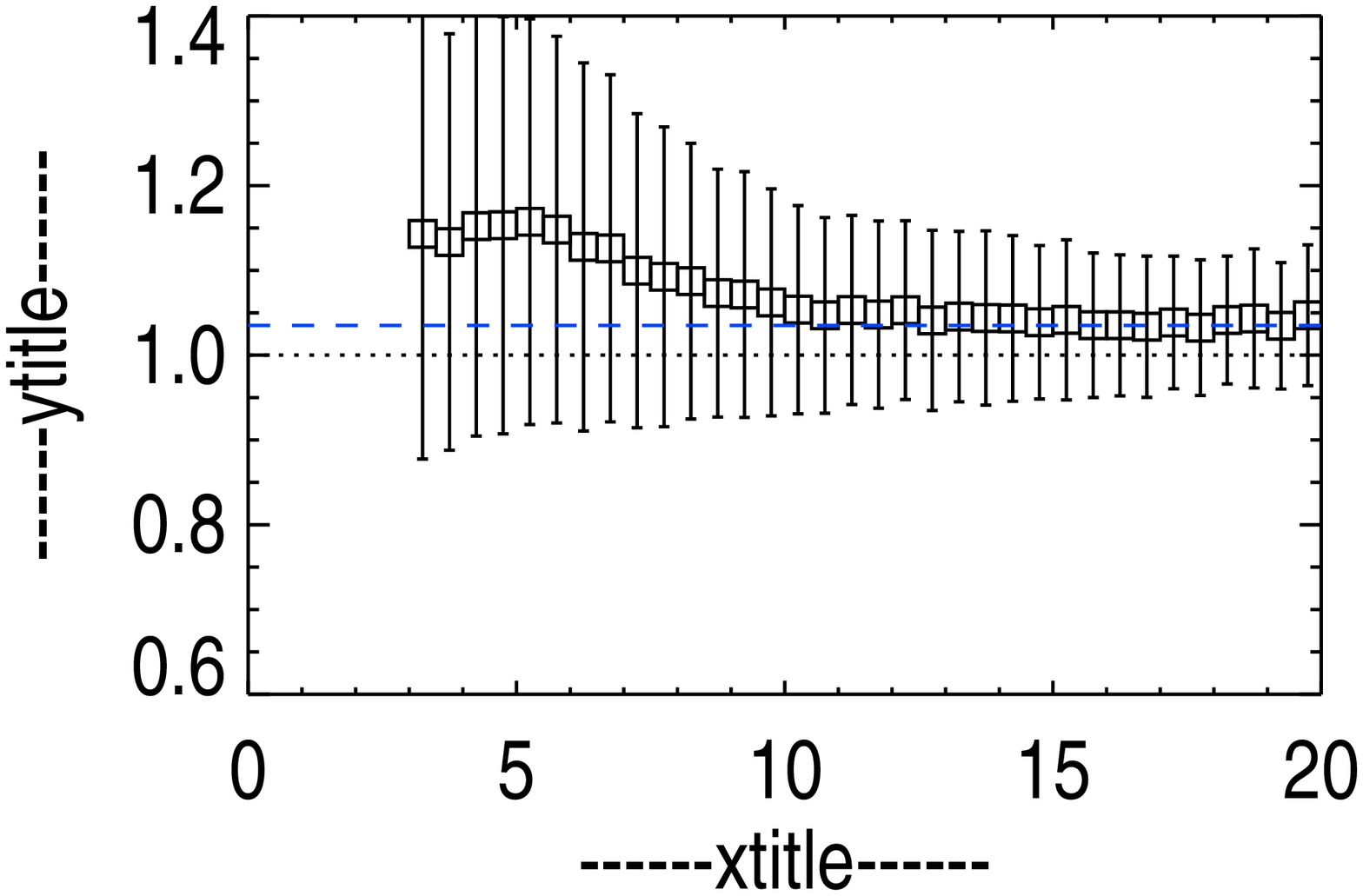}&
\psfrag{------xtitle------}{S/N $\quad S_{545}^{\rm{D}} $}
$\quad\quad$\includegraphics[width=0.17\textwidth, viewport=100 0 450 380]{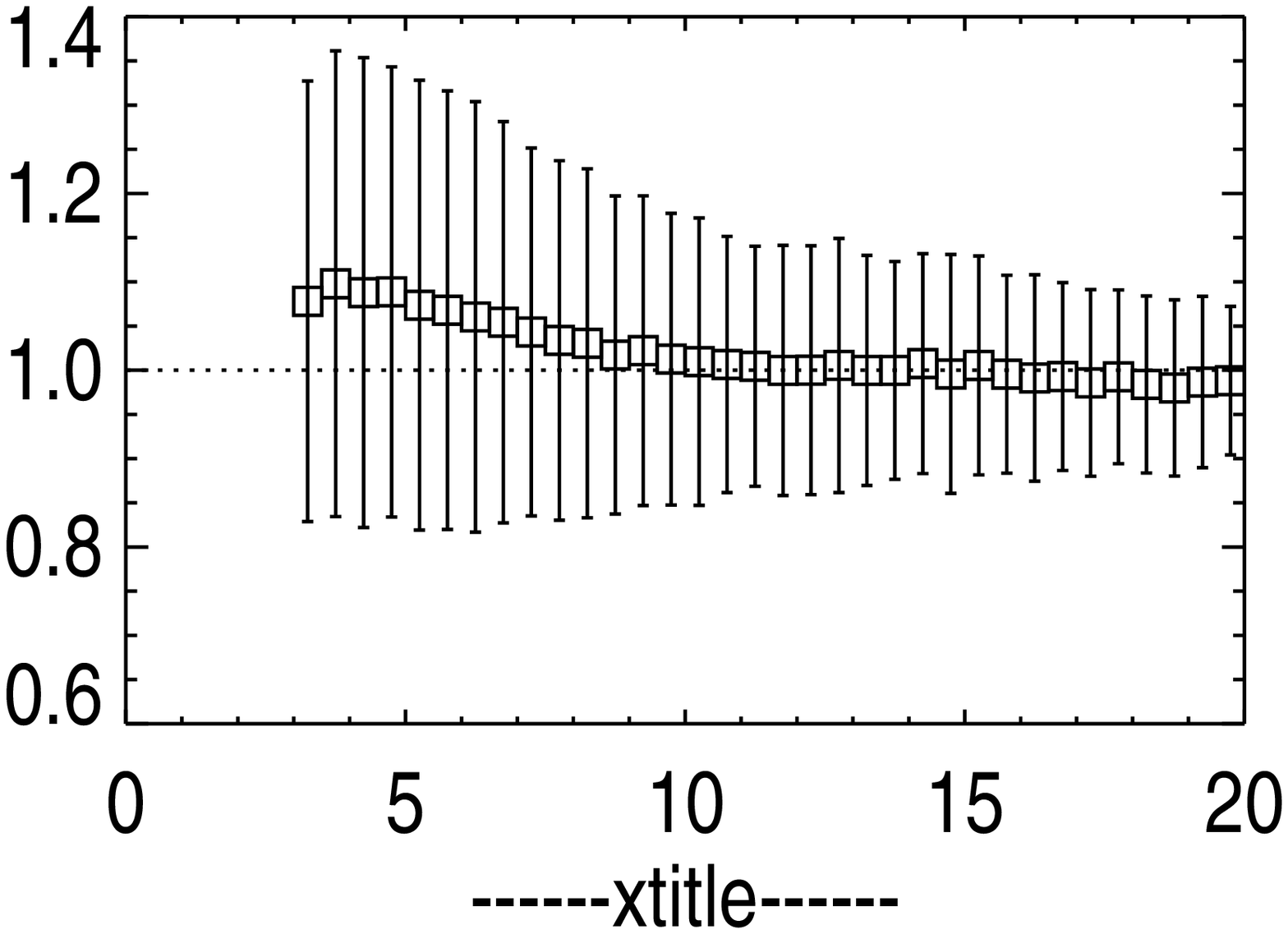} \\
& \\
\psfrag{------xtitle------}{$\rm{FWHM}_{\rm{in}}$}
\psfrag{------ytitle------}{$\rm{FWHM}_{\rm{rec}}$}
\includegraphics[width=0.21\textwidth, viewport=40 20 470 480]{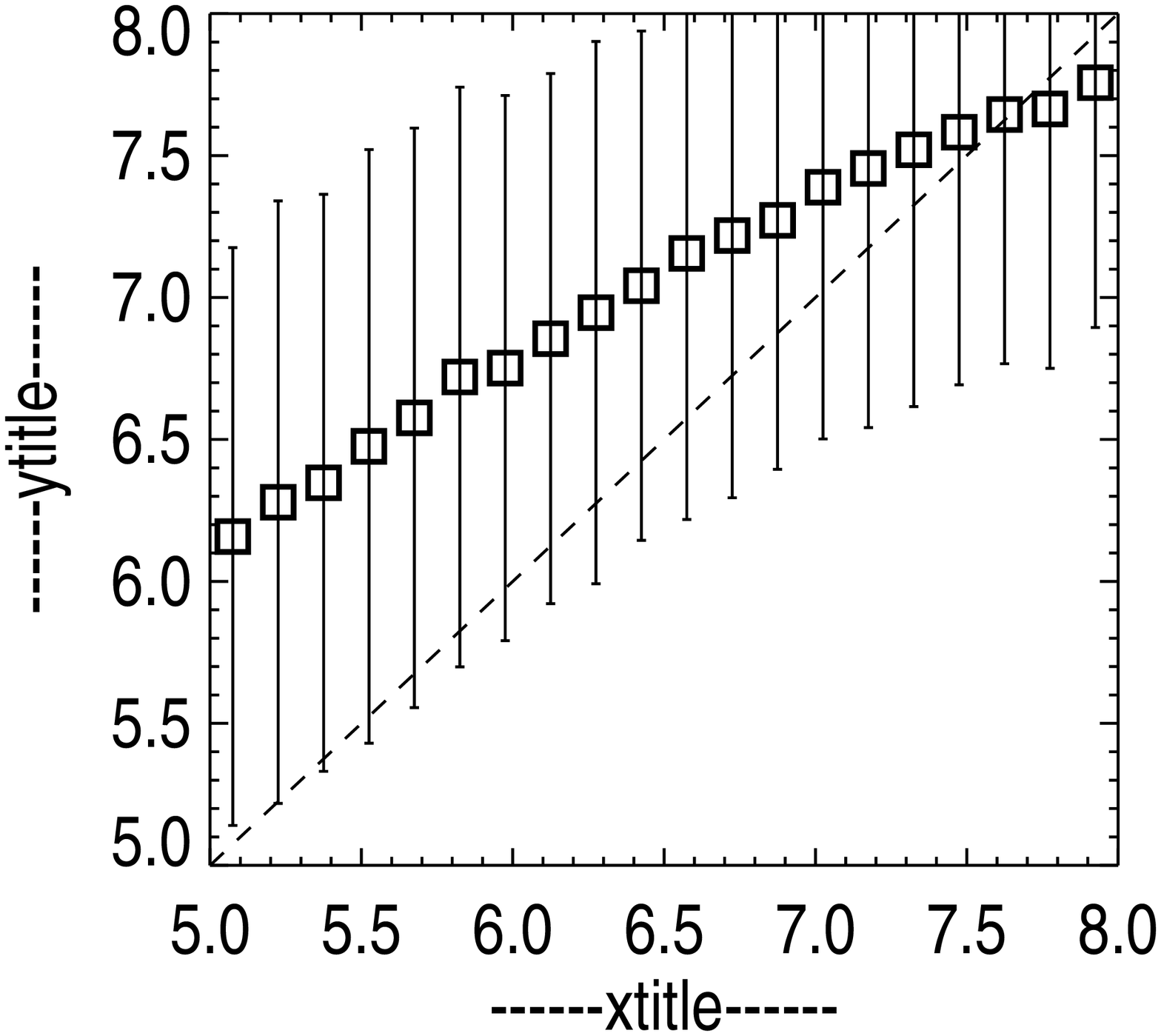}&
$\quad$\includegraphics[width=0.21\textwidth, viewport=40 20 470 480]{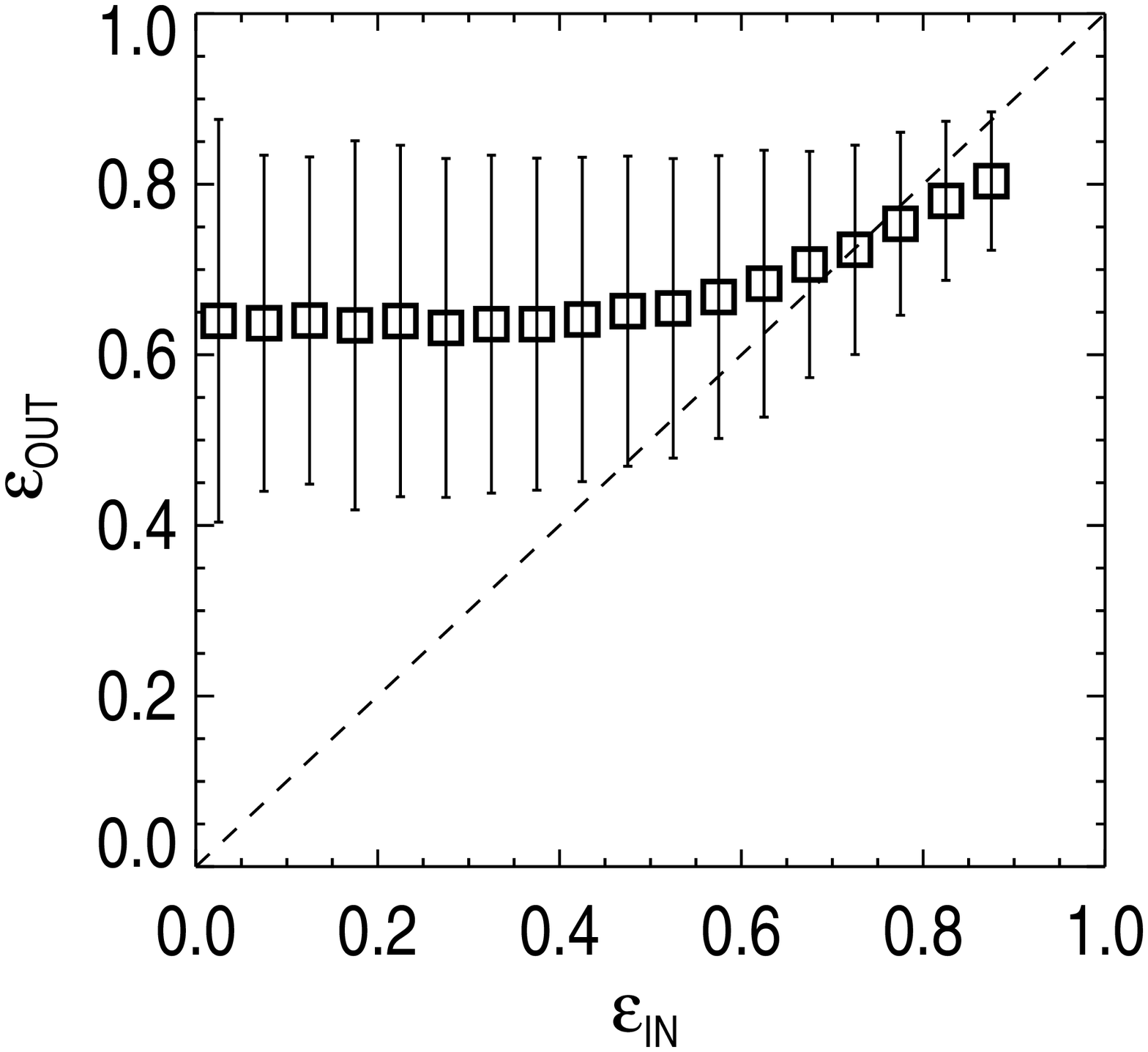}\\
\end{tabular}
\caption{FWHM and ellipticity recovery. {\it Left}: FWHM. {\it Right}:
ellipticity.  {\it Top}: ratio of recovered over injected FWHMs and
ellipticities (Rec./In.) as a function of the S/N of detection on the 545\,GHz
cleaned map.  {\it Bottom}: recovered versus injected FWHMs and ellipticities.
The dashed line gives the 1:1 relation.}
\label{fig:mcqa_accuracy_fwhm_ellipticity}
\end{figure}

We first analyse the positional accuracy of the detected sources and show the results in Fig.~\ref{fig:mcqa_accuracy_position}.
Recall that the centroid coordinates of the elliptical Gaussian are obtained through a fit on the cleaned 545\,GHz map.
Hence 68\,\% of the sources exhibit a positional offset smaller than 1{\parcm}2 and 95\,\% of them within 2{\parcm}9, 
which are not negligible values compared to the 5{\arcmin} resolution of the maps. This positional uncertainty
is mainly due to the confusion level of the CIB in which these sources are embedded.

More problematic is the efficiency of the FWHM recovery, which drives the
 computation of the aperture photometry (see Fig.~\ref{fig:mcqa_accuracy_fwhm_ellipticity}).
Recall that the FWHM is defined as the geometric mean of the minor and major FWHM, 
FWHM=$(\rm{FWHM}_{\rm{min}}\times\rm{FWHM}_{\rm{maj}})^{1/2}$.
The recovered FWHM is overestimated compared with the input FWHM
over the whole range of S/N of detection in the 545\,GHz cleaned map (see left panels), 
by an average value of 3.5\,\% at high S/N, and up to 15\,\% at low S/N (below 10). 
When looking more carefully at the distribution of recovered versus injected FWHMs, 
it appears that the largest FWHM bin, the one close to 8{\arcmin}, is better recovered than the smallest 
FWHM values, which are strongly overestimated by up to 30\,\%.
In fact, the distribution of the recovered FWHM peaks around 6{\parcm}5, 
while the input values were uniformly distributed between 5{\arcmin} and 8{\arcmin}.

In addition to this, the dispersion of the ratio between the recovered and the
injected FWHMs does not significantly decrease with the S/N of the detection, 
and lies around 7\,\%. This is larger than the level of uncertainty provided 
by the elliptical Gaussian fit, which is about 1.5\,\% at maximum.
Indeed the uncertainty on the FWHM is mainly dominated by the confusion level of the CIB. 

The same analysis is performed on the ellipticity of the sources, defined as
\begin{equation}
\varepsilon = \sqrt{1 - \left( \frac{\theta_{\rm{min}}}{\theta_{\rm{maj}}} \right)^2 } \, ,
\end{equation}
where $\theta_{\rm{min}}$ and $\theta_{\rm{maj}}$ are the minor and major axis of the ellipse, respectively.
When looking at the ratio between the recovered and injected ellipticity as a function of the S/N of the 545\,GHz flux density
(right panels of Fig.~\ref{fig:mcqa_accuracy_fwhm_ellipticity}), the estimates do not seem biased for S/N larger than 5.
However, the recovered versus injected ellipticity comparison shows that low ellipticities are systematically overestimated. 
The average ellipticity estimates are greater than 0.6 over the whole range of input ellipticity. 
Recall that an ellipticity $\varepsilon=0.6$ corresponds to a major axis 1.25 times larger than the minor axis.
Such an error of 25\,\% between minor and major FWHMs is fully compatible with the level of uncertainty of the recovered FWHM, pointed above.
Again this effect is probably explained by the CIB confusion.

We have observed that these results are totally independent of the choice of the CMB template (ideal or highly foreground-contaminated) for the cleaning processing, 
because the geometry parameters are obtained in the 545\,GHz cleaned map, which are barely impacted by the CMB cleaning.

\begin{figure*}
\hspace{-0.5cm}
\begin{tabular}{ccccc}
$\quad \quad \quad \quad \quad \quad$ 857\,GHz& 
$\quad \quad \quad \quad \quad \quad$ 545\,GHz& 
$\quad \quad \quad \quad \quad \quad$ 353\,GHz& 
$\quad \quad \quad \quad \quad \quad$ 217\,GHz& \\
\psfrag{------xtitle------}{$\, \, \, S_{\nu}^{\rm{D}}/\sigma_{\nu}^{\rm{sky}} $}
\psfrag{----ytitle----}{$S_{\nu}^{\rm{D}}\,/\, S_{\nu}^{\rm{I}}$}
$\quad$\includegraphics[width=0.17\textwidth, viewport=100 0 450 380]{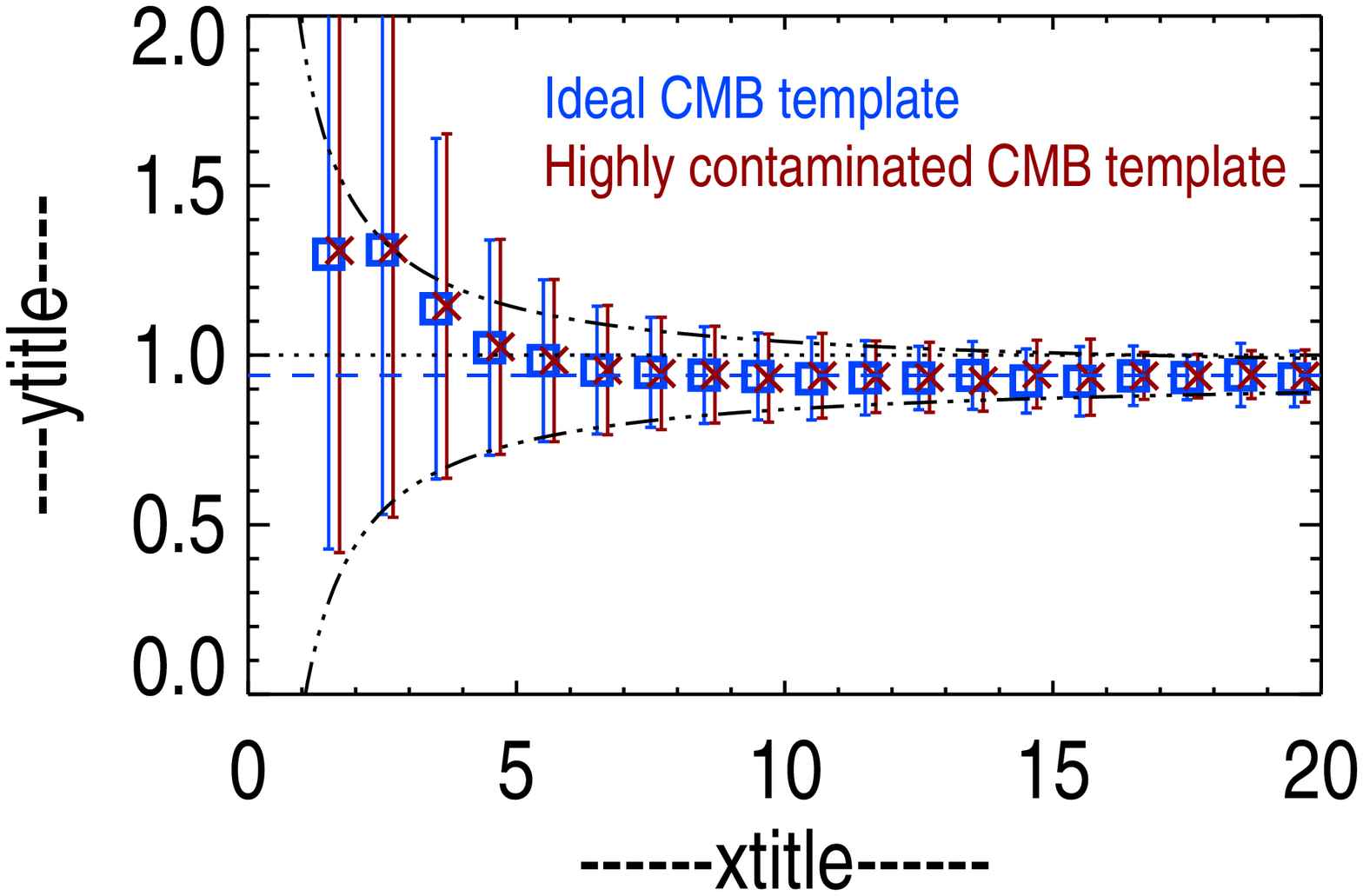}&
\psfrag{------xtitle------}{$\, \, \, S_{\nu}^{\rm{D}}/\sigma_{\nu}^{\rm{sky}} $}
\psfrag{------ytitle------}{$S_{545}^{\rm{D}}\,/\, S_{545}^{\rm{I}}$}
$\quad$\includegraphics[width=0.17\textwidth, viewport=100 0 450 380]{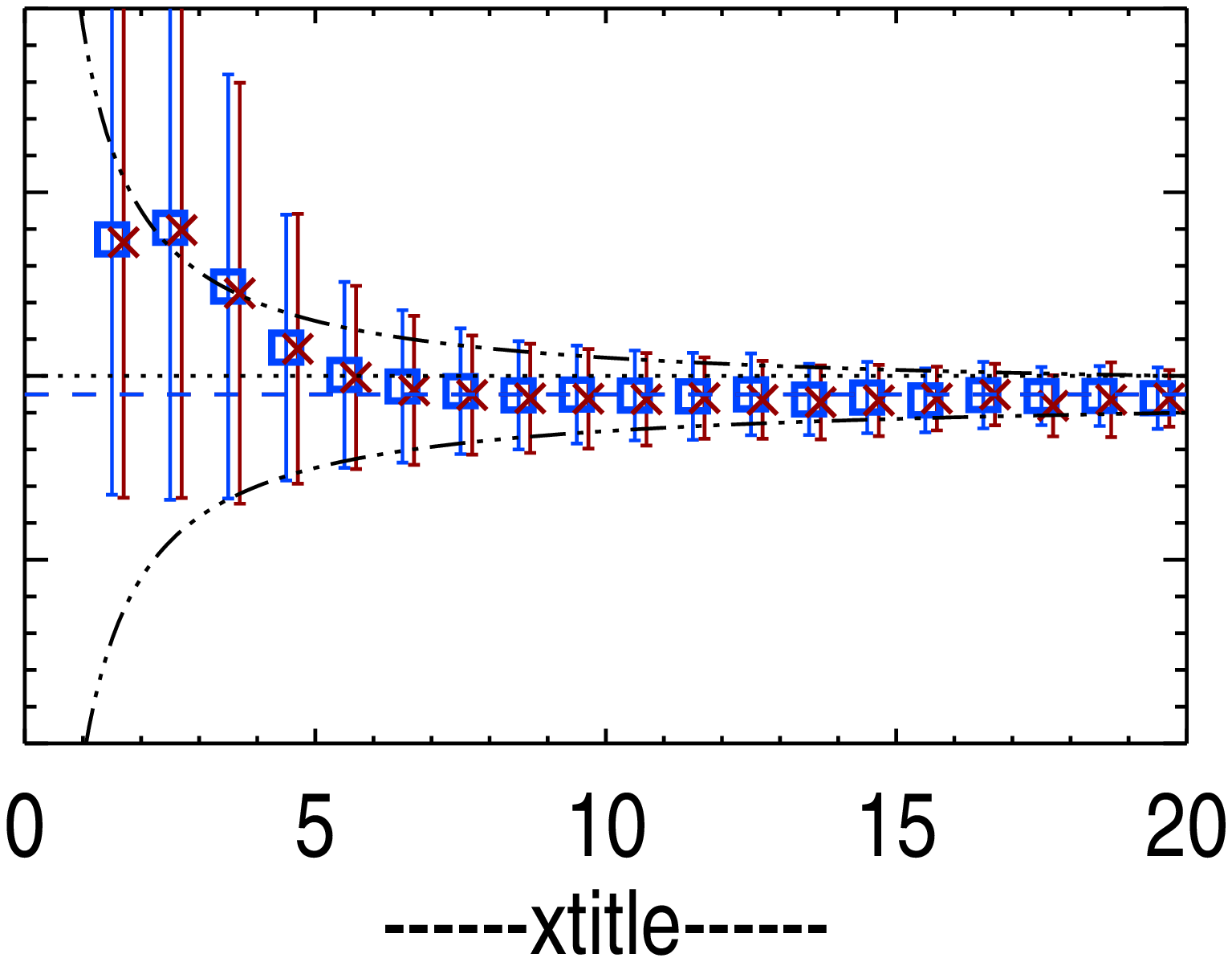}&
\psfrag{------xtitle------}{$\, \, \, S_{\nu}^{\rm{D}}/\sigma_{\nu}^{\rm{sky}} $}
\psfrag{------ytitle------}{$S_{353}^{\rm{D}}\,/\, S_{353}^{\rm{I}}$}
$\quad$\includegraphics[width=0.17\textwidth, viewport=100 0 450 380]{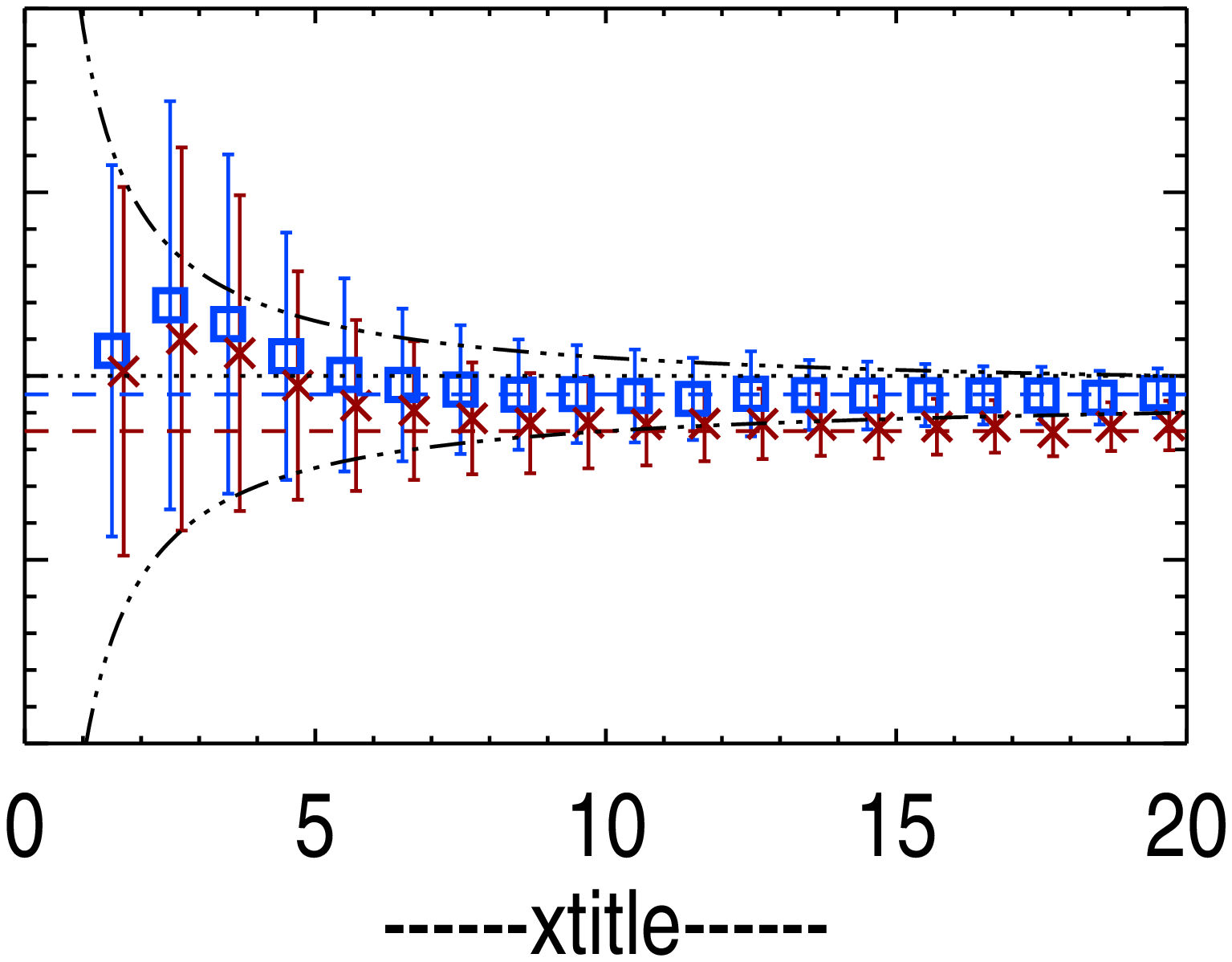}&
\psfrag{------xtitle------}{$\, \, \, S_{\nu}^{\rm{D}}/\sigma_{\nu}^{\rm{sky}} $}
\psfrag{------ytitle------}{$S_{217}^{\rm{D}}\,/\, S_{217}^{\rm{I}}$}
$\quad$\includegraphics[width=0.17\textwidth, viewport=100 0 450 380]{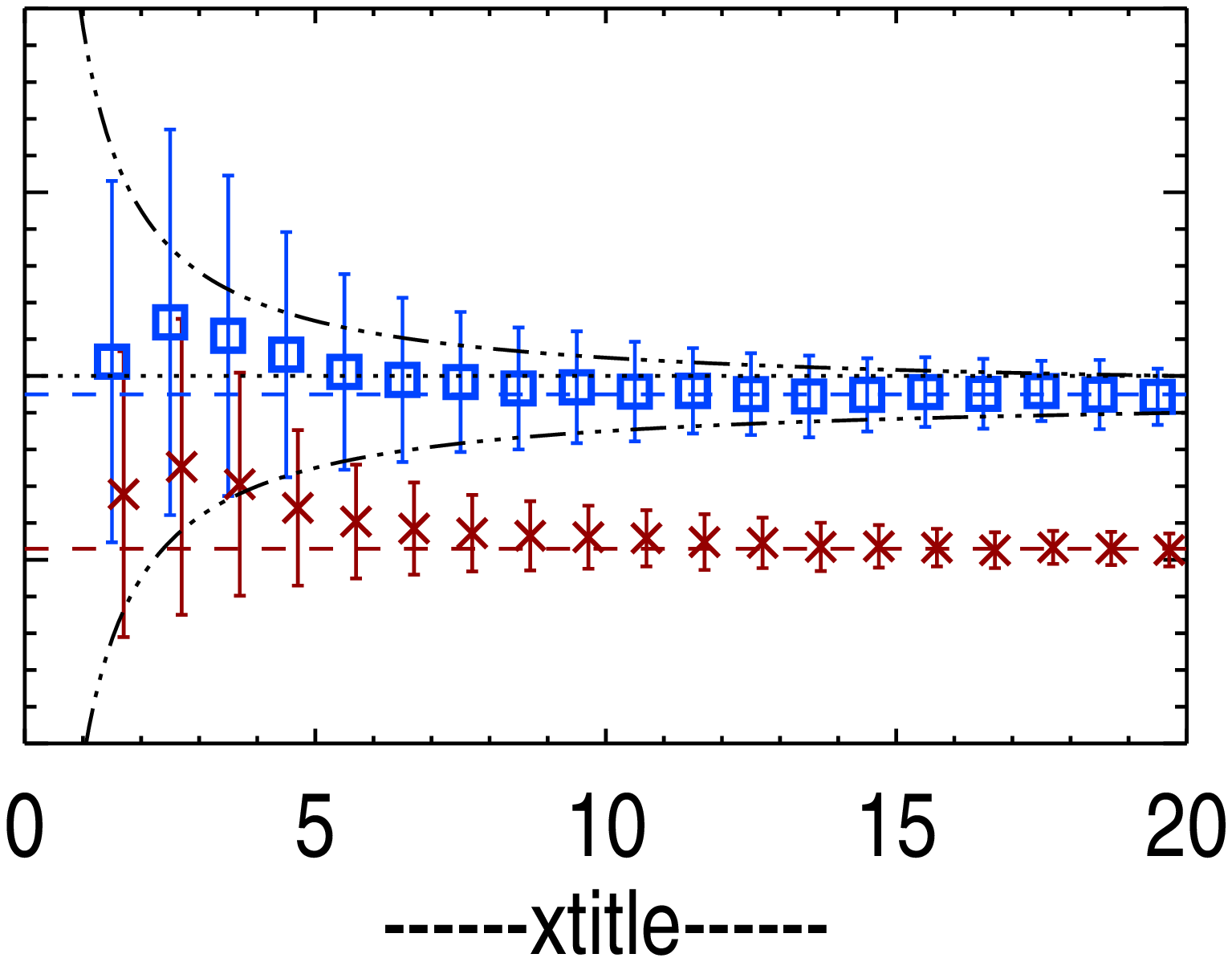}& \\
\psfrag{------xtitle------}{$S_{\nu}^{\rm{I}}$ [Jy]}
\psfrag{------ytitle------}{$S_{\nu}^{\rm{D}}$ [Jy]}
\psfrag{------ztitle------}{S/N $\quad S_{857}^{\rm{D}} $}
\includegraphics[width=0.21\textwidth, viewport=40 20 470 480]{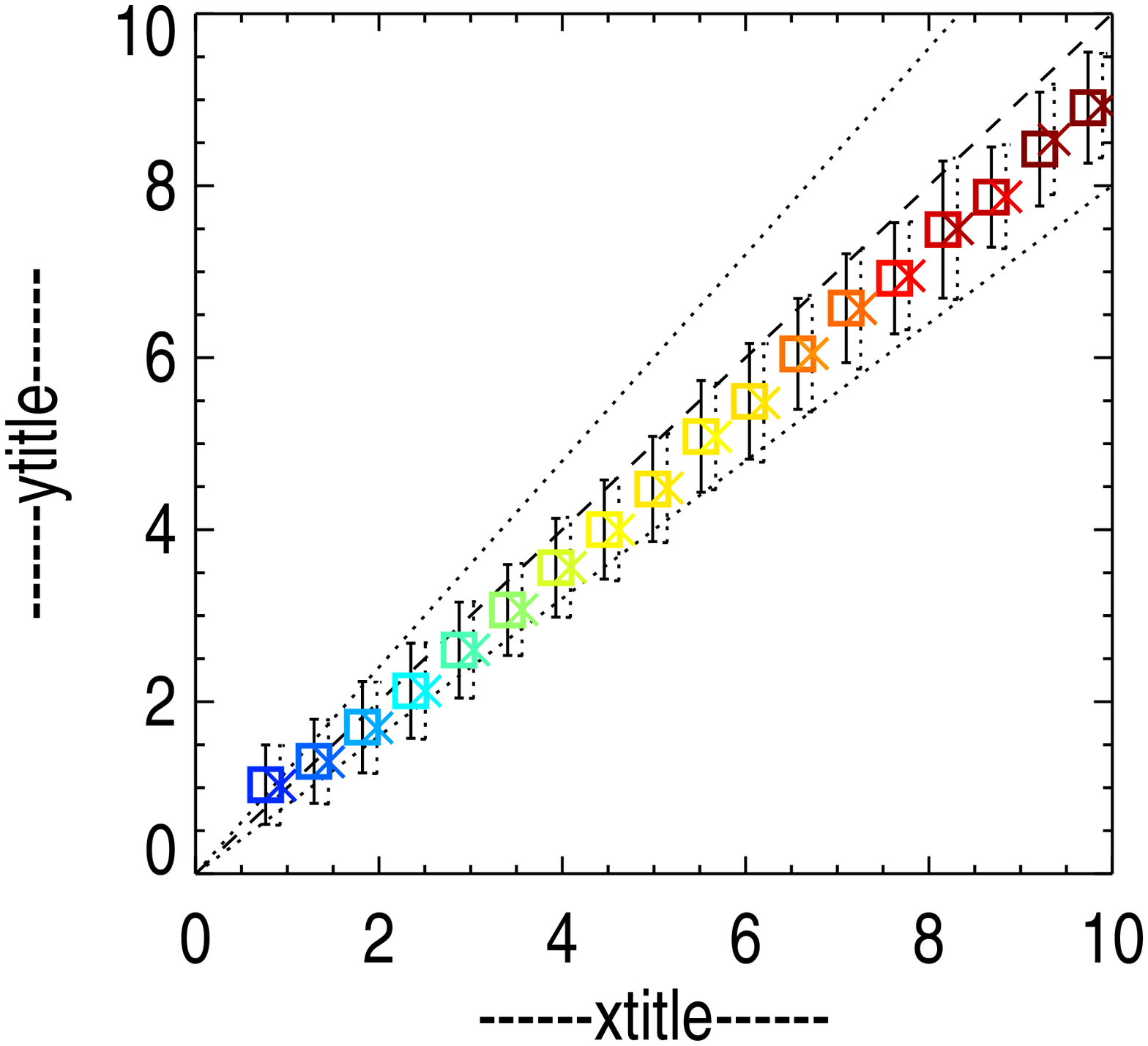}&
\psfrag{------xtitle------}{$S_{\nu}^{\rm{I}}$ [Jy]}
\psfrag{------ytitle------}{$S_{\nu}^{\rm{D}}$ [Jy]}
\psfrag{------ztitle------}{S/N $\quad S_{545}^{\rm{D}} $}
\includegraphics[width=0.21\textwidth, viewport=40 20 470 480]{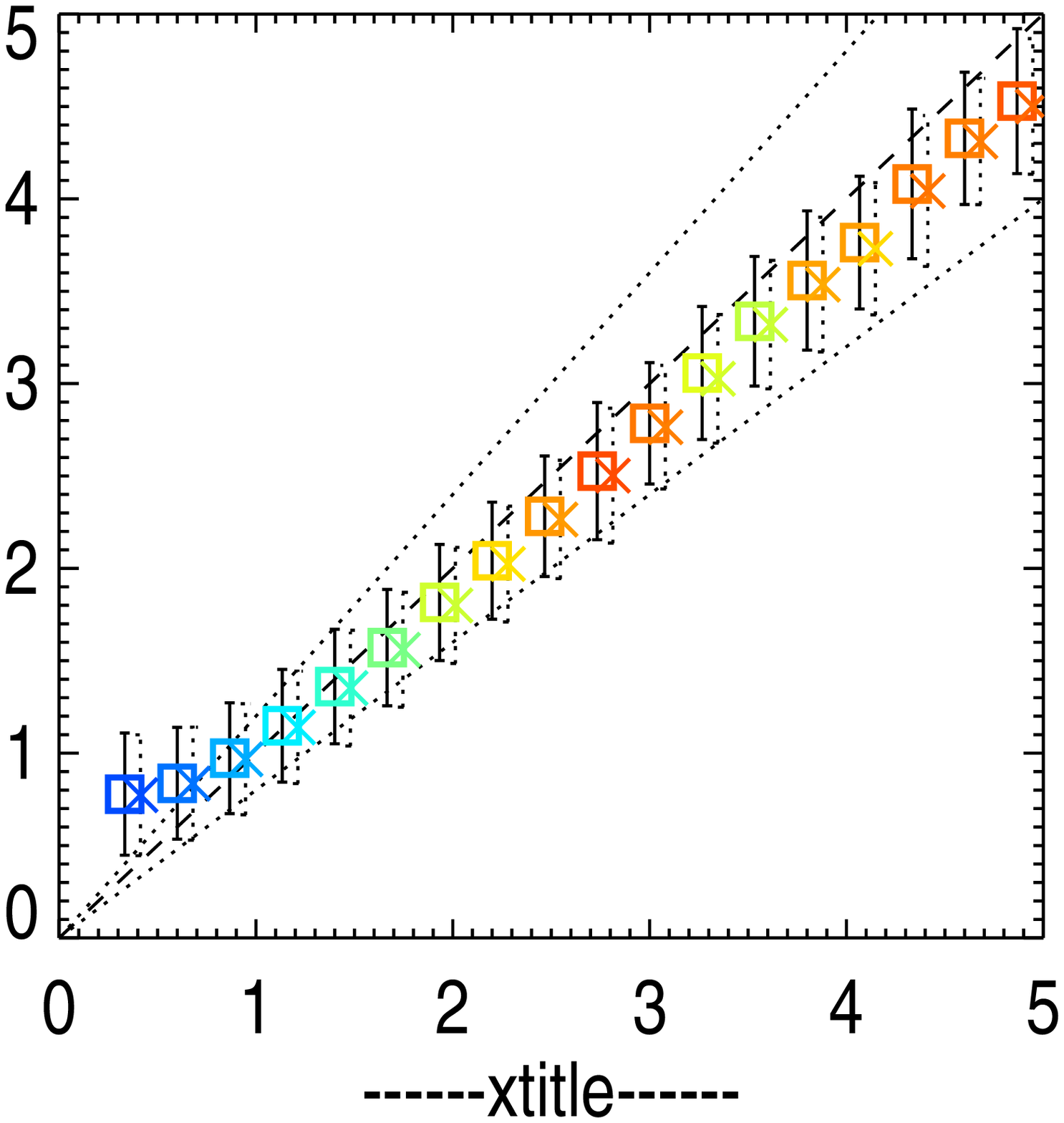}&
\psfrag{------xtitle------}{$S_{\nu}^{\rm{I}}$ [Jy]}
\psfrag{------ytitle------}{$S_{\nu}^{\rm{D}}$ [Jy]}
\psfrag{------ztitle------}{S/N $\quad S_{353}^{\rm{D}} $}
\includegraphics[width=0.21\textwidth, viewport=40 20 470 480]{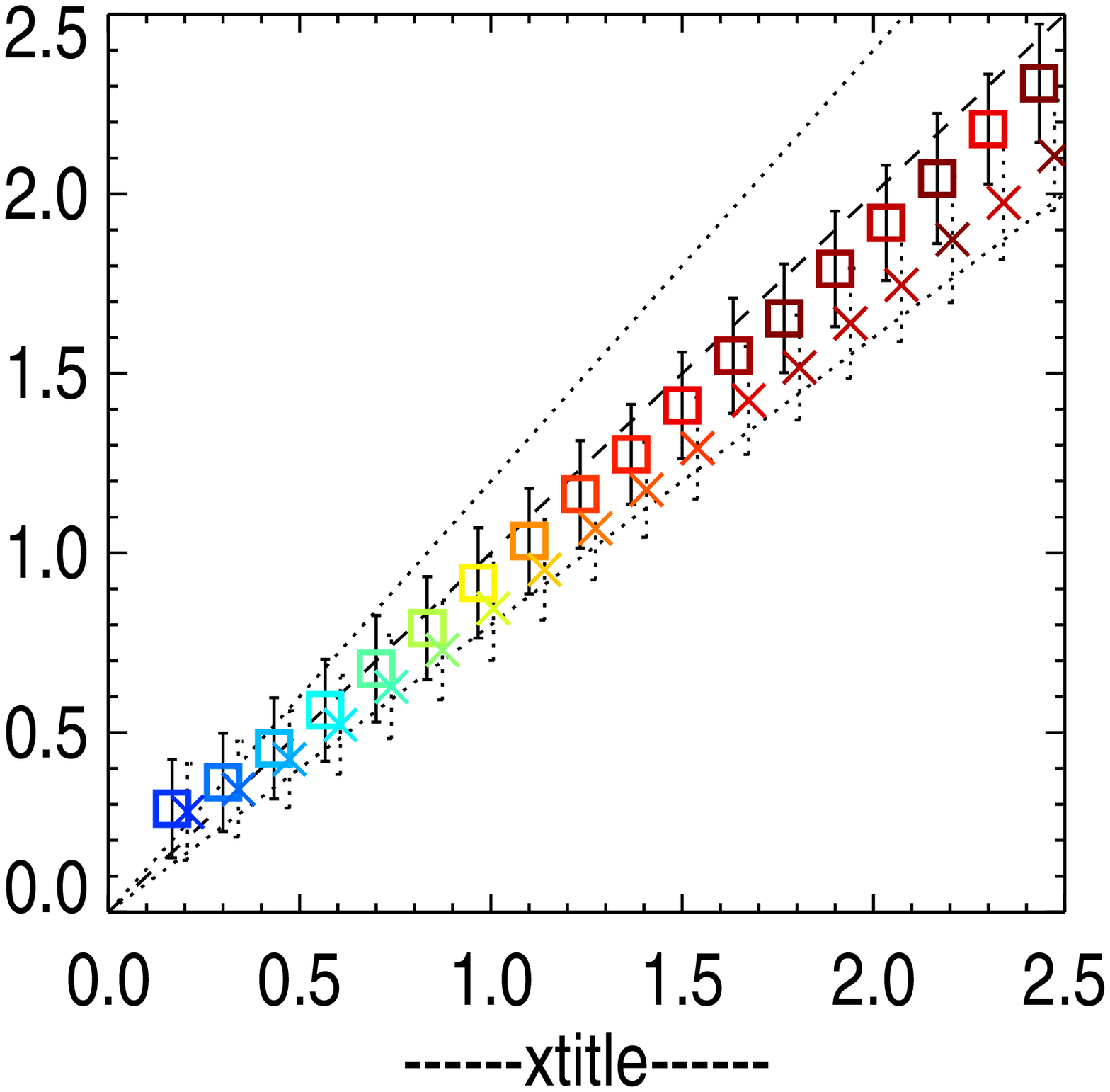}&
\psfrag{------xtitle------}{$S_{\nu}^{\rm{I}}$ [Jy]}
\psfrag{------ytitle------}{$S_{\nu}^{\rm{D}}$ [Jy]}
\psfrag{------ztitle------}{S/N $\quad S_{217}^{\rm{D}} $}
\includegraphics[width=0.21\textwidth, viewport=40 20 470 480]{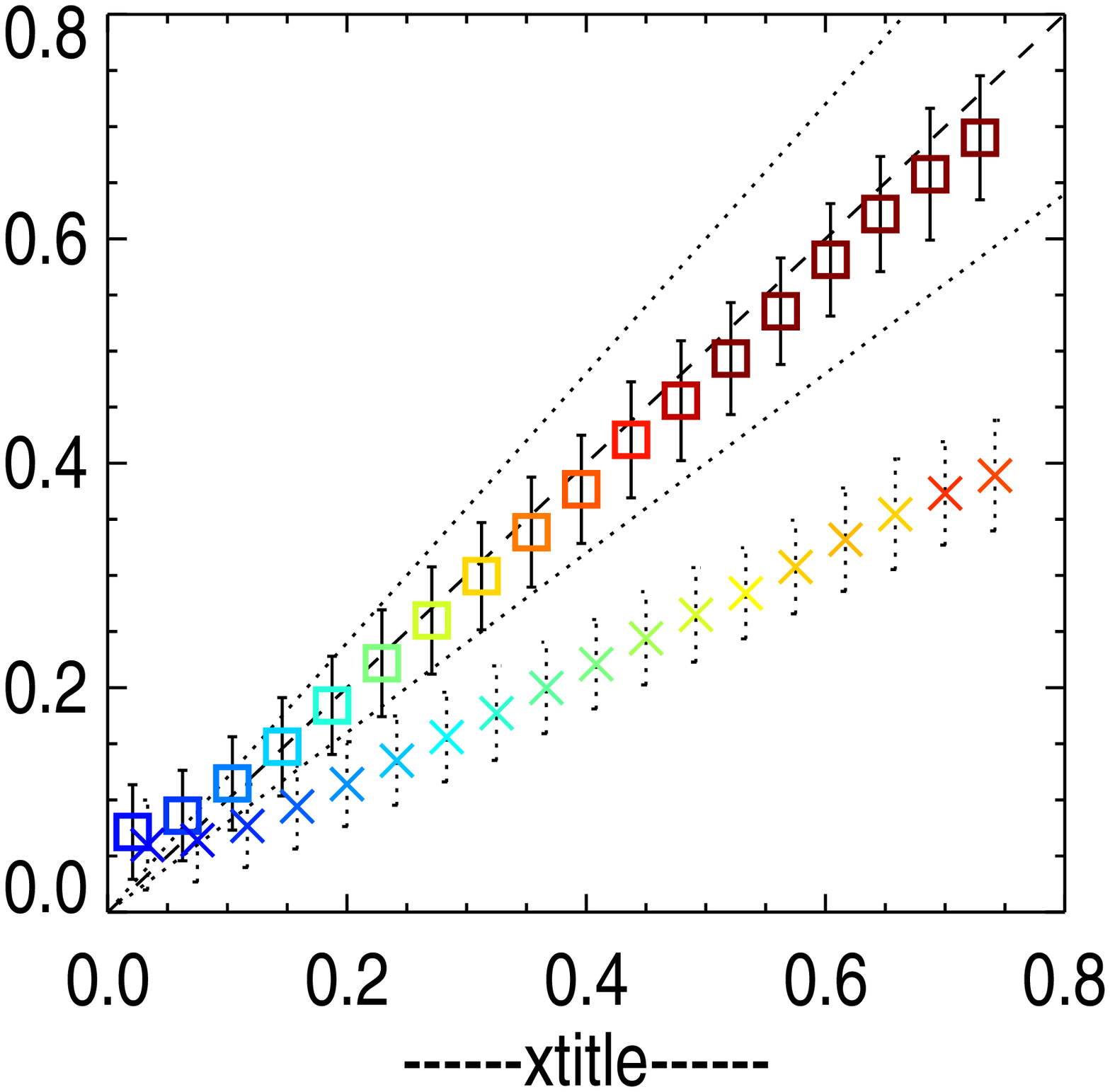}& 
\psfrag{------ztitle------}{$\quad S_{\nu}^{\rm{D}}/\sigma_{\nu}^{\rm{sky}} $}
$\quad$\includegraphics[width=0.057\textwidth, viewport=480 5 600 480]{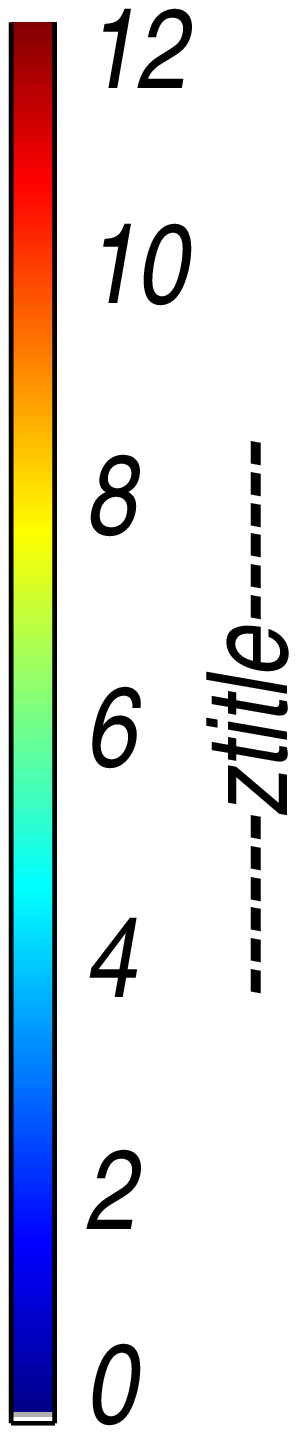} \\
\end{tabular}
\caption{Flux density recovery, from left to right, at 857, 545, 353, and 217\,GHz. 
{\it Top}: ratio of the recovered to the input flux density ($S_{\nu}^{\rm{D}}$/$S_{\nu}^{\rm{I}}$) as a function of 
the recovered flux density S/N, $S_{\nu}^{\rm{D}}/\sigma_{\nu}^{\rm{sky}}$. This is shown for two choices of the CMB template, 
i.e., ideal (blue squares) or highly contaminated by extragalactic foregrounds (red crosses). The average bias at high S/N is shown with a dashed line, while
the $\pm1\,\sigma$ envelope expected at each S/N is plotted as a dash-dot-dot-dot line.
The error bars correspond to the $\pm1\,\sigma$ standard deviation computed over the sub-sample of sources in each bin of S/N.
{\it Bottom}: recovered ($S_{\nu}^{\rm{D}}$) versus input ($S_{\nu}^{\rm{I}}$) flux density per bin of input flux density. 
Again, two cases are shown depending on the quality of the CMB template, ideal (square) or highly contaminated (crosses).
The colour scale provides the average S/N of the flux density inside each bin of input flux density.
The dotted lines show the $\pm20$\,\% limits around the 1:1 relation (dashed line).}
\label{fig:mcqa_accuracy_flux}
\end{figure*}

\subsection{Photometry quality}
\label{sec:mcqa_photometry}

We first recall that the recovered photometry, $S_{\nu}^{\rm{D}}$, is obtained on cleaned maps and suffers from the
noise and the CIB confusion, but also from the attenuation effect due to the cleaning process.
For each \Planck\ band, the ratio of the recovered to input flux density ($S_{\nu}^{\rm{D}}$/$S_{\nu}^{\rm{I}}$) 
is shown in the top row of Fig.~\ref{fig:mcqa_accuracy_flux} as a function of the S/N of the flux density, defined here as 
the ratio of the recovered flux density to the uncertainty due to CIB confusion, $S_{\nu}^{\rm{D}}/\sigma_{\nu}^{\rm{sky}}$.
This is shown for the two options of the CMB template, i.e., ideal (squares) or highly contaminated (crosses).

When assuming a very low level of foreground contamination in the CMB template, flux density
 estimates in all \Planck\ bands are recovered with a very good accuracy, as expected according to theoretical predictions of the attenuation effect
 of Fig.~\ref{fig:impact_cleaning_attenuation}. The fact that all flux density estimates appear statistically slightly underestimated by about 4\,\% for $\rm{S/N}>5$ 
is related to the quality of the source shape recovery.
 On the contrary, when the CMB template is assumed to 
 be highly contaminated by the extragalactic foregrounds, flux density estimates are more impacted 
 by the cleaning process, especially at 217\,GHz. In this band, the attenuation factor due to cleaning reaches a level of 47\,\% at high S/N, 
which is compatible with the predictions of Sect.~\ref{sec:impact_cleaning_highz}.
The attenuation at 353\,GHz is about 17\,\% at high S/N.
 
Below a S/N of around 5 two other effects appear: a much larger overestimation of the FWHM, 
up to 30\,\% at very low S/N, as discussed in Sect.~\ref{sec:mcqa_position_accuracy}; and the 
so-called flux boosting effect, which represents the tendency to
overestimate the flux densities of faint sources close to the CIB confusion because of noise upscatters being more likely
 than downscatters \citep[see][]{Hogg1998}. 
While the latter can be addressed using a Bayesian approach \citep{Coppin2005,
Coppin2006, Scott2008} for intermediate S/N (\ie, $\rm{S/N}>8$), we used this set of Monte Carlo simulations, as done by \citet{Scott2002} and
\citet{Noble2012}, to assess its impact on photometry estimates.
As observed in the bottom panels of Fig.~\ref{fig:mcqa_accuracy_flux}, flux densities of faint sources are strongly overestimated, producing
a plateau around 0.5\,Jy at 545\,GHz. This is consistent with the
confusion noise levels predicted by \citet{Negrello2004} in the
\Planck\ bands.

\begin{figure}
\hspace{-0.7cm}
\begin{tabular}{ccc}
$\quad \quad \quad \quad \quad \quad$ $S_{353}\,/\, S_{545}$& 
$\quad \quad \quad \quad \quad \quad$ $S_{545}\,/\, S_{857}$& \\
\psfrag{------xtitle------}{S/N $\quad S_{545}^{\rm{X}}$}
\psfrag{------ytitle------}{Rec./In.}
$\quad$\includegraphics[width=0.15\textwidth, viewport=100 0 450 380]{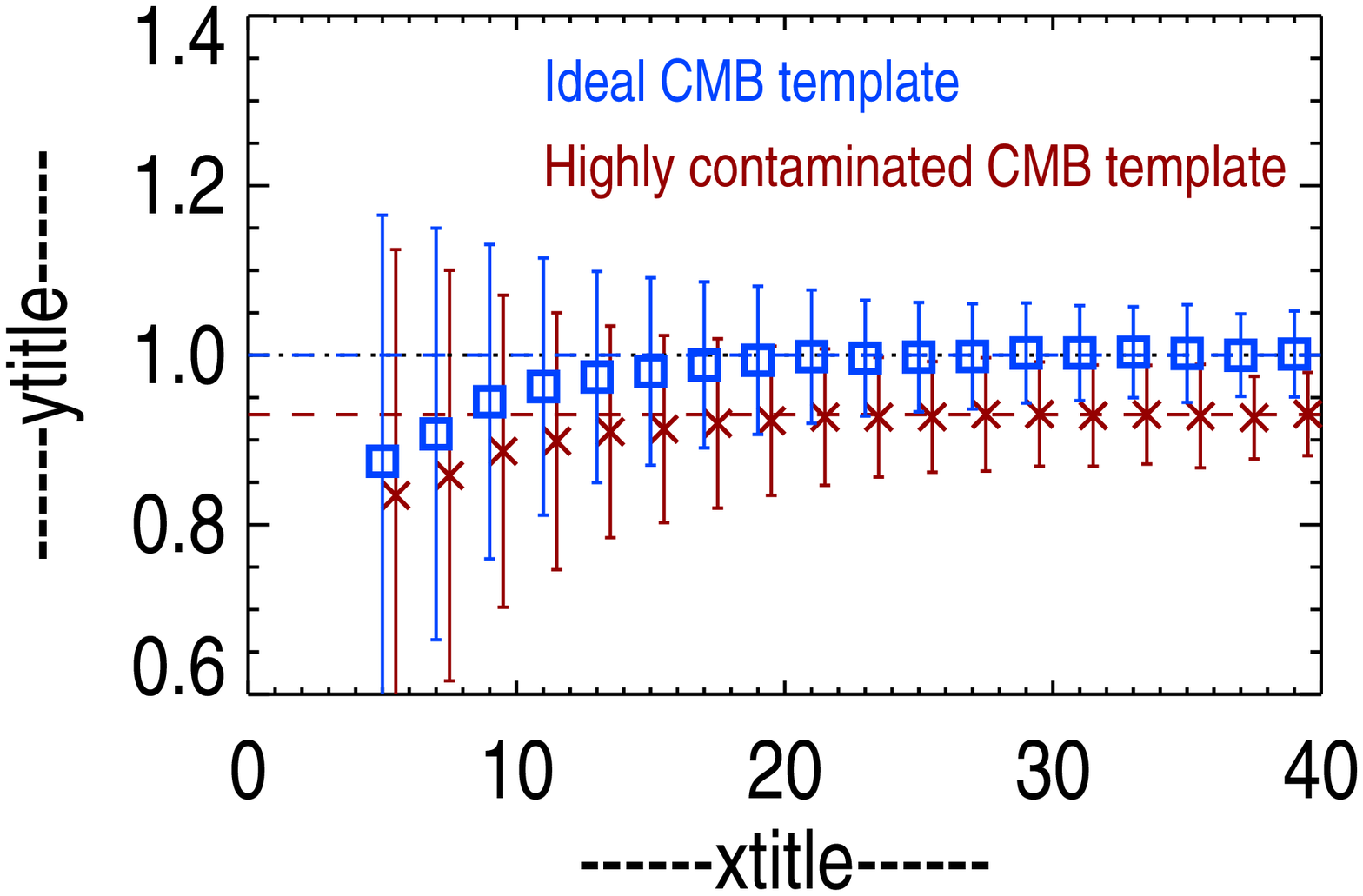}&
\psfrag{------xtitle------}{S/N $\quad S_{545}^{\rm{X}} $}
\psfrag{------ytitle------}{Rec./In.}
$\quad$\includegraphics[width=0.15\textwidth, viewport=100 0 450 380]{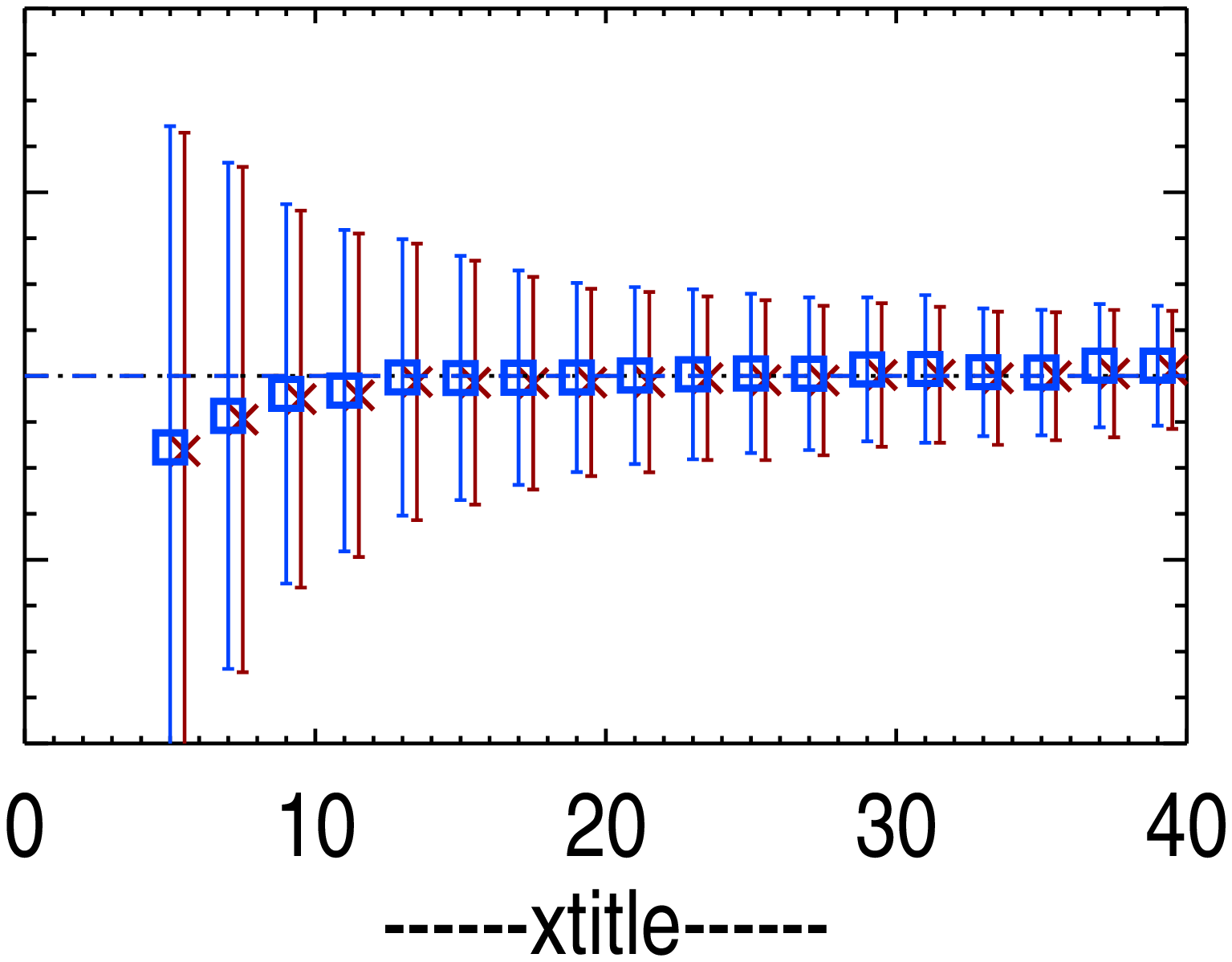}& \\
& & \\
\psfrag{------xtitle------}{\ Injected}
\psfrag{------ytitle------}{\ Recovered}
\includegraphics[width=0.19\textwidth, viewport=40 20 470 480]{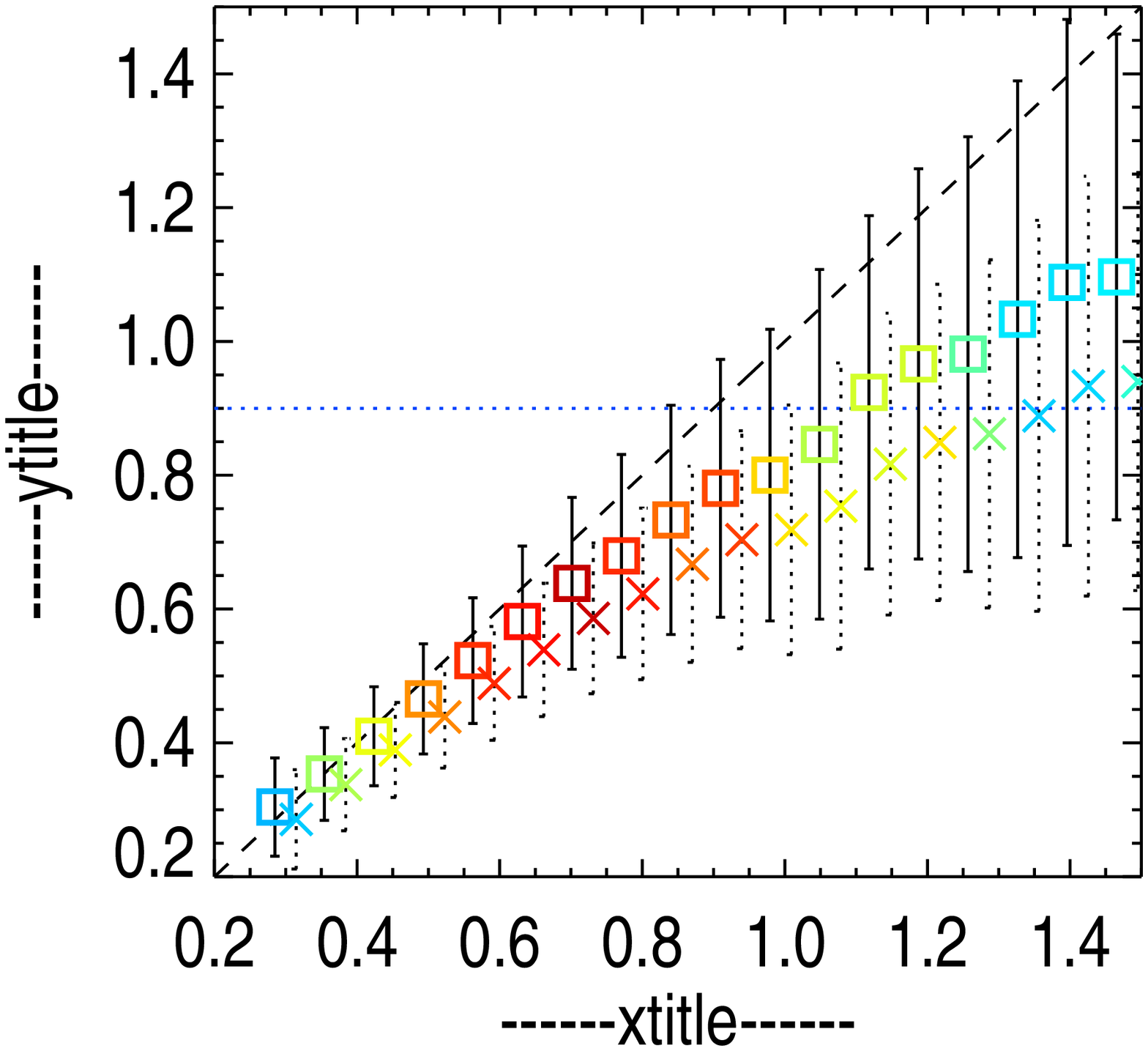}&
\psfrag{------xtitle------}{\ Injected}
\includegraphics[width=0.19\textwidth, viewport=40 20 470 480]{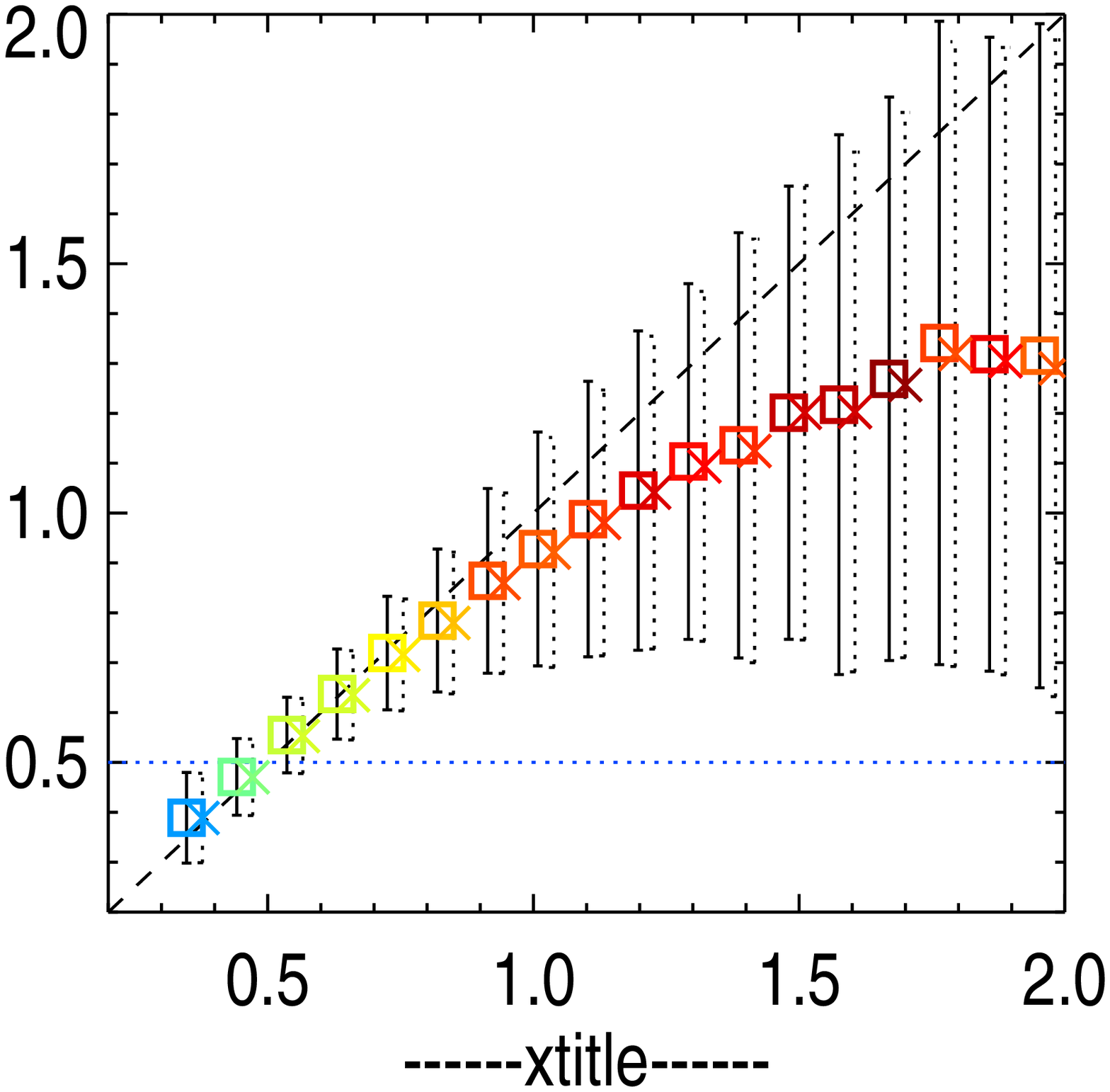}&
\psfrag{------ztitle------}{S/N $\quad S_{545}^{\rm{X}} $}
$\quad$\includegraphics[width=0.052\textwidth, viewport=480 5 600 480]{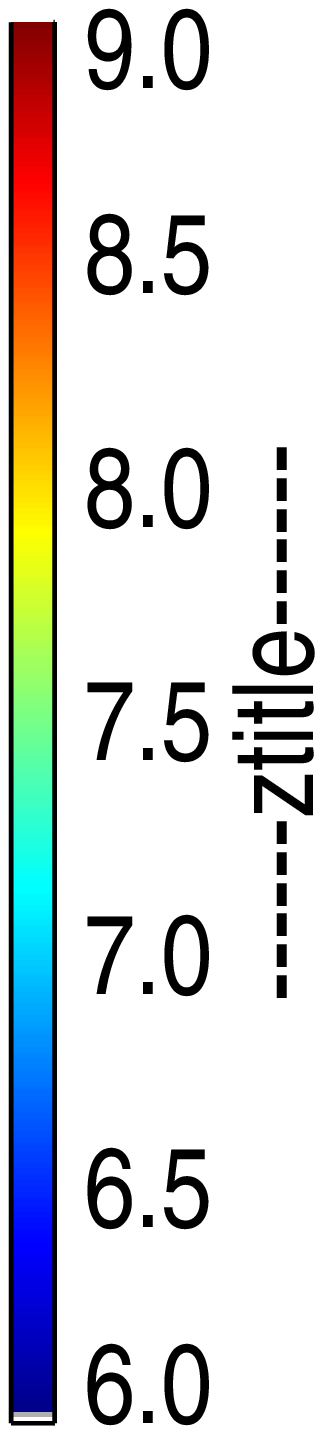} \\
\end{tabular}
\caption{Colour-colour ratio recovery. {\it Left}: $S_{353}/S_{545}$.
{\it Right}: $S_{545}/S_{857}$. 
{\it Top}: ratio of recovered over injected colour-colour ratio (Rec./In.)
as a function of the 545\,GHz excess S/N. 
This is shown for two choices of the CMB template, 
i.e., ideal (blue squares) or highly contaminated by extragalactic
foregrounds (red crosses). 
{\it Bottom}: recovered versus injected colour-colour ratio per bin of
input colour-colour ratio.  Again, two cases are shown depending on the
quality of the CMB template, ideal (squares) or highly contaminated (crosses).
The colour scale provides the average S/N of the 545\,GHz excess inside
each bin of input colour-colour ratio.  The blue dotted lines show
$S_{353}/S_{545}<0.9$ and $S_{545}/S_{857}>0.5$, which are the colour
criteria adopted for source selection. }
\label{fig:mcqa_accuracy_col_col}
\end{figure}

\begin{figure*}
\hspace{0.5cm}
\hspace{-1cm}
\begin{tabular}{cccccc}
\begin{minipage}[c]{0.03\linewidth} \vspace{-4.5cm} 50\,K \end{minipage}&
\psfrag{------xtitle------}{$ \quad S_{857}^{\rm{I}} $ [Jy]}
\includegraphics[width=0.2\textwidth, viewport=40 20 410 380]{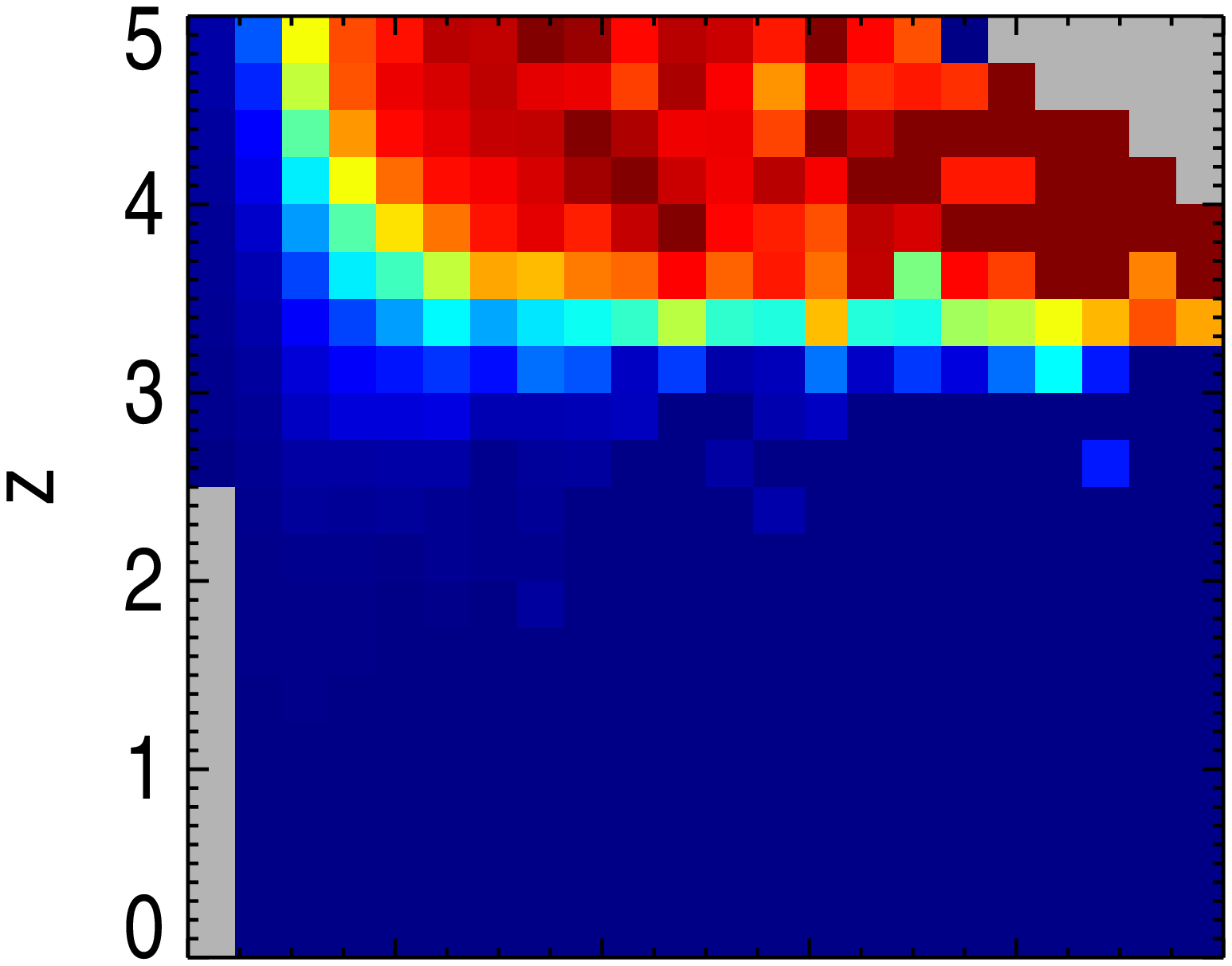}&
\psfrag{------xtitle------}{$ \quad S_{545}^{\rm{I}} $ [Jy]}
\includegraphics[width=0.2\textwidth, viewport=40 20 410 380]{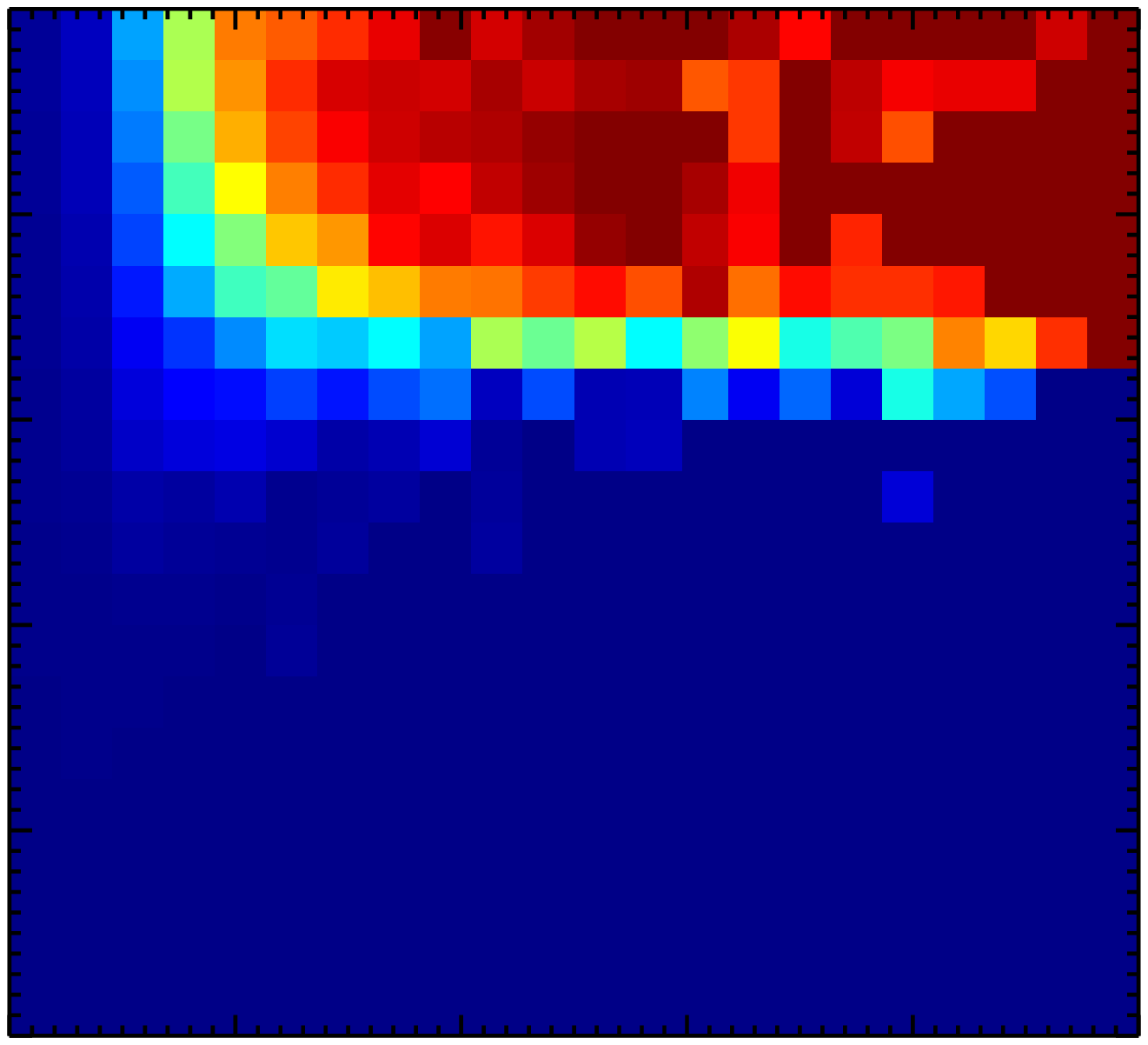}&
\psfrag{------xtitle------}{$ \quad S_{353}^{\rm{I}} $ [Jy]}
\includegraphics[width=0.2\textwidth, viewport=40 20 410 380]{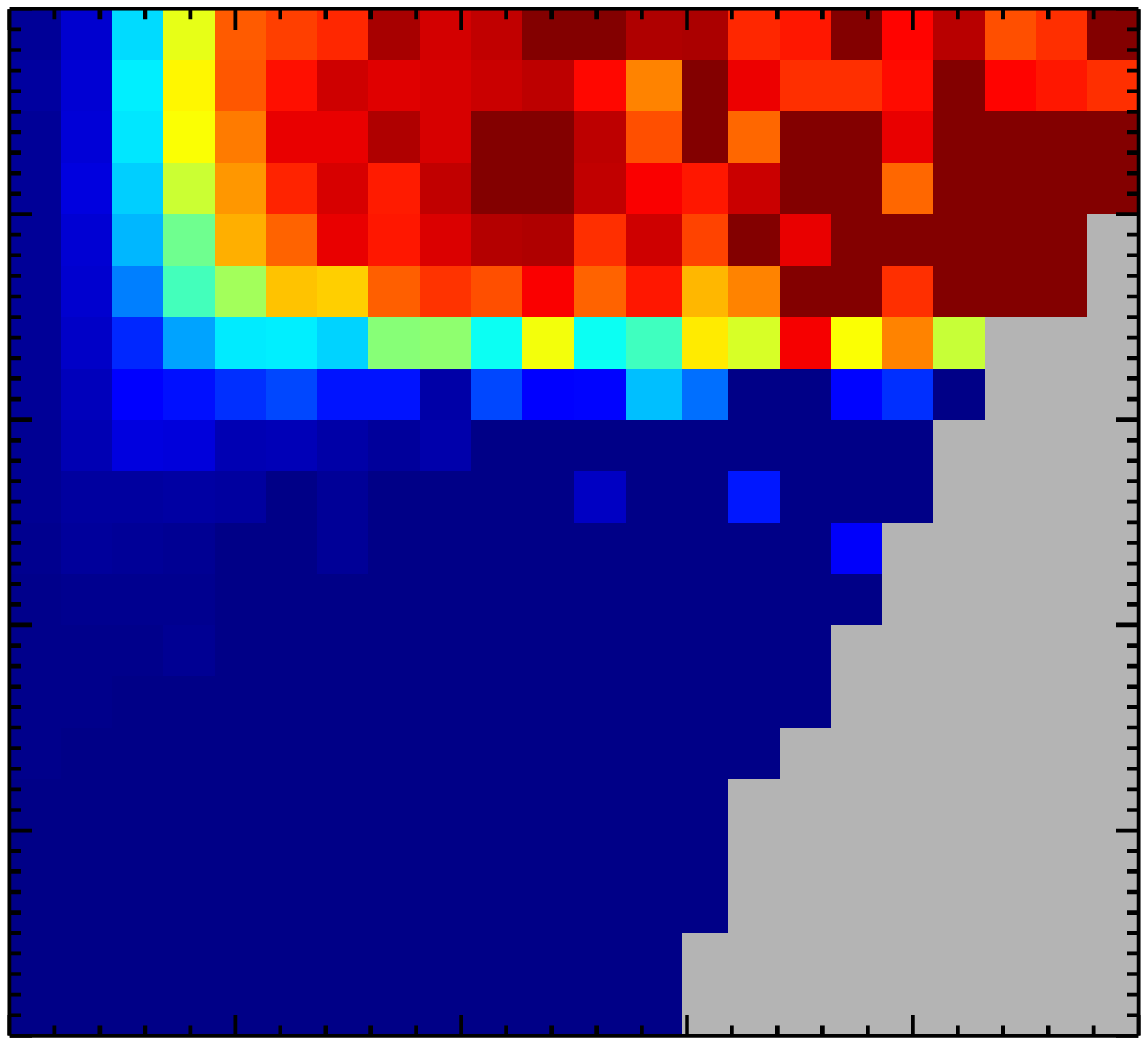}&
\psfrag{------xtitle------}{$ \quad S_{217}^{\rm{I}} $ [Jy]}
\includegraphics[width=0.2\textwidth, viewport=40 20 410 380]{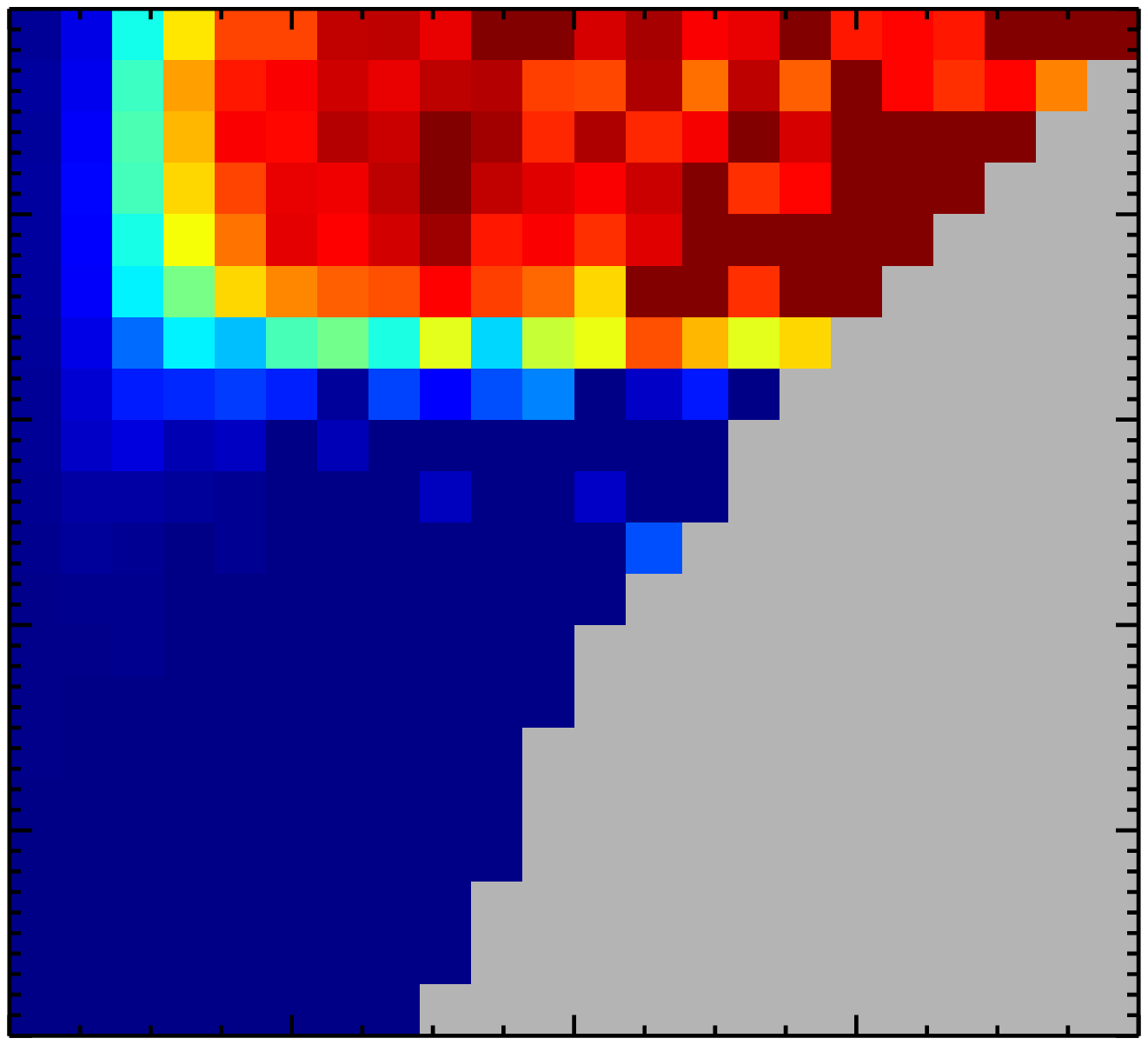}&
\psfrag{----------ytitle----------}{Completeness [\%]}
\includegraphics[width=0.057\textwidth, viewport=470 20 575 380]{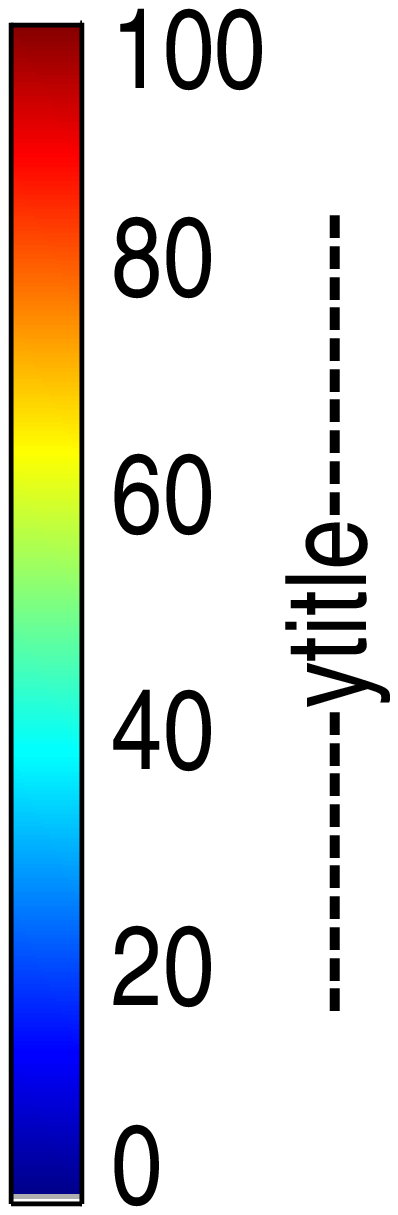}\\
\begin{minipage}[c]{0.03\linewidth} \vspace{-4.5cm} 40\,K \end{minipage}&
\psfrag{------xtitle------}{$ \quad S_{857}^{\rm{I}} $ [Jy]}
\includegraphics[width=0.2\textwidth, viewport=40 20 410 380]{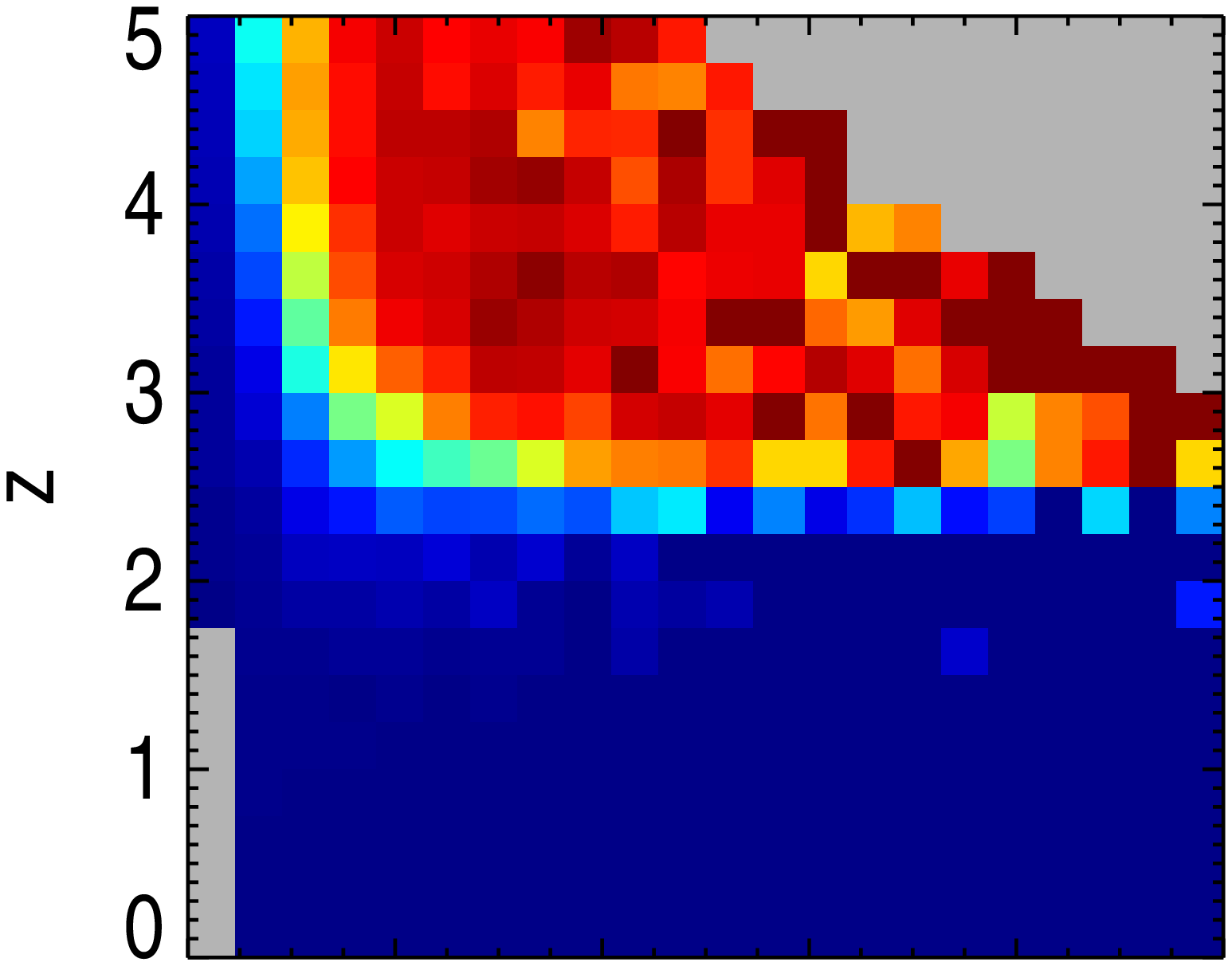}&
\psfrag{------xtitle------}{$ \quad S_{545}^{\rm{I}} $ [Jy]}
\includegraphics[width=0.2\textwidth, viewport=40 20 410 380]{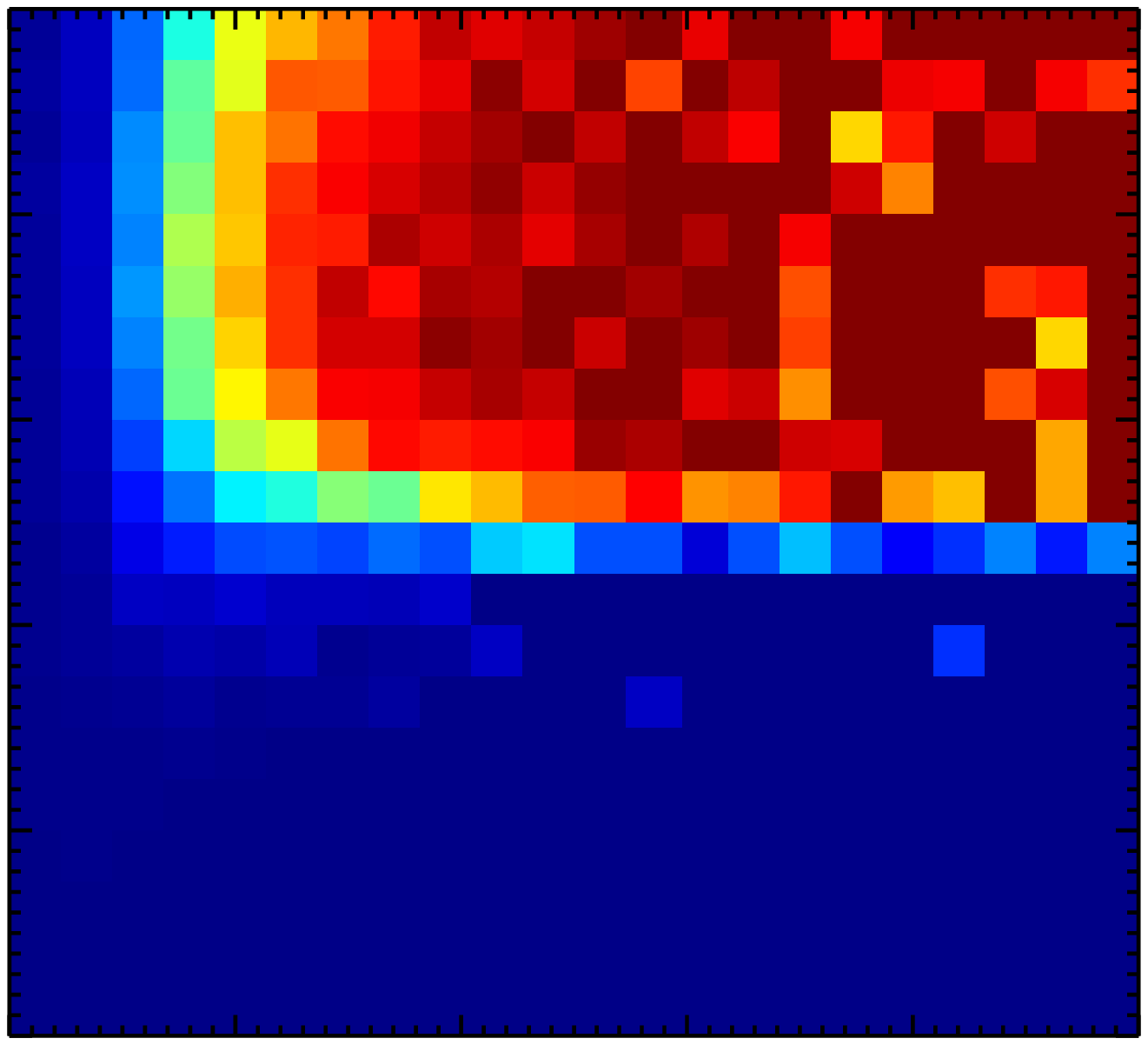}&
\psfrag{------xtitle------}{$ \quad S_{353}^{\rm{I}} $ [Jy]}
\includegraphics[width=0.2\textwidth, viewport=40 20 410 380]{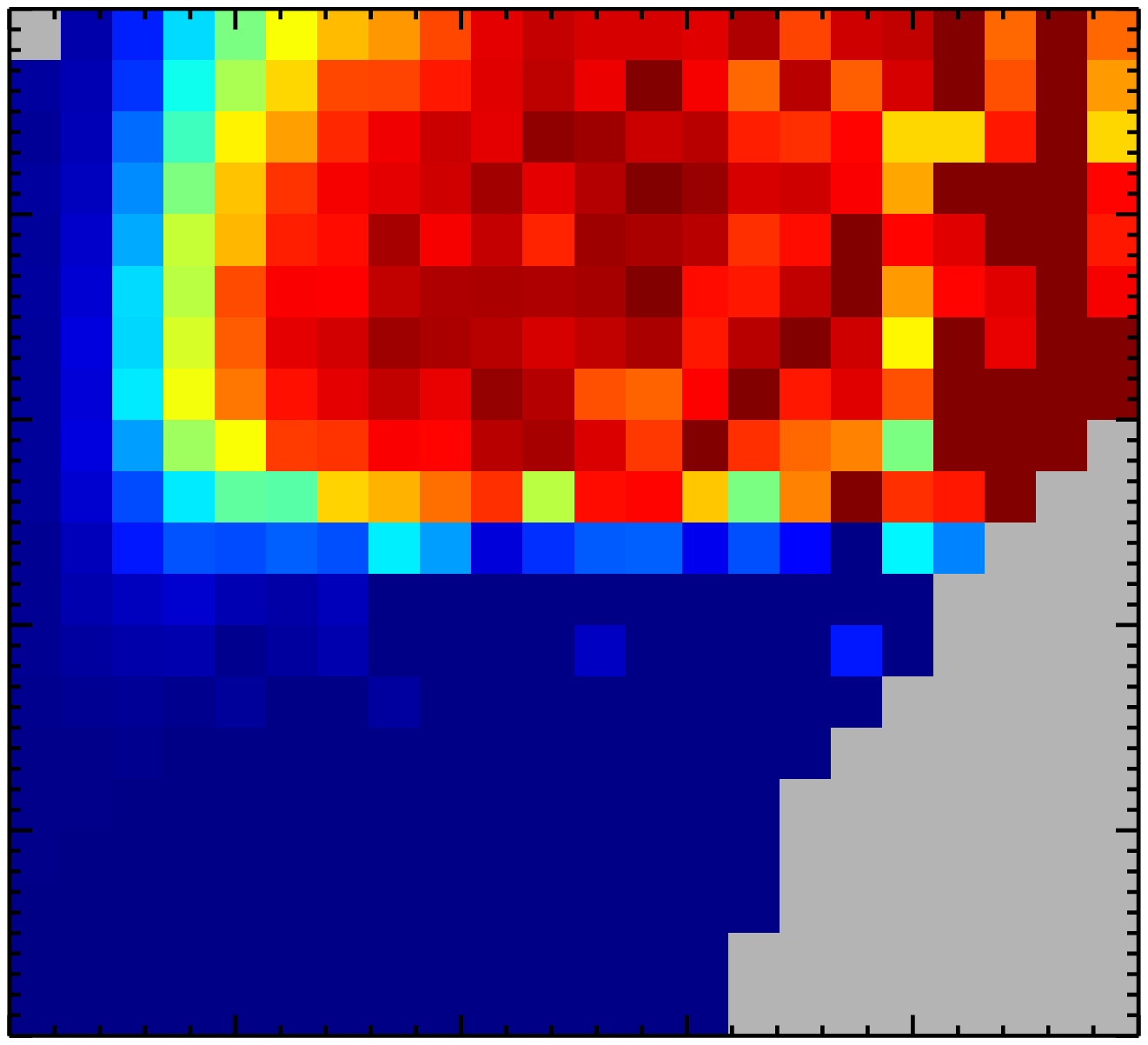}&
\psfrag{------xtitle------}{$ \quad S_{217}^{\rm{I}} $ [Jy]}
\includegraphics[width=0.2\textwidth, viewport=40 20 410 380]{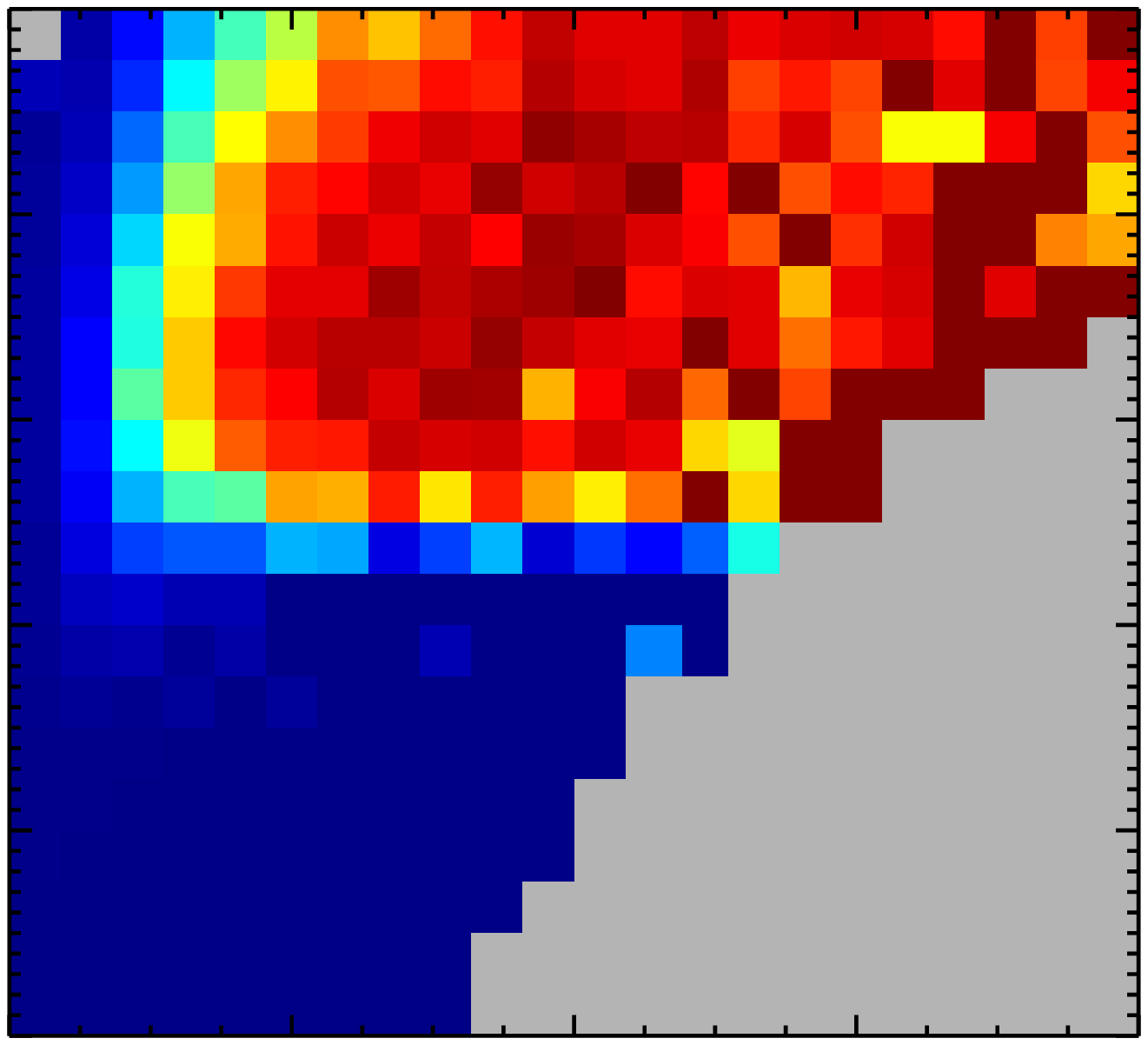}&
\psfrag{----------ytitle----------}{Completeness [\%]}
\includegraphics[width=0.057\textwidth, viewport=470 20 575 380]{mcqa_selection_redshift_bar_v7.0.ps}\\
\begin{minipage}[c]{0.03\linewidth} \vspace{-4.5cm} 30\,K \end{minipage} &
\psfrag{------xtitle------}{$ \quad S_{857}^{\rm{I}} $ [Jy]}
\includegraphics[width=0.2\textwidth, viewport=40 20 410 380]{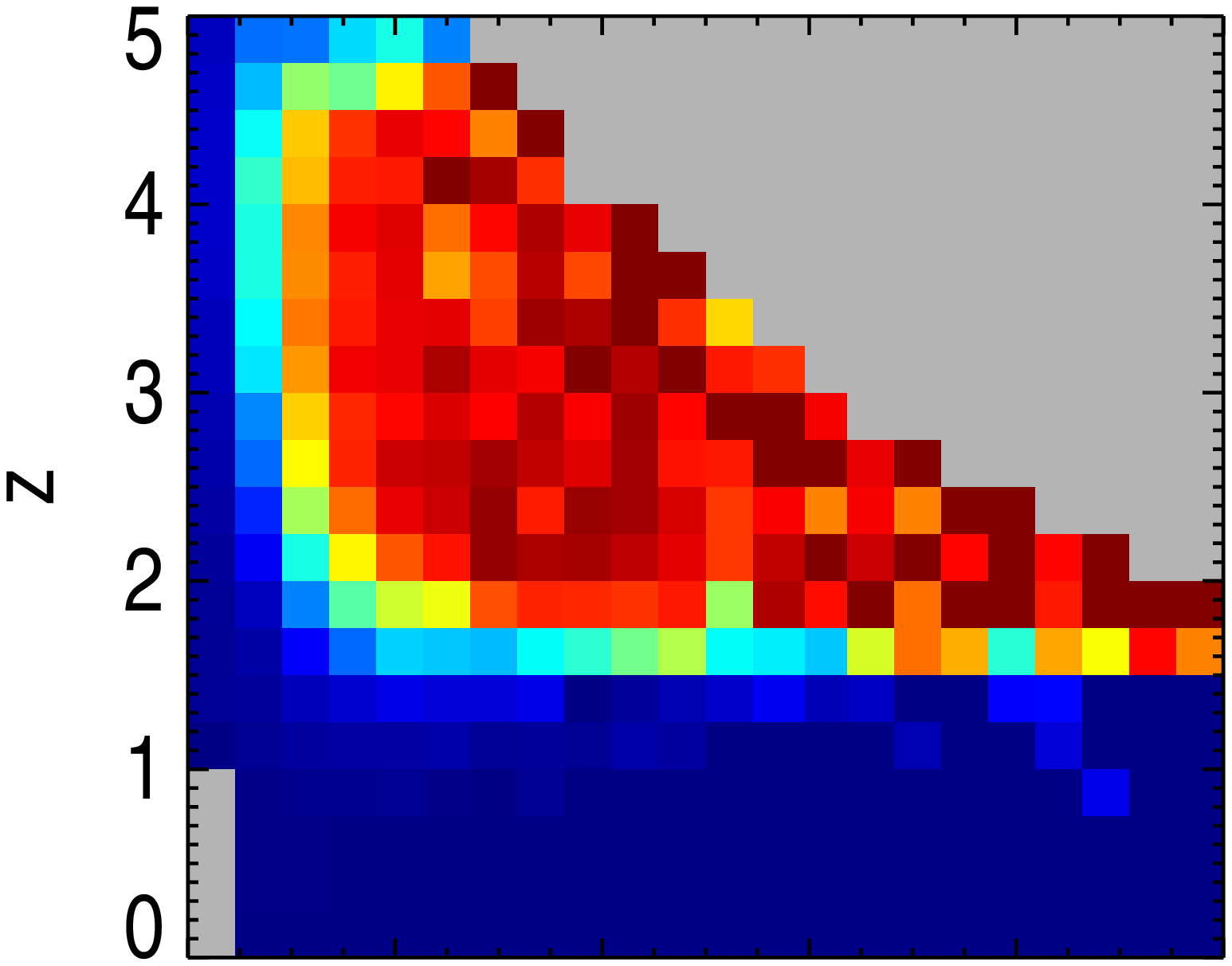}&
\psfrag{------xtitle------}{$ \quad S_{545}^{\rm{I}} $ [Jy]}
\includegraphics[width=0.2\textwidth, viewport=40 20 410 380]{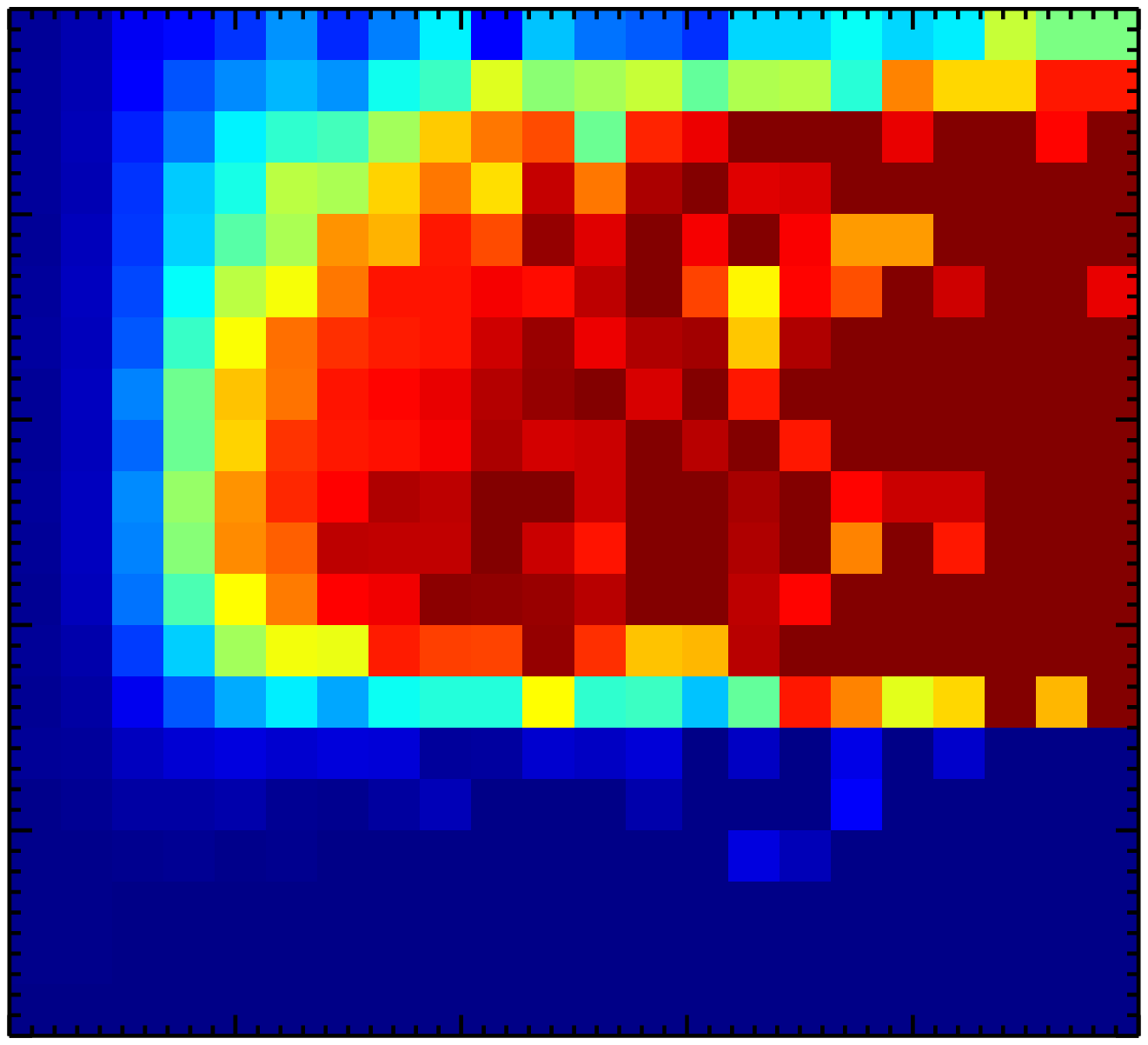}&
\psfrag{------xtitle------}{$ \quad S_{353}^{\rm{I}} $ [Jy]}
\includegraphics[width=0.2\textwidth, viewport=40 20 410 380]{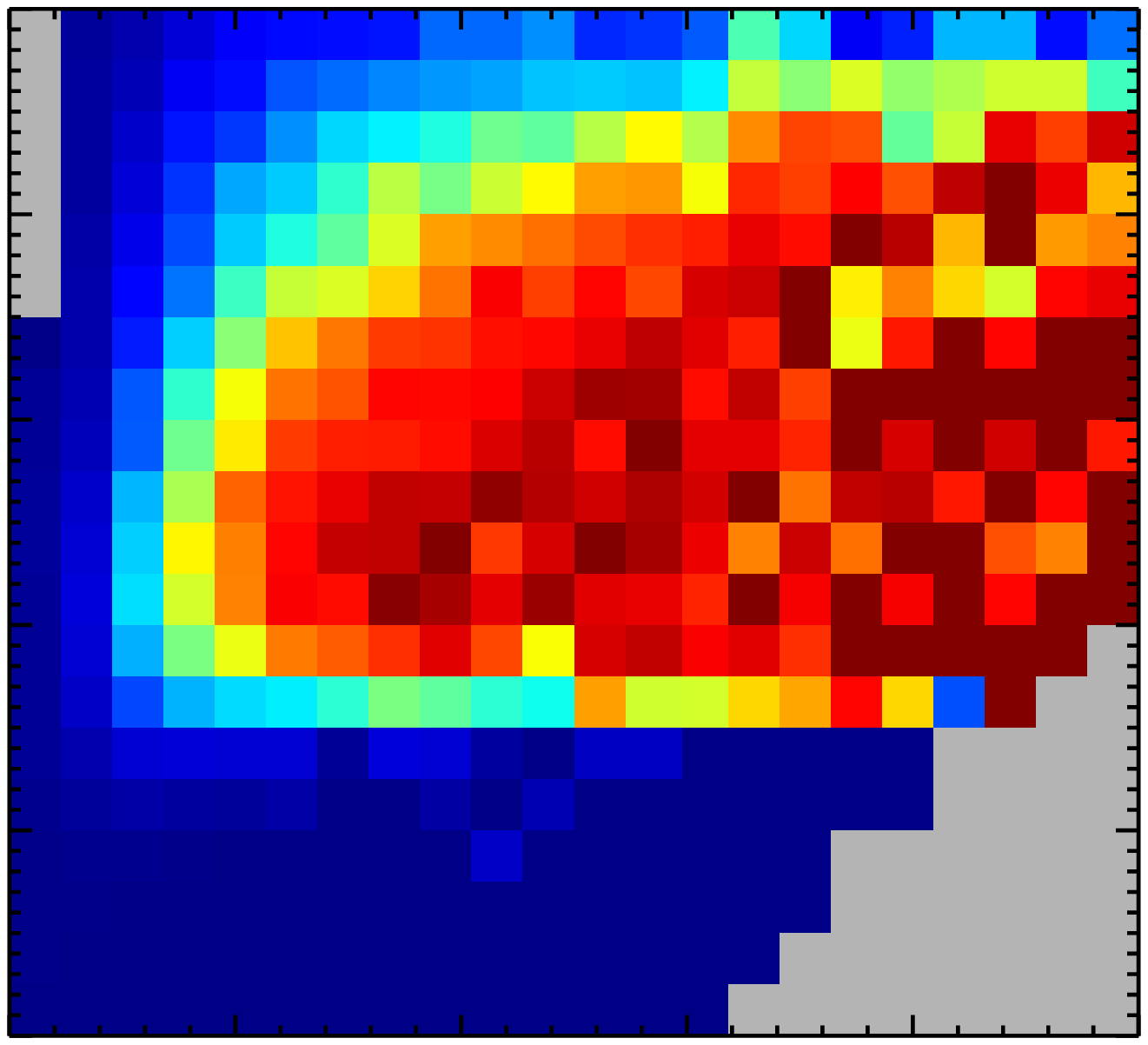}&
\psfrag{------xtitle------}{$ \quad S_{217}^{\rm{I}} $ [Jy]}
\includegraphics[width=0.2\textwidth, viewport=40 20 410 380]{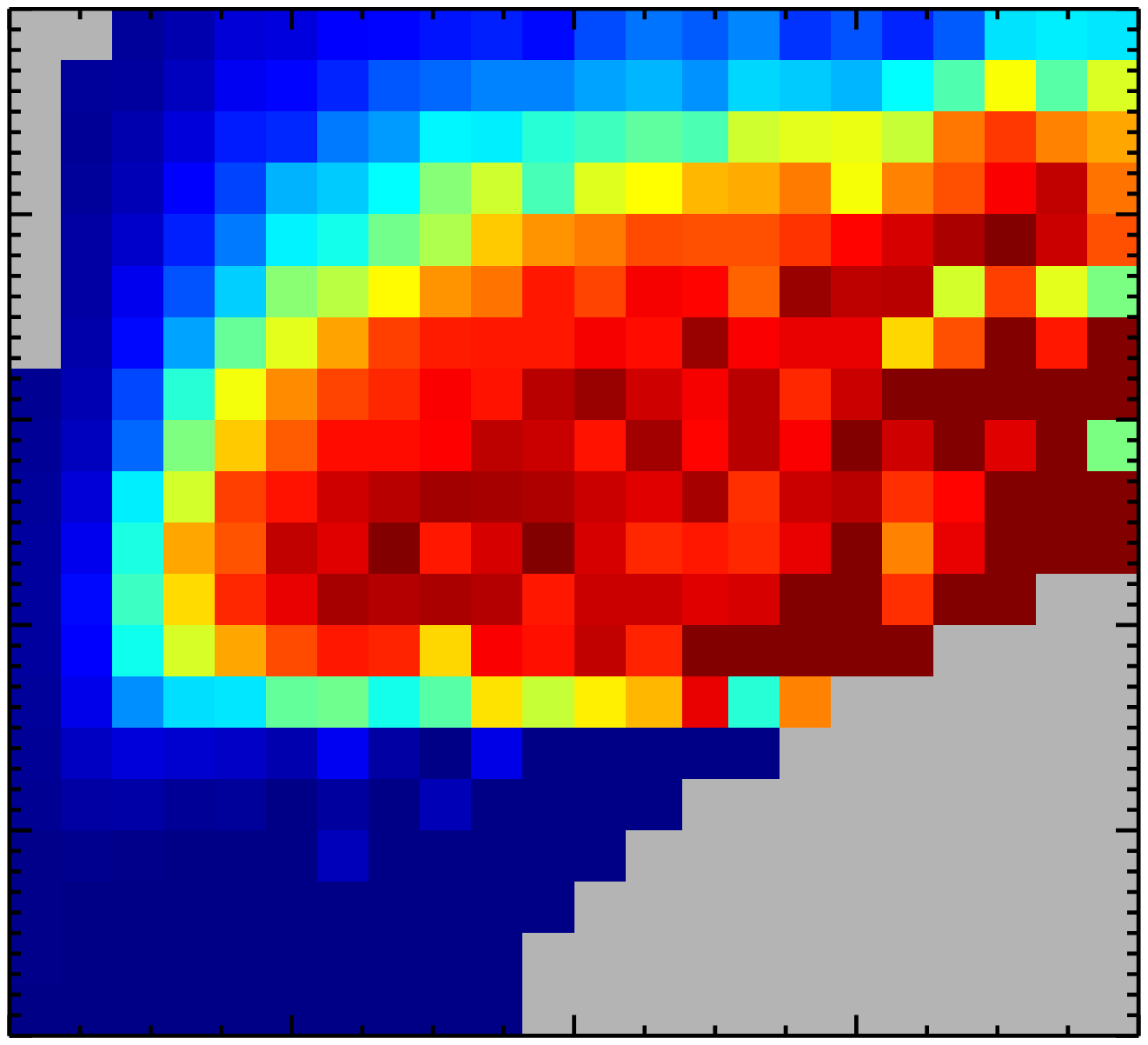}&
\psfrag{----------ytitle----------}{Completeness [\%]}
\includegraphics[width=0.057\textwidth, viewport=470 20 575 380]{mcqa_selection_redshift_bar_v7.0.ps}\\
\begin{minipage}[c]{0.03\linewidth} \vspace{-4.5cm} 20\,K \end{minipage}&
\psfrag{------xtitle------}{$ \quad S_{857}^{\rm{I}} $ [Jy]}
\includegraphics[width=0.2\textwidth, viewport=40 20 410 380]{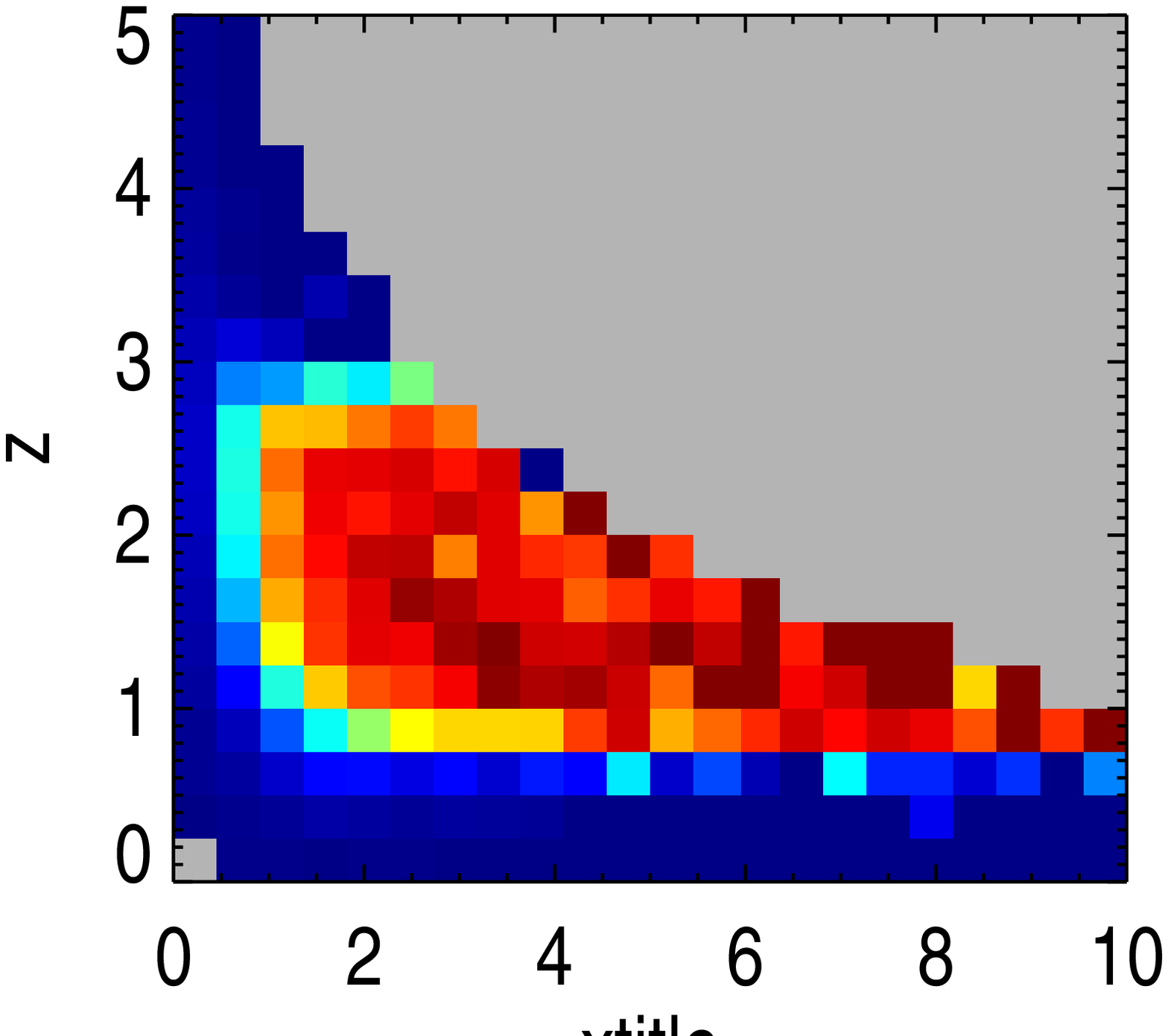}&
\psfrag{------xtitle------}{$ \quad S_{545}^{\rm{I}} $ [Jy]}
\includegraphics[width=0.2\textwidth, viewport=40 20 410 380]{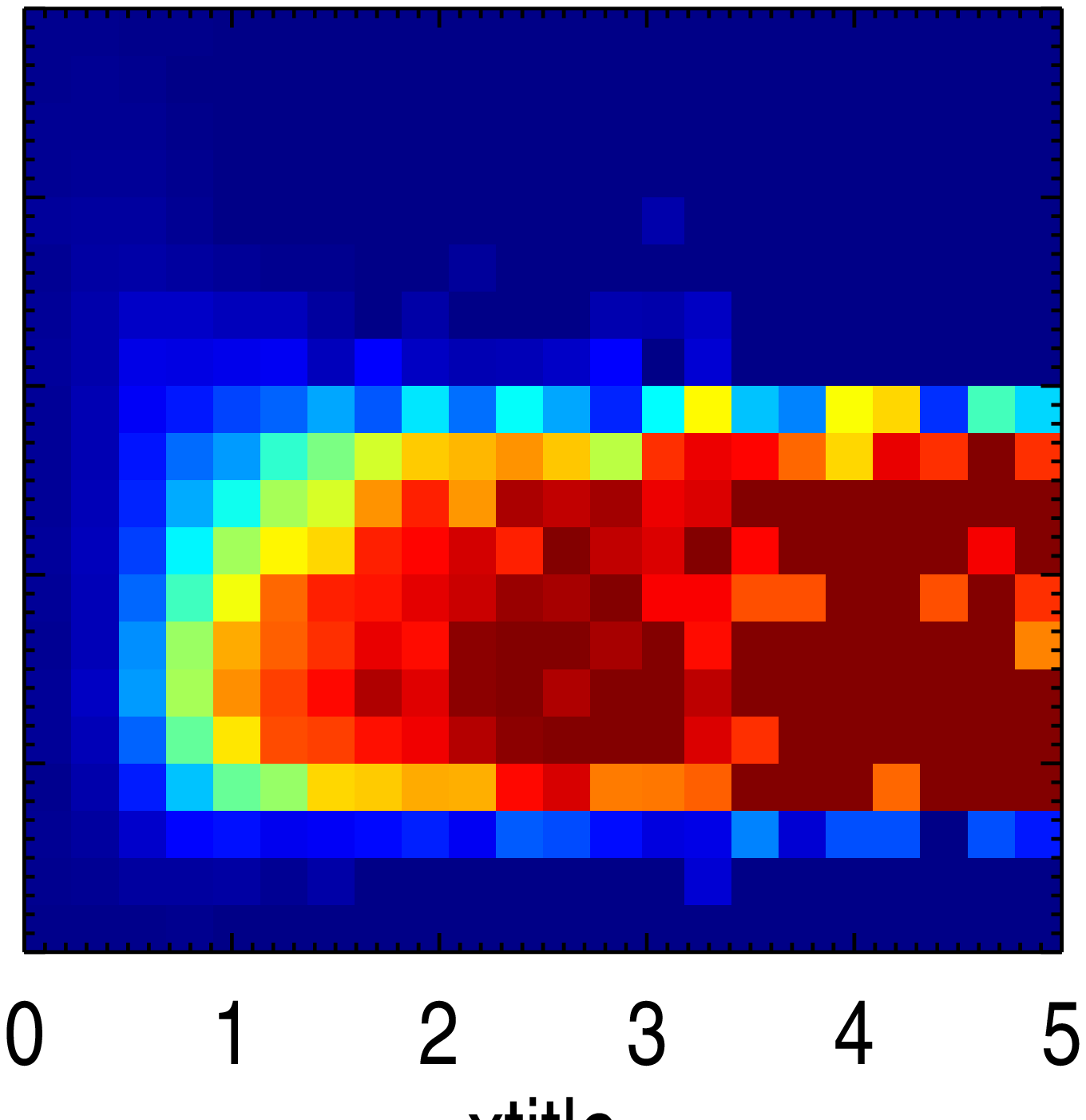}&
\psfrag{------xtitle------}{$ \quad S_{353}^{\rm{I}} $ [Jy]}
\includegraphics[width=0.2\textwidth, viewport=40 20 410 380]{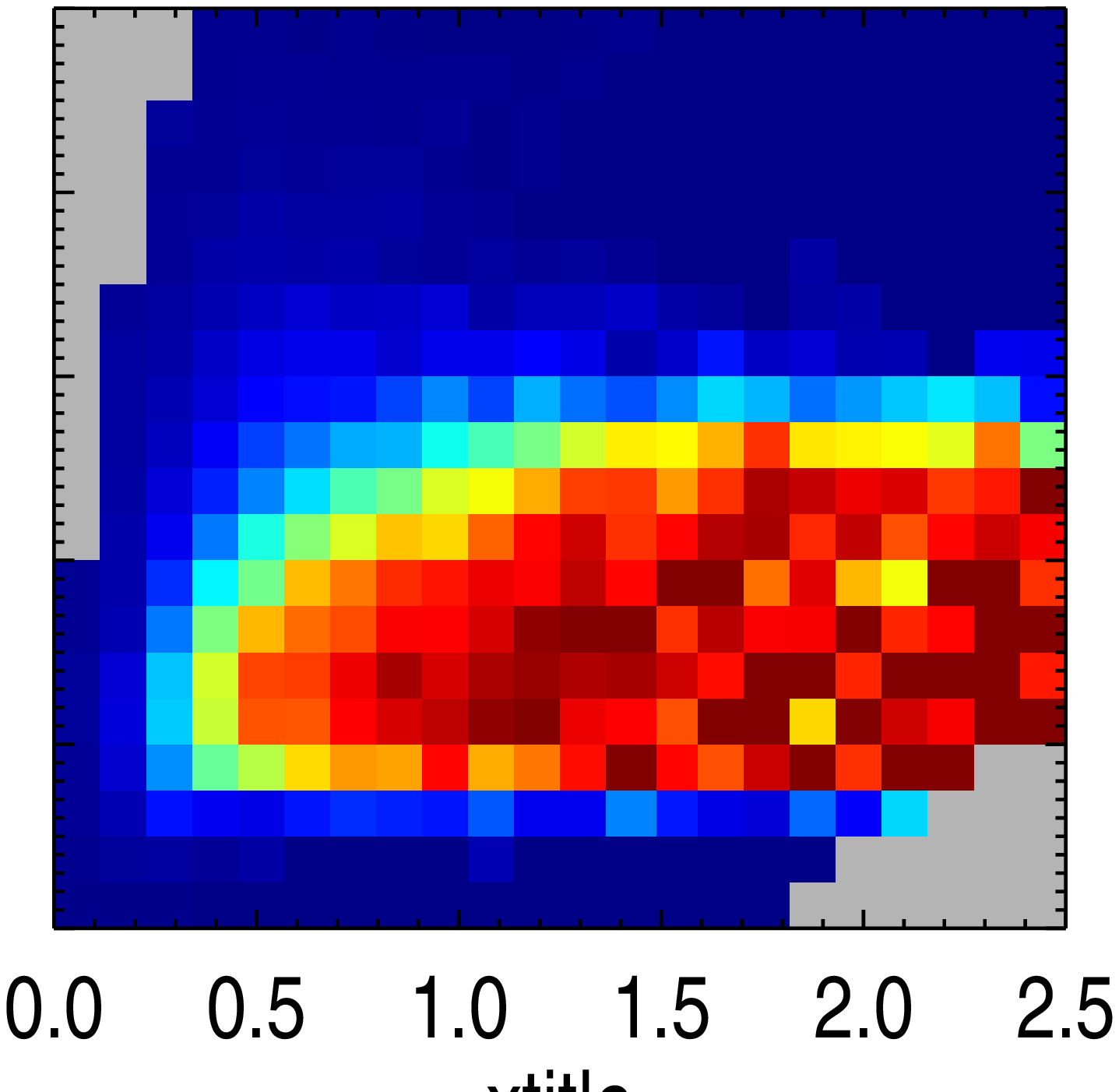}&
\psfrag{------xtitle------}{$ \quad S_{217}^{\rm{I}} $ [Jy]}
\includegraphics[width=0.2\textwidth, viewport=40 20 410 380]{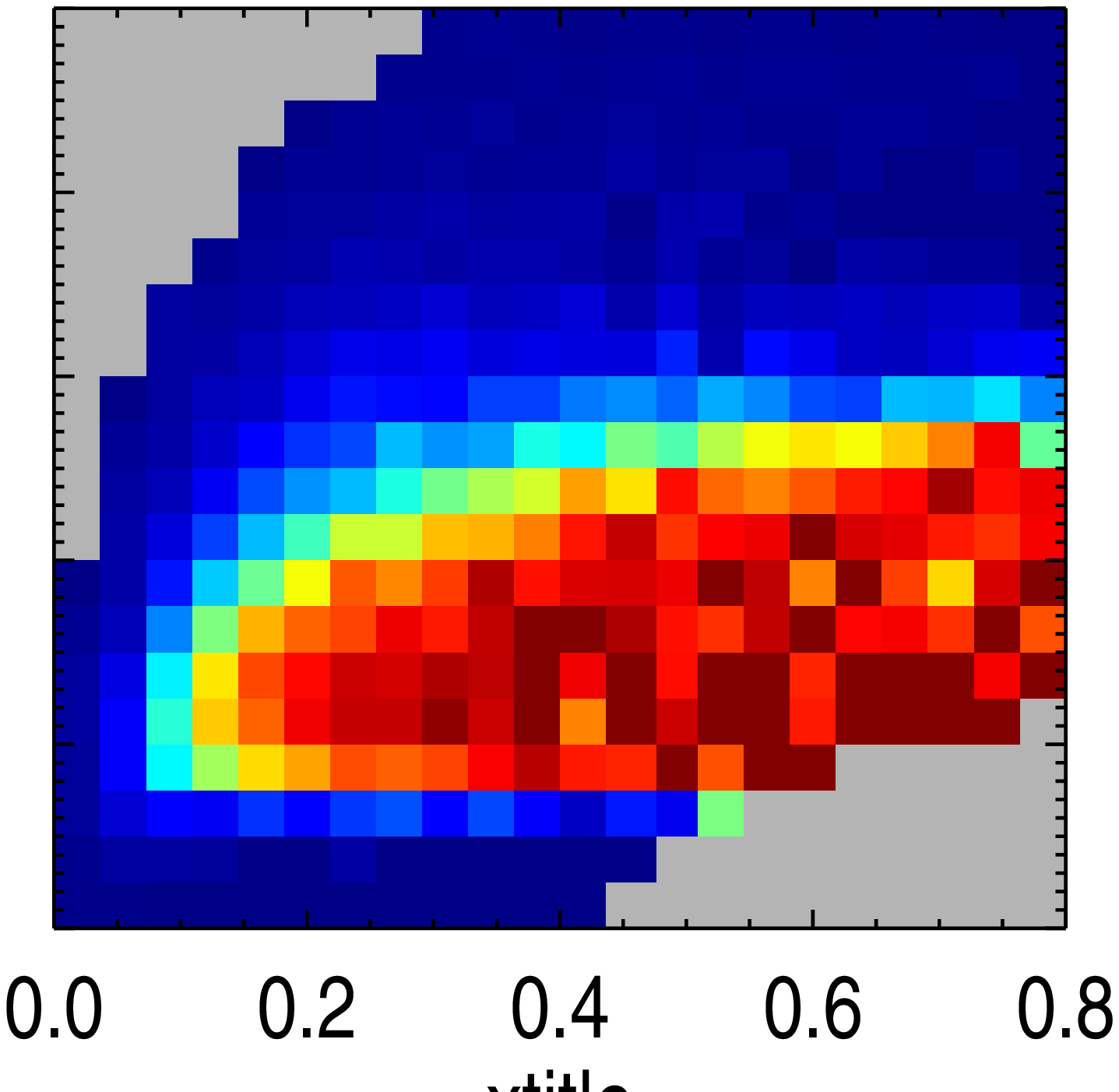}&
\psfrag{----------ytitle----------}{Completeness [\%]}
\includegraphics[width=0.057\textwidth, viewport=470 20 575 380]{mcqa_selection_redshift_bar_v7.0.ps}\\
\end{tabular}
\vspace{0.5cm}
\caption{Completeness as a function of redshift and flux density at 857, 545,
353, and 217\,GHz (from left to right) and for each category of injected
sources with dust temperatures of 20, 30, 40, and 50\,K, from top to bottom,
respectively.  Grey regions are domains without simulated data for these
flux densities and redshifts.}
\label{fig:mcqa_selection_function_redshift_fx}
\end{figure*}

Because of the complex interplay between the attenuation due to the cleaning
process, the geometry recovery, and the flux boosting effect, any simple
Bayesian approach for flux de-boosting would be difficult to implement.
For this reason, the flux density estimates of the \Planck\ high-$z$
candidates presented in this work are not corrected for flux boosting or
cleaning attenuation. However, in order to minimize the impact of flux
boosting when building the final list, we will apply a minimal threshold on
the 545\,GHz flux density estimates, which has been set to 
500\,mJy, as determined through these simulations.

\subsection{Colour selection accuracy}
\label{sec:mcqa_colour}

It is important to notice that the colour ratios of the detected sources 
are relatively well preserved by the cleaning and photometry processing,
which is crucial to ensure the quality of the colour-colour selection of
these high-$z$ candidates.  The dependence with the S/N of the detection in
the excess map of the ratio between the recovered to input colour ratios is
shown for both $S_{353}/S_{545}$ and $S_{545}/S_{857}$ in the left panels
of Fig.~\ref{fig:mcqa_accuracy_col_col}.
Note that for this analysis we include all the sources detected before
applying any colour-colour selection, in order to assess the robustness
of the latter selection.  Again, in this analysis, the ideal and highly
contaminated cases of the CMB template are explored.
 
When assuming an ideal CMB template, the recovered
$S_{353}^{\rm{D}}/S_{545}^{\rm{D}}$ ratio (top left panel of
Fig.~\ref{fig:mcqa_accuracy_redshift}) is unbiased on average for S/N larger
than 15 when compared to the injected values.  More precisely, when looking
at the recovered versus injected trend (bottom left panel), 
it appears that the higher the $S_{353}^{\rm{I}}/S_{545}^{\rm{I}}$ ratio,
the more underestimated the output colour, so that the recovered
$S_{353}^{\rm{D}}/S_{545}^{\rm{D}}$ ratio always remains below 1 (within
1$\,\sigma$) for input ratios $S_{353}^{\rm{I}}/S_{545}^{\rm{I}}<1$.
The case is even worse when assuming a highly-contaminated CMB template,
yielding an underestimate of the recovered $S_{353}^{\rm{D}}/S_{545}^{\rm{D}}$
ratio by 17\,\% to 7\,\%, from low to high S/N.  This is well explained by
the attenuation coefficient, which may differ between the 545- and 353-GHz
bands. This effect has been taken into account when setting the colour-colour
criteria in Sect.~\ref{sec:ps_detection_colcol} in a conservative way.

The recovered $S_{545}^{\rm{D}}/S_{857}^{\rm{D}}$ ratio (right panels of
Fig.~\ref{fig:mcqa_accuracy_redshift}) 
does not appear as strongly biased on average, but is still underestimated for high $S_{545}^{\rm{I}}/S_{857}^{\rm{I}}$ inputs ($> 0.8$); 
this does not impact the overall colour-colour selection, since in this case the recovered ratio still satisfies the selection criterion ($>0.5$).

We have also used these Monte Carlo simulations to check the accuracy of the colour-colour selection process.
The probability $\mathcal{P}$, introduced in Sect.~\ref{sec:ps_detection_colcol}, and based on the recovered colour ratios 
$S_{545}^{\rm{D}}/S_{857}^{\rm{D}}$ and $S_{353}^{\rm{D}}/S_{545}^{\rm{D}}$, has been compared to the
exact probability that the input colour ratios $S_{545}^{\rm{I}}/S_{857}^{\rm{I}}$ and $S_{353}^{\rm{I}}/S_{545}^{\rm{I}}$
satisfy the colour criteria. Hence requesting a probability of 0.84 to
find the true colour values inside the expected colour-colour domain
(which is equivalent to a 1$\,\sigma$ constraint on a half-bounded domain),
gives a minimal threshold of $\mathcal{P} > 0.9$ based on the recovered
colour values.  This is what has been applied to build the official list.

\subsection{Selection function}
\label{sec:mcqa_selection}

We now focus on the sample of detected sources, 
obtained after applying the S/N criteria in all bands and the colour-colour criteria of Sect.~\ref{sec:ps_detection_colcol},
in agreement with the criteria used for the true extraction.
This allows us to quantify the selection function of our detection algorithm by 
computing the completeness of the detected sources as a function of redshift, extinction, and flux density. 
Here we define the completeness as the ratio between the initial number of injected sources and
 the number of detected sources in the same bin for a given property.

In Fig.~\ref{fig:mcqa_selection_function_redshift_fx} the completeness is
presented as a function of both redshift and input flux density in all
\Planck\ bands for each category of dust temperature of the extragalactic
source, $T_{\rm{xgal}}$.  Of course, the completeness is highly dependent on
the input temperature of the extragalactic source ($T_{\rm{xgal}}$), because
of the temperature-redshift degeneracy.  Sources with a high temperature
(50\,K) are only detected when located at high redshift ($>3$), while
sources with a low temperature (20\,K) can be detected up to redshift $z=1$.
To solve for this well known degeneracy, \citet{Greve2012} have used a prior
on the temperature built on a sample of 58 unlensed and 14 lensed high-$z$
submm sources.  They state that the median temperatures of the unlensed and
lensed population of sources at $z>1$ are $T_{\rm{xgal}}=34$\,K and
$T_{\rm{xgal}}=46$\,K, respectively, and range from 15 to 80\,K, and
30 to 80\,K, respectively.  Studies have shown similar ranges of temperature
with {\it Herschel}, SCUBA-2 ad other instruments \citep{Chapin2009,Chapin2011,
Chapman2010, Magnelli2012, Symeonidis2013, Swinbank2014}.
As a confirmation, the mean temperature of the dusty star forming galaxies
discovered by SPT and confirmed with ALMA observations as strongly lensed
sources has been estimated at $T_{\rm{xgal}}=38$\,K \citep{Weiss2013}.
Furthermore, first confirmations of sources of this list have shown median
temperature of 44\,K for lensed candidates \citep{Canameras2015}, and 32\,K
for proto-cluster candidates \citep{Florescacho2015}.
However, we have to keep in mind that this degeneracy cannot be broken for
all other sources of this list without any direct measurement of the redshift.

The completeness exhibits a very sharp cut-off on the lowest redshift 
side (e.g., at $z>1.5$ for $T_{\rm{xgal}}=30$\,K), dropping suddenly
to zero below this limit.  On the high redshift side, after a plateau, it
goes back smoothly to zero, because of the impact of the attenuation due to
the cleaning, which becomes more and more important with higher redshifts.
Focusing again on the $T_{\rm{xgal}}=30$\,K case, the completeness reaches
100\,\% for strong sources ($S_{545}>3$\,Jy) and $2<z<3$.
However, the completeness drops quickly for fainter sources, reaching a
maximum of about 50\,\% at a flux density of 700\,mJy and redshifts between
1 and 3.  Our detection method therefore operates as a filter in redshift by
selecting sources peaking in the submm range.  For an average dust temperature
of $T_{\rm{xgal}}=30$\,K, this redshift window ranges from about 1.5 to 4.5.

\begin{figure}[t]
\vspace{-0.5cm}
\hspace{-0.7cm}
\psfrag{----xtitle----}{$E(B-V)_{\rm{xgal}}$}
\includegraphics[width=0.55\textwidth]{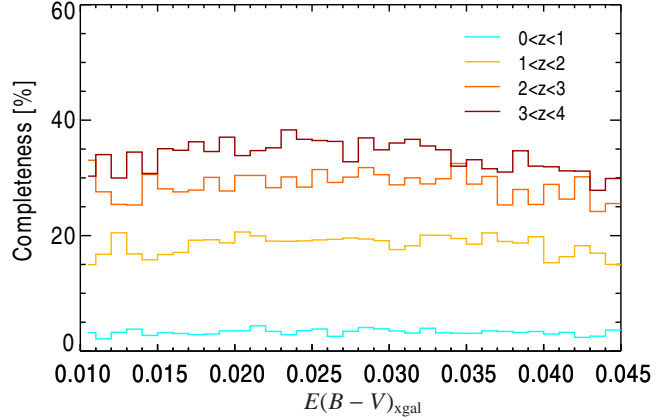}
\caption{Completeness as a function of extinction. This is shown for four bins
of redshift: $0<z<1$ (blue); $1<z<2$ (yellow); $2<z<3$ (orange); and $3<z<4$
(brown).  Only sources with input flux densities
$S_{545}^{\rm{I}}>0.5\,\rm{Jy}$ have been taken into account in this analysis.}
\label{fig:mcqa_selection_function_ebv}
\end{figure}

Finally, as shown in Fig.~\ref{fig:mcqa_selection_function_ebv}, there is no
dependence of the completeness on extinction, which implies that the cleaning
method and the presence of Galactic structures do not affect the
ability of the detection algorithm to extract high-$z$ candidates (at least
over the cleanest 26\,\% of the sky).  This does not prevent the possible
presence of some spurious detections due to Galactic cirrus, which can be
addressed by looking at the $\rm {H}_{\rm{I}}$ column density, as discussed
in Sect.~\ref{sec:ancillary_internal_xcat}.

\begin{figure*}
\center
\vspace{-0.4cm}
\includegraphics[width=0.55\textwidth, angle=90,viewport=50 0 650 500]{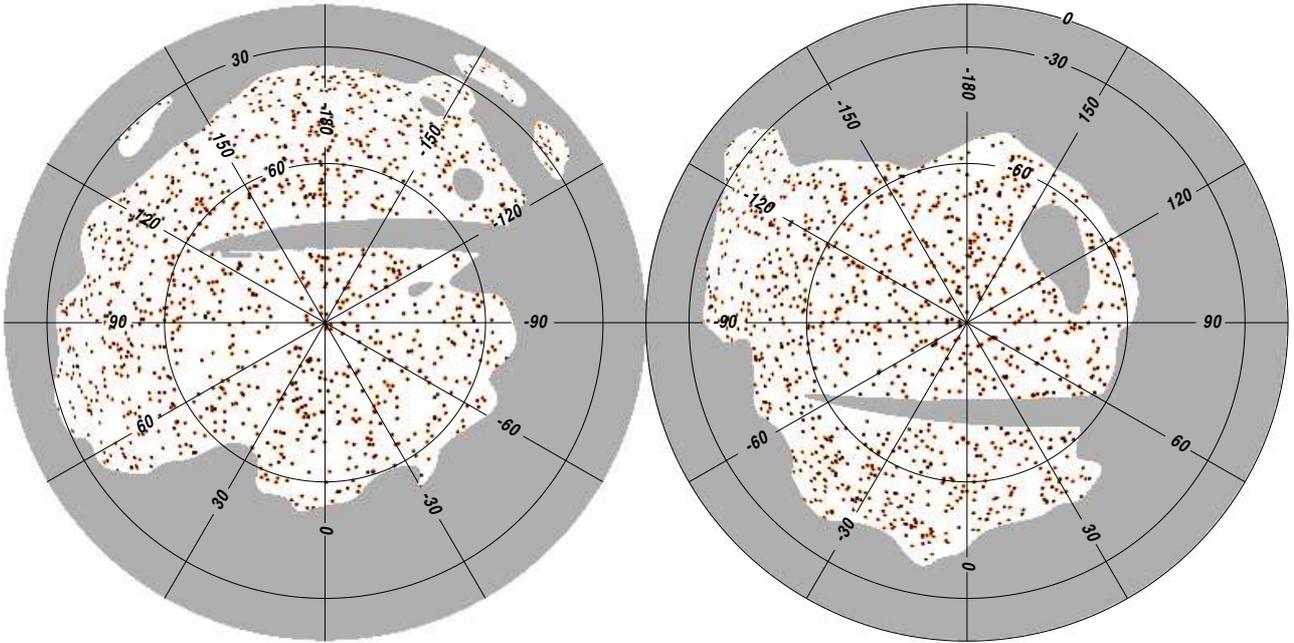} 
\caption{Distribution of the 2151 \Planck\ high-$z$ source candidates over the whole
sky in Galactic coordinates and orthographic projections.}
\label{fig:allsky_PHZ}
\end{figure*}

\section{The PHZ}
\label{sec:content}

\subsection{Building the source list}

The full procedure of CMB and Galactic cirrus cleaning is performed on the
set of \Planck\ and IRAS data, enabling us to build the 545\,GHz excess map
on which the detection criterion $S_{545}^{\rm{X}}/\sigma_{545}^{\rm{X}}>5$
is applied, combined with the requirements
$S_{\nu}^{\rm{D}}/\sigma_{\nu}^{\rm{D}}>3$ in all 857-, 545-, and
353-GHz-cleaned maps simultaneously, and the requirement
$S_{100}^{\rm{C}}/ \sigma_{100}^{\rm{C}}<3$ to reject contamination by radio
sources.  This yields a first sample of 9052 source candidates for which the
photometry in the 857-, 545-, and 353-GHz-cleaned maps
 is computed with associated uncertainties due to noise measurement and CIB
confusion.  Notice that 44 sources have been rejected during this first step
because of their clear detection at 100\,GHz, confirming
the possible contamination by radio sources as discussed in
Sect.~\ref{sec:compsep_radio}.  The colour-colour selection is performed
by requiring a probability of 90\,\% to satisfy both colour criteria,
$S_{545}^{\rm{D}}/S_{857}^{\rm{D}}>0.5$ and
$S_{353}^{\rm{D}}/S_{545}^{\rm{D}}<0.9$.
In addition to the colour-colour selection, we also apply a cut in flux
density, $S_{545}>500~\mathrm{mJy}$, to ensure a minimum bias due to the
flux boosting effect, following the prescriptions motivated by the numerical
simulations detailed in Sect.~\ref{sec:mcqa_photometry}. This leads to a
final number of 2151 high-$z$ source candidates present
in the \Planck\ List of High-redshift Source Candidates (PHZ).

The all-sky distribution of the PHZ sources is shown in
Fig.~\ref{fig:allsky_PHZ}, where it can be seen that they span the whole
northern and southern caps. The distribution shown does not exhibit any
evidence of contamination by the extended Galactic structures.

A full description of the content of the PHZ is given in
Table~\ref{tab:PHZ_listing}.  We stress that the flux densities provided in
this list have been obtained on the cleaned maps and may be strongly affected
by attenuation due to the cleaning process, depending on their SED type and
redshift, which are still unknown. For this reason
these flux density estimates have to be taken with some caution.
In order to help the user to assess the reliability of the PHZ sources,
we also provide cutouts ($1^{\circ}\times1^{\circ}$) 
of the excess map at 545\,GHz and the cleaned maps at 857, 545, 353, and
217\,GHz, available soon through the Planck Legacy
Archive\footnote{\url{http://www.cosmos.esa.int/web/planck/pla}} and the {\tt MuFFInS}\footnote{\url{http://muffins.irap.omp.eu}} ({\tt Mu}lti {\tt F}requency {\tt F}ollow-up {\tt In}ventory {\tt S}ervice) portal.

\subsection{Statistical description}
\label{sec:cat_statistics}

\begin{figure*}
\begin{tabular}{cccc}
\psfrag{------xtitle------}{$\quad$ S/N}
\psfrag{yy}{$N$}
\psfrag{legend4}{$S_{545}^{\rm{X}}/\sigma_{545}^{\rm{X}}$}
\psfrag{legend1}{$S_{857}^{\rm{D}}/\sigma_{857}^{\rm{D}}$}
\psfrag{legend2}{$S_{545}^{\rm{D}}/\sigma_{545}^{\rm{D}}$}
\psfrag{legend3}{$S_{353}^{\rm{D}}/\sigma_{353}^{\rm{D}}$}
\includegraphics[width=0.21\textwidth, viewport=60 0 470 500]{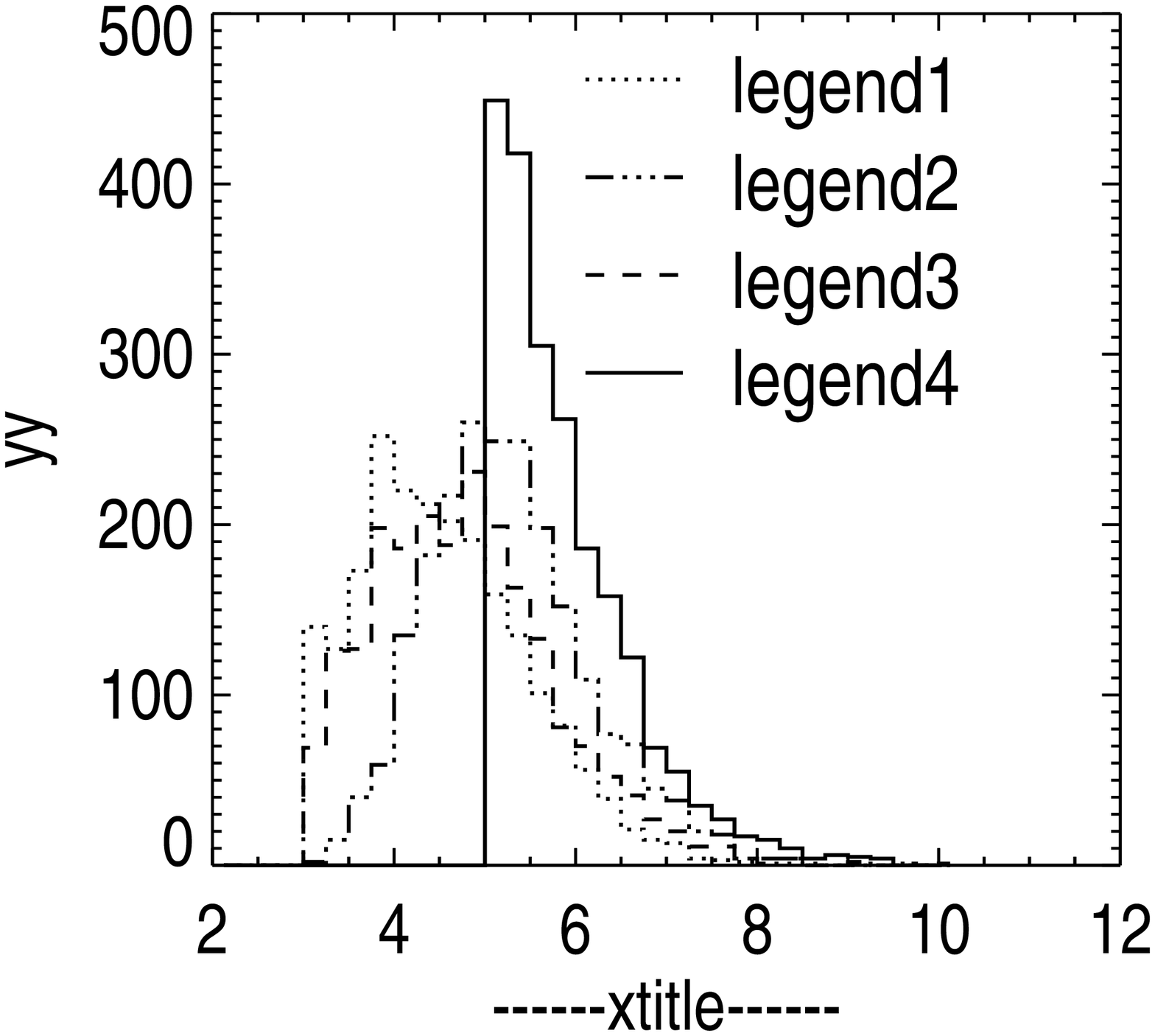}&
\includegraphics[width=0.21\textwidth, viewport=60 0 470 500]{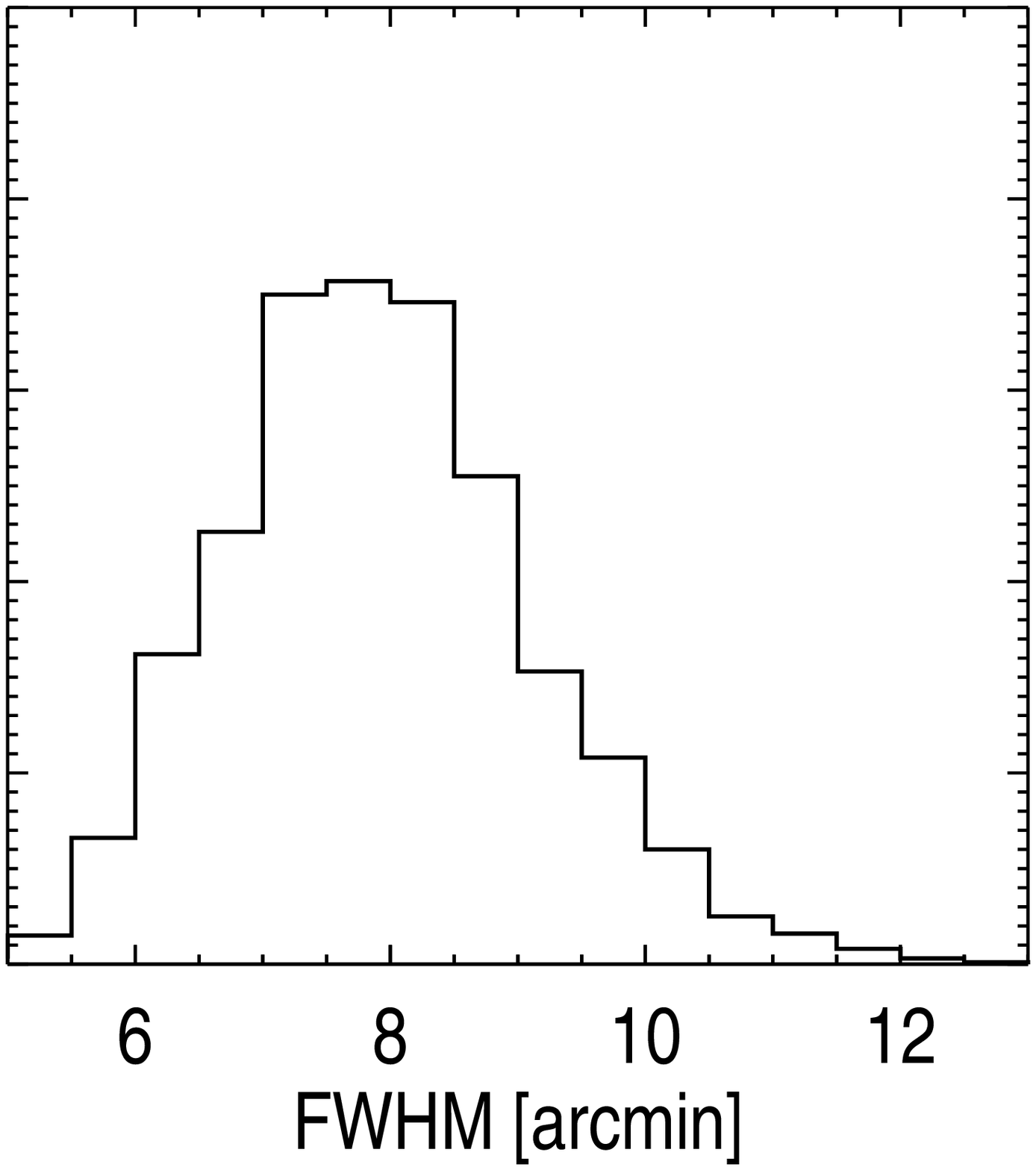}&
\includegraphics[width=0.21\textwidth, viewport=60 0 470 500]{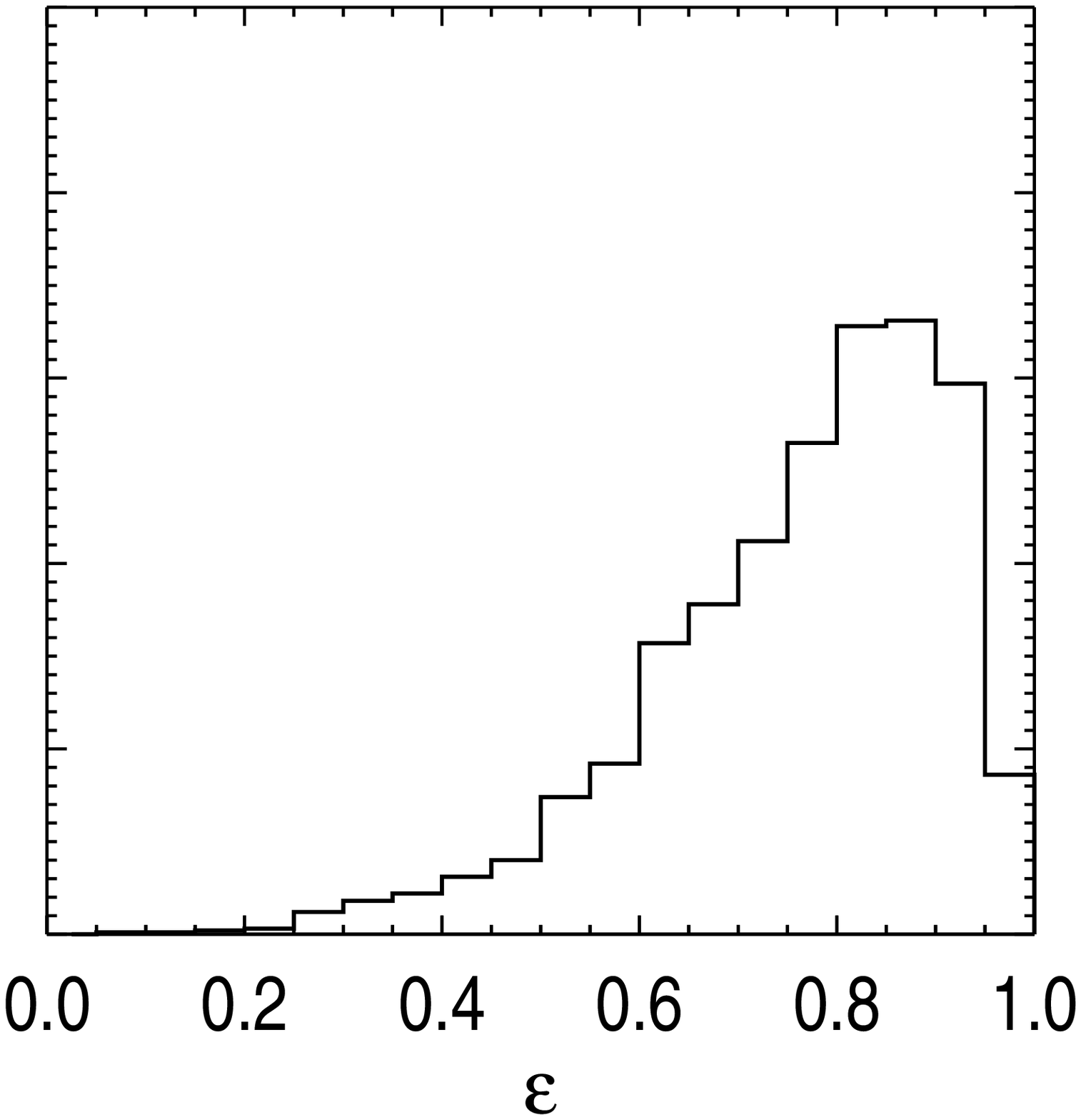}&
\psfrag{------------xtitle------------}{$E(B-V)_{\rm{xgal}}$ ($\times 10^2$)}
\includegraphics[width=0.21\textwidth, viewport=60 0 470 500]{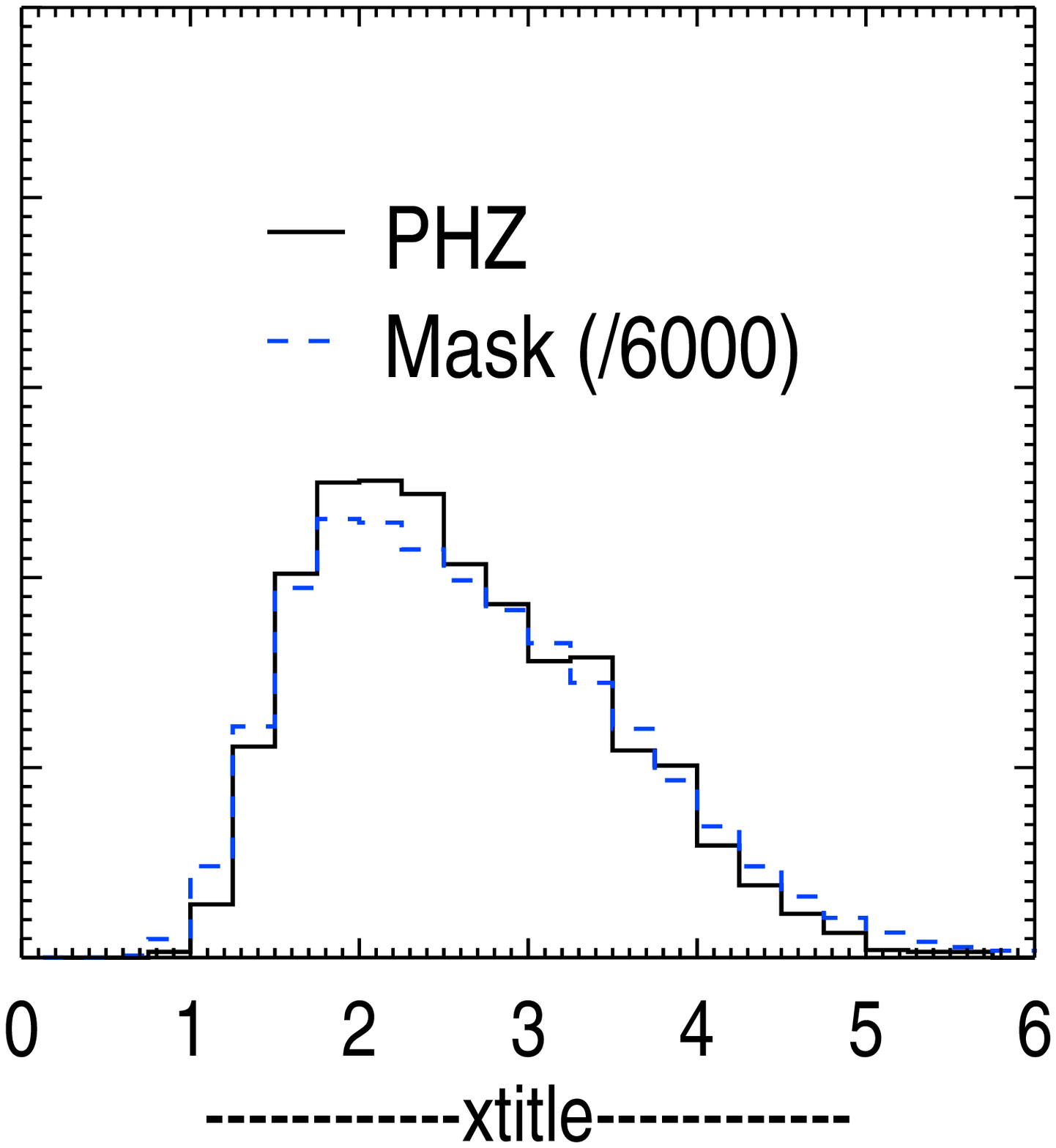} \\
\psfrag{---------xtitle---------}{$\quad \quad S_{857}^{\rm{D}}$}
\psfrag{yy}{$N$}
\includegraphics[width=0.21\textwidth, viewport=60 0 470 500]{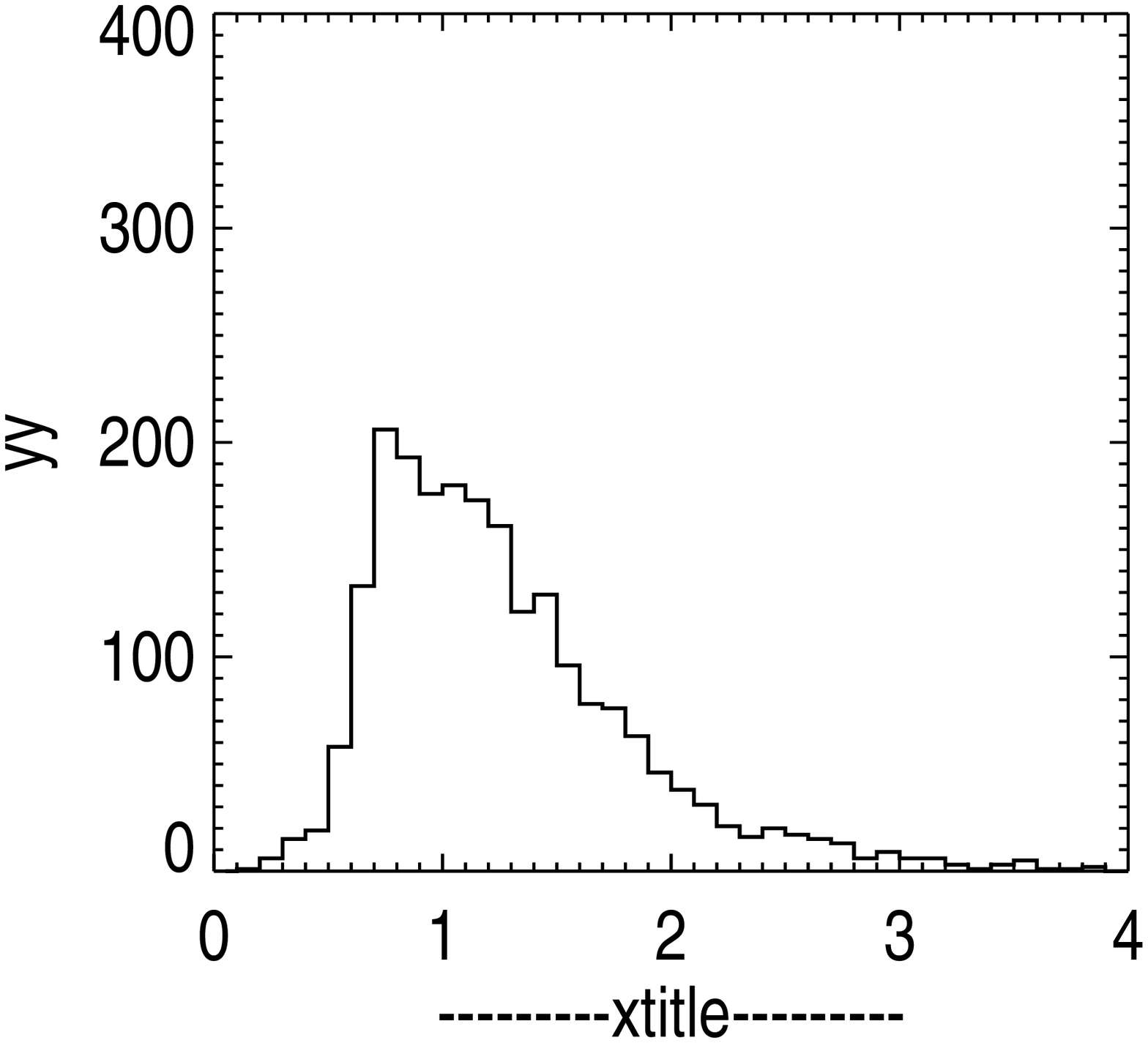}&
\psfrag{---------xtitle---------}{$\quad \quad S_{545}^{\rm{D}}$}
\includegraphics[width=0.21\textwidth, viewport=60 0 470 500]{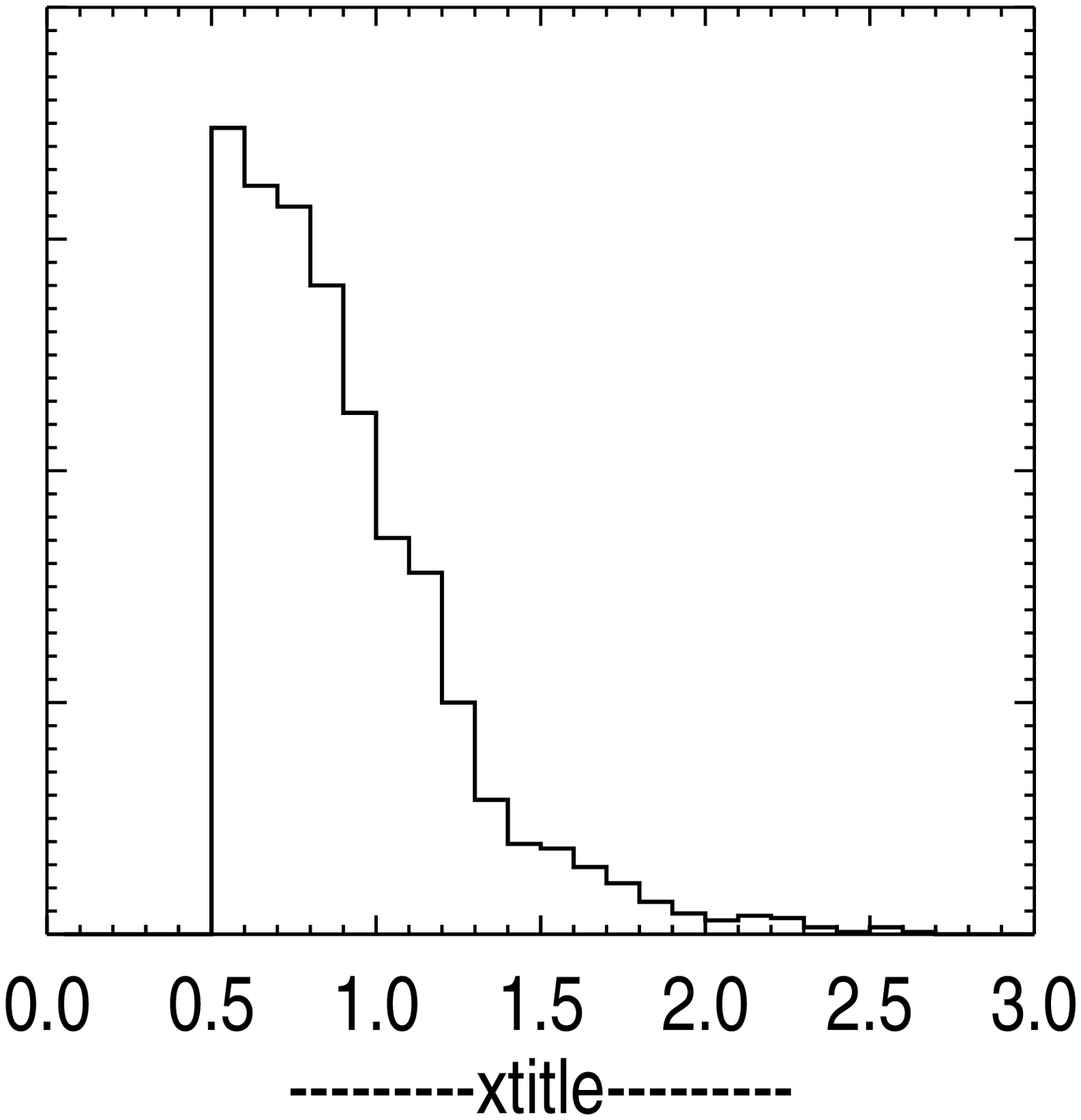}&
\psfrag{---------xtitle---------}{$\quad \quad S_{353}^{\rm{D}}$}
\includegraphics[width=0.21\textwidth, viewport=60 0 470 500]{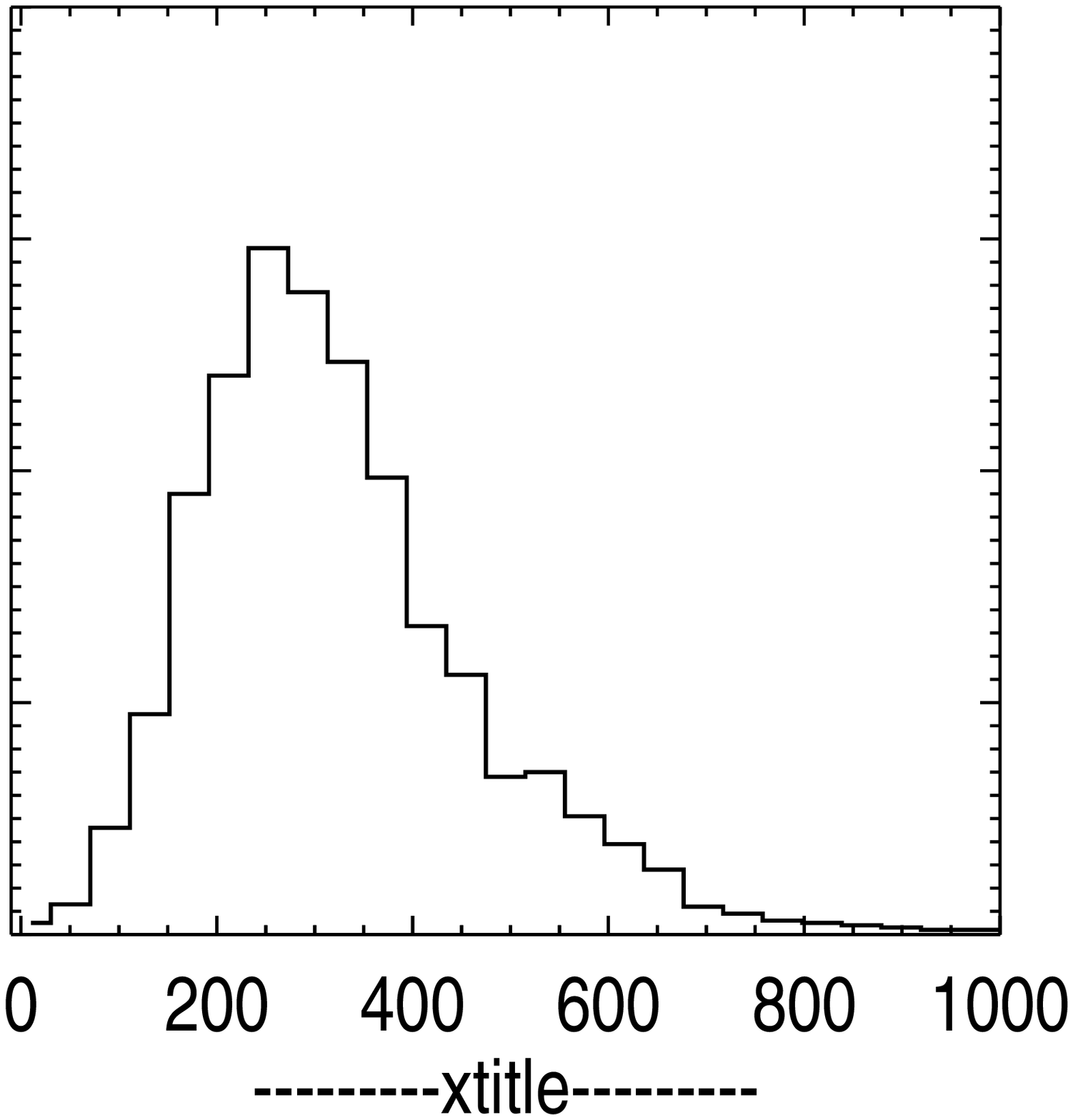}&
\psfrag{---------xtitle---------}{$\quad \quad S_{217}^{\rm{D}}$}
\includegraphics[width=0.21\textwidth, viewport=60 0 470 500]{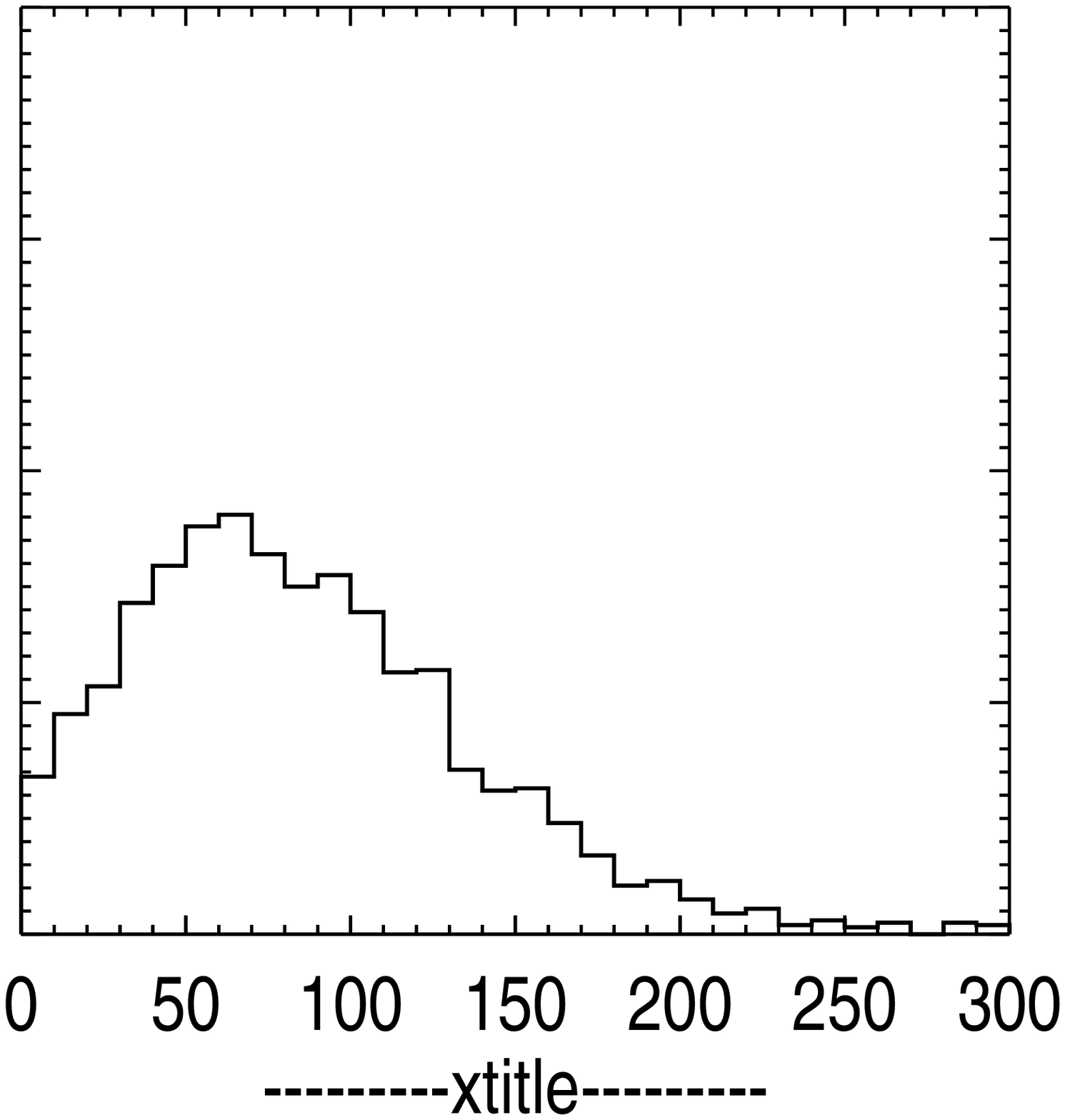} 
\end{tabular}
\caption{Distribution of the PHZ geometric properties and photometry.
{\it Top}: from left to right, S/N of the detection in the 545\,GHz excess
maps; FWHM; and ellipticity of the Gaussian elliptical fit; 
plus the local extinction $E(B-V)_{\rm{xgal}}$. The distribution of 
the extinction $E(B-V)_{\rm{xgal}}$ is also shown for the whole mask (blue
dashed line).  {\it Bottom}: flux density estimates on the cleaned 857, 545,
353, and 217\,GHz maps (from left to right, respectively).}
\label{fig:statistics}
\end{figure*}

The statistics of the main properties of the \Planck\ high-$z$ candidates
are shown in Fig.~\ref{fig:statistics}:
S/N of the detection on the excess map at 545\,GHz; FWHM and ellipticity of
the Gaussian elliptical fit; average local extinction $E(B-V)_{\rm{xgal}}$;
and flux densities in all cleaned bands.  The S/N of the detection in the
545\,GHz excess map does not extend to values larger than 10, 
peaking close to 5 (i.e., the threshold imposed by the detection criteria), 
while the S/N of the detection in the cleaned maps at 857, 545, and 353\,GHz
have 80\,\% to 90\,\% of their values below 6.
The PHZ sources are not extremely high S/N detections.  The distribution
of the FWHM peaks around 7{\parcm}9. As has been shown with Monte Carlo 
simulations (see Sect.~\ref{sec:mcqa_position_accuracy}), the FWHM are
statistically overestimated by 20\,\% at low S/N (below 10), 
which is the case for most of the detections.  This means that the actual
size distribution of PHZ sources is probably centred around 6{\parcm}3, 
leading to a real average size of 3{\parcm}8 after deconvolution by the
5{\arcmin} \Planck\ beam.  Concerning the ellipticity distribution, Monte
Carlo simulations have shown that it is artificially stretched to an average
ellipticity of 0.65, because of the confusion with the CIB in which the
PHZ sources are embedded.  However, the actual distribution peaks at even
larger ellipticities, around 0.8, suggesting that the PHZ sources are not
compact or spherical but somewhat extended objects.

The distribution of the Galactic extinction $E(B-V)_{\rm{xgal}}$ (bottom right
panel of Fig.~\ref{fig:statistics}) is similar to the statistics
of the whole mask. This is entirely consistent with what has been 
observed in Monte Carlo simulations in Sect.~\ref{sec:mcqa_selection}, i.e., 
our detection algorithm is not sensitive to the Galactic foreground level,
thanks to the efficient Galactic cirrus cleaning.
The distribution of the flux density estimates at 545\,GHz is sharply
cutoff at 500\,mJy because of the threshold applied to avoid 
too strong a flux boosting effect, and extending to 2.5\,Jy. In the other
bands the distribution peaks around 
0.8\,Jy, 250\,mJy, and 70\,mJy at 857, 353, and 217\,GHz, respectively.

\begin{figure}[t]
\vspace{-0.4cm}
\hspace{-0.7cm}
\psfrag{------xtitle------}{$S_{353}/S_{545}$}
\psfrag{------ytitle------}{$S_{545}/S_{857}$}
\includegraphics[width=0.55\textwidth]{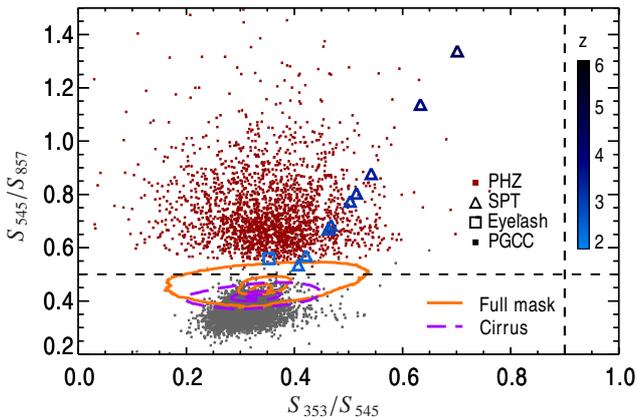} 
\caption{Colour-colour diagram of the 2151 \Planck\ high-$z$ 
source candidates (red dots), and other typical astrophysical sources.
Galactic cold clumps from the PGCC catalogue are shown
as grey dots. Triangles represent the
nine DSFGs discovered by SPT, with confirmed redshifts from 2.5 to 4.5.
The square is the ``Cosmic Eyelash'' submm galaxy, lying at $z=2.33$.
The colour of the symbols is proportional to the redshift of the object
from 2 to 6 (in blue scale). 
Contours give the pixel distribution of the full mask (orange) and towards
cirrus (purple), including 99.9\,\%, 50\,\%, and 10\,\% of the distribution.
Notice that the colours of PHZ sources are computed using flux densities
obtained on cleaned maps, $S_{857}^{\rm{D}}$, $S_{545}^{\rm{D}}$,
and $S_{353}^{\rm{D}}$.  The dashed lines show the two criteria
used to build the colour-colour selection.}
\label{fig:cc_highz}
\end{figure}

\subsection{Colour-colour domain}
\label{sec:colcol_statistics}

The distribution of the PHZ sources in the colour-colour diagram is shown
in Fig.~\ref{fig:cc_highz}, and compared to 
the loci of a few typical high-$z$ astrophysical
sources: the Galactic cold clumps of the PGCC catalogue; a subset of
nine dusty star forming galaxies (DSFG) discovered with
the South Pole Telescope \citep[SPT;][]{Vieira2010} and followed-up
with SABOCA and LABOCA \citep{Greve2012}; and the submm galaxy
SMMJ2135$-$0102, the ``Cosmic Eyelash,'' located at $z=2.33$
\citep{Swinbank2011, Ivison2010, Danielson2011}. The contours 
of the pixel distribution inside the full mask and towards Galactic
cirrus in the initial \Planck\ maps are also shown, including 99.9\,\%, 50\,\%,
and 10\,\% of the distribution.  Hence the Galactic cirrus pixels
(defined as those pixels with an extinction $E(B-V)_{\rm{xgal}}$
larger than 0.03 inside the mask), as well as the Galactic cold sources of
the PGCC, occupy very distinct domains compared with the high-$z$ candidates, 
as ensured by the colour criteria on the $S_{545}/S_{857}$ colour ratio. 
Furthermore it can be seen that the above criteria allow us to separate the
high-$z$ ($>2$) from the intermediate and low-$z$ ($<2$) component of the CIB,
which dominates the distribution of the full mask.

Comparing now to the loci of known high-$z$ objects, the PHZ sources span a
quite different domain; this is fully explained by the impact of attenuation
on the flux density estimates obtained on cleaned maps, as has been
investigated using numerical simulations in Sect.~\ref{sec:mcqa_photometry}.
The $S_{353}^{\rm{D}}/S_{545}^{\rm{D}}$ colour ratio
is especially affected by the cleaning for high redshift sources, i.e., at
high intrinsic $S_{353}/S_{545}$ colour ratio, so that the measured
$S_{353}^{\rm{D}}/S_{545}^{\rm{D}}$ ratio lies between 0.2 and 0.6 even for
redshifts as high as 4.  That is why we cannot use this colour ratio to
obtain an estimate of the redshift of the PHZ sources.
On the contrary the second colour ratio $S_{545}^{\rm{D}}/S_{857}^{\rm{D}}$
is not affected by the cleaning, up to a value of 0.8, and then slightly
underestimated by about 10\,\% for an intrinsic colour ratio of 1. 
This can then be used as a direct tracer of the redshift combined with the
dust temperature of the detected sources.  The fact that 73\,\% of the PHZ
sources exhibit a colour ratio $S_{545}^{\rm{D}}/S_{857}^{\rm{D}}$ between
0.5 and 0.8 is mainly due to the efficiency of the detection algorithm in
this colour range. The 27\,\% of sources with
$S_{545}^{\rm{D}}/S_{857}^{\rm{D}}>0.8$ represents an interesting sample of
highly redshifted or extremely cold sources.

\subsection{Redshift estimates}
\label{sec:redshift}

\begin{figure}[t]
\vspace{-0.03cm}
\hspace{-0.7cm}
\psfrag{xx}{$z$}
\psfrag{yy}{$N$}
\includegraphics[width=0.55\textwidth]{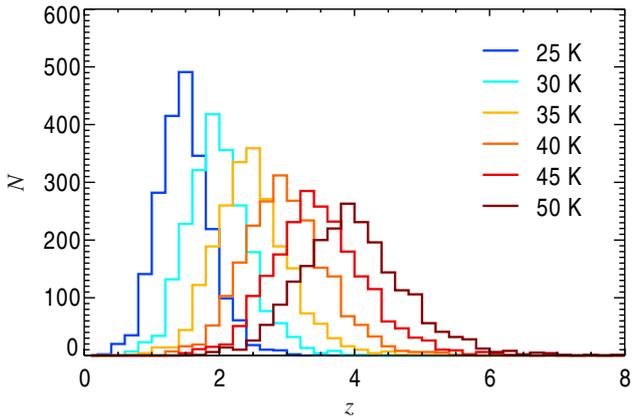}
\caption{Distribution of the submm photometric redshift estimates assuming 
six different cases for the extragalactic dust temperature, from 25 to 50\,K.}
\label{fig:z_stat}
\end{figure}

\begin{figure}
\hspace{-0.7cm}
\psfrag{yy}{$N$}
\psfrag{------xtitle------}{$L$ [$10^{14}\, \rm{L}_{\odot}]$}
\psfrag{---------xtitle---------}{SFR [$10^{4}\, \rm{M}_{\odot} \rm{yr}^{-1}]$}
\includegraphics[width=0.55\textwidth]{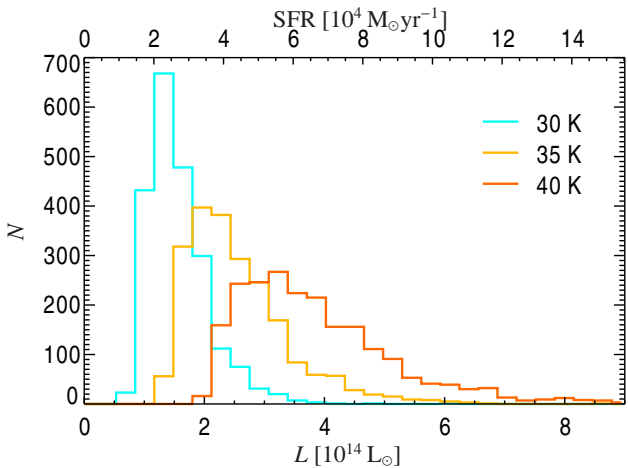} 
\caption{Distribution of the FIR bolometric luminosity and SFR for three assumptions of the extragalactic dust temperature, $T_{\rm{xgal}}=30$, 35, and 40\,K.}
\label{fig:lbol_sfr_stat}
\end{figure}

We performed a photometric redshift determination for each source, assuming 
 simple SED modelling given by a modified blackbody emission with a dust spectral index
$\beta_{\rm{xgal}}=1.5$ and six different cases of the dust temperature, namely $T_{\rm{xgal}}= 25$, 30, 35, 40, 45, and 50\,K.
In order to take into account the impact of the cleaning algorithm introduced in Sect.~\ref{sec:impact_cleaning_highz}, 
 we built a grid of attenuated flux densities modelled for each value of the redshift ($0<z<8$) and the dust temperature.
 A $\chi^2$ analysis based on this grid yields the best fit of the redshift together with 1$\,\sigma$ lower and upper limits.
 The accuracy of the redshift estimate processing has been analysed on Monte Carlo simulations (see Appendix~\ref{sec:redshift_accuracy}).
The average uncertainties associated with these photometric redshift estimates are about 0.5, given a specific dust temperature.
Of course the degeneracy between the redshift and the dust temperature may induce much larger uncertainties on those sources
without spectroscopic data.
 
The distribution of redshift estimates can be seen in Fig.~\ref{fig:z_stat} for each case of the extragalactic dust temperature.
For an average dust temperature of 35\,K, which is consistent with the latest analyses \citep[e.g.,][]{Greve2012, Weiss2013, Magnelli2014, Swinbank2014}, 
the distribution exhibits a median value of $z=2.5$, with 95\,\% of sources lying between 1.5 and 3.7. This is in perfect agreement 
with the outcomes of the Monte Carlo analysis of Sect.~\ref{sec:mcqa}. Because of the degeneracy 
between the redshift and the dust temperature, this redshift range shifts to $2.6<z<5.7$ for the highest temperature, $T_{\rm{xgal}}=50$\,K.

\subsection{FIR luminosities and SFRs}
\label{sec:lum_mass_sfr}

Given the redshift estimates, we derive for each source the FIR bolometric luminosity associated with the six different assumptions made on the
dust temperature. This is computed as the
integral of the redshifted modified blackbody emission between 300\,GHz and 37.5\,THz. 
Following the prescription of \citet{kennicutt1998} and assuming that the contribution from the AGN is negligible for these objects,
 we finally derive an estimate of the star formation rate as SFR [$\rm{M}_{\odot} \rm{yr}^{-1}$] = $1.7\times10^{-10}L_{\rm{FIR}}$[$\rm{L}_{\odot}$].
The distributions of bolometric luminosity and SFR are shown in Fig.~\ref{fig:lbol_sfr_stat}, 
for three options of the dust temperature, 
$T_{\rm{xgal}}=30$, 35, and 40\,K, lying in the most probable range of temperature expected for dusty submm galaxies.

The FIR bolometric luminosity distribution peaks around
$2\times10^{14}$\,$\rm{L}_{\odot}$ (assuming $T_{\rm{xgal}}=35$\,K), with an
associated SFR around 3200\,$\rm{M}_{\odot} \rm{yr}^{-1}$, which is not
really compatible with the expected luminosities of single 
submm galaxies at high-$z$, typically
$10^{11}$--$3\times10^{13}$\,$\rm{L}_{\odot}$
\citep{kovacs2006, Chapin2011, Geach2013, Swinbank2014, Casey2014}.
Only strongly lensed galaxies may reach such high apparent luminosities,
because of the magnification.  The brightest strongly lensed dusty galaxies
detected by SPT exhibit intrinsic FIR luminosities ranging between
1.9 and $6.9\times10^{13}$$\mu^{-1}\rm{L}_{\odot}$, where $\mu$ is the unknown 
magnification factor \citep{Vieira2013, Hezaveh2013}, which represents the
lowest tail of our sample distribution.  \citet{Canameras2015} reported
intrinsic FIR luminosities of
(0.5--1.7)$\times10^{14}$\,$\mu^{-1}\rm{L}_{\odot}$ 
towards 11 high-$z$ strongly lensed star-forming galaxies selected using
\Planck\ data and confirmed with {\it Herschel\/} (see
Appendix~\ref{sec:herschel}). Focusing now on the four sources of the latter
sample with a counterpart in the final PHZ, we observe that these sources
exhibit an apparent FIR luminosity about 3 to 5 times larger in \Planck\ than
in {\it Herschel}.

\begin{figure*}
\begin{tabular}{cccc}
$\quad \quad \quad \quad \quad \quad \quad$ 857\,GHz& 
$\quad \quad \quad \quad \quad \quad \quad$ 545\,GHz& 
$\quad \quad \quad \quad \quad \quad \quad$ 353\,GHz& 
$\quad \quad \quad \quad \quad \quad \quad$ 217\,GHz\\
\psfrag{-----------------ytitle-----------------}{$\,$log$(dN/dS S^{5/2}\, [\rm{Jy}^{1.5} \rm{sr}^{-1}]$}
\psfrag{--------xtitle--------}{log$(S_{850})\, [\rm{Jy}]$}
\includegraphics[width=0.21\textwidth, viewport=60 0 450 500]{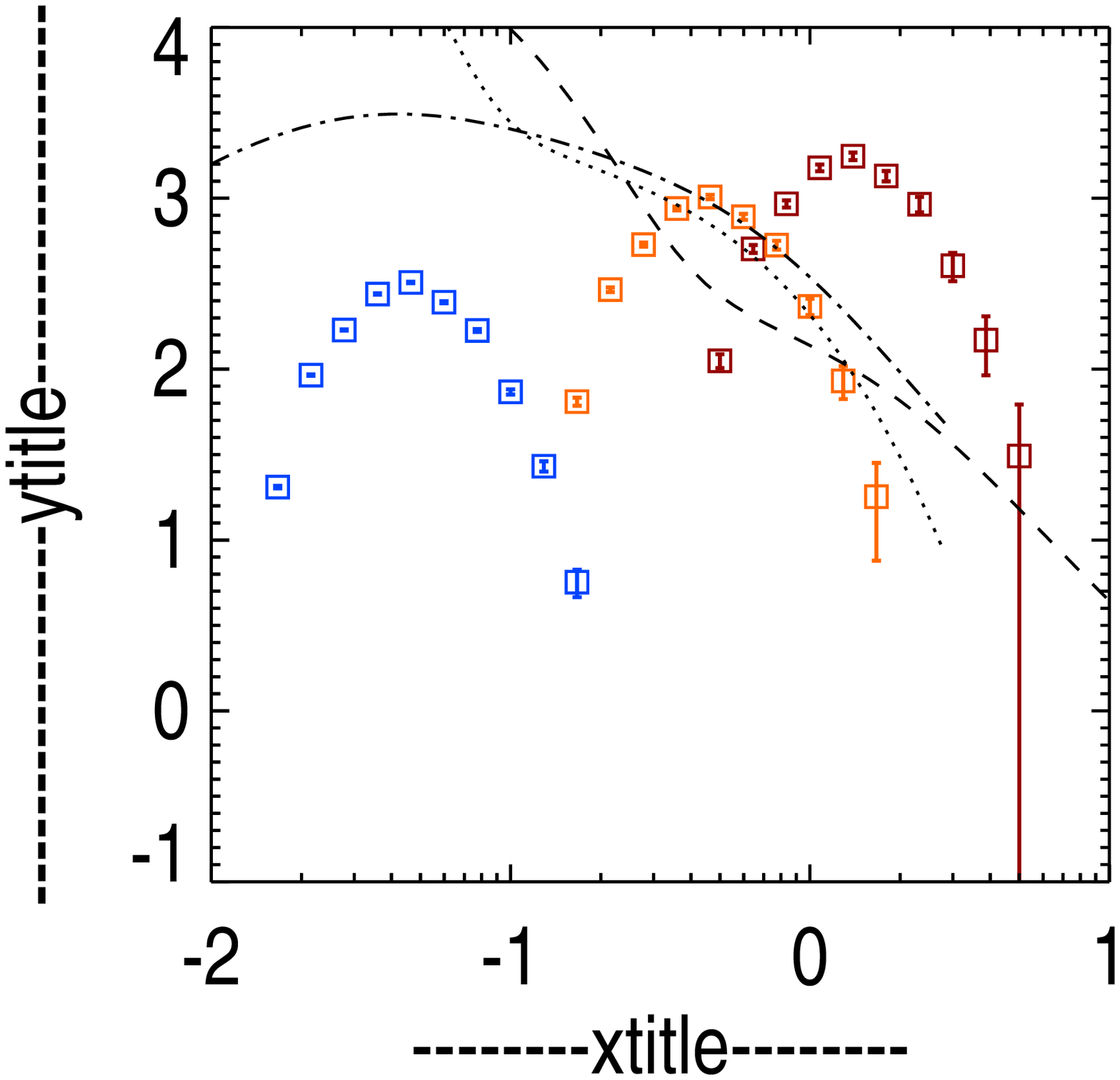}&
\psfrag{--------xtitle--------}{log$(S_{545})\, [\rm{Jy}]$}
\includegraphics[width=0.21\textwidth, viewport=60 0 450 500]{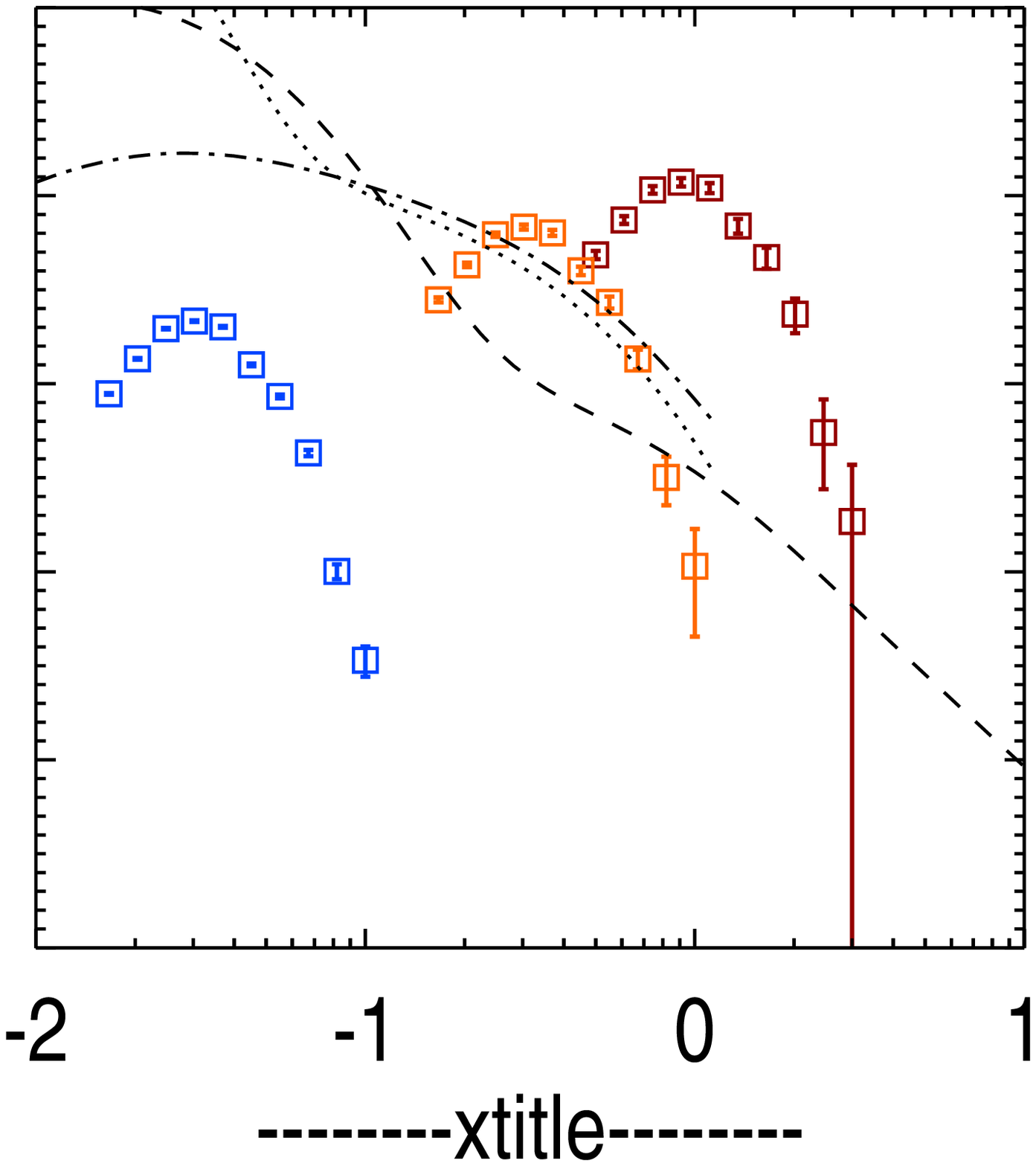}&
\psfrag{--------xtitle--------}{log$(S_{353})\, [\rm{Jy}]$}
\includegraphics[width=0.21\textwidth, viewport=60 0 450 500]{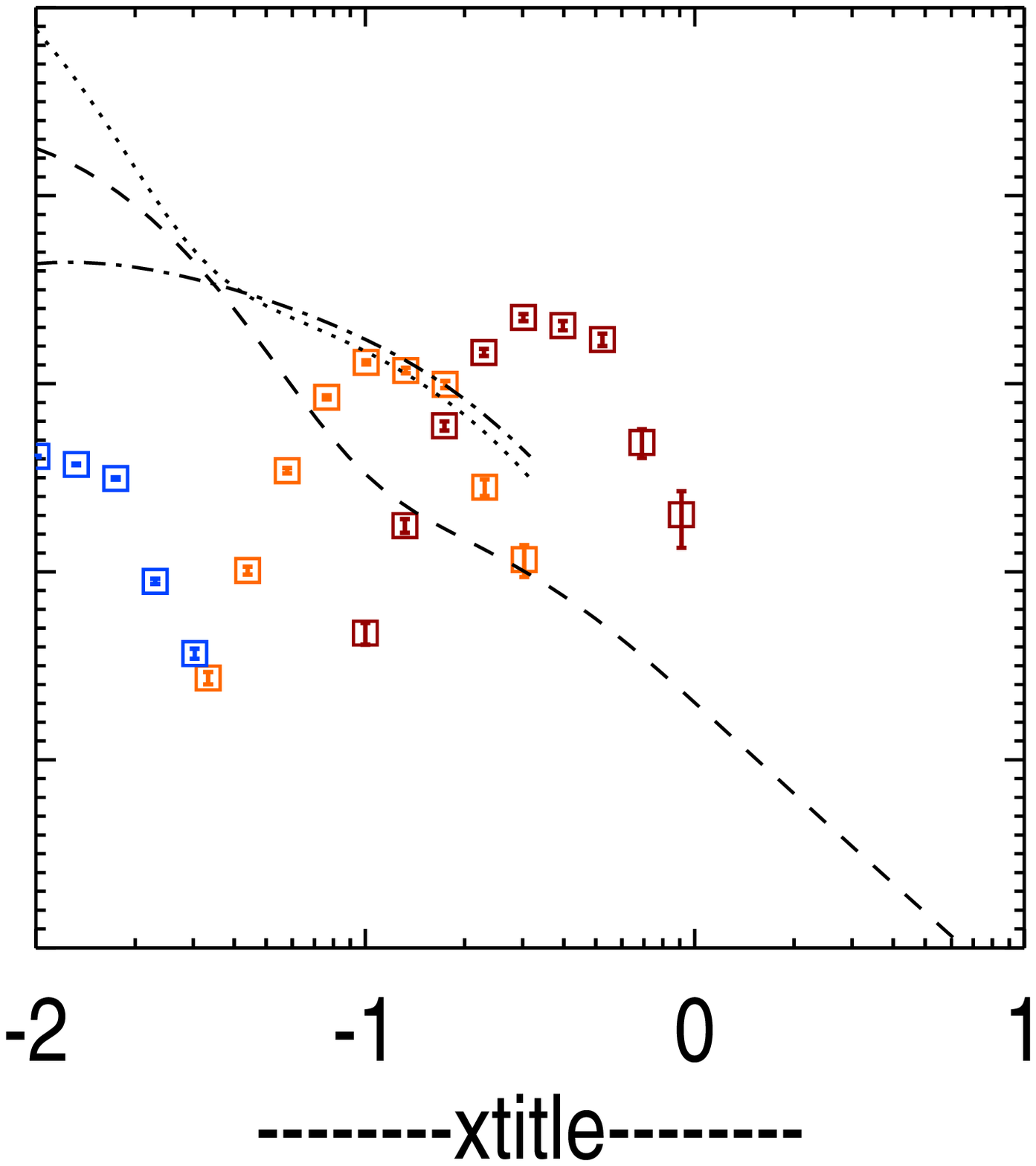}&
\psfrag{--------xtitle--------}{log$(S_{217})\, [\rm{Jy}]$}
\includegraphics[width=0.21\textwidth, viewport=60 0 450 500]{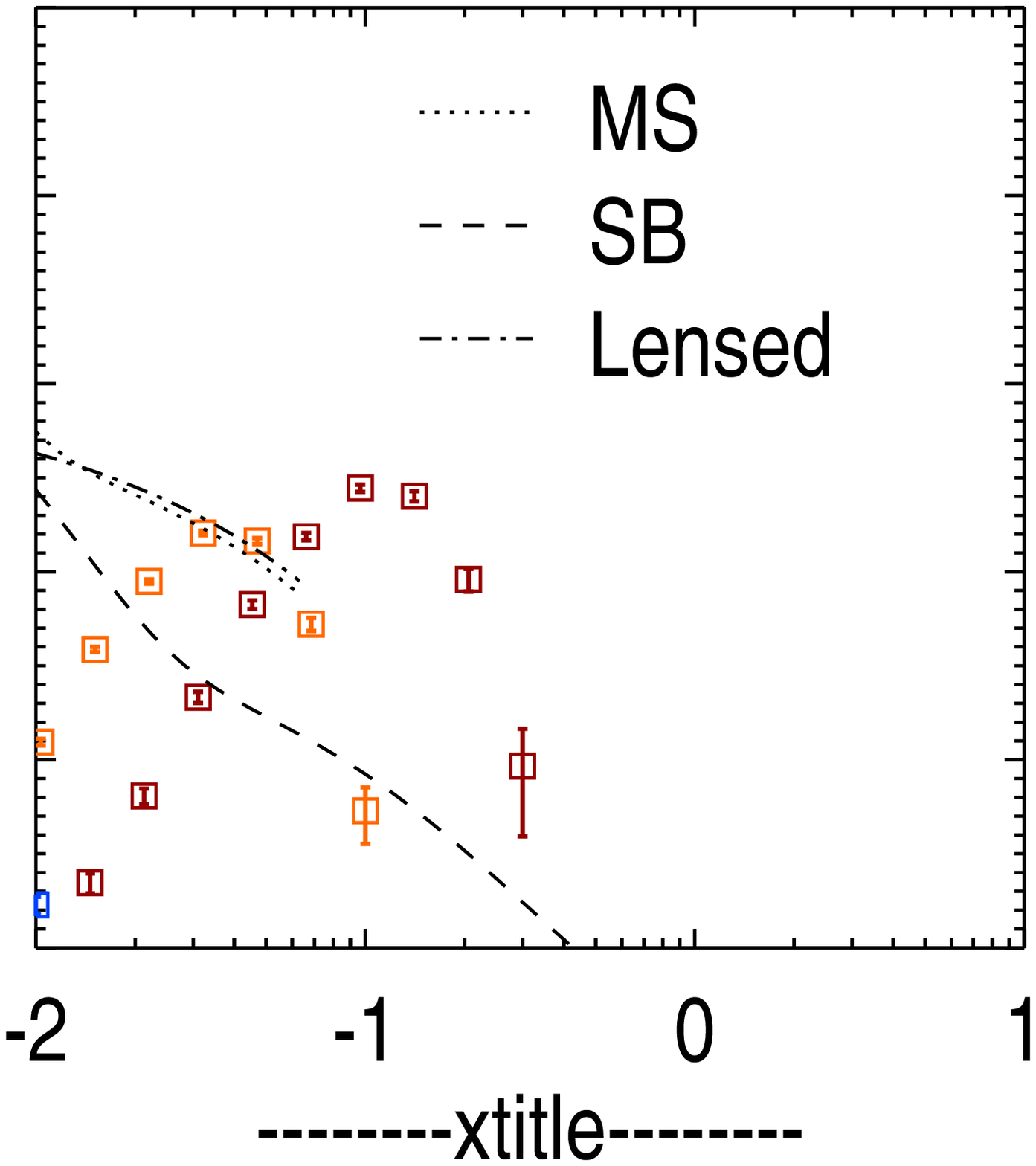} 
\end{tabular}
\caption{Number counts of PHZ sources in the 857-, 545-, 353-, and 217-GHz
bands (from left to right, respectively), where the flux densities have been
computed in the cleaned maps and may be affected by the attenuation effect.
Three cases are presented, depending on the assumption made about the intrinsic
number of objects composing the PHZ source: 
$n=1$ (red); $n=3$ (orange); and $n=30$ (blue). The predictions of the
\citet{Bethermin2012} model integrated between redshift 1.5 and 4
are shown for three populations of sources: main sequence (dotted line);
starburst (dashed line); and lensed (dot-dash line).}
\label{fig:numbercounts}
\end{figure*}

Assuming now that the \Planck\ PHZ sources are composed of multiple
galaxies, the range of FIR luminosities derived above may be compared to
recent estimates obtained by integrating the submm emission of 
galaxy members towards proto-cluster candidates at high-$z$, e.g., about
$10^{13}$\,$\rm{L}_{\odot}$ at redshift $1\,{<}\,z\,{<}\,1.5$
\citep{Brodwin2013}, or (0.5--7)$\times10^{13}$\,$\rm{L}_{\odot}$ at
$z\simeq2$ \citep{Clements2014}.  Using the dedicated {\it Herschel\/}
follow-up of 228 \Planck\ candidates \citep{planck2014-XXVII} 
described in Appendix~\ref{sec:herschel}, it also appears that the 
\Planck\ FIR luminosity estimates are about 2 to 3 times larger than the
integrated luminosities of the galaxy members identified with {\it Herschel\/}
inside the elliptical Gaussian profiles of the \Planck\ PHZ sources. 
Despite the precaution we made by applying a flux density threshold at
500\,mJy at 545\,GHz, the flux boosting effect can still reach 20\,\% for
flux density estimates around 0.5\,Jy; this may explain a fraction of the
discrepancy between \Planck\ and {\it Herschel}, but not all.
This remaining discrepancy suggests that the \Planck\ estimates
integrated over a 5{\arcmin} beam include a component that is barely traced
by SPIRE because of confusion. As characterized by
\citet{Viero2015}, this effect is even stronger for sources at high redshift,
and can reach 50\,\% of enhancement when going from {\it Herschel}-SPIRE
resolution to \Planck\ resolution.  Hence \Planck\ flux densities allow us
to recover an estimate of the overall budget of the submm emission at high-$z$,
by including a population of faint sources contributing to the \Planck\ flux,
but undetected in {\it Herschel}'s higher resolution data.

\subsection{Number counts}
\label{sec:nb_counts}

The reliability of the flux density estimates in the cleaned maps has already been discussed above.
It is impacted by the overestimation of the extension of the sources, but also by the CIB fluctuations, 
and more seriously by the attenuation 
due to the cleaning process, which may strongly affect the flux density estimates 
(depending on the dust temperature, the redshift of the sources, and the level of foreground contamination of the CMB template). 
A theoretical approach has shown that the flux densities at 353 and 217\,GHz 
can be underestimated on average by about 10\,\% and 40\,\%, respectively, while the 857- and 545-GHz bands are not affected.
The numerical analysis of Sect.~\ref{sec:mcqa} pointed out an additional bias of 3.5\,\%.
However, these biases are both compensated at low flux densities by the flux boosting effect.
 We stress that an exact correction for this attenuation effect for each individual source 
could only be carried out by knowing its SED and redshift. 

\begin{figure}[t]
\vspace{-.3cm}
\psfrag{-----------------ytitle-----------------}{$\quad$log$(dN/d\rm{log}S_{\nu} \, [\rm{sr}^{-1}]$}
\psfrag{--------xtitle--------}{log$(S_{353})\, [\rm{Jy}]$}
\psfrag{legend1}{\tiny{$Q=1$}}
\psfrag{legend2}{\tiny{$Q=1/b$}}
\psfrag{legend3}{\tiny{$Q=1/b^2$}}
\includegraphics[width=0.5\textwidth]{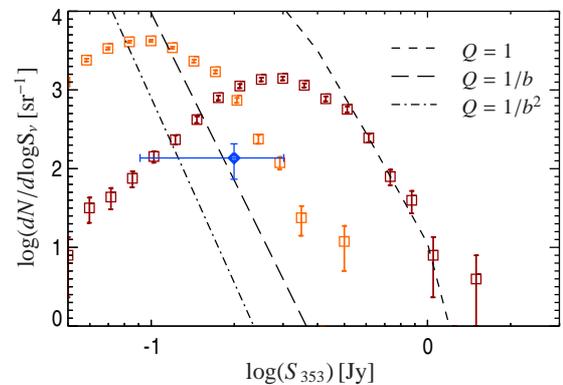}
\caption{PHZ number counts at 353\,GHz (red squares), compared to the
``clumps'' number counts analytical predictions
of \citet{Negrello2005} for three assumptions of the amplitude of the 3-point
correlation function, $Q$, where $b$ is the clustering bias parameter.
Two cases of the PHZ number counts are presented, depending on the assumption
made about the intrinsic number of clumps composing the PHZ source: 
$n=1$ (red); and $n=3$ (orange). The blue diamond gives the number counts
estimate reported by \citet{Clements2014} for proto-cluster candidates
selected from the \Planck\ ERCSC catalogue. }
\label{fig:dndlogs_353}
\end{figure}

Despite this warning, it is interesting to perform a crude number counts
analysis on the PHZ sources.  The number counts are shown in
Fig.~\ref{fig:numbercounts} for all channels. The population of PHZ sources
appear extremely bright compared to the predictions of \citet{Bethermin2012}
for three types of individual galaxies: main sequence (MS); starburst (SB);
and lensed sources. For this analysis the models of \citet{Bethermin2012}
have been integrated in the range of redshift $1.5<z<4$, according to the
expected detection range of our algorithm.  Three versions of the PHZ number
counts are shown, depending of the assumed number of individual objects
composing the \Planck\ source, namely $n=1$, 3, and 30. 
If we assume the \citet{Bethermin2012} model represent the PHZ contents,
then the ``$n=3$'' counts being closest to the model suggests that the PHZ
candidate sources typically include multiple galaxies.
The PHZ number counts at 353\,GHz may also be compared with analytical
predictions by \citet{Negrello2005} that explore the impact of clustering
when building number counts with large beams such as those of \Planck\ or
{\it Herschel}.  Those authors considered three scenarios for the clustering,
associated with the 3-point correlation function
(see Fig.~\ref{fig:dndlogs_353}). It appears that the PHZ distribution is
more or less consistent with the modelling that is the most realistic of
the three, assuming no evolution with redshift for the amplitude of the
3-point correlation function.  Again, two versions of the PHZ number counts
are shown, depending on the assumed typical number of internal objects
composing the \Planck\ source, i.e., $n=1$ or $n=3$. This may suggest that
a fraction of the PHZ sources are combinations of objects located along the
line of sight either by chance or because they belong to the same cosmic
filament.

\subsection{Cross-check with \Planck\  catalogues}
\label{sec:ancillary_internal_xcat}

\begin{table}[t]
\caption{Number of associations within 5{\arcmin} between the PHZ and the
 \Planck\ catalogues PCCS2, PSZ2, and PGCC.
Matches with the PCCS2 catalogue are divided into two zones corresponding to
the quantified-reliability zone (0) and filament zone (1). 
Notice that the zones 2 and 3, corresponding to Galactic regions, are already
rejected by the mask used in this work \citep[see][]{planck2014-a35}.
Concerning the PGCC, the cross-correlations are divided into the three
categories of flux quality (FQ) 1 to 3 \citep[see][]{planck2014-a37}.}
\label{tab:catxchecks}
\nointerlineskip
\setbox\tablebox=\vbox{
\newdimen\digitwidth 
\setbox0=\hbox{\rm 0} 
\digitwidth=\wd0 
\catcode`*=\active 
\def*{\kern\digitwidth} 
\newdimen\signwidth 
\setbox0=\hbox{+} 
\signwidth=\wd0 
\catcode`!=\active 
\def!{\kern\signwidth} 
\newdimen\pointwidth 
\setbox0=\hbox{.} 
\pointwidth=\wd0 
\catcode`?=\active 
\def?{\kern\pointwidth} 
\halign{\tabskip=0pt\hbox to 1.5 in{#\leaderfil}\tabskip=1.0em&
\hfil#\hfil&
\hfil#\hfil&
\hfil#\hfil\tabskip=0pt\cr
\noalign{\doubleline}
\omit\hfill Catalogue \hfill& \omit\hfill Option \hfill& Mask& PHZ\cr
\noalign{\vskip 3pt\hrule\vskip 4pt}
 PCCS2 857& zone 0& 2447& 10\cr 
 \omit\hfill& zone 1& *297& *1\cr 
 \noalign{\vskip 3pt\hrule\vskip 4pt}
 PCCS2 545& zone 0& *818& 12\cr 
 \omit\hfill& zone 1& *114& $\dots$\cr 
 \noalign{\vskip 3pt\hrule\vskip 4pt}
 PCCS2 353& zone 0& *779& 17\cr 
 \omit\hfill& zone 1& **47& $\dots$\cr 
 \noalign{\vskip 3pt\hrule\vskip 4pt}
 PCCS2 217&   zone 0& *831& *1\cr 
 \omit\hfill& zone 1& ***4& $\dots$\cr 
 \noalign{\vskip 3pt\hrule\vskip 4pt}
 PCCS2 143&   zone 0& *683& *1\cr 
 \omit\hfill& zone 1& **$\dots$& $\dots$\cr 
 \noalign{\vskip 3pt\hrule\vskip 4pt}
 PCCS2 100&   zone 0& *520& *2\cr 
 \omit\hfill& zone 1& **$\dots$& $\dots$\cr 
\noalign{\vskip 3pt\hrule\vskip 4pt}
 PCCS2 70&    zone 0& *232& $\dots$\cr 
 \omit\hfill& zone 1& **$\dots$& $\dots$\cr 
 \noalign{\vskip 3pt\hrule\vskip 4pt}
 PCCS2 44&    zone 0& *189& $\dots$\cr 
 \omit\hfill& zone 1& **$\dots$& $\dots$\cr 
 \noalign{\vskip 3pt\hrule\vskip 4pt}
 PCCS2 30&    zone 0& *367& *1\cr 
 \omit\hfill& zone 1& **$\dots$& $\dots$\cr 
 \noalign{\vskip 3pt\hrule\vskip 4pt}
 PCCS2 857$\times$545$\times$353& zone 0& *407& *2\cr 
 \omit\hfill& zone 1& **51& $\dots$\cr 
 \noalign{\vskip 3pt\hrule\vskip 4pt}
 PCCS2 70$\times$44$\times$30& zone 0& *157& $\dots$\cr
 \omit\hfill& zone 1& **$\dots$& $\dots$\cr 
\noalign{\vskip 3pt\hrule\vskip 4pt}
 PSZ2& & *548& *3\cr 
\noalign{\vskip 3pt\hrule\vskip 4pt}
PGCC& FQ 1& **31& *3\cr
 \omit\hfill& FQ 2& **43& 16\cr
 \omit\hfill& FQ 3& **13& $\dots$\cr
 \noalign{\vskip 3pt\hrule\vskip 4pt}
}}
\endPlancktable
\end{table}

We performed a cross-check between the 2151 sources of the 
PHZ and the other catalogues made available 
with this \Planck\ 2015 release (see Table~\ref{tab:catxchecks}):
the Planck Catalogue of Compact Sources \citep[PCCS2;][]{planck2014-a35}; the Planck Catalogue of SZ sources
\citep[PSZ2;][]{planck2014-a36}; and Planck Catalogue of Galactic Cold Clumps \citep[PGCC;][]{planck2014-a37}.

We counted only three associations between the PHZ and the PSZ2,
which confirms the different astrophysical nature of these two populations
of objects. Sources from the PSZ2 catalogue are virialized galaxy clusters
traced by their Sunyaev-Zeldovich signal due to the hot intergalactic gas, 
while sources from the PHZ are traced by their dust submm emission coming
from the high-$z$ galaxies located inside the \Planck\ beam.  The probable
nature of the PHZ sources will be discussed in Sect.~\ref{sec:conclusion}.

The cross-match with the PCCS2 has been performed with the
catalogues extracted in all nine individual \Planck-HFI and LFI bands,
but also with two band-merged catalogues: the HFI band-merged catalogue is
defined as the PCCS2 sources with simultaneous detections in the 857, 545,
and 353\,GHz HFI bands; and the LFI band-merged catalogue requires detection
in all LFI bands, i.e., 70, 44, and 30\,GHz \citep[see][]{planck2014-a35}.
The HFI and LFI band-merged catalogues trace two different populations, 
dusty submm sources and radio sources, respectively. 
As shown in Table~\ref{tab:catxchecks}, the overlap between the PCCS2
and the PHZ is extremely small.
Taking into account the redundancy between bands, a total of 35 sources
are present in both catalogues, while no radio sources (from the LFI
band-merged catalogue) and only two dusty submm sources 
(from the HFI band-merged catalogue) are found in the PHZ sample.
Notice that the sources of the PCCS2 bands are divided into two categories,
depending on their reliability, namely
high reliability sources (zone 0) or unvalidated sources (zones 1, 2, and 3),
where the 0-3 zones correspond to quantified-reliability zone, filament zone,
Galactic zone, and Galactic filaments zone, respectively.
Matches between the PHZ and PCCS2 only happen in the
quantified-reliability zone, suggesting that the PHZ sources are quite clean
from the cirrus contamination traced by the PCCS2 masks.
When looking at the individual low-frequency matches between PHZ
and PCCS2 sources, the dust emission signature in the HFI
bands is clear, but may be associated with radio emission
observed in the LFI bands.  The PHZ is thus seen to be complementary to
the PCCS2, by picking out the faintest and coldest objects at high latitude.

The PGCC catalogue has been built over the whole sky, but focuses on the 
Galactic objects by rejecting any possible associations with extragalactic
sources.  This purification step was performed using three independent
methods \citep[see][]{planck2014-a37}: 
cross-correlation with well-characterized catalogues of extragalactic sources;
identification with galaxies in optical data; and colour-colour selection.
Among the 87 PGCC sources lying in the high-latitude mask used in this work,
19 are found to be correlated with PHZ sources within 5{\arcmin}.
These 19 cross-matched sources exhibit very low temperature in the PGCC
catalogue (with a median around 9\,K),
and are associated with low $\ion{H}{i}$ column densities (amongst the
lowest 10\,\% of the PGCC catalogue). On the PHZ side, 
these 19 sources exhibit a similar distribution of flux density at 545\,GHz as
the whole PHZ, with extinction values spanning the full mask statistics,
suggesting that the PHZ population does not consist of the faintest component
of the PGCC population. Despite this, it is still hard to determine if these
sources are Galactic or extragalactic, and they are flagged in both catalogues
accordingly.
 
\begin{figure}[t]
\vspace{-0.2cm}
\hspace{-0.7cm}
\psfrag{-----ytitle-----}{$\rm{N}_{\rm{H}}$\,[${\rm cm}^{-2}$]}
\psfrag{--xtitle--}{$T$\,[K]}
\includegraphics[width=0.55\textwidth]{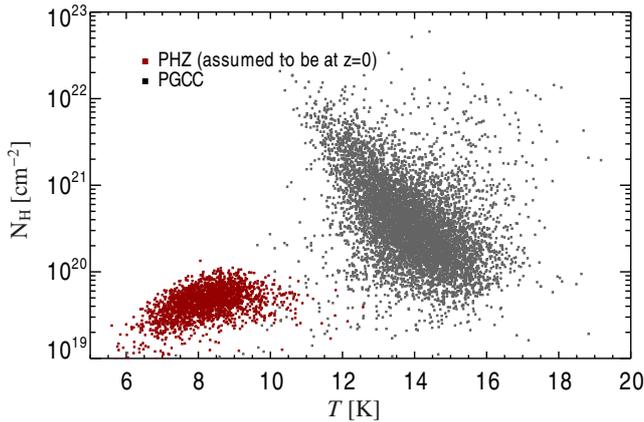} 
\caption{\ion{H}{i} column density versus temperature relation for PGCC sources
(grey dots) and PHZ sources (red dots) when they are assumed to be at $z=0$.}
\label{fig:nht_highz}
\end{figure}

This analysis can be used to disentangle the possible contamination of the
PHZ by cirrus. Because of the degeneracy between redshift and temperature,
the PHZ sources can be interpreted as ``cold'' or ``red'' sources.
For the analysis here, we assume that
each PHZ source is located inside the Galaxy, i.e., $z=0$. We derive its
temperature from the flux density estimates at 857, 545, and 353\,GHz, 
assuming a dust spectral index of 2, as is observed for dense regions with
temperature below 10\,K.  We compute the column density of each source by
applying the same recipe as for the PGCC sources \citep[see][]{planck2014-a37}.
Hence the PHZ source candidates, assumed to lie at $z=0$, exhibit temperatures
around 8\,K and mean column densities of about $5\times10^{19}\,{\rm cm}^{-2}$.
The relation between the temperature (assuming a dust spectral index of 2)
and the column density of the PGCC sources and the PHZ sources assumed to be
Galactic objects is shown in \ref{fig:nht_highz}.
For PGCC sources, the lower the temperature, the higher the column density,
as expected for the dense Galactic medium.   However, the opposite trend is
observed for the PHZ sources, which are located in a very distinct domain
compared to the PGCC.  Similarly, the $E(B-V)_{\rm{}xgal}$ distribution of
the PHZ sources has been shown to perfectly follow the distribution inside
the full mask (see top right panel of Fig.~\ref{fig:statistics}), 
without showing any bias towards denser regions associated with cirrus.
This reinforces the fact that the PHZ source candidates are not linked to
dense Galactic structures located in cirrus, but lie at high redshift instead, 
and represent a complementary sample of sources to the PGCC catalogue.

\section{Discussion and conclusions}
\label{sec:conclusion}

We have applied an original multi-frequency detection algorithm on
the \Planck-HFI plus 3\,THz IRAS data set to build the List of \Planck\
High-redshift Source Candidates (the PHZ), comprising 2151 objects selected
by their dust emission excess in the 545-GHz band, over the 25.8\,\%
cleanest part of the sky.  We have fully characterized our detection
algorithm using Monte Carlo simulations.  This has enabled us to assess
the quality of the flux densities provided in this list, and, more
specifically, the impact of the attenuation due to the cleaning process, which
tends to statistically underestimate the flux densities by 4\,\% to 40\,\%,
depending on the frequency.  However, we have demonstrated the robustness
of the colour-colour selection process, which allows us to efficiently reject
Galactic cold clumps, low-$z$ dusty sources, and contaminants such as radio
galaxies or low-$z$ galaxy clusters exhibiting strong SZ signatures.
The algorithm has been shown to preferentially detect dusty sources located
at redshifts between 1.5 and 4, depending on their intrinsic temperature
(ranging from 20 to 40\,K), reaching a completeness levels of about
50\,\%, 80\,\%, and 100\,\% for sources with $S_{\rm{545}}=1$, 2, and 3\,Jy,
respectively. 

Despite the reliability of the high-$z$ dusty signature for all the PHZ
sources, the astrophysical nature of these candidates is still uncertain.
They could first of all be statistical fluctuations of the CIB, i.e., chance
alignments of field galaxies along the line of sight
\citep{Negrello2005, Negrello2010, Chiang2013, Chiang2014}. 
Given the flux density threshold of 500\,mJy applied at 545\,GHz, all the
PHZ detections have been obtained at more than 3.7 and 3.3 times the
confusion noise estimated for a Poisson plus clustering contribution with
two different correlation models \citep{Negrello2004}.  Assuming a Gaussian
distribution for the Poisson plus clustering fluctuations as a first guess, 
the associated probabilities to find such CIB fluctuations at a 5{\arcmin}
scale become $0.012$ and $0.061$\,$\rm{deg}^{-2}$, respectively, to be
compared with the density of the PHZ sources which is about
$0.21$\,$\rm{deg}^{-2}$.  Hence the PHZ source density is 17.5 and 3.5
times larger than chance alignment expectations derived in the two clustering
cases of \citet{Negrello2004}.  While it has been shown with other
{\it Herschel\/} analysis that this chance alignment may be larger than
expected, the population of the PHZ sources is still hard to explain by
chance alignment alone, even if this cannot be fully rejected yet for some
fraction of the candidates.

First hints about the nature of the \Planck\ high-$z$ candidates have been
obtained with {\it Herschel\/} follow-up observations. \citet{Negrello2007}
and \citet{Bethermin2012} predicted that a small fraction of the very bright
sources at high redshift ($z>2$) are expected to be lensed dusty starburst
galaxies.  Hence the source H-ATLAS J114637.9$-$001132, simultaneously
detected in the {\it Herschel\/} H-ATLAS survey field, in the ERCSC catalogue
\citep{planck2011-1.10}, and in a previous incarnation of the \Planck\ list
of high-$z$ candidates, was confirmed to be a gravitationally lensed galaxy
at $z\,{=}\,3.3$ \citep{Fu2012, Herranz2013}.
Similarly the source HLS J091828.6+514223, discovered in the {\it Herschel\/}
Lensing Survey \citep[][]{Egami2010} and independently detected in \Planck\
data, was confirmed to be a strongly lensed galaxy at $z=5.2$
\citep{Combes2012}.  Furthermore a dedicated {\it Herschel\/} follow-up
programme on a sub-sample of 228 \Planck\ high-$z$ source candidates
\citep{planck2014-XXVII}, described in more detail in
Appendix~\ref{sec:herschel}, provided unique information on the nature of
this sample.  While 3\,\% of the {\it Herschel\/} fields show clear evidence
of single bright sources inside the \Planck\ beam, further follow-up
observations in optical, Far-IR and the submm of 11 candidates confirmed that
these objects are \Planck-discovered strongly lensed galaxies. They exhibit
flux densities at 350\,$\mu$m larger than 350\,mJy and up to 1\,Jy,
with spectroscopic redshifts ranging from 2.2 to 3.6 \citep{Canameras2015}. 
Compared to the properties of the recent discoveries by {\it Herschel\/}
and the South-Pole Telescope (SPT) of large sets of strongly gravitationally
lensed submm galaxies with flux densities between 100 and 200\,mJy
\citep[e.g.,][]{Negrello2010, Vieira2013, Wardlow2013}, these \Planck\
high-$z$ lensed sources are amongst the brightest lensed galaxies
in the submm range.

Complementary to this population of strongly lensed galaxies,
\citet{planck2014-XXVII} states that more than 93\,\% of the \Planck\
high-$z$ sources followed-up with {\it Herschel\/} are overdensities of around
10 red sources on average, with SEDs peaking at 350\,$\mu$m. This confirms,
on a small sub-sample of sources, what was suggested by the number counts
analysis performed on the whole list (see Sect.~\ref{sec:nb_counts}), i.e., 
PHZ sources are preferentially structures of multiple sources instead of
single red objects.  This statement is in agreement with the predictions by
\citet{Negrello2005} on the detectability of such overdensities 
of high-$z$ dusty star forming galaxies
in the submm, and with recent works \citep[e.g.,][]{Gobat2011, Santos2011,
Santos2013, Santos2014, Clements2014} providing the first observations.

The first newly discovered PHZ proto-cluster candidate with spectroscopic
confirmation is the source PHZ~G095.50$-$61.59, which consists of two systems
at $z=1.7$ and $z=2.0$ \citep[][]{Florescacho2015}. 
Spectroscopic redshifts have been obtained towards four and eight galaxies,
associated with each one of the two structures, respectively, within a
comoving radius of 1\,Mpc, consistent with sizes of local cluster and recently
discovered proto-clusters at $z>1.5$
\citep{Castellano2007, Andreon2011, Gobat2013}.  With an integrated SFR of
2000--3000\,$\rm{M_{\odot}}\rm{yr}^{-1}$ over the \Planck\ beam and a mass
of $4.5\times10^{14}$\,$\rm{M}_{\odot}$, this object fits into the galaxy
cluster category.  Despite the fact that this source has turned out to be a
line of sight combination of two structures, it nevertheless has acted as a
pointer towards high-$z$ objects.  This indicates that the PHZ will be
useful for finding such structures, even if a fraction of the sources are
multiple objects; the reason is that the selection process ensures that
something along the line of sight has to be red, i.e., has to have the colours of 
star-forming galaxies.

Considering the above option of a proto-cluster population, it is interesting
to compare the expected surface density of massive halos at high redshift
with the one of the PHZ sources, i.e., $0.21$\,$\rm{deg}^{-2}$.
From the \citet{Tinker2010} halo model we derive a surface density of dark
matter halos with $M>10^{14}\,{\rm M}_{\odot}$ at $z>2$ of about
$0.5$\,$\rm{deg}^{-2}$.  Given the detection efficiency of our algorithm
(depending on the redshift and flux density), and the fact that only a
fraction of these dark matter halos may be observed during their star-forming
phase, the total number of PHZ source candidates and the expected numbers
of massive high-$z$ galaxy clusters are about the same order of magnitude. 

Moreover the submm photometric redshift distribution of the PHZ sources,
likely ranging from $z=1.5$ to 4, corresponds to the expected redshifts
of the star-formation peak activity of such proto-cluster objects. The fact
that no associations have been found between the PHZ and the \Planck\
Sunyaev-Zeldovich Catalogue (PSZ2) also reveals that the population traced
by the PHZ does not exhibit any clear feature in the SZ effect, 
which means that these objects may still be in a very early stage of their
evolution and not virialized yet.  It is interesting to notice that the
PHZ number counts are compatible with predictions 
of clump number counts made earlier by \citet{Negrello2005}. 

This \Planck\ list of high-$z$ candidates opens a new window on the brightest
and rarest structures at high redshift, which remain unaccessible to other
detection methods.  It is the largest list of proto-cluster candidates at
$z>2$, detected in a homogeneous way over more than 25\,\% of the sky. It is
a unique and powerful sample of particular interest for structure formation
studies.  The full characterization of the PHZ sample is challenging and
it will require a huge effort to follow-up these objects and constrain their
nature.  A comparison with detailed structure formation models could then
be performed in order to reveal what can be learned from this population of 
high-$z$ objects about the early ages of our Universe.

\begin{acknowledgements}
The Planck Collaboration acknowledges the support of: ESA; CNES and
CNRS/INSU-IN2P3-INP (France); ASI, CNR, and INAF (Italy); NASA and DoE (USA);
STFC and UKSA (UK); CSIC, MINECO, JA, and RES (Spain); Tekes, AoF, and CSC
(Finland); DLR and MPG (Germany); CSA (Canada); DTU Space (Denmark);
SER/SSO (Switzerland); RCN (Norway); SFI (Ireland); FCT/MCTES (Portugal);
ERC and PRACE (EU). A description of the Planck Collaboration and a list of
its members, indicating which technical or scientific activities they have
been involved in, can be found at
\url{http://www.cosmos.esa.int/web/planck/planck-collaboration}.
\end{acknowledgements}

\bibliographystyle{aat}
\bibliography{PIP_XXXIX.bbl}

\appendix

\section{Cleaning with the \Planck\ CMB 8{\arcmin} map}
\label{sec:cmbclean}

Complementary to the Monte Carlo analysis performed in Sect.~\ref{sec:mcqa}
to study the impact of the CMB template quality on the detection and
photometry processing, we used the \Planck\ 143\,GHz map as a CMB template
to assess the level of extragalactic foregrounds included in the {\tt SMICA}
CMB component map, and its possible impact on the PHZ.
All \Planck, IRIS, and {\tt SMICA} CMB component map have been first smoothed
at a common resolution 8{\arcmin} in order to be compatible with the
143\,GHz map.  On this alternative set of maps, we applied the full
processing of cleaning, detection, photometry, and colour selection, to build
two new lists of high-$z$ source candidates at 8{\arcmin}, using either the
{\tt SMICA} CMB component map or the \Planck\ 143\,GHz map as a CMB template,
counting 1121 and 1038 high-$z$ source candidates, respectively.

The two catalogues have about 80\,\% of their sources in common. The 20\,\%
of non-matches correspond to sources with S/N close to the detection
thresholds, which is explained by the fact that the {\tt SMICA} CMB component
map and the \Planck\ 143\,GHz map do not exhibit the same noise properties.

\begin{figure}
\center
\hspace{-1cm}
\begin{tabular}{ccc}
\psfrag{------ytitle------}{$S_{857}^{\rm{D\, 8{\arcmin}\, 143\,\rm{GHz}}}$\,[Jy] }
\psfrag{------xtitle------}{$S_{857}^{\rm{D\, 8{\arcmin}}}$\,[Jy]}
\includegraphics[width=0.20\textwidth, viewport=60 0 490 500]{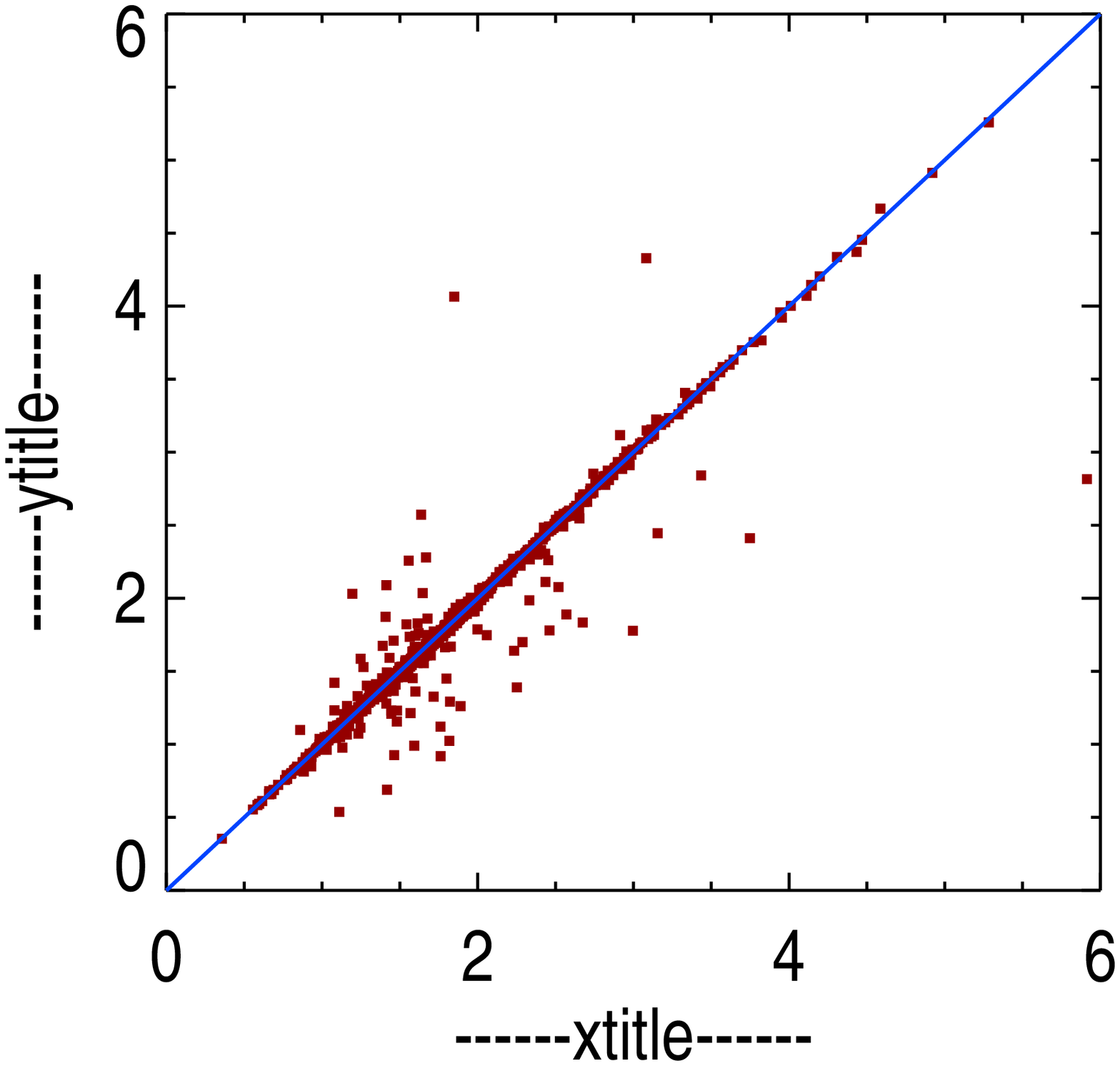}&
\quad&
\psfrag{------ytitle------}{$S_{545}^{\rm{D\, 8{\arcmin}\, 143\,\rm{GHz}}}$\,[Jy]}
\psfrag{------xtitle------}{$S_{545}^{\rm{D\, 8{\arcmin}}}$\,[Jy]}
\includegraphics[width=0.20\textwidth, viewport=60 0 490 500]{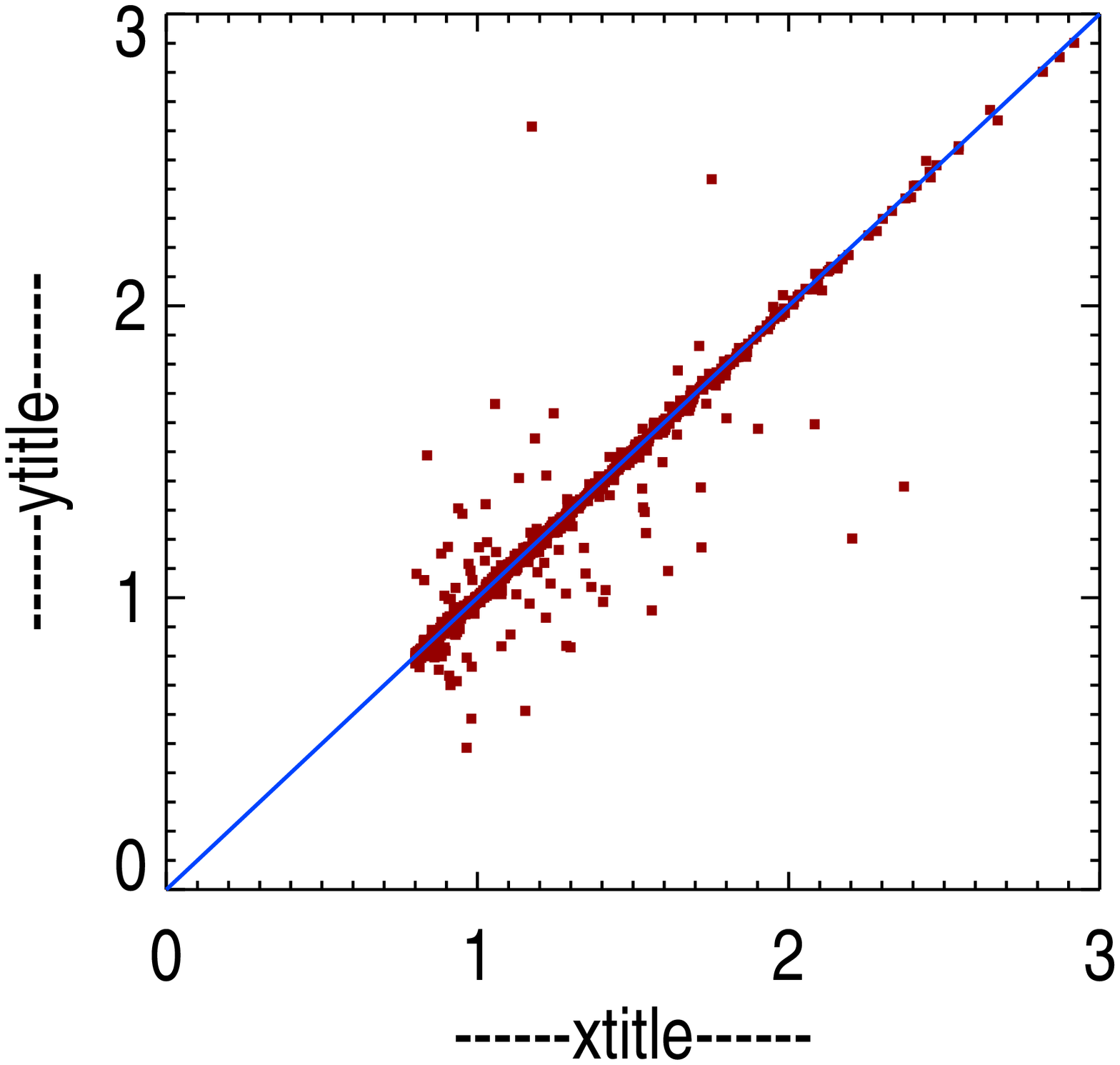} \\
\psfrag{------ytitle------}{$S_{353}^{\rm{D\, 8{\arcmin}\, 143\,\rm{GHz}}}$\,[Jy]}
\psfrag{------xtitle------}{$S_{353}^{\rm{D\, 8{\arcmin}}}$\,[Jy]}
\includegraphics[width=0.20\textwidth, viewport=60 0 490 500]{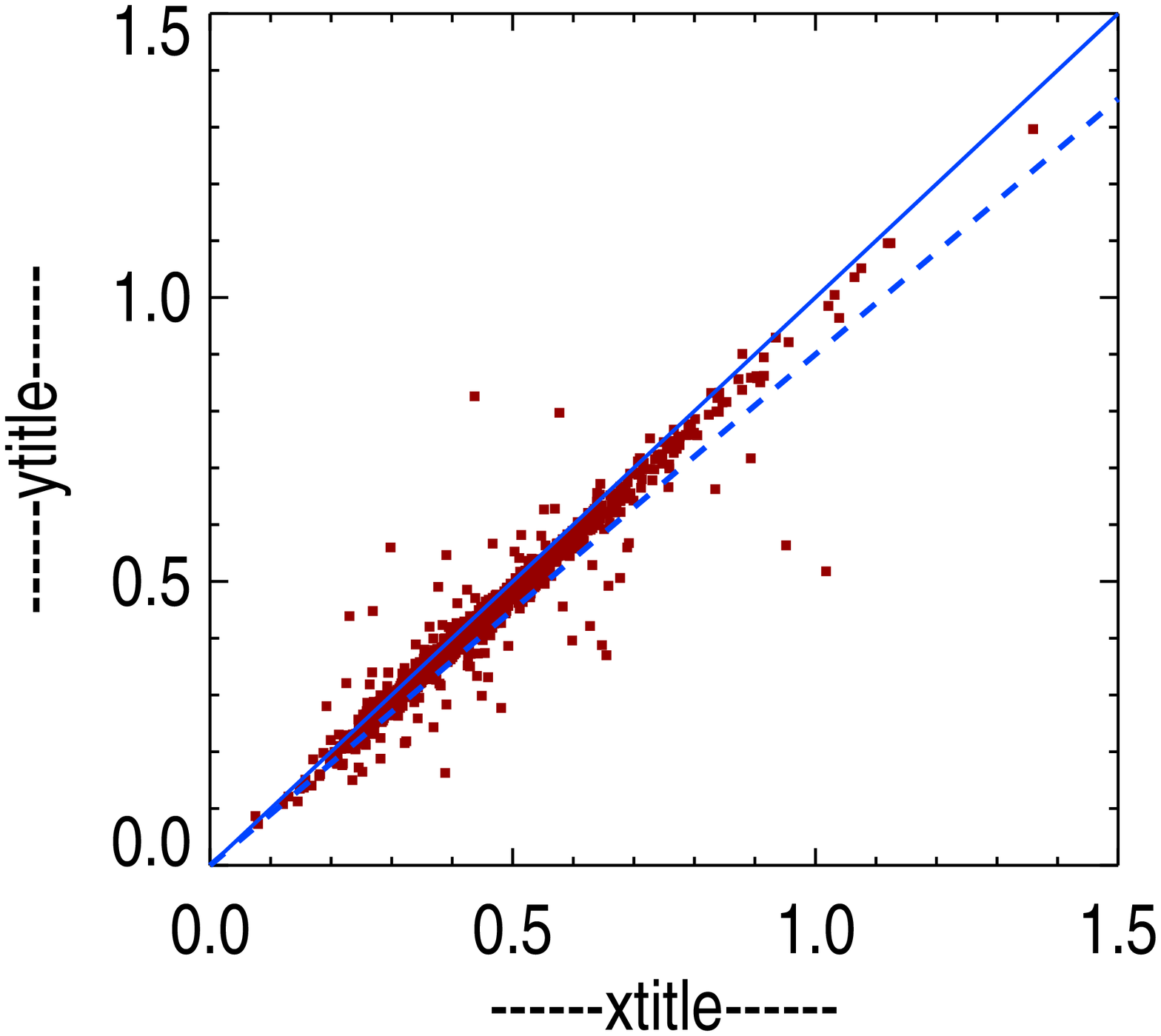}& &
\psfrag{------ytitle------}{$S_{217}^{\rm{D\, 8{\arcmin}\, 143\,\rm{GHz}}}$\,[Jy]}
\psfrag{------xtitle------}{$S_{217}^{\rm{D\, 8{\arcmin}}}$\,[Jy]}
\includegraphics[width=0.20\textwidth, viewport=60 0 490 500]{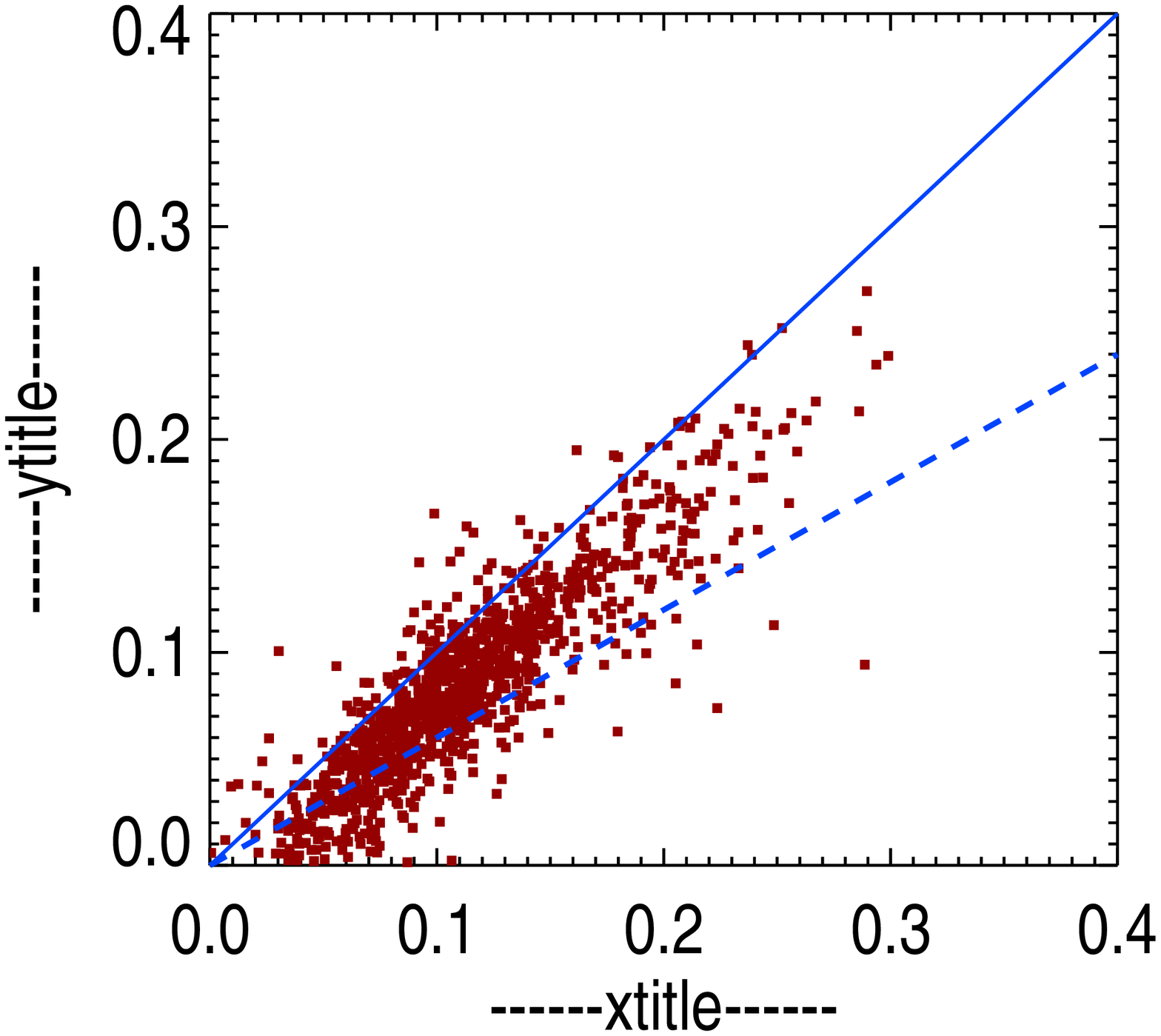} \\
\end{tabular}
\caption{Comparison of the flux densities of the PHZ source candidates
(at 8{\arcmin}) computed on the cleaned maps obtained using the \Planck\
143\,GHz map ($S_{\nu}^{\rm{D\, 8{\arcmin}\, 143}}$) or the {\tt SMICA} CMB
component map ($S_{\nu}^{\rm{D\, 8{\arcmin}}}$) as a CMB template during the
cleaning processing.  The blue line provides the 1:1 reference. The dashed
line gives the expected attenuation coefficients of 10\,\% at 353\,GHz and
40\,\% at 217\,GHz due to extragalactic foreground contamination of the CMB
template.}
\label{fig:xcheck_cmbclean}
\end{figure}

The level of extragalactic foreground contamination in the {\tt SMICA} CMB
template can be seen by comparing the flux densities toward to 1121 sources
of the list based on the {\tt SMICA} CMB component map at 8{\arcmin} and
obtained on the two versions of the cleaned maps at 857, 545, 353,
and 217\,GHz maps, as shown in Fig.~\ref{fig:xcheck_cmbclean}.
The flux density estimates of both cases are fully consistent in the 857- and
545-GHz bands, as is expected for the range of redshift of the PHZ sources
($1<z<4$, see Sect.~\ref{sec:impact_cleaning_highz}). 
The attenuation becomes important in the 353- and 217-GHz bands. 
The flux densities obtained using the {\tt SMICA} CMB component map appear
statistically larger than when using the \Planck\ 143\,GHz map, 
which confirms that they are less affected by the attenuation. However, they
do not entirely follow the statistical expectation of unattenuated flux
density estimates shown in blue dashed line of Fig.~\ref{fig:xcheck_cmbclean}.
This discrepancy may come from the diversity of the SEDs that have been
assumed to follow a modified blackbody emission law with a dust spectral
index of 1.5 in our modelling.  It can also be due to a residual of
extragalactic foregrounds in the {\tt SMICA} CMB component map, yielding up
to 5\,\% of attenuation in the 353-GHz band, instead of the 
10\,\% expected in the worse case. Unfortunately this residual emission is
hard to quantify, and has to be included in the photometric uncertainties.

It should be noted that earlier versions of the PHZ, which were used to select
targets for follow-up observations, such as the {\it Herschel\/} follow-up
described in Appendix~\ref{sec:herschel}, were all built using the 143\,GHz map
as a CMB template, because no CMB component maps were available at this time
at 5{\arcmin} resolution.  However, the 8\arcm\ and 5\arcm\ PHZ lists do not
exactly cover the same population. Only 458 objects match both lists within
5{\arcmin}.  This is explained by the fact that compact sources detected at
5{\arcmin} may be diluted in an 8{\arcmin} beam, yielding no detection in the
latter case.  On the other hand, extending structures integrated within a
8{\arcmin} beam may not exhibit any 5{\arcmin} features, yielding no detection
in the 5{\arcmin} list.

\section{Redshift estimate accuracy}
\label{sec:redshift_accuracy}

\begin{figure*}
\hspace{-0.4cm}
\begin{tabular}{ccccc}
$\quad \quad$ $T_{\rm{xgal}}=20\,$K& 
$\quad \quad$ $T_{\rm{xgal}}=30\,$K& 
$\quad \quad$ $T_{\rm{xgal}}=40\,$K& 
$\quad \quad$ $T_{\rm{xgal}}=50\,$K& \\
\\
\psfrag{------xtitle------}{Injected $z$}
\psfrag{------ytitle------}{Recovered $z$}
\includegraphics[width=0.23\textwidth, viewport=0 0 500 500]{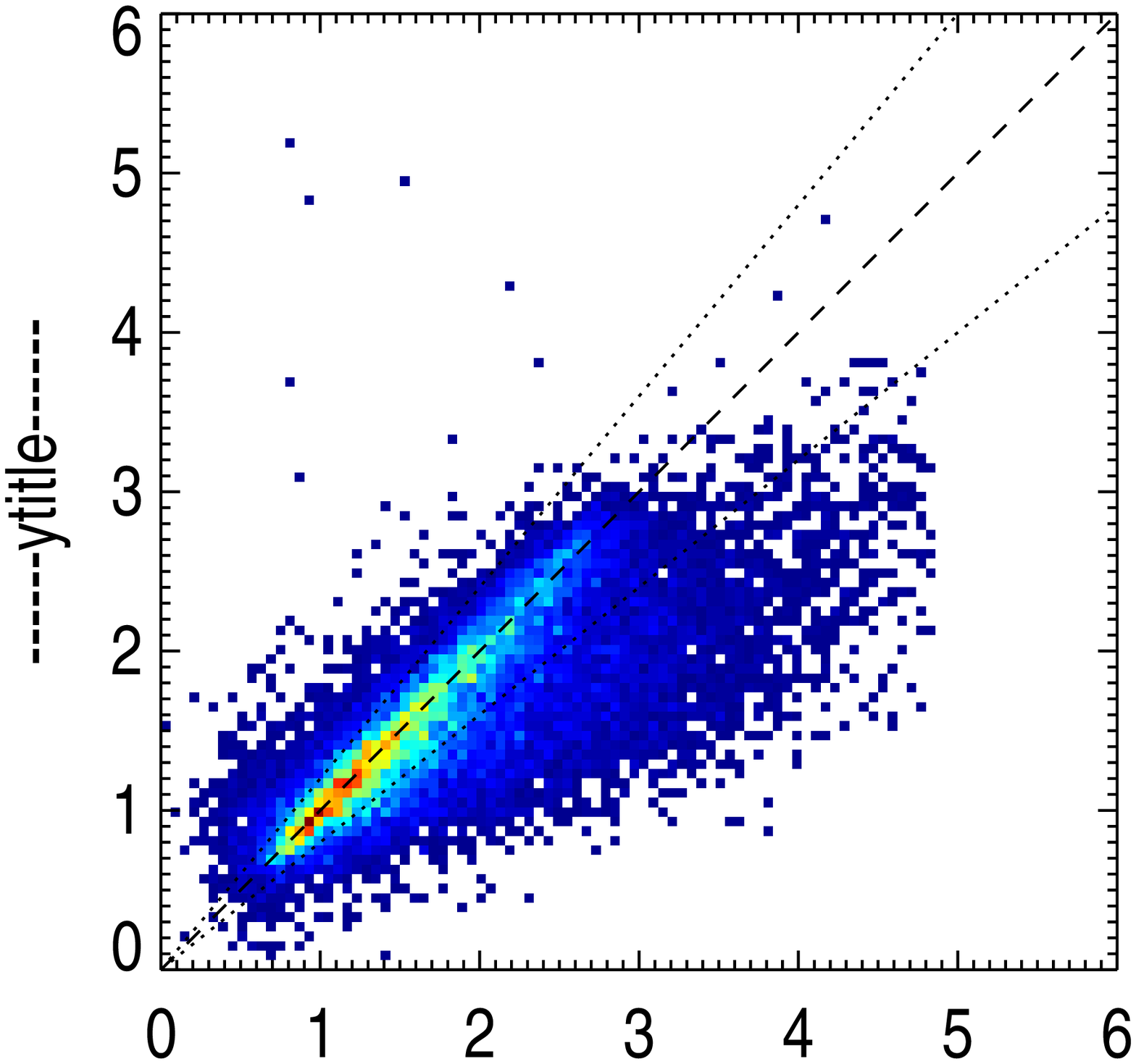}&
\psfrag{------xtitle------}{Injected $z$}
\includegraphics[width=0.23\textwidth, viewport=0 0 500 500]{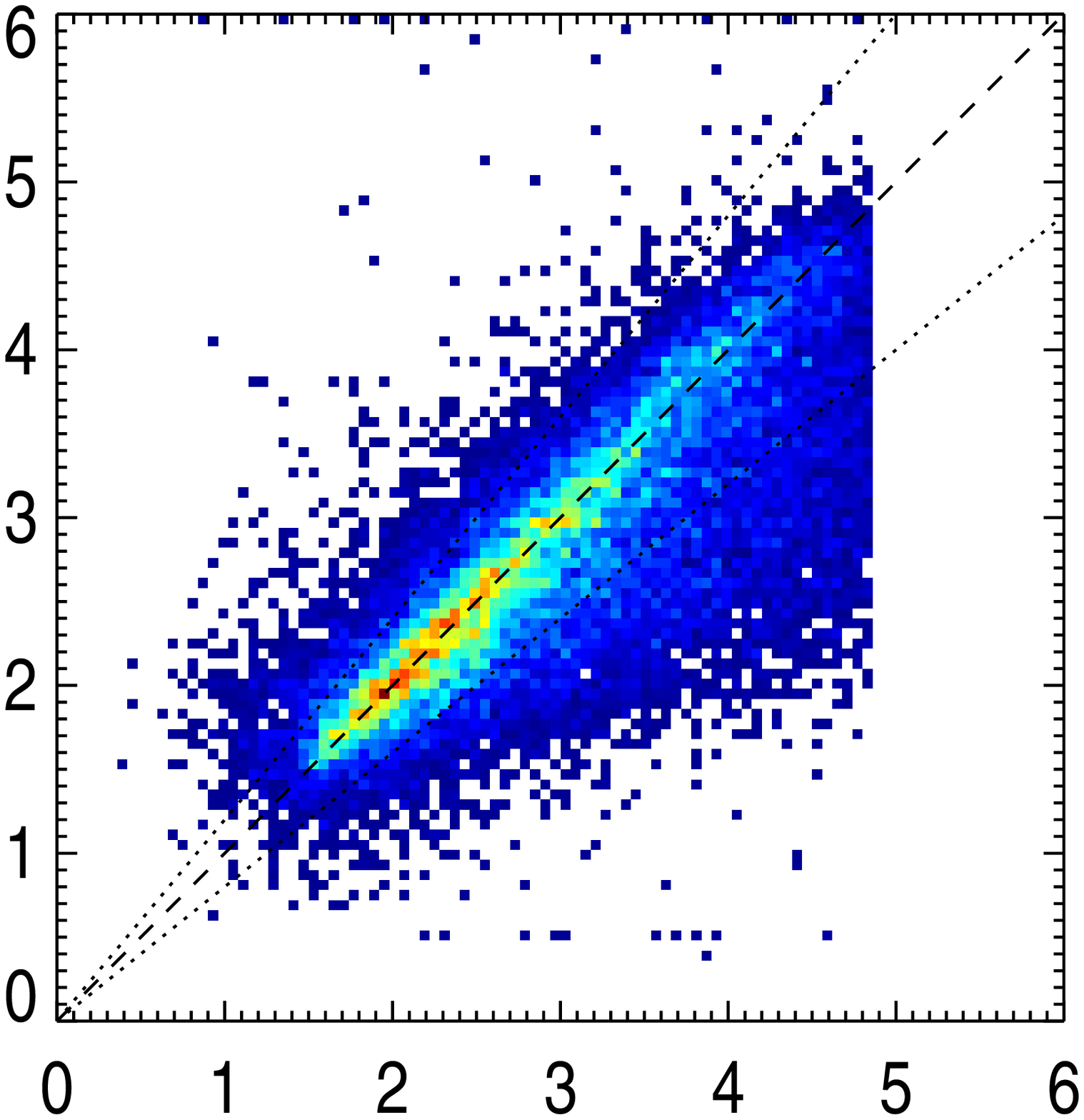}&
\psfrag{------xtitle------}{Injected $z$}
\includegraphics[width=0.23\textwidth, viewport=0 0 500 500]{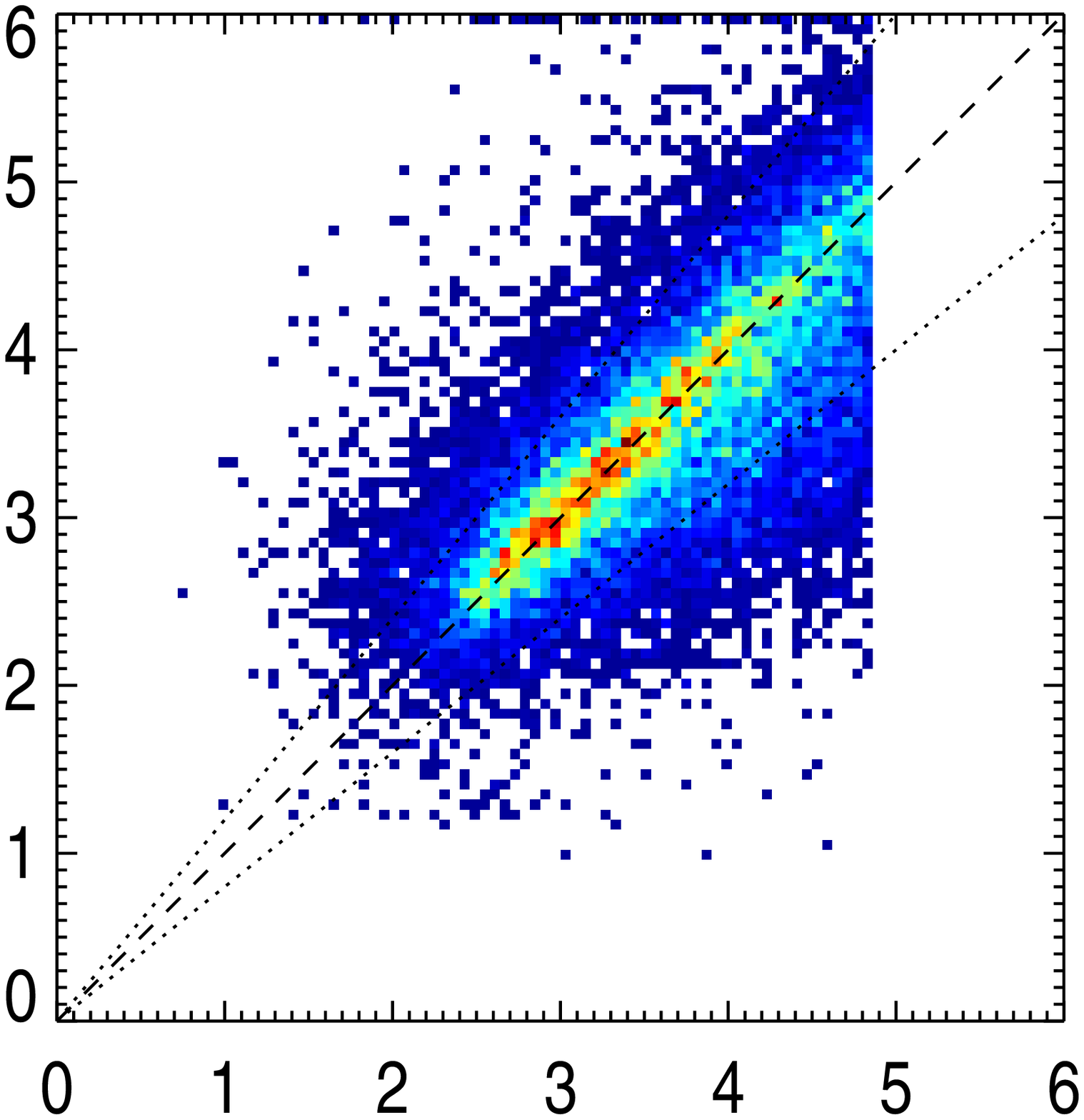}&
\psfrag{------xtitle------}{Injected $z$}
\includegraphics[width=0.23\textwidth, viewport=0 0 500 500]{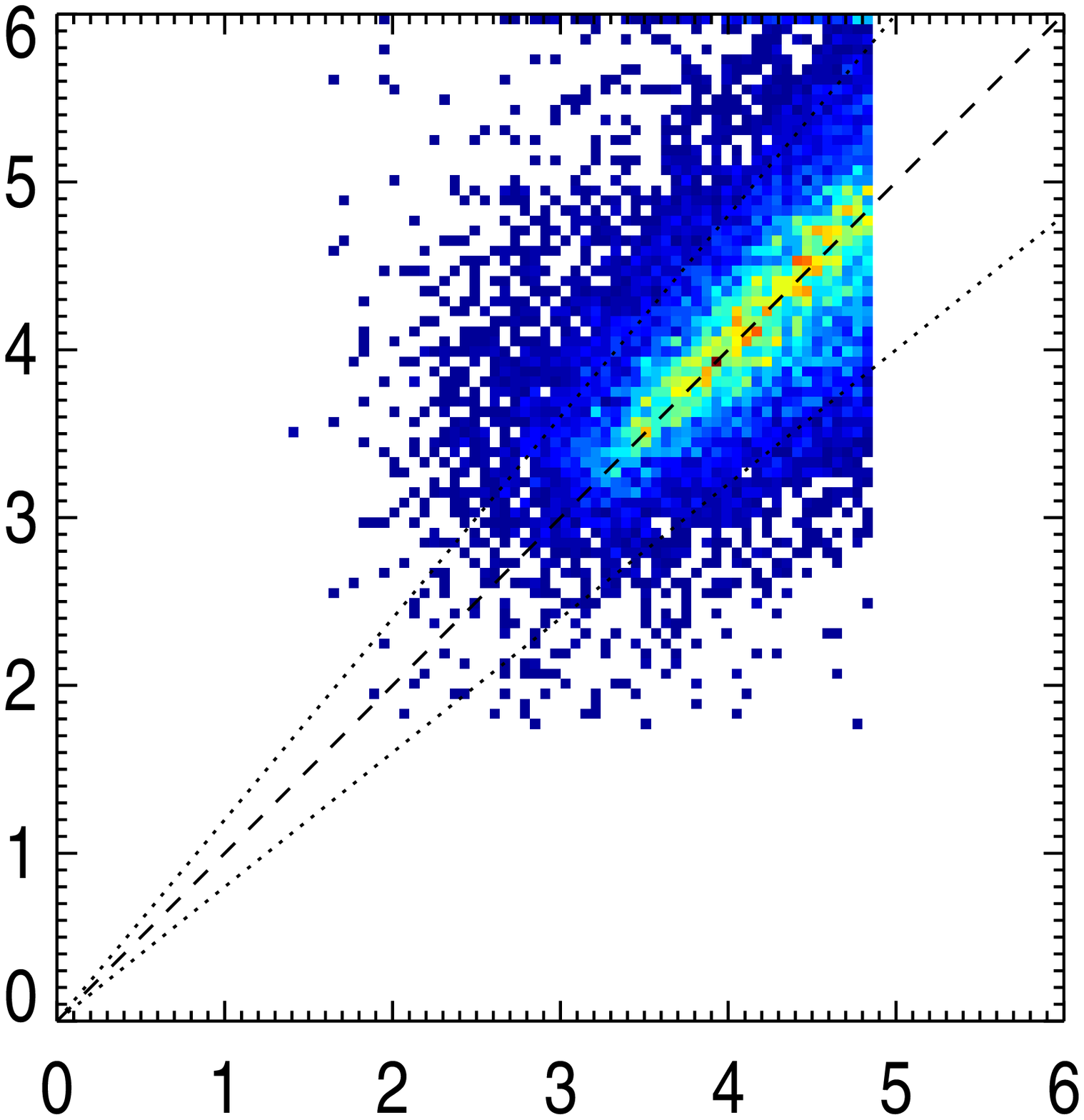} \\
\psfrag{------xtitle------}{Injected $z$}
\psfrag{------ytitle------}{Recovered $z$}
\includegraphics[width=0.23\textwidth, viewport=0 0 500 500]{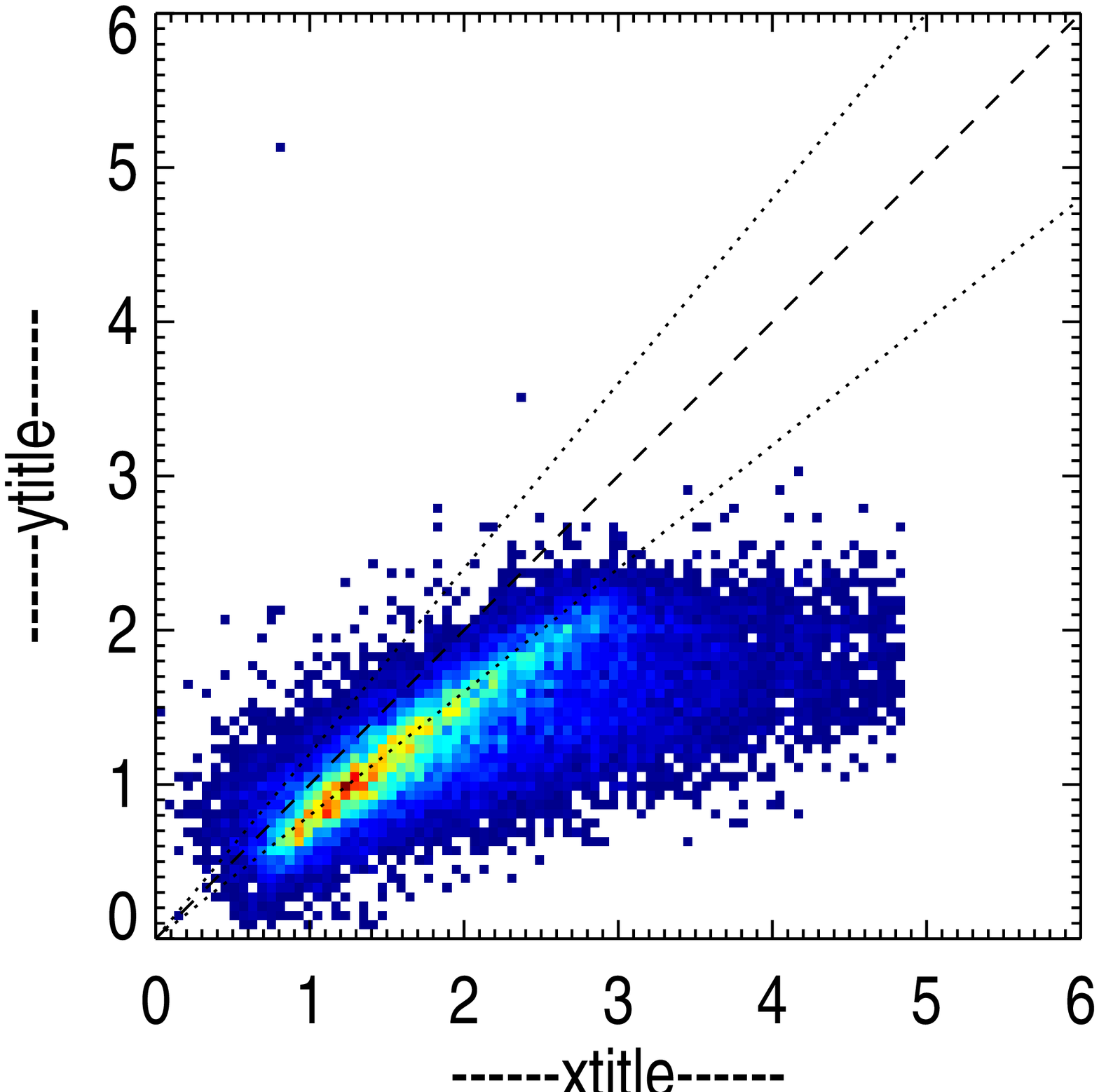}&
\psfrag{------xtitle------}{Injected $z$}
\includegraphics[width=0.23\textwidth, viewport=0 0 500 500]{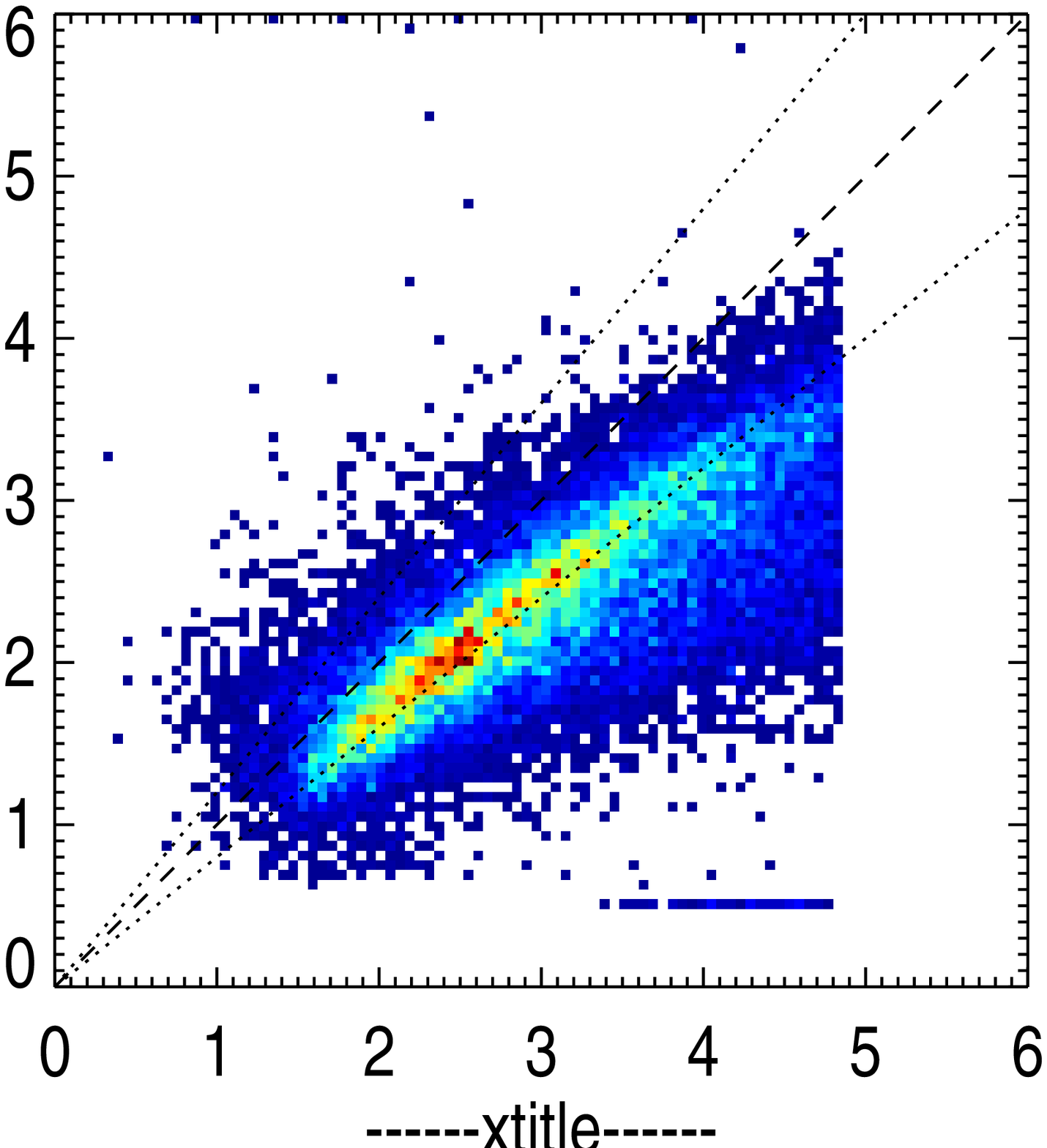}&
\psfrag{------xtitle------}{Injected $z$}
\includegraphics[width=0.23\textwidth, viewport=0 0 500 500]{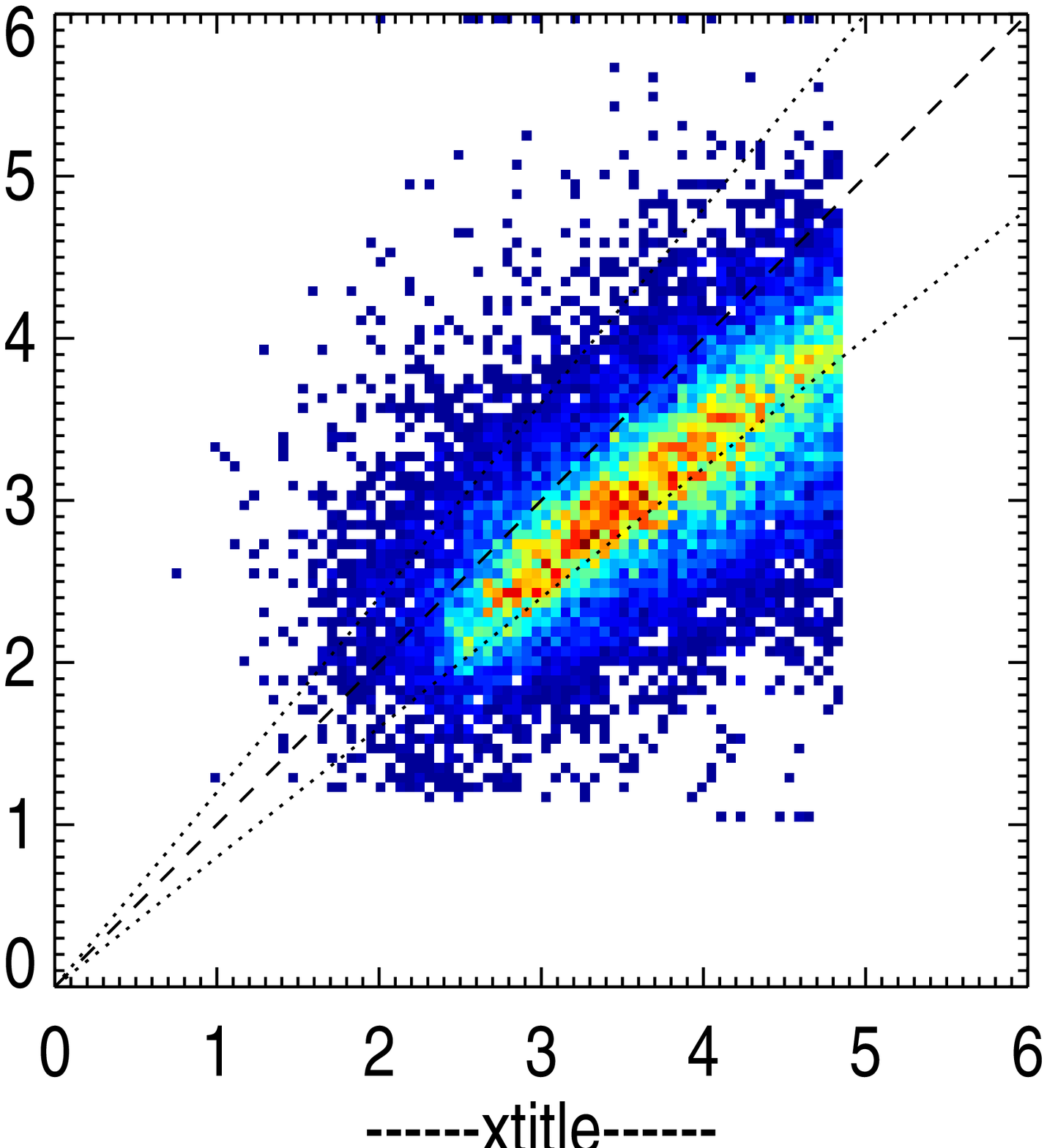}&
\psfrag{------xtitle------}{Injected $z$}
\includegraphics[width=0.23\textwidth, viewport=0 0 500 500]{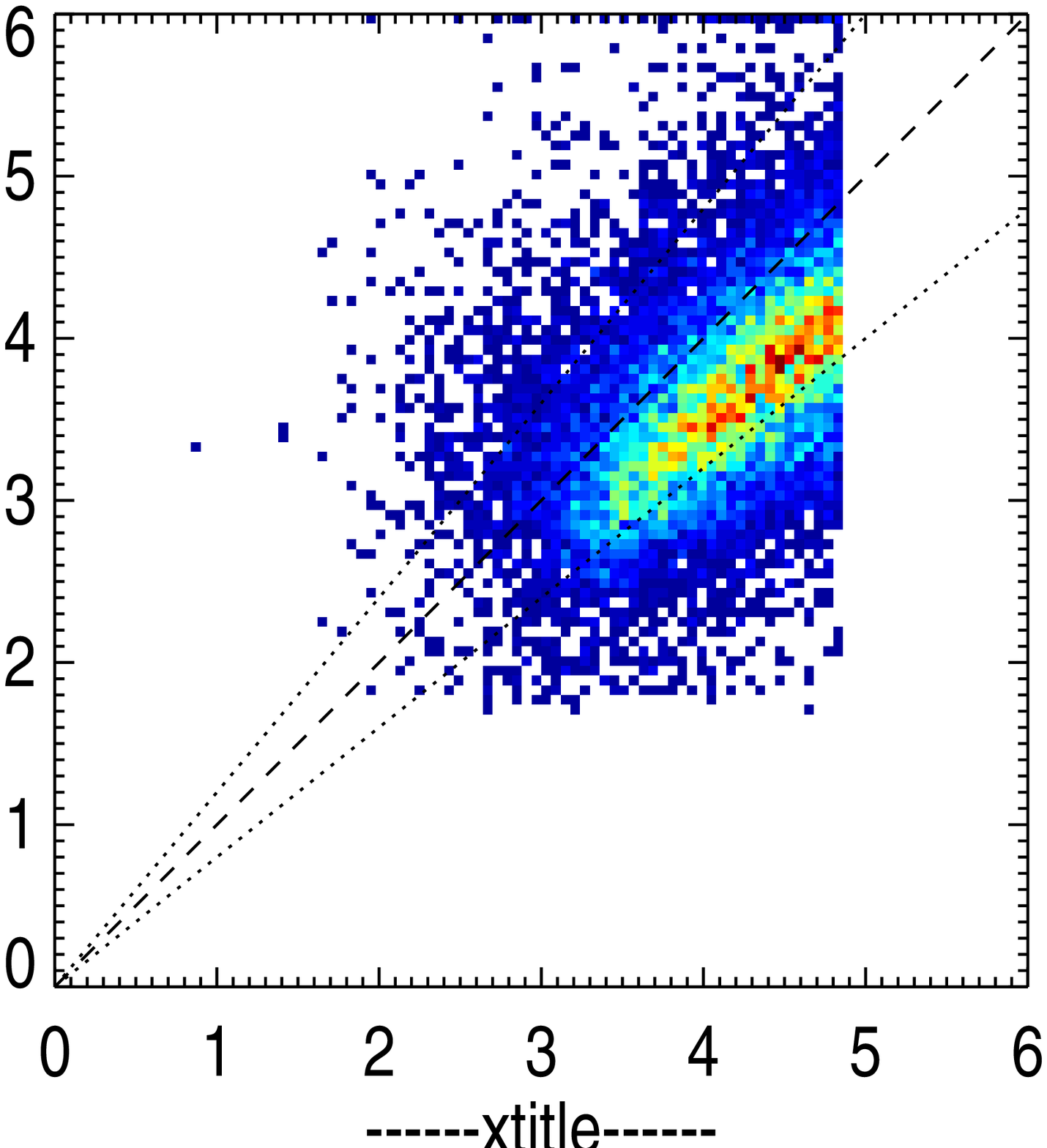} \\
\end{tabular}
\caption{Photometric redshift recovery, for injected dust temperatures
$T_{\rm{xgal}}=$ 20, 30, 40, and 50\,K (from left to right).
Redshift estimates have been obtained assuming the exact dust temperature for
each source, and an ideal CMB template.  The 2-D histogram of the recovered
redshift estimates versus the injected redshifts are shown for two cases of
the quality of the CMB template used for cleaning:
ideal (top) or highly contaminated (bottom).  The dotted lines show the
$\pm20$\,\% limits around the 1:1 relation (dashed line).}
\label{fig:mcqa_accuracy_redshift}
\end{figure*}

We have tested the accuracy of the photometric redshift estimate processing (see Sect.~\ref{sec:redshift}) using the Monte Carlo simulations
 presented in Sect.~\ref{sec:mcqa_method}. 
 We applied the same SED-fitting algorithm based on the recovered flux densities at 857, 545, 353, and 217\,GHz for each injected and detected source 
 of the mock catalogue. In order to check the impact of the cleaning process and the photometric accuracy on these redshift estimates, we
have compared the recovered redshift estimates with the input values injected in the Monte Carlo simulations, assuming the correct injected temperature.

As shown in Fig.~\ref{fig:mcqa_accuracy_redshift}, the photometric redshift estimates are not reliable over the full range of redshift. 
Even if the CMB template is assumed to be ideal for both the simulations and the SED modelling used to fit the redshift, 
the photometric redshift estimates are systematically high for the lowest detectable redshifts, 
and are underestimated for the largest detectable redshifts, for each range of dust temperature. However, in the intermediate range of redshift, 
where most of the sources are detected, the accuracy is about 10\,\%, which is sufficiently accurate for our purpose.

When assuming an ideal CMB template in computing the theoretical attenuation
coefficients for each \Planck\ band before the SED fitting,
Fig.~\ref{fig:mcqa_accuracy_redshift} shows that when the estimate
is actually highly contaminated by extragalactic foregrounds (bottom panels),
the associated photometric redshift estimates are statistically underestimated
by 15 to 20\,\%. This last number gives the maximum impact due to the
contamination of the CMB template on the redshift estimates.

This simple analysis, of course, does not take into account all other
uncertainties impacting any photometric redshift estimate, 
such as the degeneracy between the redshift and the dust temperature,
or the SED assumption.  For all these reasons, the photometric redshift
estimates delivered in this list are provided as basic estimates only, and 
should be used with caution.

\section{The \textit{Herschel} sub-sample}
\label{sec:herschel}

\begin{figure*}
\begin{tabular}{cccc}
\psfrag{------xtitle------}{$\quad$S/N $S_{545}^{\rm{X}}$}
\includegraphics[width=0.21\textwidth, viewport=60 0 470 500]{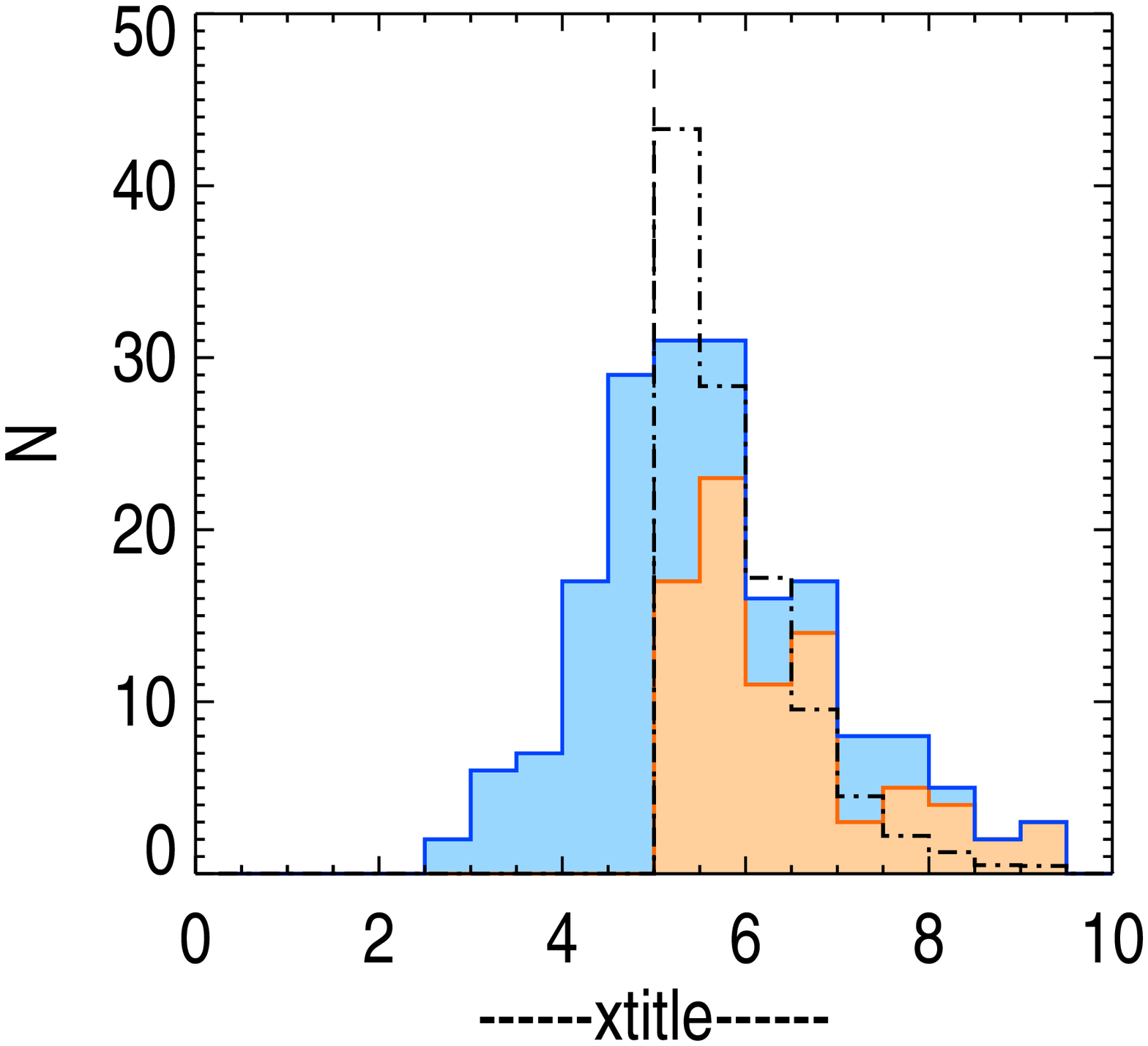}&
\psfrag{------xtitle------}{$\quad$S/N $S_{857}^{\rm{D}}$}
\includegraphics[width=0.21\textwidth, viewport=60 0 470 500]{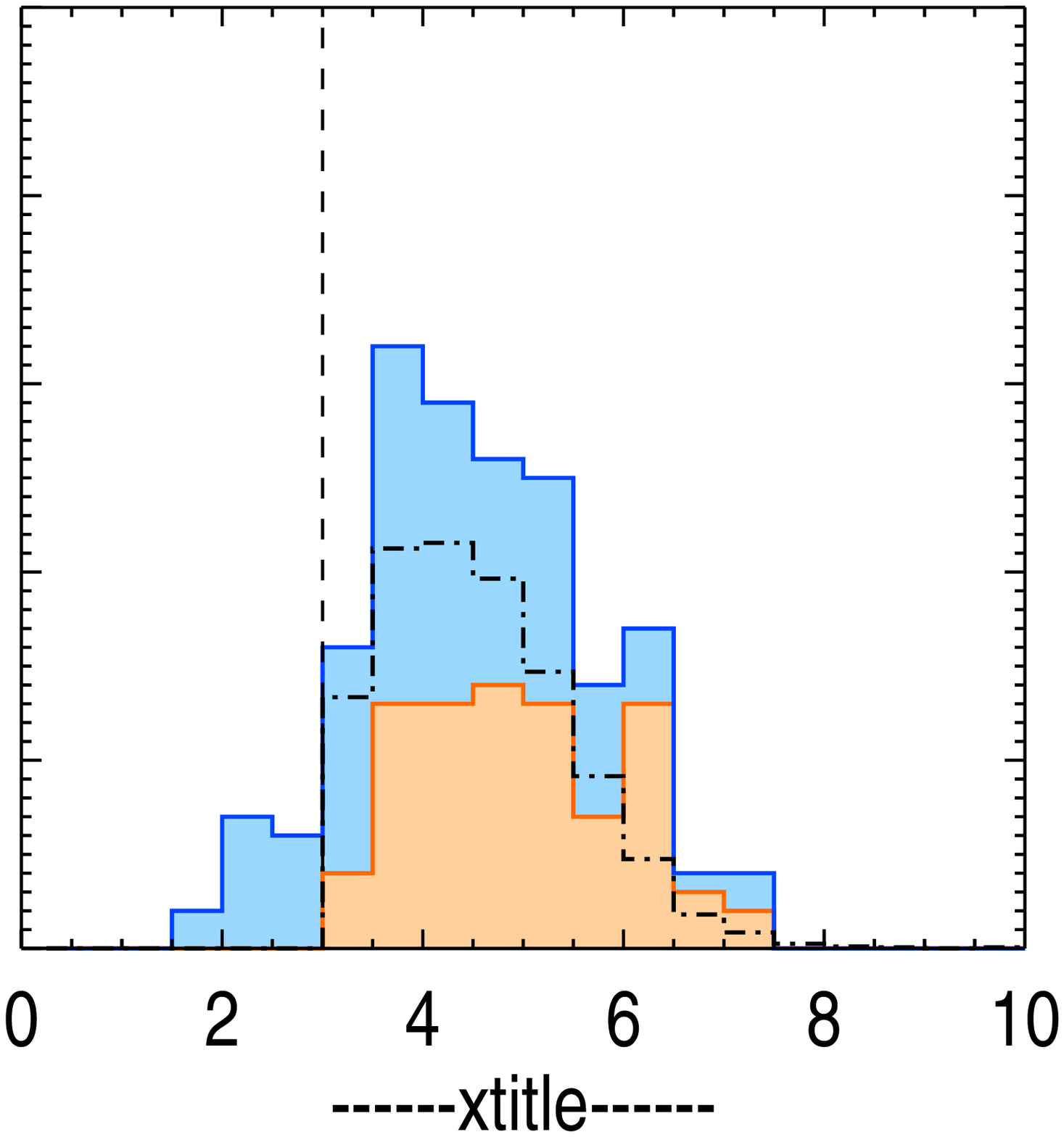}&
\psfrag{------xtitle------}{$\quad$S/N $S_{545}^{\rm{D}}$}
\includegraphics[width=0.21\textwidth, viewport=60 0 470 500]{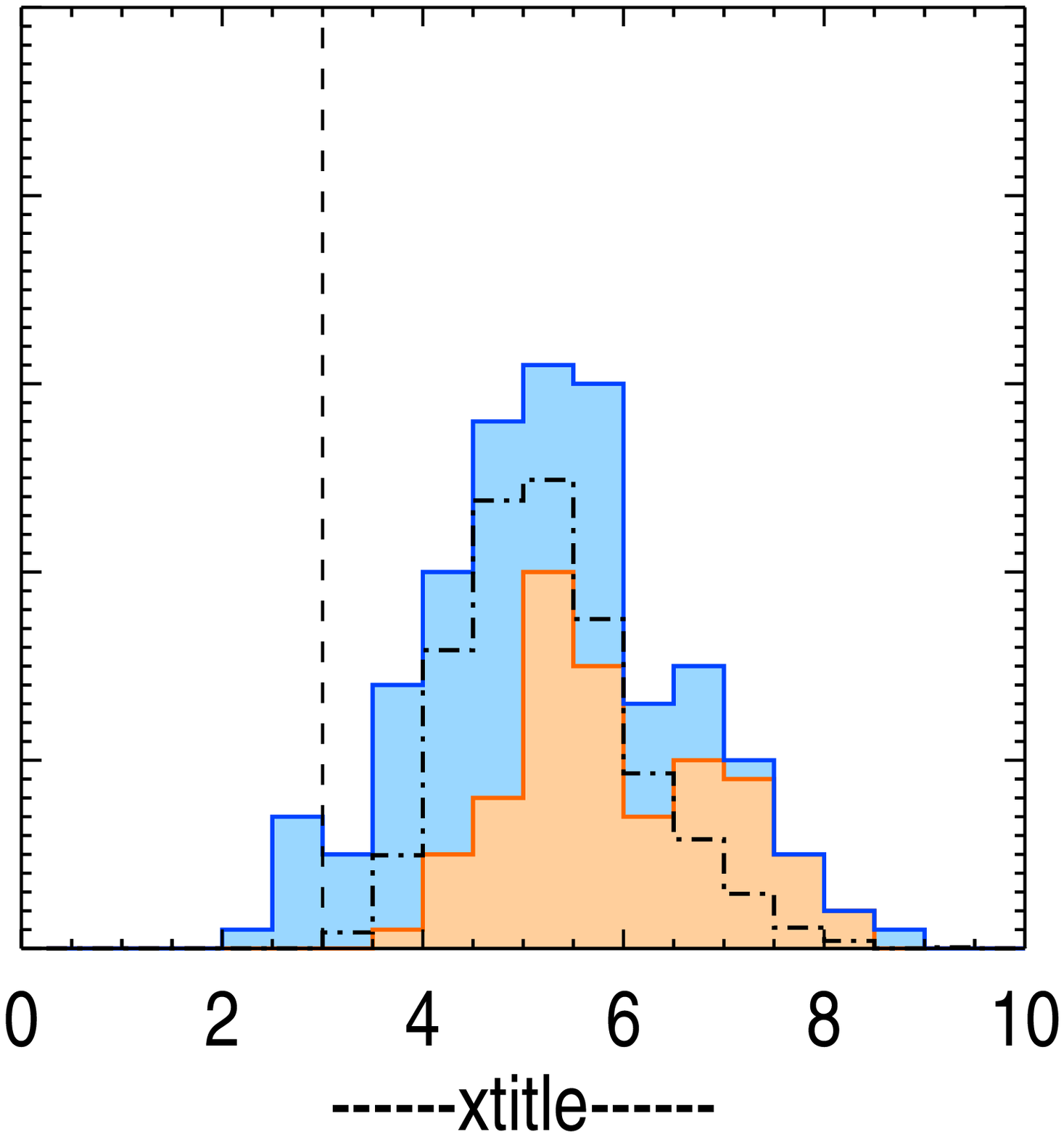}&
\psfrag{------xtitle------}{$\quad$S/N $S_{353}^{\rm{D}}$}
\includegraphics[width=0.21\textwidth, viewport=60 0 470 500]{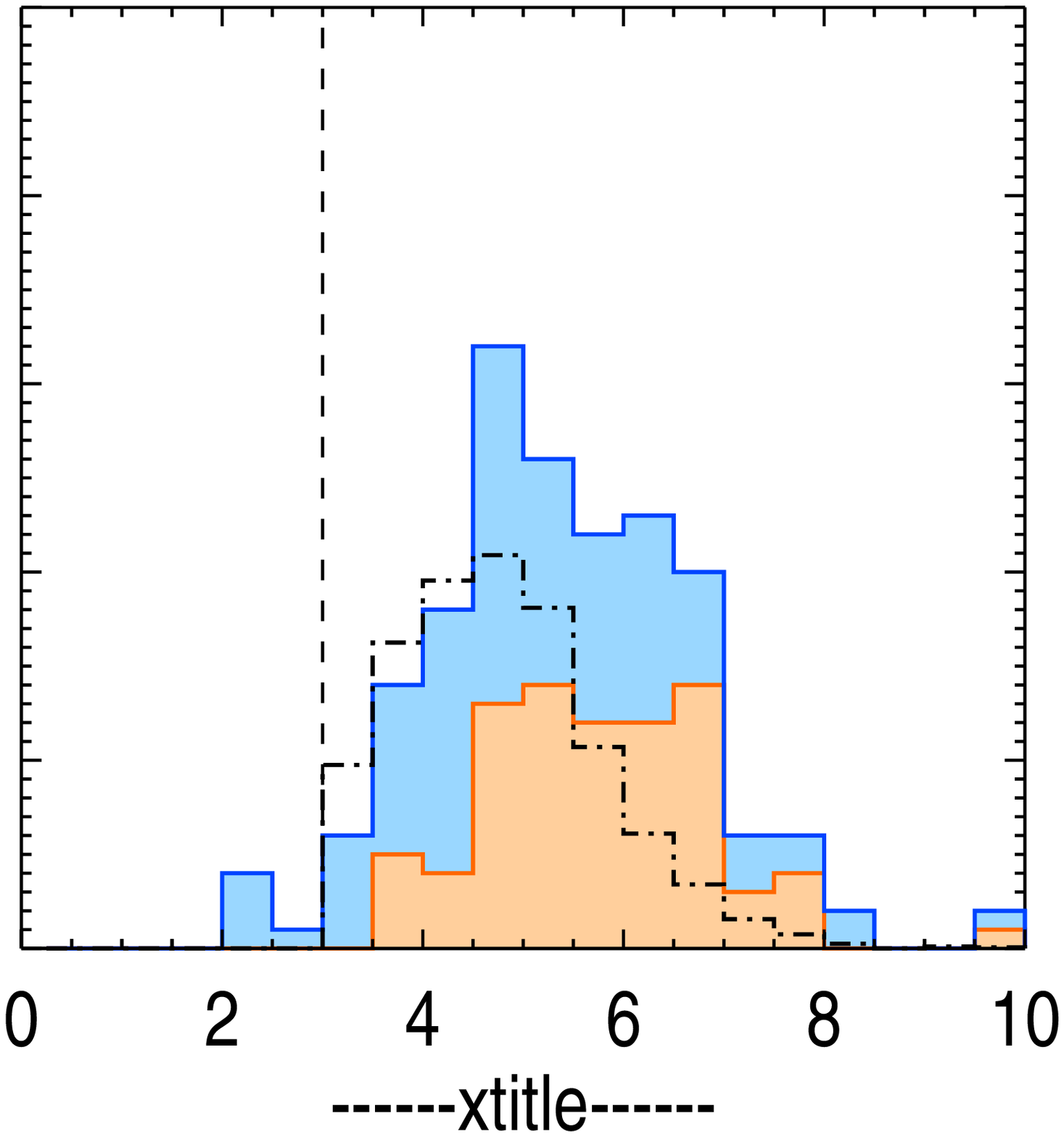}\\
\end{tabular}
\caption{Distribution of the S/N in the excess and cleaned \Planck\ maps at
the coordinates of the 182 sources of the {\it Herschel\/} sample that exhibit a
detection in the deep list obtained with a 1$\,\sigma$ threshold in all bands.
The distribution of the 83 sources followed-up by {\it Herschel\/} and present
in the PHZ is shown in orange, while the rest of the sample is shown in blue.
The distribution of the full PHZ is shown in dot-dashed black line, scaled
by a factor 1/20.  The dashed lines show the S/N thresholds required in all
bands for a detection.}
\label{fig:herschel_sn}
\end{figure*}

\begin{figure*}
\center
\hspace{-1cm}
\begin{tabular}{ccc}
\psfrag{------xtitle------}{$\quad$S/N $S_{545}^{\rm{X}}$}
\includegraphics[width=0.21\textwidth, viewport=60 0 470 500]{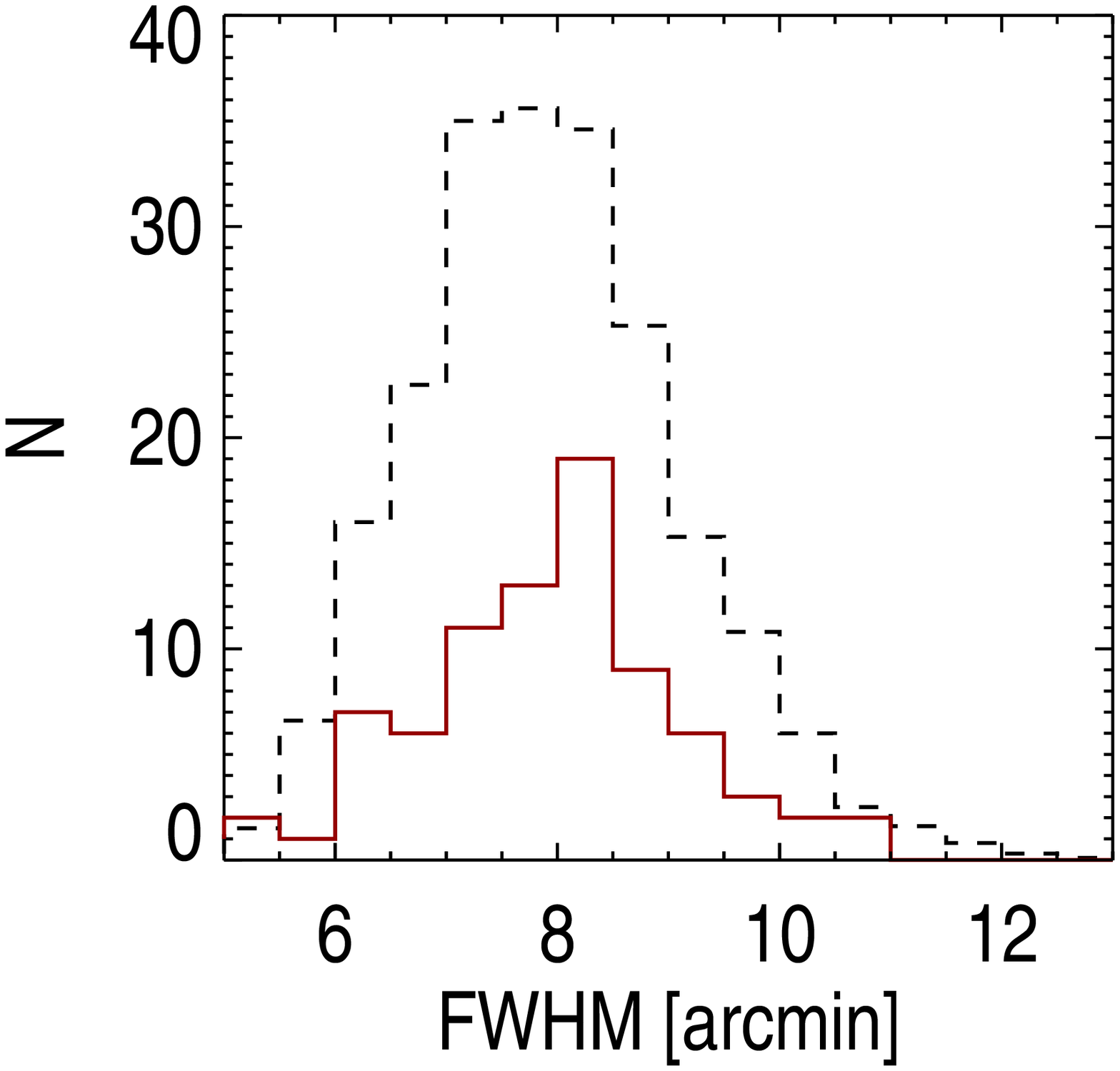}&
\includegraphics[width=0.21\textwidth, viewport=60 0 470 500]{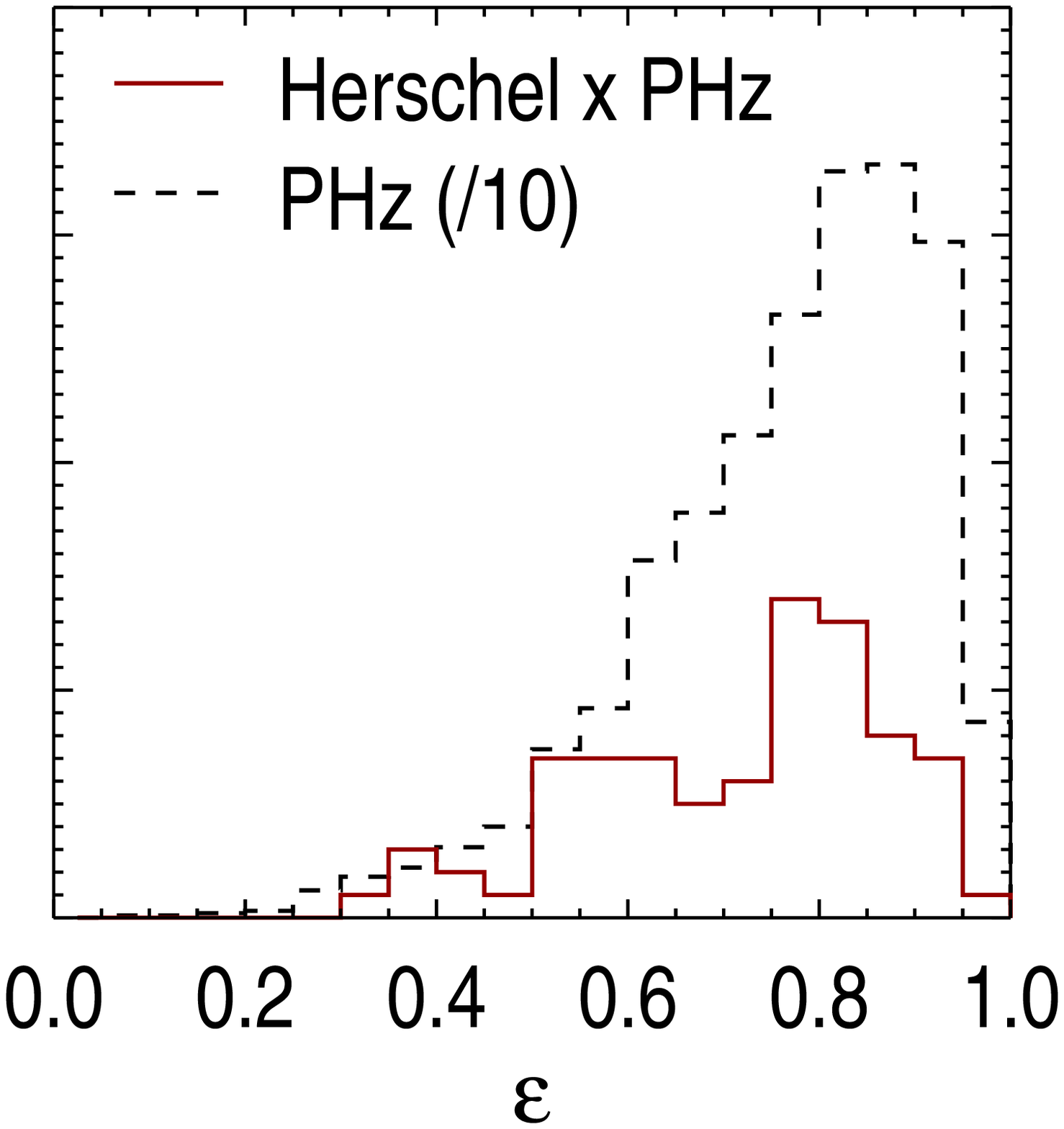}&
\includegraphics[width=0.21\textwidth, viewport=60 0 470 500]{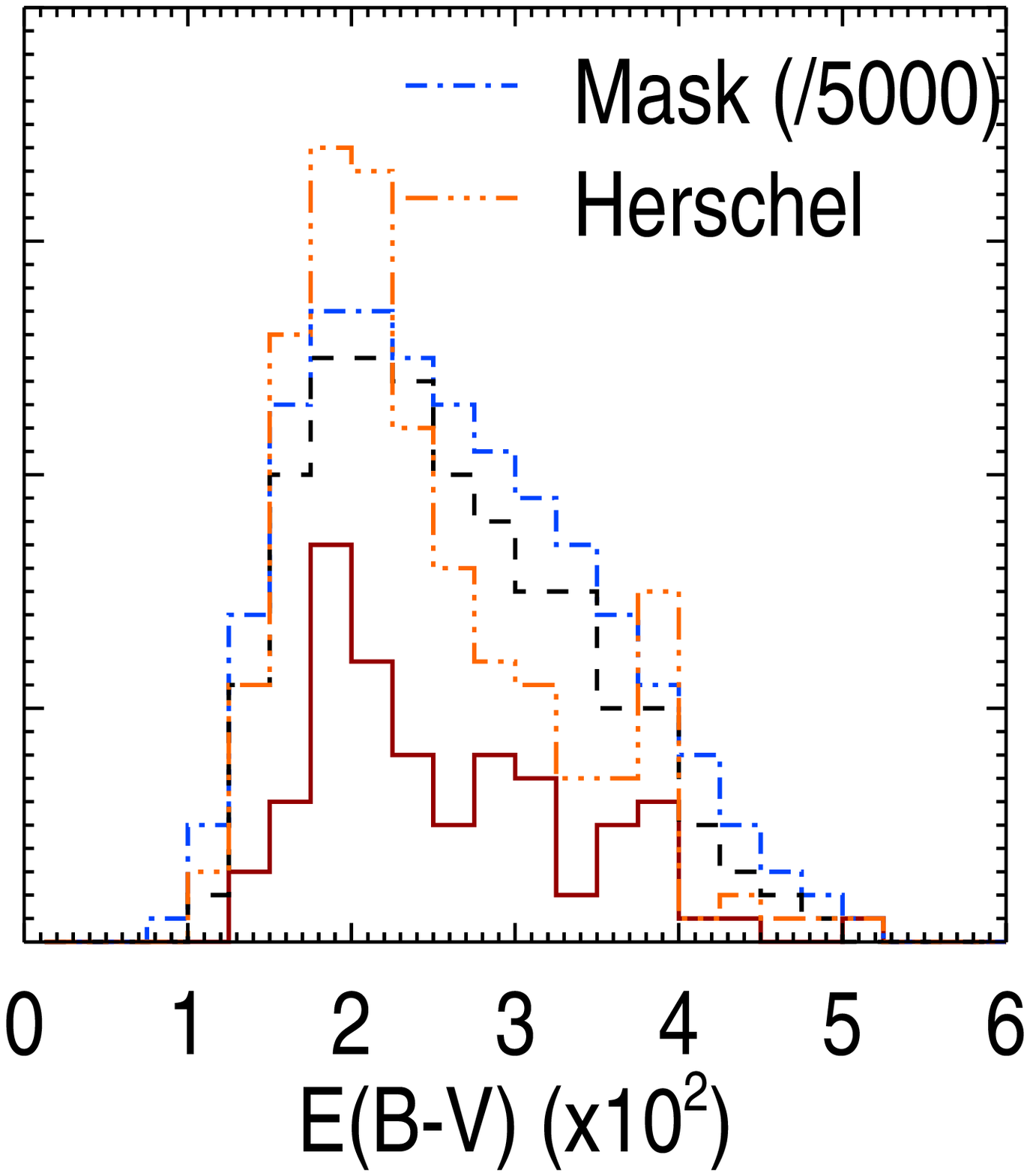} \\
\end{tabular}
\caption{Distribution of the properties for the {\it Herschel\/} sub-sample of
83 sources still present in the PHZ. {\it Left}: FWHM. {\it Middle}:
ellipticity of the Gaussian elliptical fit. {\it Right}: local extinction
$E(B-V)_{\rm{xgal}}$.  The distribution of the PHZ is repeated (dashed line)
with a factor 1/10. The distribution of the extinction $E(B-V)_{\rm{xgal}}$ is
also shown for the whole mask (dashed blue line), and for the whole
{\it Herschel\/} sub-sample (orange).}
\label{fig:statistics_herschel}
\end{figure*}

\begin{figure*}
\center
\hspace{-3cm}
\begin{tabular}{cccc}
\quad \quad \quad Sources not in PHZ& \quad \quad Sources in PHZ& & \\
\\
\includegraphics[width=0.23\textwidth, viewport=60 0 490 500]{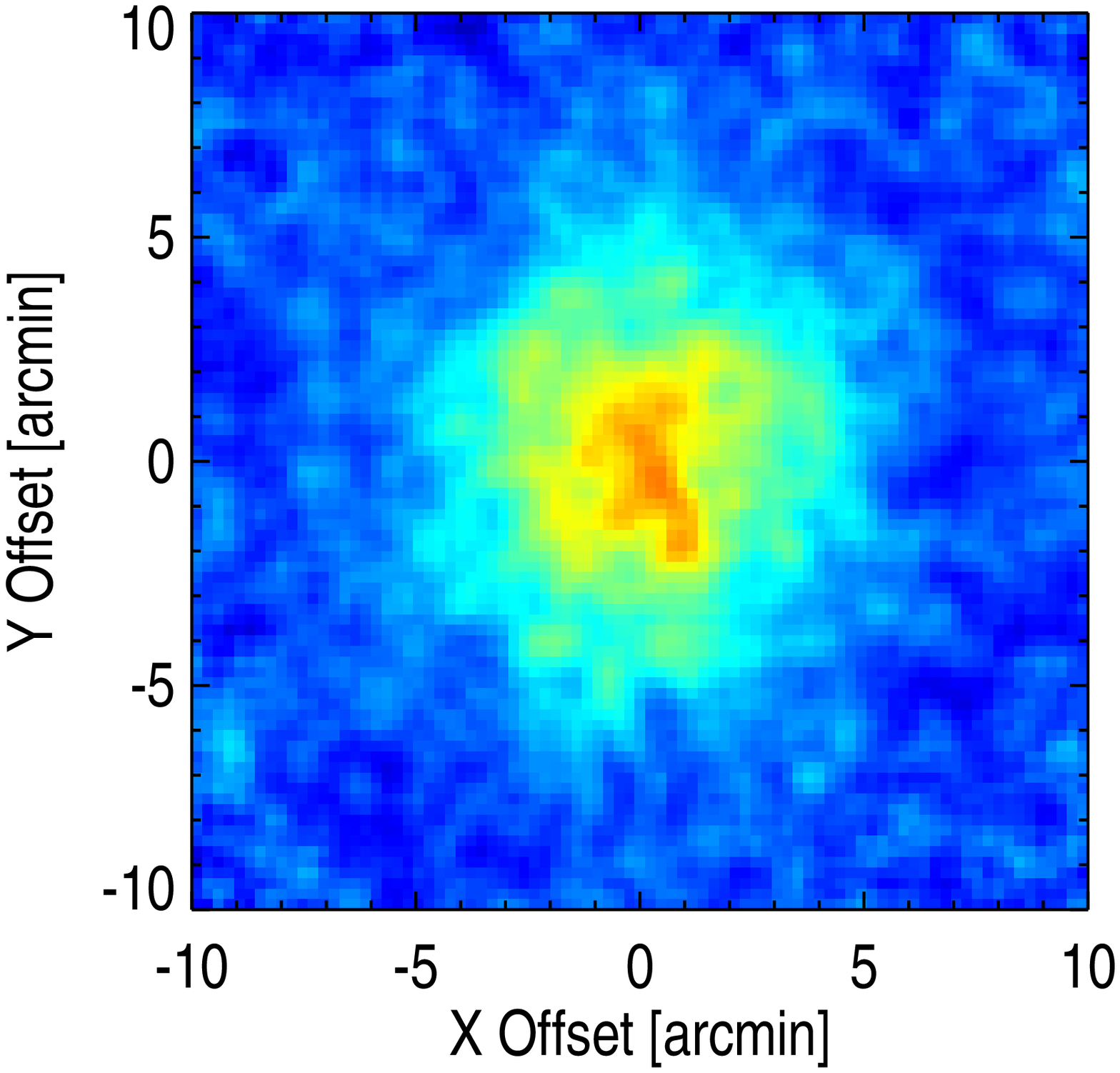}&
\includegraphics[width=0.23\textwidth, viewport=60 0 490 500]{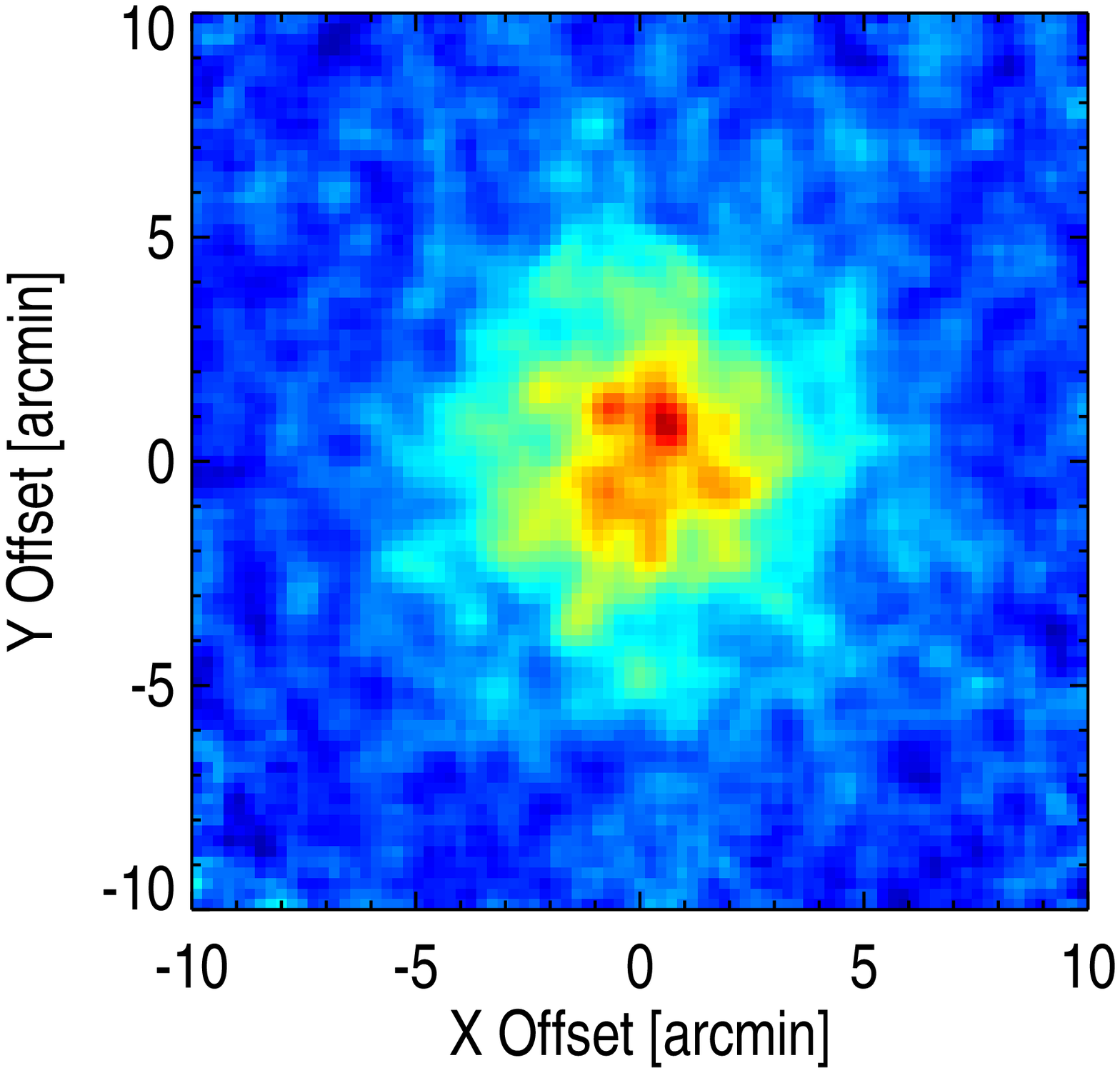}&
\psfrag{----ytitle----}{\tiny{$\rm{MJy}\,\rm{sr}^{-1}$}}
\includegraphics[width=0.0735\textwidth, viewport=0 -53 120 47]{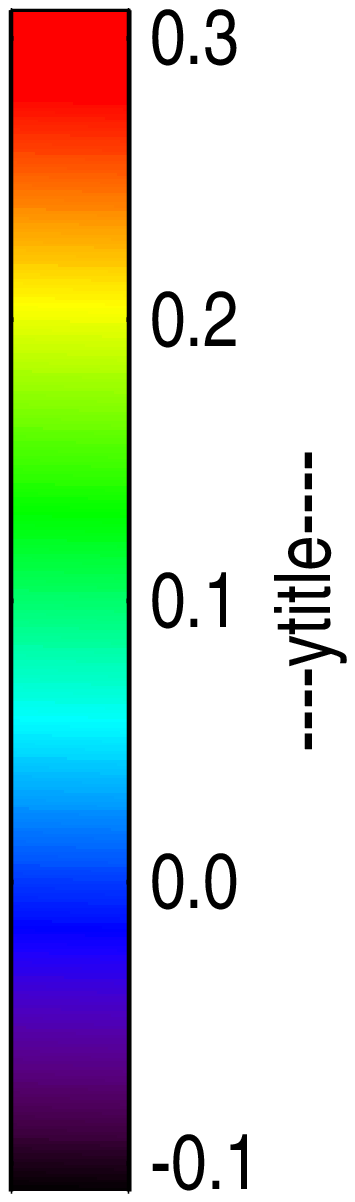}&
\psfrag{------xxxlegendxxx------}{{\tiny Sources in PHZ}}
\psfrag{------yyylegendyyy------}{{\tiny Sources not in PHZ}}
\psfrag{----------ytitle----------}{\tiny{$500\,\mu \rm{m} \, \rm{[MJy}\,\rm{sr}^{-1}\rm{]}$}}
\includegraphics[width=0.22\textwidth, viewport=30 -15 360 360]{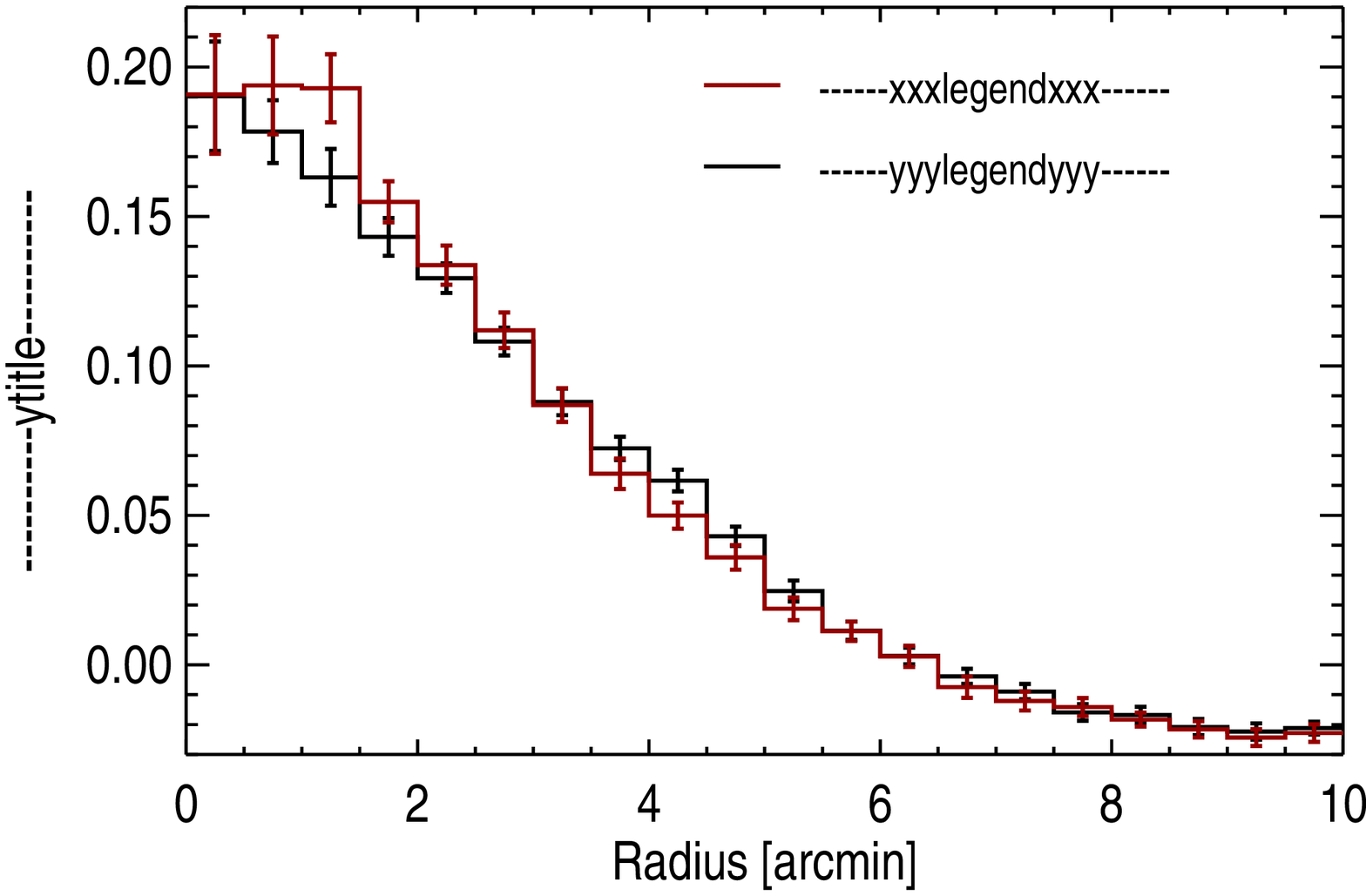}\\
\includegraphics[width=0.23\textwidth, viewport=60 0 490 500]{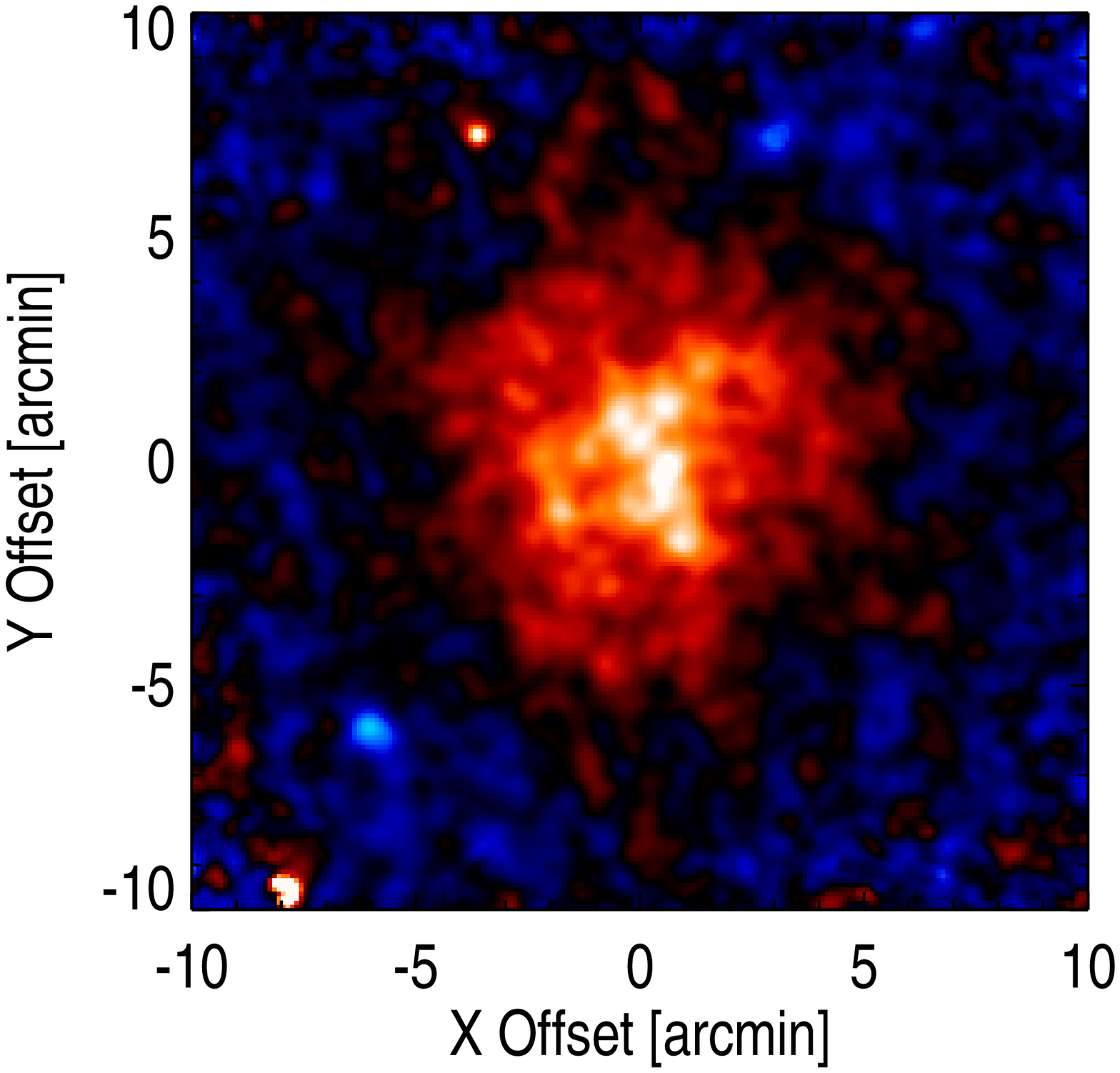}&
\includegraphics[width=0.23\textwidth, viewport=60 0 490 500]{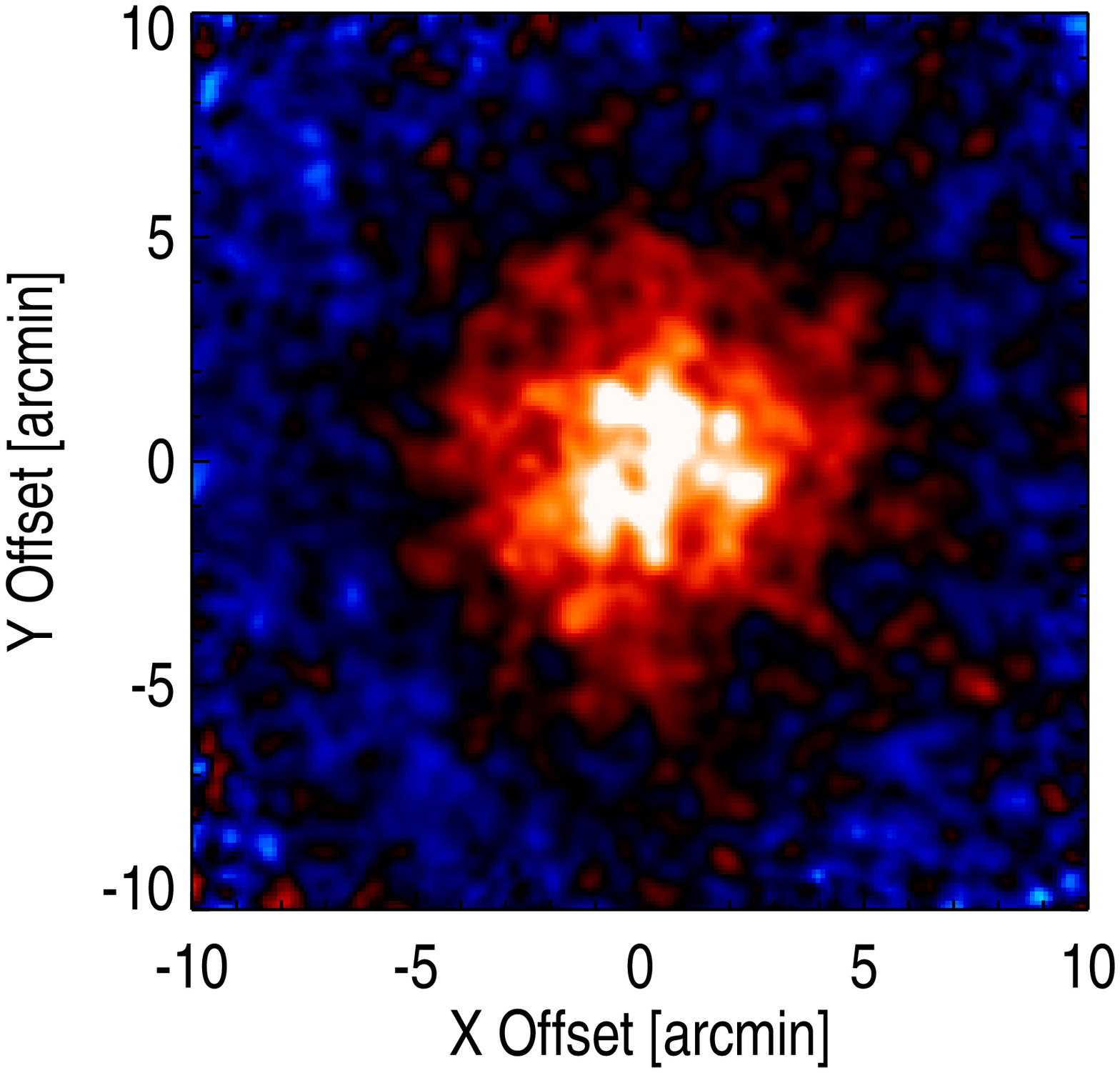}&
\psfrag{----ytitle----}{\tiny{$\rm{MJy}\,\rm{sr}^{-1}$}}
\includegraphics[width=0.0735\textwidth, viewport=0 -53 120 47]{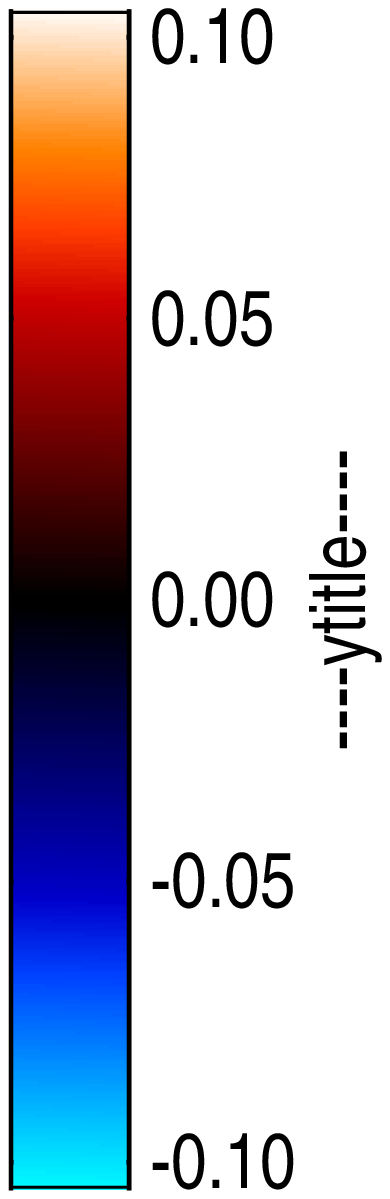}&
\psfrag{------xxxlegendxxx------}{{\tiny Sources in PHZ}}
\psfrag{------yyylegendyyy------}{{\tiny Sources not in PHZ}}
\psfrag{--------------ytitle--------------}{\tiny{RX $500\,\mu \rm{m} \, \rm{[MJy}\,\rm{sr}^{-1}\rm{]}$}}
\includegraphics[width=0.22\textwidth, viewport=30 -15 360 360]{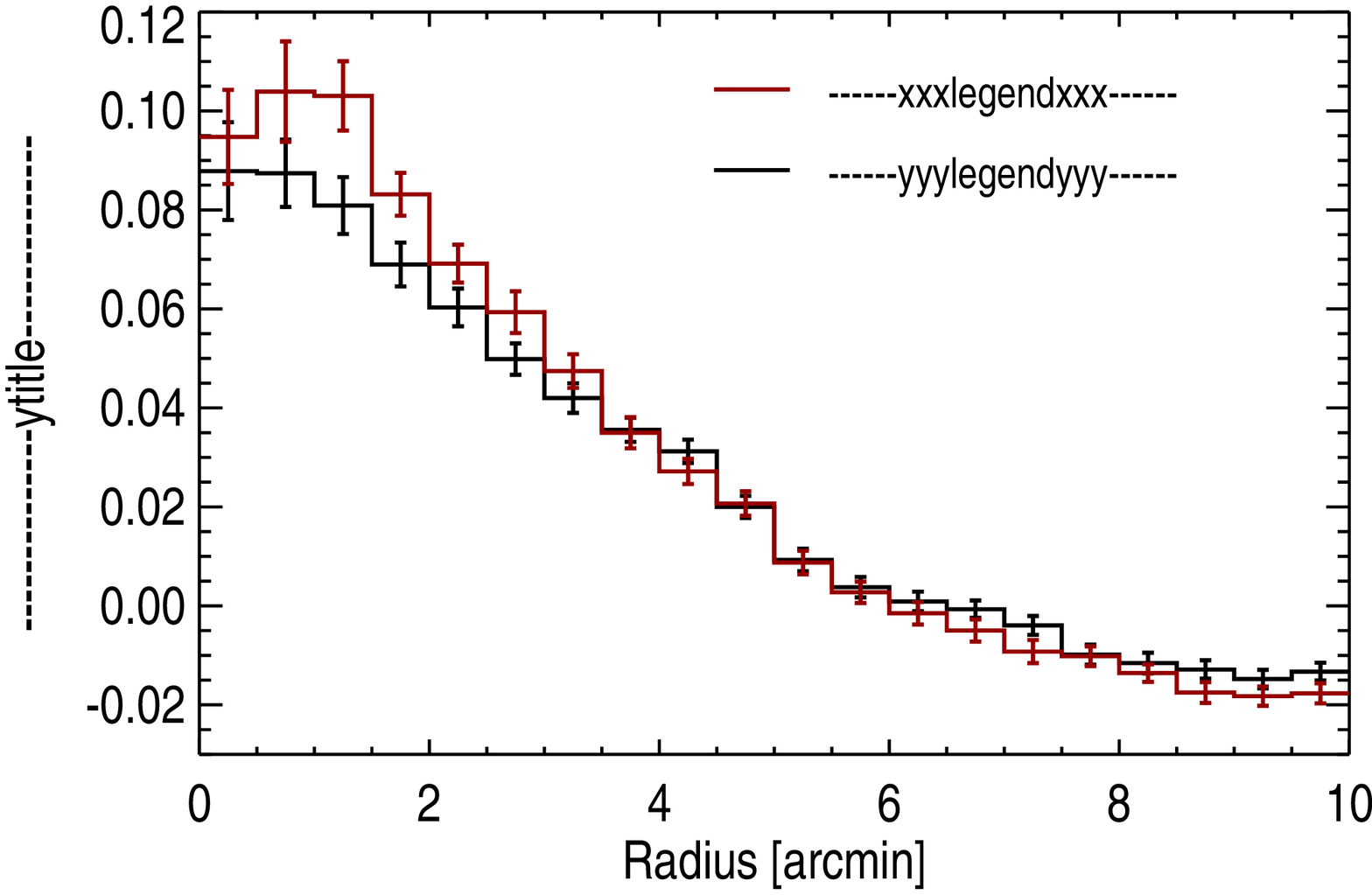}\\
\end{tabular}
\caption{Stacking analysis on two sub-samples of sources followed-up with
{\it Herschel}, depending on their presence in the final PHZ.
{\it Top:} stacked maps and radial profiles obtained on the 500\,$\mu$m
{\it Herschel}-SPIRE intensity maps.
{\it Bottom:} stacked maps and radial profiles obtained on the 500\,$\mu$m
{\it Herschel}-SPIRE red excess maps.
Error bars have been obtained via bootstrapping.}
\label{fig:stacking_herschel}
\end{figure*}

A dedicated follow-up of the \Planck\ high-$z$ candidates has been carried 
out with the {\it Herschel}-SPIRE instrument, culminating in
three accepted programmes during the OT1 (10 sources, PI: Montier),
OT2 (70, PI: Dole), and Must-Do (106, HPASSS, PI: Dole) calls.
A total of 228 sources were selected from the \Planck\ data: 204 sources
were selected using an algorithm similar to the one described in this work, 
but applied at 8{\arcmin} resolution on earlier versions of the \Planck\
data before the completion of the full mission, and 24 others were selected
from the Planck Catalogue of Compact Sources \citep[PCCS;][]{planck2013-p05}.
From this sample, 25 (16 from PHZ plus 9 from PCCS) sources are now
outside the mask defined for Galactic extinction, 
and 83 (82 from PHZ plus 1 from PCCS) sources remain in the final PHZ.
Without including the sub-sample of sources selected separately from the
PCCS, 120 of the observed sources are not in the final PHZ list,
which is explained by two main factors: improvement of the data quality; and 
evolution of the detection method. The S/N of the \Planck\
maps has been improved thanks to the completion of the full mission
and a better control of the systematics, so that previous detections may
now fall at lower S/N. To characterize this effect, we produced a
larger list of \Planck\ sources by relaxing the S/N criteria of the
detection to 1 in all bands (excess and cleaned maps), and we find
associations in this deep list for almost 90\,\% (182 sources) of the
203 sources of the {\it Herschel\/} sample present inside the mask. 
The S/N distribution of the {\it Herschel\/} sample is shown as a blue
histogram in Fig.~\ref{fig:herschel_sn}, while the sub-sample of sources
present in the PHZ is given in orange. It appears that most of the
sources of the {\it Herschel\/} sample that have not been selected in the
final PHZ exhibit a S/N close to the threshold criteria in at least
one band, so that they are rejected when simultaneously constraining
detections in all bands.
Hence only 10\,\% of the sources fail in more than one band.
Furthermore the detection algorithm has been improved compared to the first
incarnations of the method, especially when applying the colour-colour
criteria. We now use a probability to reject sources not satisfying the
colour-colour criteria, while a simple threshold-cut on each colour was
applied before.  This enabled us to improve the robustness of the final
product. All of these investigations show that sources of the {\it Herschel\/}
sample are not likely to be spurious if they happen not to be included in the
PHZ, but are simply at lower significance.

The statistics of the FWHM, ellipticity and extinction of the sub-sample of
83 candidates followed-up with {\it Herschel\/} and present in the final
PHZ are shown in Fig.~\ref{fig:statistics_herschel}. They spans the same
range of properties as the full list (dashed line). However, this
{\it Herschel\/} sub-sample is characterized by statistically higher S/N,
smaller FWHM, smaller ellipticities, and lower extinctions than the full PHZ.
This can be explained by the process of selection applied to obtain robust
target lists for the three various {\it Herschel\/} calls, which tended to
bias the selection towards cleaner regions of the high-latitude sky, and to 
preferentially pick high S/N compact sources, i.e., with small sizes and
regular shapes.

Another way to probe the reliability of the \Planck\ candidates followed-up
with {\it Herschel\/} but not present in the final PHZ is to compare,
via a stacking analysis, the statistical properties of two sub-samples,
namely sources included or not included in the final PHZ.
Thus we have performed the stacking of the {\it Herschel}-SPIRE
$20{\arcmin}\times20{\arcmin}$ cutouts at 500\,$\mu$m, 
over the 83 sources included in the PHZ on the one hand, and over the
120 sources no longer included in the PHZ on the other hand.
The resulting stacked maps and the associated profiles are shown in the
first row of Fig.~\ref{fig:stacking_herschel}.
The overdensity of {\it Herschel\/} sources appears slightly more compact
for the sub-sample of sources still included in the PHZ, with a radial
profile presenting a plateau within about 2{\arcmin}.
This is consistent with the fact that the PHZ has been built at
5{\arcmin} resolution, while the initial selection of the {\it Herschel\/}
sample was based on a first list built at 8{\arcmin} resolution. However,
the overdensity of sources is still clearly identified in both sub-samples.
Furthermore, we have performed on {\it Herschel}-SPIRE maps a similar process
as the one applied on \Planck\ maps to show the red excess at 500\,$\mu$m.
For each source, the background colour is first estimated between the 250
and 500\,$\mu$m maps on a region defined outside the \Planck\ peak emission
at 545\,GHz.  This background colour is used to extrapolate the 250\,$\mu$m
map at 500\,$\mu$m, and this is then removed from the original 500\,$\mu$m map, 
yielding the red excess map at 500\,$\mu$m:
\begin{equation}
M_{500}^{\rm{RX}} = M_{500} -
 \left\langle \frac{M_{500}}{M_{250}} \right\rangle_{\rm{bkg}} M_{250}\, ,
\end{equation}
where $M_{250}$ and $M_{500}$ are the {\it Herschel}-SPIRE maps at 250 and
500\,$\mu$m, and $\left\langle\right\rangle_{\rm{bkg}}$ 
means the average over the background region defined above. Positive pixels
in this kind of red excess map are associated with colours redder than the
background, and potentially associated with higher redshift structures, 
while negative pixels are bluer and mostly associated with lower redshift
structures. By stacking the red excess maps over the two {\it Herschel\/}
sub-samples, we get the stacked maps shown in the second row of
Fig.~\ref{fig:stacking_herschel}, which exhibit a clear excess of red colours
for both samples. The radial profile obtained with the sample of
{\it Herschel\/} sources included in the PHZ presents a red excess larger
in the central part compared to the other sample.
This is linked to the left panel of Fig.~\ref{fig:statistics_herschel},
where it can be seen that the 83 {\it Herschel\/} sources included in the
PHZ exhibit larger S/N in the \Planck\ excess and cleaned maps 
than the other sources. This analysis demonstrates firstly that the PHZ
represents a sample of sources with a larger reliability than the
initial selection made for the {\it Herschel\/} follow-up, and secondly
that the sub-sample of sources without any counter-parts in the PHZ
are not spurious detections, but simply have lower significance, as already
stressed above.

From the 11 sources of the {\it Herschel\/} sub-sample confirmed as strongly
lensed star-forming galaxies \citep{Canameras2015}, four sources (over the
five previously selected with a similar algorithm used in this work) are
present in the final PHZ.  Two other sources, confirmed at redshifts 2.2
and 2.4, did not pass the colour-colour criteria, while a third one 
exhibits a S/N on the 545\,GHz excess map just below the required threshold
of 5.  The last five sources, which have been selected from the PCCS catalogue,
have no counter part in the PHZ.
Additionally, the first spectroscopically confirmed \Planck-discovered
proto-cluster candidate, PHZ~G095.50$-$61.59 \citep{Florescacho2015}, 
exhibits one of the smallest S/N values for the 545\,GHz excess in the PHZ.

Finally, it is worth remarking that because the {\it Herschel}-SPIRE follow-up
of the \Planck\ high-$z$ source candidates and the final PHZ
are not fully consistent (in a statistical sense), it is hard to draw
definitive conclusions about the nature of the PHZ sources based on the
{\it Herschel\/} analysis.

\section{List description}
\label{sec:phz_description}

In this last appendix we present a description of the PHZ.
Table~\ref{tab:PHZ_listing} gives the names, units and, explanation
of the contents of each column.

\begin{table*}[!hp]
\vspace{-0.5cm}
\caption{Columns in the PHZ.}
\label{tab:PHZ_listing}
\nointerlineskip
\setbox\tablebox=\vbox{
\newdimen\digitwidth 
\setbox0=\hbox{\rm 0} 
\digitwidth=\wd0 
\catcode`*=\active 
\def*{\kern\digitwidth} 
\newdimen\signwidth 
\setbox0=\hbox{+} 
\signwidth=\wd0 
\catcode`!=\active 
\def!{\kern\signwidth} 
\newdimen\pointwidth 
\setbox0=\hbox{.} 
\pointwidth=\wd0 
\catcode`?=\active 
\def?{\kern\pointwidth} 

\vspace{-0.65cm}
\halign{\hbox to 2.2 in{#\leaderfil}\tabskip=1.0em&
\hfil#\hfil&
#\hfil\tabskip=0pt\cr
\noalign{\doubleline}
\omit \hfil Column Name \hfil& Unit& \hfil Description\cr
\noalign{\vskip 3pt\hrule\vskip 4pt}
\multicolumn{3}{c}{Identification}\cr
\noalign{\vskip 4pt}
 {\tt NAME}& \dots & Source name\cr
 {\tt SNR\_X545}& \dots & S/N in the 545\,GHz excess map\cr
 {\tt SNR\_D857}& \dots & S/N in the 857\,GHz cleaned map\cr
 {\tt SNR\_D545}& \dots & S/N in the 545\,GHz cleaned map\cr
 {\tt SNR\_D353}& \dots & S/N in the 353\,GHz cleaned map\cr
\noalign{\vskip 3pt\hrule\vskip 4pt}
\multicolumn{3}{c}{Source position}\cr
\noalign{\vskip 4pt}
 {\tt GLON}& [deg]& Galactic longitude based on morphology fitting\cr
 {\tt GLAT}& [deg]& Galactic latitude based on morphology fitting\cr
 {\tt RA}& [deg]& Right ascension (J2000) in degrees\cr
 {\tt DEC}& [deg]& Declination (J2000) in degrees\cr
\noalign{\vskip 3pt\hrule\vskip 4pt}
\multicolumn{3}{c}{Morphology}\cr
\noalign{\vskip 4pt}
{\tt GAU\_MAJOR\_AXIS}& [arcmin]& FWHM along the major axis of the elliptical Gaussian\cr
 {\tt GAU\_MAJOR\_AXIS\_SIG}& [arcmin]& 1$\,\sigma$ uncertainty of the FWHM along the major axis\cr
 {\tt GAU\_MINOR\_AXIS}& [arcmin]& FWHM along the minor axis of the elliptical Gaussian\cr
 {\tt GAU\_MINOR\_AXIS\_SIG}& [arcmin]& 1$\,\sigma$ uncertainty of the FWHM along the minor axis\cr
 {\tt GAU\_POSITION\_ANGLE}&[rd]& Position angle of the elliptical Gaussian, defined as the clockwise angle \cr
 \omit \hfil & & between the Galactic plane orientation and the orientation of the major axis  \cr  
 {\tt GAU\_POSITION\_ANGLE\_SIG}& [rd]& 1$\,\sigma$ uncertainty of the position angle \cr
\noalign{\vskip 3pt\hrule\vskip 4pt}
\multicolumn{3}{c}{Photometry on cleaned maps}\cr
\noalign{\vskip 4pt}
 {\tt FLUX\_CLEAN\_857}& [Jy]& Flux density of the source at 857\,GHz\cr
 {\tt FLUX\_CLEAN\_857\_SIG\_SKY}& [Jy]& 1$\,\sigma$ uncertainty at 857\,GHz due to sky confusion\cr
 {\tt FLUX\_CLEAN\_857\_SIG\_DATA}& [Jy]& 1$\,\sigma$ uncertainty at 857\,GHz due to measurement error\cr
 {\tt FLUX\_CLEAN\_857\_SIG\_GEOM}& [Jy]& 1$\,\sigma$ uncertainty at 857\,GHz due to elliptical Gaussian fit accuracy\cr
 {\tt FLUX\_CLEAN\_545}& [Jy]& Flux density of the source at 545\,GHz\cr
 {\tt FLUX\_CLEAN\_545\_SIG\_SKY}& [Jy]& 1$\,\sigma$ uncertainty at 545\,GHz due to sky confusion\cr
 {\tt FLUX\_CLEAN\_545\_SIG\_DATA}& [Jy]& 1$\,\sigma$ uncertainty at 545\,GHz due to measurement error\cr
 {\tt FLUX\_CLEAN\_545\_SIG\_GEOM}& [Jy]& 1$\,\sigma$ uncertainty at 545\,GHz due to elliptical Gaussian fit accuracy\cr
 {\tt FLUX\_CLEAN\_353}& [Jy]& Flux density of the source at 353\,GHz\cr
 {\tt FLUX\_CLEAN\_353\_SIG\_SKY}& [Jy]& 1$\,\sigma$ uncertainty at 353\,GHz due to sky confusion\cr
 {\tt FLUX\_CLEAN\_353\_SIG\_DATA}& [Jy]& 1$\,\sigma$ uncertainty at 353\,GHz due to measurement error\cr
 {\tt FLUX\_CLEAN\_353\_SIG\_GEOM}& [Jy]& 1$\,\sigma$ uncertainty at 353\,GHz due to elliptical Gaussian fit accuracy\cr
 {\tt FLUX\_CLEAN\_217}& [Jy]& Flux density of the source at 217\,GHz\cr
 {\tt FLUX\_CLEAN\_217\_SIG\_SKY}& [Jy]& 1$\,\sigma$ uncertainty at 217\,GHz due to sky confusion\cr
 {\tt FLUX\_CLEAN\_217\_SIG\_DATA}& [Jy]& 1$\,\sigma$ uncertainty at 217\,GHz due to measurement error\cr
 {\tt FLUX\_CLEAN\_217\_SIG\_GEOM}& [Jy]& 1$\,\sigma$ uncertainty at 217\,GHz due to elliptical Gaussian fit accuracy\cr
\noalign{\vskip 3pt\hrule\vskip 4pt}
\multicolumn{3}{c}{Physical Properties}\cr
\noalign{\vskip 4pt}
 {\tt PROB\_COLCOL}& \dots & Colour-colour selection probability\cr
 {\tt EBV\_MEAN}& \dots & Mean extinction $E(B-V)_{\rm{xgal}}$ within the source PSF\cr
 {\tt EBV\_APER}& \dots & Aperture estimate of the extinction $E(B-V)_{\rm{xgal}}$ within the source PSF\cr
 {\tt EBV\_APER\_SIG}& \dots & 1$\,\sigma$ uncertainty of the aperture extinction $E(B-V)_{\rm{xgal}}$ within the source PSF\cr
 {\tt ZPHOT\_[25,30,35,40,45,50]K}& \dots & Submm photometric redshift estimate with $T_{\rm{xgal}}=25$, 30, 35, 40, 45, and 50\,K\cr
 {\tt ZPHOT\_[25,30,35,40,45,50]K\_LOW}& \dots & Lower limit of the 68\,\% confidence level\cr
 {\tt ZPHOT\_[25,30,35,40,45,50]K\_UP}& \dots & Upper limit of the 68\,\% confidence level\cr
 {\tt ZPHOT\_[25,30,35,40,45,50]K\_CHI2}& \dots & Reduced $\chi^2$ of the best fit\cr
 {\tt LFIR\_[25,30,35,40,45,50]K}& [$\rm{L}_{\odot}$]& FIR luminosity estimate with $T_{\rm{xgal}}=25$, 30, 35, 40, 45, and 50\,K\cr
 {\tt LFIR\_[25,30,35,40,45,50]K\_LOW}& [$\rm{L}_{\odot}$]& Lower limit of the 68\,\% confidence level\cr
 {\tt LFIR\_[25,30,35,40,45,50]K\_UP}& [$\rm{L}_{\odot}$]& Upper limit of the 68\,\% confidence level\cr
 {\tt SFR\_[25,30,35,40,45,50]K}& [$\rm{M}_{\odot} \rm{yr}^{-1}$]& Star Formation Rate estimate with $T_{\rm{xgal}}=25$, 30, 35, 40, 45, and 50\,K\cr
 {\tt SFR\_[25,30,35,40,45,50]K\_LOW}& [$\rm{M}_{\odot} \rm{yr}^{-1}$]& Lower limit of the 68\,\% confidence level\cr
 {\tt SFR\_[25,30,35,40,45,50]K\_UP}& [$\rm{M}_{\odot} \rm{yr}^{-1}$]& Upper limit of the 68\,\% confidence level\cr
\noalign{\vskip 3pt\hrule\vskip 4pt}
\multicolumn{3}{c}{Flags}\cr
\noalign{\vskip 4pt}
 {\tt XFLAG\_PCCS\_857}& [0,1]& 1 if present in the PCCS2 857-GHz band\cr
 {\tt XFLAG\_PCCS\_545}& [0,1]& 1 if present in the PCCS2 545-GHz band\cr
 {\tt XFLAG\_PCCS\_353}& [0,1]& 1 if present in the PCCS2 353-GHz band\cr
 {\tt XFLAG\_PCCS\_217}& [0,1]& 1 if present in the PCCS2 217-GHz band\cr
 {\tt XFLAG\_PCCS\_143}& [0,1]& 1 if present in the PCCS2 143-GHz band\cr
 {\tt XFLAG\_PCCS\_100}& [0,1]& 1 if present in the PCCS2 100-GHz band\cr
 {\tt XFLAG\_PCCS\_70}& [0,1]& 1 if present in the PCCS2 70-GHz band\cr
 {\tt XFLAG\_PCCS\_44}& [0,1]& 1 if present in the PCCS2 44-GHz band\cr
 {\tt XFLAG\_PCCS\_30}& [0,1]& 1 if present in the PCCS2 30-GHz band\cr
 {\tt XFLAG\_PCCS\_HFI}& [0,1]& 1 if present in the PCCS2 857-, 545-, and 353-GHz bands\cr
 {\tt XFLAG\_PCCS\_LFI}& [0,1]& 1 if present in the PCCS2 70-, 44-, and 30-GHz bands\cr
 {\tt XFLAG\_PCCS\_SZ}& [0,1]& 1 if present in the PSZ2\cr
 {\tt XFLAG\_PGCC}& [0,1]& 1 if present in the PGCC\cr
 {\tt XFLAG\_HERSCHEL}& [0,1]& 1 if present in the {\it Herschel\/} follow-up programme\cr
 \noalign{\vskip 3pt\hrule\vskip 4pt}
}}
\endPlancktable
\end{table*}

\end{document}